%% file: hyperk-loi.tex
\documentclass[aps,preprint,superscriptaddress,twoside,11pt]{revtex4-1}

\pdfoutput=1

\usepackage{graphicx}%
\usepackage{dcolumn}%
\usepackage{bm}%
\usepackage{hyperref}%
\usepackage[usenames]{color}%

\begin{document}

\title{\Large Letter of Intent: \\
       The Hyper-Kamiokande Experiment \\
       --- Detector Design and Physics Potential ---}

\newcommand{\INSTGFU}{\affiliation{Gifu University, Department of Physics, Gifu, Gifu 501-1193, Japan}}
\newcommand{\INSTKEK}{\affiliation{High Energy Accelerator Research Organization (KEK), Tsukuba, Ibaraki, Japan}}
\newcommand{\INSTKBE}{\affiliation{Kobe University, Department of Physics, Kobe, Hyogo 657-8501, Japan}}
\newcommand{\INSTKYT}{\affiliation{Kyoto University, Department of Physics, Kyoto, Kyoto 606-8502, Japan}}
\newcommand{\INSTMYG}{\affiliation{Miyagi University of Education, Department of Physics, Sendai, Miyagi 980-0845, Japan}}
\newcommand{\INSTNGY}{\affiliation{Nagoya University, Solar Terrestrial Environment Laboratory, Nagoya, Aichi 464-8602, Japan}}
\newcommand{\INSTOKY}{\affiliation{Okayama University, Department of Physics, Okayama, Okayama 700-8530, Japan}}
\newcommand{\INSTTHK}{\affiliation{Tohoku University, Research~Center~for~Neutrino~Science, Sendai 980-8578, Japan}}
\newcommand{\INSTTKI}{\affiliation{Tokai University, Department of Physics, Hiratsuka, Kanagawa 259-1292, Japan}}
\newcommand{\INSTTKY}{\affiliation{University of Tokyo, Department of Physics, Bunkyo, Tokyo 113-0033, Japan}}
\newcommand{\INSTERI}{\affiliation{University of Tokyo, Earthquake Research Institute, Bunkyo, Tokyo 113-0032, Japan}}
\newcommand{\INSTKAM}{\affiliation{University of Tokyo, Institute for Cosmic Ray Research, Kamioka Observatory, Kamioka, Gifu 506-1205, Japan}}
\newcommand{\INSTRCCN}{\affiliation{University of Tokyo, Institute for Cosmic Ray Research, Research~Center~for~Cosmic~Neutrinos, Kashiwa, Chiba 277-8582, Japan}}
\newcommand{\INSTIPMU}{\affiliation{University of Tokyo, Institute~for~the~Physics~and~Mathematics~of~the~Universe, Kashiwa, Chiba 277-8583, Japan}}

\INSTGFU
\INSTKEK
\INSTKBE
\INSTKYT
\INSTMYG
\INSTNGY
\INSTOKY
\INSTTHK
\INSTTKI
\INSTTKY
\INSTERI
\INSTKAM
\INSTRCCN
\INSTIPMU

\author{K.~Abe}\INSTKAM\INSTIPMU
\author{T.~Abe}\INSTTKY
\author{H.~Aihara}\INSTTKY\INSTIPMU
\author{Y.~Fukuda}\INSTMYG
\author{Y.~Hayato}\INSTKAM\INSTIPMU
\author{K.~Huang}\INSTKYT
\author{A.~K.~Ichikawa}\INSTKYT
\author{M.~Ikeda}\INSTKYT
\author{K.~Inoue}\INSTTHK\INSTIPMU
\author{H.~Ishino}\INSTOKY
\author{Y.~Itow}\INSTNGY
\author{T.~Kajita}\INSTRCCN\INSTIPMU
\author{J.~Kameda}\INSTKAM\INSTIPMU
\author{Y.~Kishimoto}\INSTKAM\INSTIPMU
\author{M.~Koga}\INSTTHK\INSTIPMU
\author{Y.~Koshio}\INSTKAM\INSTIPMU
\author{K.~P.~Lee}\INSTRCCN
\author{A.~Minamino}\INSTKYT
\author{M.~Miura}\INSTKAM\INSTIPMU
\author{S.~Moriyama}\INSTKAM\INSTIPMU
\author{M.~Nakahata}\INSTKAM\INSTIPMU
\author{K.~Nakamura}\INSTKEK\INSTIPMU
\author{T.~Nakaya}\INSTKYT\INSTIPMU
\author{S.~Nakayama}\INSTKAM\INSTIPMU
\author{K.~Nishijima}\INSTTKI
\author{Y.~Nishimura}\INSTKAM
\author{Y.~Obayashi}\INSTKAM\INSTIPMU
\author{K.~Okumura}\INSTRCCN
\author{M.~Sakuda}\INSTOKY
\author{H.~Sekiya}\INSTKAM\INSTIPMU
\author{M.~Shiozawa}\email[]{masato@suketto.icrr.u-tokyo.ac.jp}\INSTKAM\INSTIPMU
\author{A.~T.~Suzuki}\INSTKBE
\author{Y.~Suzuki}\INSTKAM\INSTIPMU
\author{A.~Takeda}\INSTKAM\INSTIPMU
\author{Y.~Takeuchi}\INSTKBE\INSTIPMU
\author{H.~K.~M.~Tanaka}\INSTERI
\author{S.~Tasaka}\INSTGFU
\author{T.~Tomura}\INSTKAM
\author{M.~R.~Vagins}\INSTIPMU
\author{J.~Wang}\INSTTKY
\author{M.~Yokoyama}\INSTTKY\INSTIPMU

\collaboration{Hyper-Kamiokande working group}

\date{\today}

\begin{abstract}

We propose the Hyper-Kamiokande (Hyper-K) detector as a next generation 
underground water Cherenkov detector.  It will    
serve as a far detector of a long baseline neutrino oscillation 
experiment envisioned for the upgraded J-PARC, and as  
a detector capable of observing -- far beyond the sensitivity 
of the Super-Kamiokande (Super-K) detector -- 
proton decays, atmospheric neutrinos, and neutrinos from astronomical origins.
The baseline design of Hyper-K is based on the highly successful Super-K,  
taking full advantage of a well-proven technology.

Hyper-K consists of two cylindrical tanks lying side-by-side, 
the outer dimensions of each tank being $48 \ ({\rm W}) \times 54\ ({\rm H}) \times 250 \ ({\rm L})\ {\rm m}^3$.
The total (fiducial) mass of the detector is 0.99 (0.56) million metric tons, 
which is about 20 (25) times larger than that of Super-K.
A proposed location for Hyper-K is about 8 km south of Super-K 
(and 295 km away from J-PARC) at an underground depth of 1,750 meters water equivalent (m.w.e.).
The inner detector region of the Hyper-K detector is viewed 
by 99,000 20-inch PMTs, 
corresponding to the PMT density of 
$20\%$ photo-cathode coverage (one half of that of Super-K).

Hyper-K presents unprecedented potential for precision measurements 
of neutrino oscillation parameters and discovery reach for $CP$ violation 
in the lepton sector.
With a total exposure of 5 years (one year being equal to $10^7$ sec) to a  
$2.5$-degree off-axis neutrino beam
produced by the 1.66 MW J-PARC proton synchrotron, it is expected
that the $CP$ phase $\delta$ can be determined to better than 18 degrees 
for all possible values of $\delta$
and $CP$ violation can be established with a statistical significance 
of $3\sigma$ for $74\%$ of the 
$\delta$ parameter space if $\sin^2 2\theta_{13}>0.03$ and the mass hierarchy 
is known.
If $\sin^2 2\theta_{13}$ is as large as 0.1
the mass hierarchy can be determined 
with more than 3$\sigma$ statistical significance for 46\% 
of the $\delta$ parameter space.
In addition, a high statistics data sample of atmospheric neutrinos
will allow us to extract the information on the mass hierarchy 
and the octant of $\theta_{23}$.
With a full 10 year duration of data taking, the significance for 
the mass hierarchy determination is expected to reach $3\sigma$ or greater 
if $\sin^2\theta_{23}>0.4$. 

Hyper-K can extend the sensitivity to nucleon decays 
beyond what was achieved by Super-K by an order of magnitude or more. 
The sensitivities to the partial lifetime of protons
for the decay modes of $p \rightarrow e^+ \pi^0$ and
$p \rightarrow \bar{\nu} K^+$ 
are expected to exceed $1 \times 10^{35}$ years
and $2 \times 10^{34}$ years, respectively.
This is the only known, realistic detector option capable of reaching such a  
sensitivity for the $p \rightarrow e^+\pi^0$ mode.

The scope of studies at Hyper-K also covers high precision measurements 
of solar neutrinos, observation of both supernova burst neutrinos and supernova relic neutrinos,
dark matter searches, and possible detection of solar flare neutrinos.
The prospects for neutrino geophysics using Hyper-K are also mentioned.

\end{abstract}

\maketitle

\tableofcontents
\newpage

\color{black}
\input{summary/summary.tex}
\newpage

\color{black}
\input{introduction/physics.tex}

\input{introduction/watercherenkov.tex}

\newpage

\color{black}
\input{experimental-setup/site.tex}
\input{experimental-setup/water.tex}

\input{experimental-setup/sensor.tex}
\input{experimental-setup/calibration.tex}

\newpage

\color{black}
\input{physics-cpv/cpv.tex}
\newpage
\input{physics-atmnu/atmnu.tex}

\newpage
\newpage
\input{physics-pdecay/pdecay.tex}
\newpage
\input{physics-solarnu/solarnu.tex}

\newpage
\input{physics-supernova/supernova.tex}

\input{physics-astro/astrophys.tex}

\newpage
\input{physics-radiography/radiography.tex}

\color{black}
\clearpage

\pagestyle{plain}

\bibliography{hyperk-loi}

\end{document}

%% file: summary/summary.tex
\section*{Executive Summary}
\vspace{4cm}
\begin{figure}[htbp]
  \begin{center}
    \includegraphics[scale=0.4]{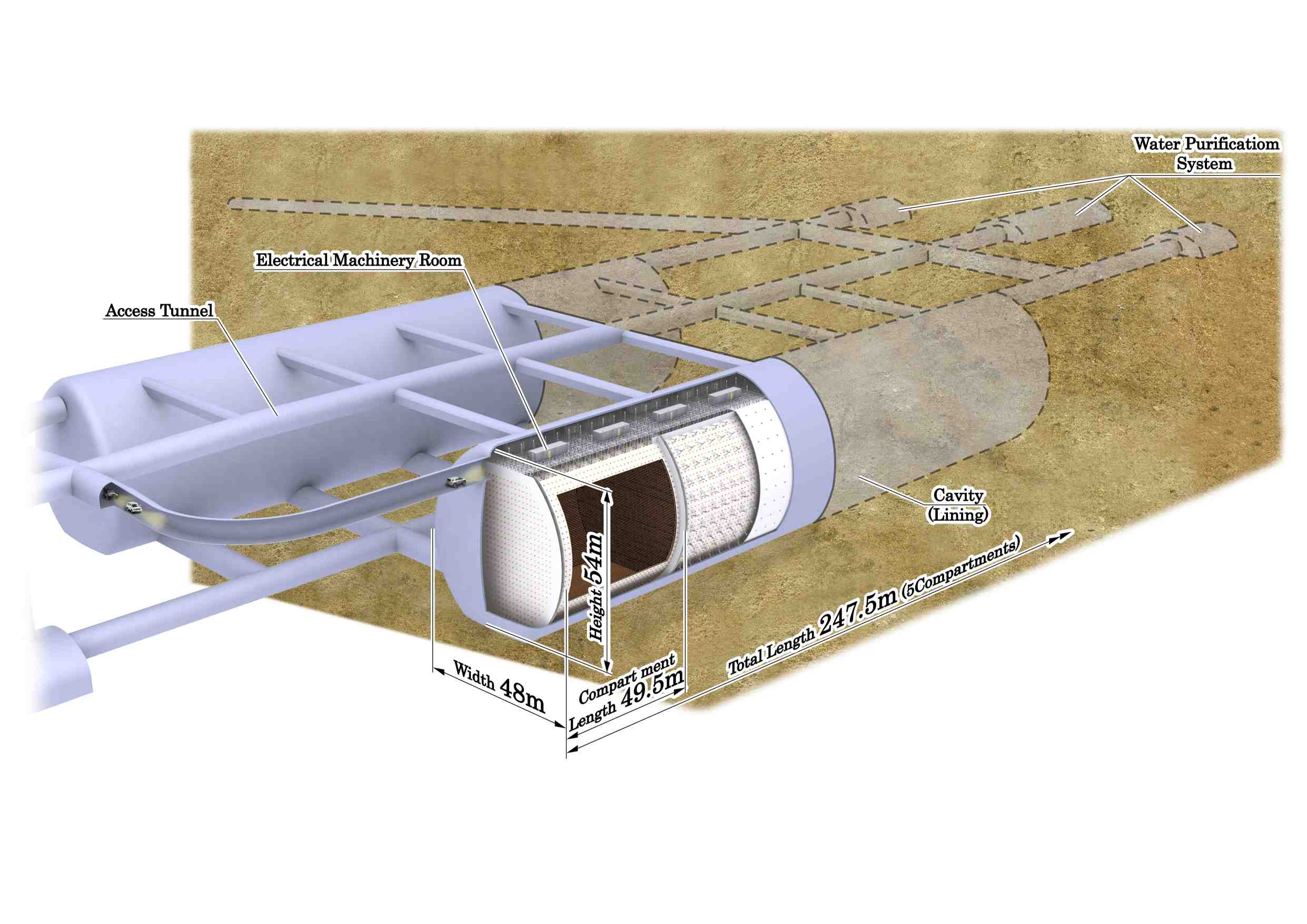}
  \end{center}
  \caption{Schematic view of the Hyper-Kamiokande detector.}
  \label{fig:HK-schematic-view}
\end{figure}
\newpage

The Hyper-Kamiokande (Hyper-K or HK) detector is proposed as 
a next generation underground water Cherenkov detector that serves as a far
detector of a long baseline neutrino oscillation experiment for the J-PARC 
neutrino beam and as a detector capable of observing proton decays,
atmospheric and solar neutrinos, and neutrinos from other astrophysical
origins.
The baseline design of Hyper-K is based on the well-proven technologies 
employed and tested at Super-Kamiokande (Super-K or SK).
Hyper-K consists of two cylindrical tanks lying side-by-side, 
the outer dimensions of each tank being 
$48 \ ({\rm W}) \times 54\ ({\rm H}) \times 250\ ({\rm L})\ {\rm m}^3.$
The total (fiducial) mass of the detector is 0.99 (0.56) million metric tons, 
which is about 20 (25) times larger than that of Super-K.
A proposed location for Hyper-K is about 8~km south of Super-K 
(and 295~km away from J-PARC) and 1,750 meters water equivalent 
(or 648~m of rock) deep.
The inner detector region is viewed by 99,000 20-inch PMTs, 
corresponding to the PMT density of 
$20\%$ photo-cathode coverage (one half of that of Super-K).
The schematic view of the Hyper-K detector is illustrated in 
Fig. ~\ref{fig:HK-schematic-view}. 
Table~\ref{tab:detector-parameters} 
summarizes the baseline design parameters of the Hyper-K detector.
\begin{table}[htdp]
\caption{Detector parameters of the baseline design.}
\begin{center}
\begin{tabular}{lll} \hline \hline
Detector type & & Ring-imaging water Cherenkov detector \\ \hline 
Candidate site & Address & Tochibora mine \\
& & Kamioka town, Gifu, JAPAN \\
& Lat. & $36^\circ21'08.928''$N \\
& Long. & $137^\circ18'49.688''$E \\
& Alt. & 508 m \\
& Overburden & 648 m rock (1,750 m water equivalent)  \\
& Cosmic Ray Muon flux & 1.0 $\sim$ 2.3 $\times$ 10$^{-6}$ sec$^{-1}$cm$^{-2}$  \\
& Off-axis angle for the J-PARC $\nu$ & $2.5^\circ$ (same as Super-Kamiokande)  \\
& Distance from the J-PARC & 295 km (same as Super-Kamiokande)  \\ \hline 
Detector geometry & Total Volume & 0.99 Megaton  \\
 & Inner Volume (Fiducial Volume) & 0.74 (0.56) Megaton  \\
 & Outer Volume & 0.2 Megaton  \\ \hline 
Photo-multiplier Tubes & Inner detector & 99,000 20-inch $\phi$ PMTs \\
& & 20\% photo-coverage \\ 
& Outer detector & 25,000 8-inch $\phi$ PMTs \\ \hline 
Water quality & light attenuation length & $>100$ m @ 400 nm  \\
 & Rn concentration & $<1$ mBq/m$^3$ \\ \hline 
\hline 

\end{tabular}
\end{center}
\label{tab:detector-parameters} 
\end{table}

Hyper-K provides rich neutrino physics programs.
In particular, it has unprecedented potential for precision measurements 
of neutrino oscillation parameters and discovery reach for $CP$ violation 
in the lepton sector.
With a total exposure of 5 years (1 year being equal to $10^7$ sec) 
to a $2.5$-degree off-axis neutrino beam
produced by the 1.66 MW J-PARC proton synchrotron, it is expected
that the $CP$ phase $\delta$ can be determined to better than 18 degrees 
for all  values of $\delta$ 
and that $CP$ violation can be established with a statistical significance 
of $3\sigma$ for $74\%$ of the 
$\delta$ parameter space if $\sin^2 2\theta_{13}>0.03$ and the mass hierarchy 
is known.
If the mass hierarchy is unknown, the sensitivity to the $CP$ violation 
is somewhat reduced due to degeneracy.
For a large value of $\sin^2 2\theta_{13}$, it is also possible to determine 
the mass hierarchy for some of $\delta$ with this program alone.
If $\sin^22\theta_{13} = 0.1$, the mass hierarchy can be determined 
with more than 3$\sigma$ statistical significance for 46\% 
of the $\delta$ parameter space.
The recent result of $\sin^2 2\theta_{13}>0.03$ obtained by the T2K 
experiment~\cite{Abe:2011sj} boosts the expectation of discovery 
of $CP$ violation by Hyper-K.

The high statistics data sample of atmospheric neutrinos obtained 
by Hyper-K will allow us to extract information on the mass hierarchy 
and the octant of $\theta_{23}$.
With a full 10 year period of data taking, the significance for the mass 
hierarchy determination is expected to reach $3\sigma$ or greater if
$\sin^2 2\theta_{13}>0.04$ and $\sin^2\theta_{23}>0.4$. 
If $\sin^2 2\theta_{23}$ is less than 0.99, it is possible to identify 
the octant of $\theta_{23}$, i.e. discriminate $\sin^2\theta_{23}<0.5$ 
from $>0.5.$

Hyper-K extends the sensitivity to nucleon decays far beyond that of Super-K.
The sensitivity to the partial lifetime of protons for the decay mode  
$p\rightarrow e^+\pi^0$,
the mode considered to be most model independent, is expected to be 
$1.3 \times 10^{35}$ years.
It is the only realistic detector option known today able to reach this sensitivity.
The sensitivity for the decay mode $p \rightarrow \bar{\nu} K^+$, 
the mode favored by super symmetry (SUSY) models, reaches $2.5 \times 10^{34}$ 
years, and therefore Hyper-K would discover proton decay if some of the SUSY models are correct.

\begin{table}[htdp]
\caption{Physics targets and expected sensitivities of the Hyper-Kamiokande
         experiment.
         $\sigma_{SD}$ is the WIMP-proton spin dependent cross section.}
\begin{center}
\begin{tabular}{lll} \hline \hline
Physics Target & Sensitivity & Conditions \\
\hline \hline
Neutrino study w/ J-PARC $\nu$~~ && 1.66 MW $\times$ 5 years (1 year $\equiv 10^7$ sec)\\
$-$ $CP$ phase precision & $<18^\circ$ & @ $s^22\theta_{13}(\equiv\sin^22\theta_{13})>0.03$ and\\
&& mass hierarchy (MH) is known \\
$-$ $CPV$ $3\sigma$ discovery coverage & 74\% (55\%) & @ $s^22\theta_{13}=0.1$, MH known(unknown) \\
 & 74\% (63\%) & @ $s^22\theta_{13}=0.03$, MH known(unknown) \\
 & 66\% (59\%) & @ $s^22\theta_{13}=0.01$, MH known(unknown) \\
\hline
Atmospheric neutrino study && 10 years observation\\
$-$ MH determination & $> 3\sigma$ CL & @ $0.4<s^2\theta_{23}$ and $0.04<s^22\theta_{13}$ \\
$-$ $\theta_{23}$ octant determination & $>90\%$ CL & @ $s^22\theta_{23}<0.99$ and $0.04<s^22\theta_{13}$ \\\hline
Nucleon Decay Searches && 10 years data \\
$-$ $p\rightarrow e^+ + \pi^0$ & $1.3 \times 10^{35}$ yrs (90\% CL)~~ &\\
 & $5.7 \times 10^{34}$ yrs ($3\sigma$ CL) &\\
$-$ $p\rightarrow \bar{\nu} + K^+$ & $2.5 \times 10^{34}$ yrs (90\% CL) &\\
 & $1.0 \times 10^{34}$ yrs ($3\sigma$ CL) &\\ \hline
Solar neutrinos && \\
$-$ $^8$B $\nu$ from Sun & 200 $\nu$'s / day & 7.0 MeV threshold (total
	 energy) w/ osc.\\
$-$ $^8$B $\nu$ day/night accuracy & $<1\%$ & 5 years, only stat. w/ SK-I BG $\times 20$\\ \hline
Astrophysical objects &&\\
$-$ Supernova burst $\nu$ & 170,000$\sim$260,000 $\nu$'s & @ Galactic center (10 kpc)\\ 
 & 30$\sim$50 $\nu$'s & @ M31 (Andromeda galaxy) \\ 
$-$ Supernova relic $\nu$ & 830 $\nu$'s / 10 years & \\
$-$ WIMP annihilation at Sun & & 5 years observation\\
 & $\sigma_{SD}=10^{-39}$cm$^2$ & @ $M_{\rm WIMP}=10$ GeV, $\chi\chi\rightarrow b\bar b$ dominant\\
 & $\sigma_{SD}=10^{-40}$cm$^2$ & @ $M_{\rm WIMP}=100$ GeV, $\chi\chi\rightarrow W^+ W^-$ dominant\\
 \hline \hline
\end{tabular}
\end{center}
\label{tab:targets}
\end{table}

Hyper-Kamiokande functions as an astrophysical neutrino observatory.
If a core collapse supernova explosion occurs halfway across our galaxy, 
the Hyper-K detector would detect approximately 170,000$\sim$260,000 neutrinos
as a $\sim 10$ second long burst. This very large statistical sample 
should at last reveal the detailed mechanism of supernova explosions.
For instance, the onset time of the explosion can be determined 
with an accuracy of 0.03 milliseconds, 
which is a key information to study the first physical process of the explosion
($p + e^- \rightarrow n + \nu_e$), 
allowing examination of the infall of the core and the ability to see the 
precise moment when a new neutron star or black hole is born.
The sharp risetime of the burst in Hyper-K can also be used to make a measurement of the 
absolute mass of neutrinos. Because of non-zero masses, their arrival times will  
depend on their energy.  The resulting measurement of the absolute neutrino mass 
would have a sensitivity of $0.5-1.3\ {\rm eV}/c^2$,  and does not depend on
whether the neutrino is a Dirac or Majorana particle.
Hyper-K is also capable of detecting supernova explosion neutrinos from 
galaxies outside of our own Milky Way;
about 7,000-10,000 neutrinos from
the Large Magellanic Cloud and
30-50 even from the Andromeda galaxy.

Detection of supernova relic neutrinos (SRN) is of great interest because the history 
of heavy element synthesis in the universe is encoded in the SRN energy spectrum.
With gadolinium added to water, a neutron produced by the inverse beta process 
($\bar{\nu}_e + p \rightarrow e^+ + n$), which is the predominant interaction mode for the SRNs, 
can be tagged by detecting gammas from the Gd($n$,$\gamma$s)Gd reaction.
Doing so greatly reduces backgrounds and opens up the SRN energy window,  
improving the detector's response to this important signal.
Our study shows that Hyper-K with $0.1\%$ by mass of gadolinium dissolved in
the water is able to detect as many as 830 SRNs in the energy range of 10-30~MeV 
for 10 years of livetime.
This large sample will enable us to explore the evolution of the universe.

Dark matter can be searched for in Hyper-K as was done in Super-K.
Neutrinos emitted by weakly interacting massive particles (WIMPs) annihilating  
in the Sun, Earth, and galactic halo can be detected using 
the upward-going muons observed in Hyper-K.
A sensitivity to the WIMP-proton spin dependent cross section would reach 
$10^{-39}\ (10^{-40})\ {\rm cm}^{2}$ for a WIMP mass of 
10 (100) GeV and  5 years of livetime.

Table \ref{tab:targets} summarizes the physics potential of Hyper-K.
This document serves to define the scope of the Hyper-K project,
describe a baseline design of the detector, and make a physics case 
for its construction.
The required R\&D items, cost, and schedule for constructing the Hyper-K 
detector will be provided in a separate documentation later. 

The Science Council of Japan announced the
``Japanese Master Plan of Large Research Projects'' for the first time on 
17 March 2010 \cite{master-plan-2010}. 
The plan includes 43 projects selected from 7 fields of science: 
Humanities and Social Sciences; Life Sciences; 
Energy, Environmental and Earth Sciences; Material and Analytical Sciences; 
Physical Sciences and Engineering; Space Sciences; and Information Sciences. 
There are a total of five large research facility projects listed in 
physical sciences.
Among them is 
``Nucleon Decay and Neutrino Oscillation Experiments with Large 
Advanced Detectors.'' 
This is a project to advance neutrino physics/astronomy and to search for 
nucleon decays using a large water Cherenkov detector that is approximately 
20 times larger in volume than Super-Kamiokande and/or a large liquid argon 
detector.
The Hyper-K detector is one of the leading options for this project.

We are also well aware of other detector technology options such as 
liquid argon and of activities 
in other regions such as LBNE \cite{LBNE} in the US, 
and LAGUNA \cite{LAGUNA} in Europe.
It is our firm intention to make the Hyper-K project completely open to the 
international community 
and contribute to the world-wide effort to make a strong neutrino physics 
program.

%% file: introduction/physics.tex
\section{Introduction}\label{section:introduction}
\subsection{Physics case}\label{section:intro-physics}

The goal of particle physics is to discover and understand the fundamental
laws of nature.
The Standard Model (SM), which is the current paradigm of elementary particles 
and their interactions, gives a successful account of the
experimental data to date \cite{Nakamura:2010zzi}.
Yet, deeper insights are still needed to answer 
more profound questions.
For instance,
why does there exist a gauge  structure of 
$SU(3)_C \otimes SU(2)_L \otimes U(1)_Y$ among  the strong, weak and electromagnetic interactions?
Why is there a three generation structure of fundamental fermions
and what are the origins of the masses and generation mixings of quarks and leptons?
To address these questions physics beyond the SM (BSM)  is required.

The discovery of neutrino oscillations by the Super-Kamiokande (Super-K or SK) 
experiment in 1998 \cite{Fukuda:1998mi} 
opened a new window to explore BSM physics.
Evidence of neutrino oscillations is the only experimental proof known today that shows the existence of BSM physics at work.
The mixing parameters of neutrinos, though not yet fully determined, were found
to be remarkably different from those of quarks,
which suggests the presence of an unknown flavor symmetry waiting to be explored.
Extremely small masses of neutrinos compared with those of 
their charged partners
lead to the preferred scenario of a seesaw mechanism, 
in which small neutrino masses are 
a reflection of the ultra-high energy scale of BSM physics.
Furthermore, recent theoretical works point to the intriguing possibility that
$CP$ asymmetry originating from flavor mixing among the three generations of 
neutrinos might have played an important role in creating the observed 
matter-antimatter asymmetry in the universe.
Therefore, to explore the full picture of neutrino masses 
and mixings and to observe $CP$ asymmetry
in the neutrino sector
are among the most important and urgent subjects in today's 
elementary particle physics world.
The indication, obtained recently by the T2K experiment \cite{Abe:2011sj},  
that the mixing angle parameter between the first 
and third generation neutrinos is sizable ($\sin^2 2\theta_{13}>0.03$) has 
further enhanced  prospects for 
Hyper-Kamiokande (Hyper-K or HK) to discover $CP$ asymmetry.

Since 1970's, Grand Unified Theories (GUT or GUTs), which 
unify the strong and electroweak interactions 
and describe them arising from larger gauge 
symmetries like \(SU(5)\),
have been extensively developed.
Because leptons and quarks are often placed in the same multiplets, most 
GUTs allow baryon number violating interactions
\cite{Pati:1973rp,Georgi:1974sy,Langacker:1980js}.
Baryon number violating 
nucleon decays would constitute an extremely sensitive probe 
of BSM physics
and the search for such a signal remains one of the major endeavors in
high energy physics.
So far nucleon decays have escaped detection even by
the world's largest nucleon decay detector,
Super-K \cite{Shiozawa:1998si,Hayato:1999az,Kobayashi:2005pe,:2009gd}, 
and so it must now be pursued by its successor, Hyper-K.

\subsubsection{Neutrino oscillations and $CP$ violation}\label{section:intro-neutrino}
The neutrino mixing matrix \(U\) \cite{Pontecorvo:1967fh,Maki:1962mu}
-- often called the Pontecorvo-Maki-Nakagawa-Sakata (PMNS) or
Maki-Nakagawa-Sakata (MNS) mixing matrix --
which translates neutrino mass eigenstates
into flavor eigenstates
\((\nu_e,\nu_\mu,\nu_\tau)^T = U(\nu_1,\nu_2,\nu_3)^T\),
is parameterized by three mixing angles 
($\theta_{12}, \theta_{23}, \theta_{13}$), one Dirac $CP$ phase $\delta$,
and two Majorana $CP$ phases ($\alpha_{21}, \alpha_{31}$)
in the three flavor neutrino framework \cite{Nakamura:2010zzi}
as
\begin{eqnarray}
U &=&
\left(
\begin{array}{ccc}
U_{e1}   & U_{e2}   & U_{e3} \\
U_{\mu1} & U_{\mu2} & U_{\mu3} \\
U_{\tau1} & U_{\tau2} & U_{\tau3} \\
\end{array}
\right)  \nonumber \\
& = &
\left(
\begin{array}{ccc}
1 & 0   & 0 \\
0 & c_{23} & s_{23} \\
0 & -s_{23} & c_{23} \\
\end{array}
\right)
\left(
\begin{array}{ccc}
c_{13} & 0   & s_{13}e^{-i\delta} \\
0 & 1 & 0 \\
-s_{13}e^{i\delta} & 0 & c_{13} \\
\end{array}
\right)
\left(
\begin{array}{ccc}
c_{12} & s_{12}  & 0 \\
-s_{12} & c_{12} & 0 \\
0 & 0 & 1 \\
\end{array}
\right)
\cdot
\left(
\begin{array}{ccc}
1 & 0 & 0 \\
0 & e^{i\frac{\alpha_{21}}{2}} & 0 \\
0 & 0 & e^{i\frac{\alpha_{31}}{2}} \\
\end{array}
\right)
\end{eqnarray}
where \(c_{ij}\) and \(s_{ij}\) represent sin\(\theta_{ij}\) and
cos\(\theta_{ij}\), respectively.
If massive neutrinos are Dirac particles, the Majorana phases are absorbed 
and only the Dirac phase can be responsible
for $CP$ violation in the lepton sector.
Neutrino oscillation frequencies are determined by mass parameters,
$\Delta m^2_{21} \equiv m^2_2 - m^2_1$ and 
$\Delta m^2_{32} \equiv m^2_3 - m^2_2$,
where $m_1, m_2,$ and $m_3$ are the three mass eigenvalues.
Because oscillation probabilities do not depend on the Majorana phases, 
the number of relevant oscillation parameters is six in total and the 
$CP$ violation effect can be seen only via the Dirac phase $\delta$
in neutrino oscillation experiments.
Among these six parameters, 
\(\theta_{12}\) and $\Delta m^2_{21}$ have been measured
by solar and reactor neutrino experiments 
\cite{Abe:2011xx,Aharmim:2008kc,Aharmim:2009gd,:2008ee}.
\(\theta_{23}\) and $|\Delta m^2_{32}|$ (only its absolute value)
have been measured by 
atmospheric neutrino \cite{Ashie:2005ik,Ashie:2004mr}
and accelerator neutrino experiments
\cite{Ahn:2006zza,Adamson:2011ig}.

For many years, much effort has been devoted to measuring the last unknown mixing angle
$\theta_{13}$ 
by long baseline experiments \cite{Yamamoto:2006ty}, 
atmospheric neutrino studies \cite{Wendell:2010md}, as well as 
solar \cite{Aharmim:2009gd}
and reactor experiments \cite{KamLAND11}.
Among them, the world's most stringent upper limit has been
\(\sin^22\theta_{13}<0.15\),
which was obtained by the short baseline reactor neutrino experiment CHOOZ
\cite{Apollonio:1999ae}.
The quest has now been advanced by
the T2K experiment announcing in June 2011 that a positive indication
of $\nu_\mu \rightarrow \nu_e$ oscillations has been observed with the result that 
zero $\theta_{13}$ is disfavored at the 2.5$\sigma$ level
\cite{Abe:2011sj}.
The significance becomes more than $3\sigma$ 
\cite{Fogli:2011qn}
when a global fit is performed
using available data from MINOS, Super-K atmospheric $\nu$, solar, and
reactor experiments in combination with the T2K observation.
This indication -- or evidence as it is called by the authors of \cite{Fogli:2011qn} -- 
would be a critical 
milestone in the overall neutrino physics program 
because a nonzero $\theta_{13}$ enables us to explore the $\delta$
parameter.
The magnitude of $CP$ violation in $\nu_l \rightarrow \nu_{l'}$
and $\bar \nu_l \rightarrow \bar \nu_{l'}$ oscillations is determined
by the rephasing invariant:
\begin{equation}
J_{CP} = \frac{1}{8}\cos\theta_{13}\sin2\theta_{12}\sin2\theta_{23}\sin2\theta_{13}\sin\delta.
\end{equation}
Thus, the size of $CP$ violation effect depends on the size of 
both $\theta_{13}$ and $\delta$ values.
The value of the $CP$ phase  
is a question of general interest because
(1) the $\delta$ is last remaining oscillation parameter to be measured 
in the MNS matrix, and  
(2) a leptonic $CP$ asymmetry is a leading candidate  
for understanding the origin of the existing matter-dominated (over antimatter) universe.
One of the goals of the Hyper-K experiment is 
to measure the $\delta$ parameter and thereby 
discover leptonic $CP$ violation by using a J-PARC muon neutrino beam.
We describe the expected sensitivity in Sec. \ref{sec:cp}.

\begin{figure}[htbp]
  \begin{center}
    \includegraphics[scale=0.6]{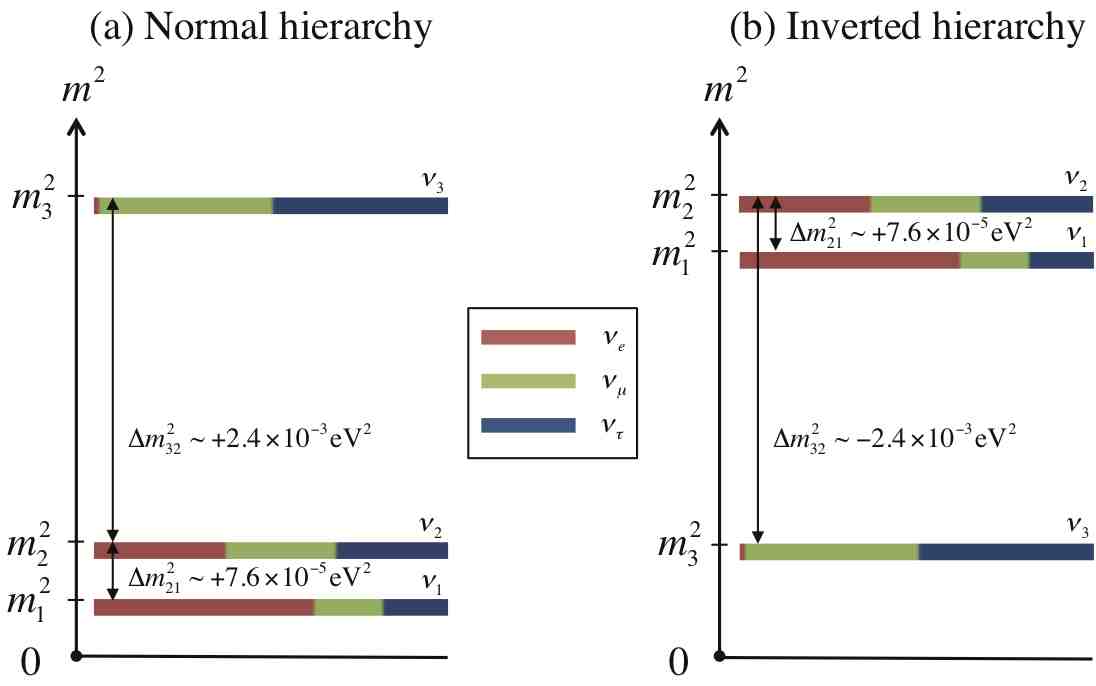}
  \end{center}
  \caption{Two mass hierarchy cases.
  \label{fig:hierarchy-illustrate}
           }
\end{figure}
The mass hierarchy or 
the sign of $\Delta m^2_{32}$ is not yet known: whether 
$m_3$ is the heaviest mass eigenvalue (normal hierarchy, $\Delta m^2_{32} > 0$)
or
$m_3$ is the lightest one (inverted hierarchy, $\Delta m^2_{32} < 0$)
remains to be experimentally determined in the future.
Two hierarchy cases are illustrated in Fig. \ref{fig:hierarchy-illustrate}.
Such an experimental determination would help in understanding  
the origin of neutrino masses, an explanation of which is expected to relate to
BSM physics. 
Determination of the mass hierarchy is also important in terms of 
neutrino-less double beta decay rates.
In the case of inverted hierarchy, the effective neutrino mass
$m_{\beta\beta}$ is expected to be larger than 10 meV; this 
could be reachable by various proposed detectors designed to  
test the Majorana nature of neutrinos \cite{DBreview-Elliott}.
The suggested $\sin^2\theta_{13}$ of a few  $\times 10^{-2}$
would also open the possibility of mass hierarchy 
determination by measuring atmospheric $\nu_e$ appearance via a resonance effect 
arising from passing through the Earth's matter.
High statistics atmospheric neutrino data would also provide 
additional information on $\delta$ and the $\theta_{23}$ octant,
as discussed in detail
in Sec. \ref{section:atmnu}.
If we are fortunate to observe the neutrino burst from a supernova explosion,
the mass hierarchy could be determined by observing time variation of the 
neutrino spectrum (see Sec. \ref{sec:supernova} for the details).

\subsubsection{Nucleon decays}\label{section:intro-ndecay}

\begin{figure}[bp]
  \begin{center}
    \includegraphics[scale=0.4]{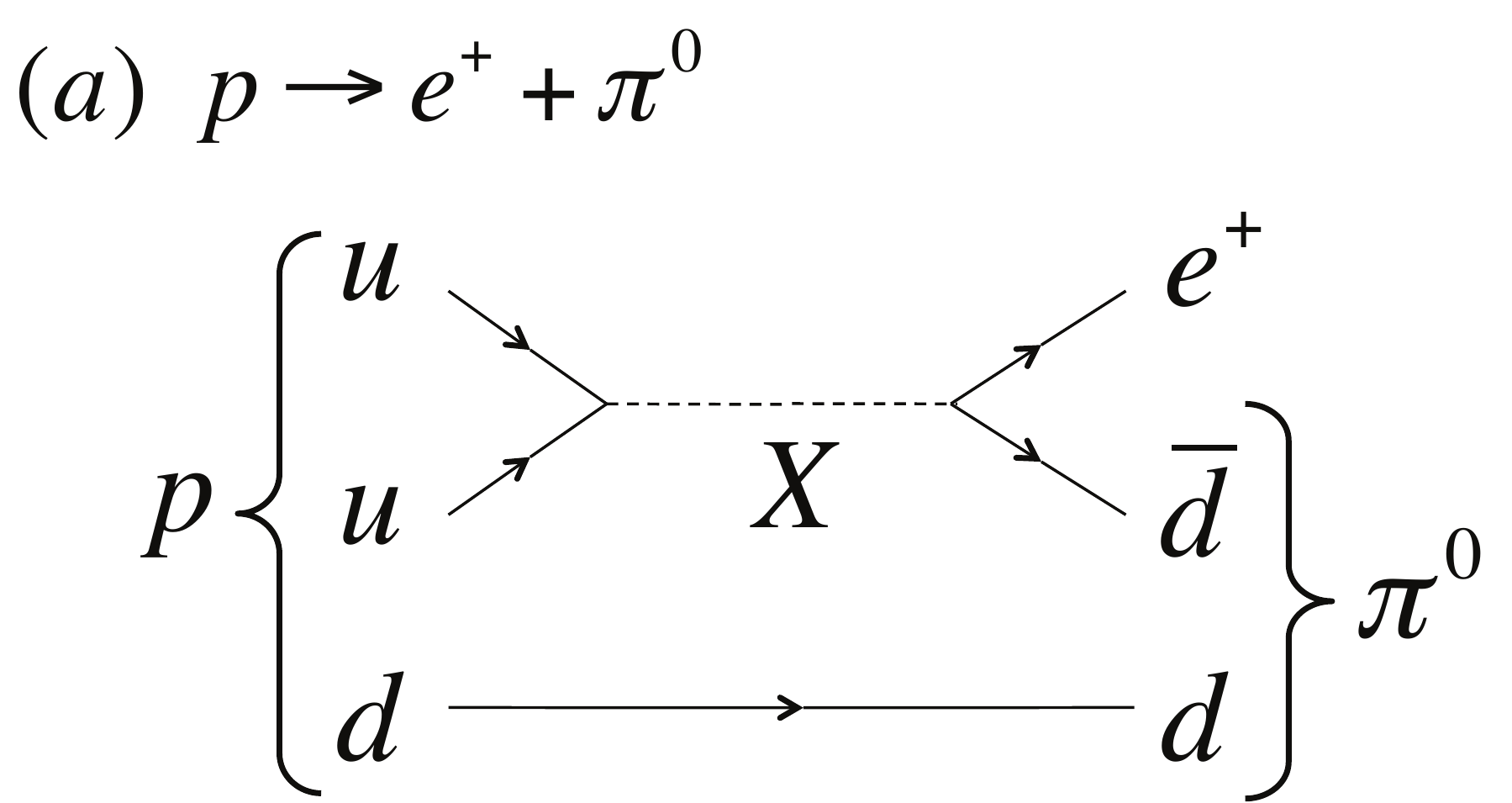}
    \includegraphics[scale=0.4]{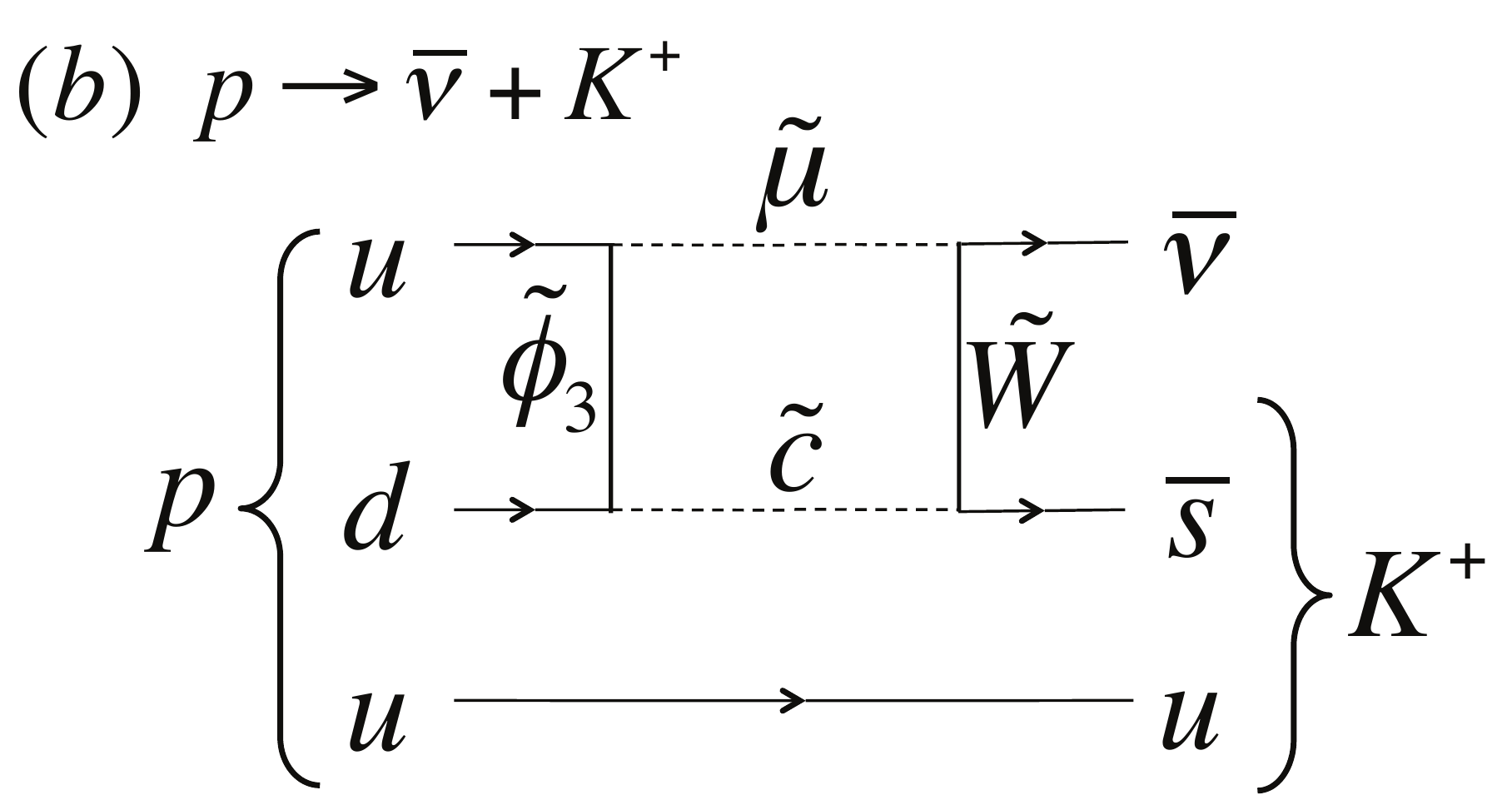}
  \end{center}
  \caption{Examples of Feynman diagrams for particular nucleon
           decay modes.  The left panel shows a proton decaying into a 
           positron and a neutral pion via exchange of a super-heavy
           gauge boson $X$.
           The right one shows the SUSY-favored decay mode, 
           proton decay into a neutrino and a $K^+$ mediated
           by a colored Higgs and other SUSY particles.
  \label{fig:pdk-FD}
           }
\end{figure}
Motivations for unifying the strong, weak, and electromagnetic forces
are very strong.  Grand Unifies Theories naturally explain
the experimental observation that electrical charges are quantized as
$Q_{\rm proton}+Q_{e^-}=0$ with better than $10^{-21}$ accuracy.
GUTs provide a simple organization of quarks and leptons by which the  
quantum numbers of quarks and leptons may be naturally understood.
GUTs are also supported by the noted unification of three gauge 
coupling constants at the very high energy scale of 
$M_{\rm GUT}\sim10^{16}$ GeV.
They have further support from the high energy physics scale implied 
by the tiny neutrino mass scale observed in neutrino oscillation experiments.

In the SM, protons are practically stable particles because
the decay rate due to the chiral anomaly is unobservably small
\cite{'tHooft:1976up}.  However, GUTs generally yield a prediction
of spontaneous proton or bound nucleon decay with experimentally
testable decay rates.  
Therefore nucleon decay experiments are well motivated
to test various proposed unified models, detect
evidence of a new paradigm, and then to pin down the details
of the successful theory by measuring the branching ratio of each nucleon
decay mode.

The favored decay mode in prototypical GUTs based on $SU(5)$ symmetry 
is $p \rightarrow e^{+}+\pi^{0}$ mediated by super-heavy
gauge bosons as shown in Fig. \ref{fig:pdk-FD}. 
On the other hand, GUTs incorporating supersymmetry (SUSY-GUTs)
suppress the decay mode $p \rightarrow e^{+} + \pi^{0}$ 
while favoring 
$p \rightarrow \overline{\nu} + K^{+}$ 
via dimension five operator interactions 
involving exchange of a heavy color triplet Higgsino
as shown in the right panel of Fig. \ref{fig:pdk-FD}. 
Proton lifetimes predicted by some GUTs and SUSY-GUTs are listed in 
Table \ref{tab:predict}.

\begin{table}[tb]
\caption{Proton lifetimes predicted by various GUTs and SUSY-GUTs.
\label{tab:predict}}
\vspace{0.4cm}
\begin{center}
\begin{tabular}{lll}
\hline\hline
Model        & Decay Mode & Lifetime Prediction (years)\\ 
\hline\hline
Minimal $SU(5)$~~~~~~~~~~~~~~~~ & $p \rightarrow e^{+} \pi^{0}$~~~~~~~~  & $10^{28.5} \sim 10^{31.5}$ ~\cite{pl}~~~~  \\ 
\hline
Minimal $SO(10)$ & $p \rightarrow e^{+} \pi^{0}$  & $10^{30} \sim 10^{40}$ ~\cite{dl}\\ 
\hline
Minimal SUSY $SU(5)$ & $p \rightarrow \overline{\nu} K^{+}$& $\leq 10^{30}$ ~\cite{hm} \\ 
\hline
SUGRA $SU(5)$ & $p \rightarrow \overline{\nu} K^{+}$ & $10^{32} \sim 10^{34}$ ~\cite{tg} \\ 
\hline
SUSY $SO(10)$ & $p \rightarrow \overline{\nu} K^{+}$ & $10^{32} \sim 10^{34}$ ~\cite{vl} \\ 
\hline
SUSY $SO(10)$ & $p \rightarrow e^{+} \pi^{0}$  & $< 5.3 \times 10^{34}$ \cite{Babu:2010ej} \\  
\hline\hline
\end{tabular}
\end{center}
\end{table}

For three decades now, the experimental pursuit of nucleon decay has been led by water Cherenkov
detectors.  
For example, increasingly stringent limits on the $p \rightarrow e^+ + \pi^0$ decay mode
have been set by a series of experiments employing this technology: 
$2.6 \times 10^{32}$ years by Kamiokande in 1989 \cite{Hirata:1989kn},
$8.5 \times 10^{32}$ years by IMB in 1999 \cite{McGrew:1999nd}, and
$8.2 \times 10^{33}$ years by Super-Kamiokande in 2009 \cite{:2009gd}.
Due to these null observations, minimal $SU(5)$ is now an excluded model.
In addition, minimal SUSY $SU(5)$ is also considered to be ruled out
\cite{Carone:1995kp}
by the results of $p \rightarrow \overline{\nu} K^{+}$ searches,
for which water Cherenkov detectors also provide best limits
\cite{Kobayashi:2005pe}. 
It is certain that a water Cherenkov detector
with 25 times the fiducial volume of Super-Kamiokande
will provide further stringent tests of various GUTs.  
Such a large detector will have a good chance to discover nucleon decay phenomena and would 
then have the potential to provide important constraints on the emergent unified theory 
by measuring the relative rates of a variety of nucleon decay modes.
The expected sensitivities for various nucleon decay modes
are described in Sec. \ref{section:pdecay}.

\subsubsection{Astrophysical objects}\label{section:intro-astro}

Hyper-Kamiokande is also capable of observing neutrinos from various astrophysical objects. One of advantages of the detector is that its
energy threshold is as low as several MeV; this enables us to reconstruct neutrinos from the Sun and supernova on an event-by-event basis.

For 40 years, many efforts have been made to computationally simulate core collapse supernova explosions, but the simulated bounce shock waves often stalled or required special assumptions to be made.  It would seem that the details of the supernova explosion mechanism are still lacking. For instance, one important effect is thought to be the matter reheating process from neutrinos trapped in the iron core. High statistics observations of neutrinos from a supernova would provide precious inside information of the explosion mechanism. If a core collapse supernova explosion were to take place near the center of our Galaxy, Hyper-K would observe as many as 250,000 neutrino interactions. There is also a good chance to explore the mechanism of neutron star or black hole formation through the fine-grained, sub-millisecond time structure details of such a burst.  Meanwhile, while waiting for a nearby explosion to occur, the continuous flux of supernova neutrinos from all past core collapse explosions will guarantee a steady accumulation of valuable astrophysical data.

Thanks to its good low energy performance for upward-going muons, Hyper-K has a larger effective area for upward-going muons below 30~GeV than do cubic kilometer-scale neutrino telescopes (see Fig.~\ref{fig:aeff} in Sec.~\ref{section:astro}). Additionally, fully contained events in Hyper-K have energy, direction, and flavor reconstruction and resolutions as good as those in Super-K.  This high performance will be useful for further background suppression or studies of source properties. For example, the detector is extremely sensitive to the energy range of neutrinos from annihilations of light (below 100~GeV) WIMP dark matter, a region which is suggested by recent direct dark matter search experiments. 

Sensitivities and expected physics outputs from astrophysics studies are discussed in Sec.~\ref{section:solar} and \ref{section:astro}.

%% file: introduction/watercherenkov.tex
\subsection{Water Cherenkov technique}\label{section:intro-watercherenkov}
\begin{figure}[htbp]
  \begin{center}
    \includegraphics[scale=0.45]{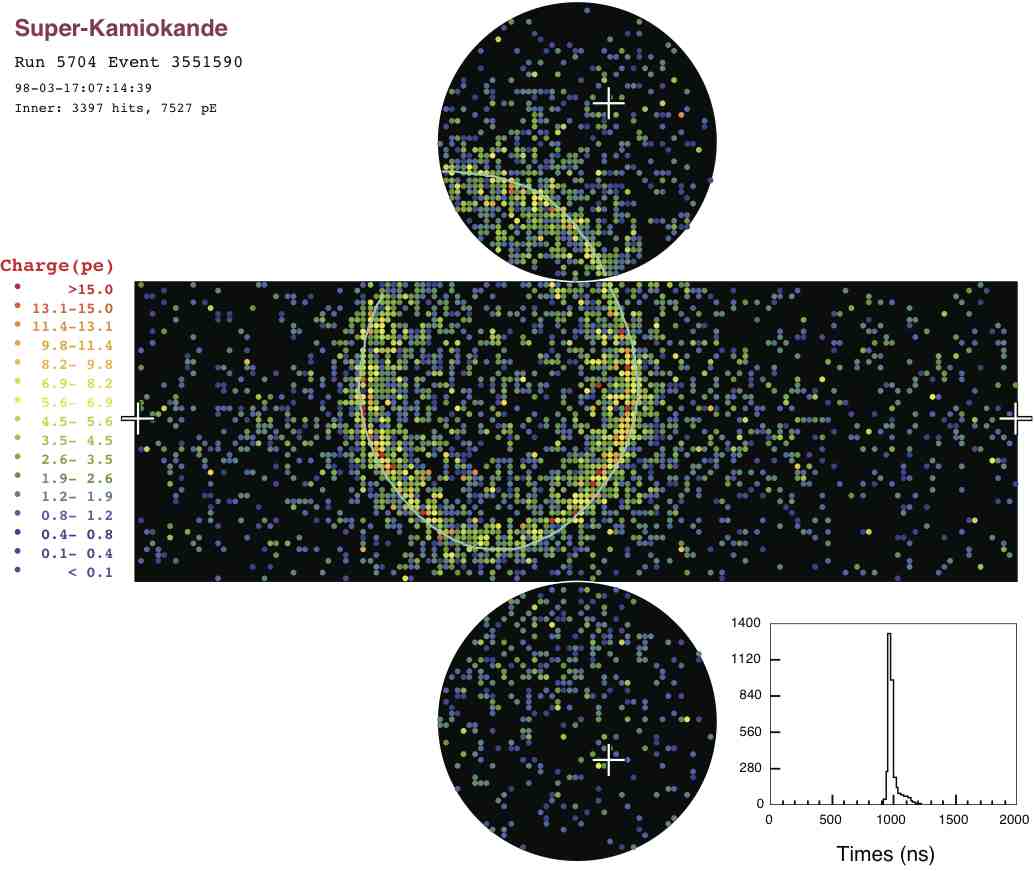}
    \includegraphics[scale=0.45]{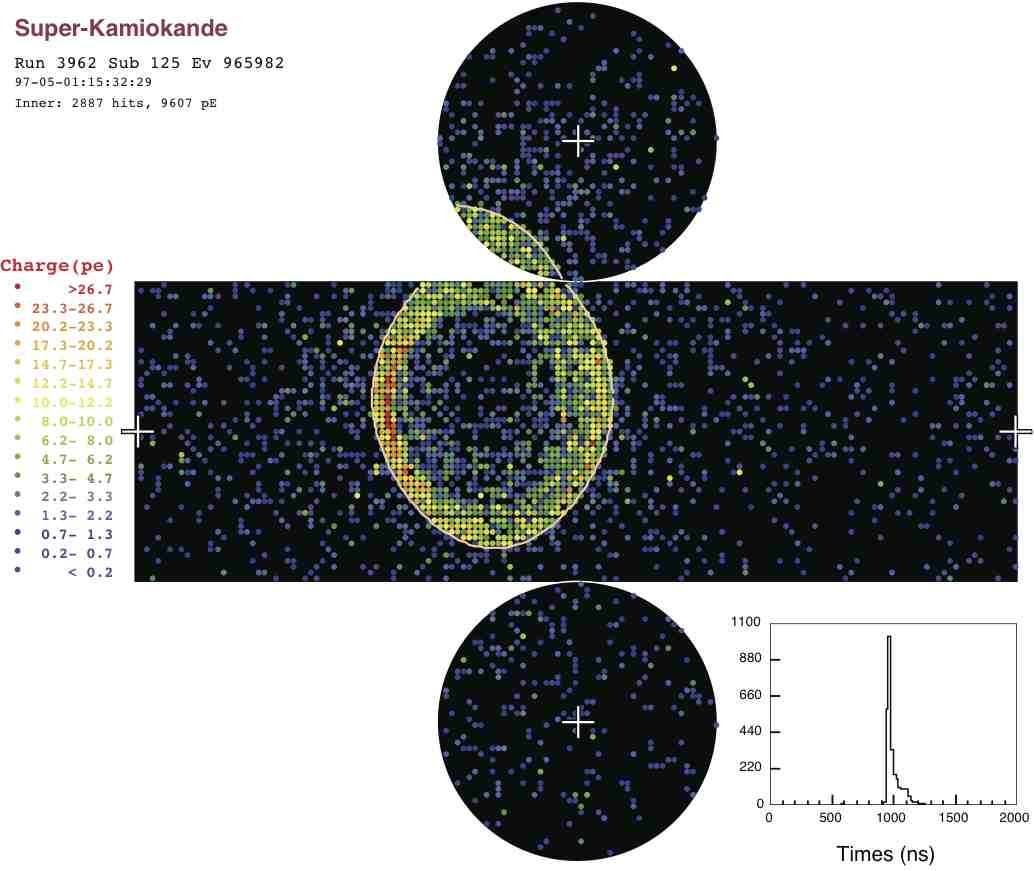}\\
    \includegraphics[scale=0.45]{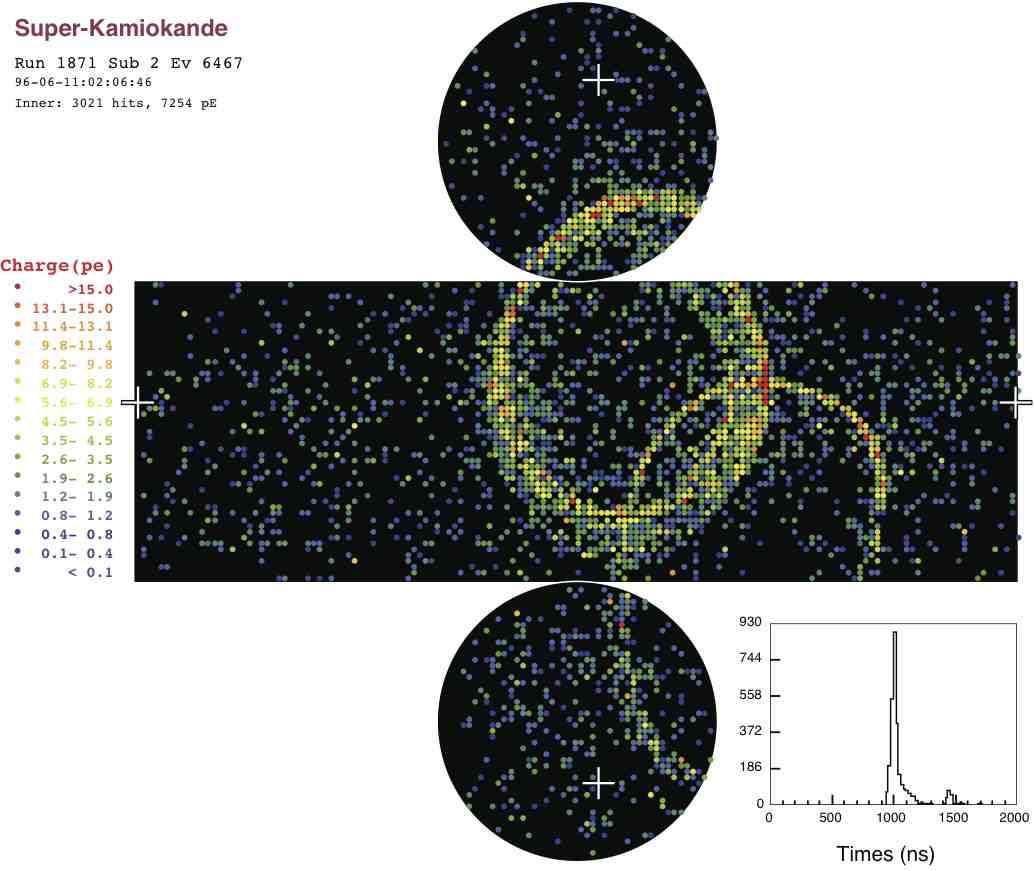}
  \end{center}
  \caption{Event displays of observed data from Super-Kamiokande's phase-I period.
           Shown are unrolled views of a single-ring electron-like event (top left), 
           a single-ring muon-like event (top right), and
           a two-ring event (bottom).
           Colored points indicate the number of detected photo-electrons
           in each photomultiplier tube.
           }
  \label{fig:display}
\end{figure}

The technique of ring imaging water Cherenkov detection,
which the Hyper-Kamiokande detector is based on,
has been successfully employed by several neutrino and nucleon decay experiments
over the last few decades.
In Japan, the Kamiokande detector (1983-1996, 3 kiloton)
and Super-Kamiokande (in operation since 1996, 50 kiloton)
have achieved several 
important scientific results, notably detection of
extraterrestrial neutrinos from the Sun 
\cite{Hirata:1989zj}
and Supernova 1987a 
\cite{Hirata:1987hu, Hirata:1988ad}, 
and discovery of neutrino flavor mixing and neutrino mass 
\cite{Fukuda:1998mi,Fukuda:2002pe}.
In the K2K long baseline neutrino oscillation experiment,
Super-K and a one kiloton water Cherenkov detector (1KT)  
provided indispensable data on the neutrino beam flux and its energy spectrum 
at the neutrino production site (using 1KT) and a location 250 km farther away (using Super-K)
\cite{Ahn:2006zza}. 
Super-K again is playing the role of the far detector in the ongoing 
T2K experiment which reported an indication of $\nu_\mu \rightarrow \nu_e$
oscillations in June 2011 \cite{Abe:2011sj}.

\begin{table}[htdp]
\caption{Expected detector performance of Hyper-Kamiokande.
         Further details can be found in Sec. \ref{section:physics}}
\begin{center}
\begin{tabular}{cc} \hline \hline
 & Resolution or Efficiency \\ \hline \hline
Vertex resolution @ 500MeV/$c$ & 28 cm (electron) / 23 cm (muon) \\
 ~~~~~~~~~~~~~~~~~~~@ 5GeV/$c$ & 27 cm (electron) / 32 cm (muon) \\ \hline 
Particle ID @ 500MeV/$c$ & $98.5\pm0.6$ \% (electron) / $99.0\pm0.2$ \% (muon) \\
 ~~~~~~~~~~~~@ 5GeV/$c$ & $99.8\pm0.2$ \% (electron) / $100^{+0.0}_{-0.4}$ \% (muon) \\ \hline 
Momentum resolution @ 500MeV/$c$ & 5.6 \% (electron) / 3.6 \% (muon) \\
 ~~~~~~~~~~~~~~~~~~~~~~~~~~@ 5GeV/$c$ & 2.0 \% (electron) / 1.6 \% (muon) \\ \hline 
Electron tagging from 500MeV/$c$ $\mu^+$ decays & 98 \% \\
 ~~~~~~~~~~~~~~~~~~~from 5GeV/$c$ $\mu^+$ decays & 58 \% \\ \hline 
\hline
J-PARC $\nu_e$ signal efficiency  & 64 \% (nominal) / 50 \% (tight) \\ \hline 
J-PARC $\nu_\mu$ CC background rejection  &  $>$99.9 \% \\ \hline 
J-PARC $\nu$ $\pi^0$ background rejection &  95 \% (nominal) / 97.6 \% (tight) \\ \hline 
\hline 
$p\rightarrow e^+ + \pi^0$ efficiency (w/ $\pi^0$ intra-nuclear scattering) &  45 \% \\ 
atmospheric $\nu$ background &  1.6 events/Mton/year \\ \hline 
$p\rightarrow \bar{\nu} + K^+$ efficiency by prompt $\gamma$ tagging method & 7.1 \% \\
atmospheric $\nu$ background & 1.6 events/Mton/year \\ \hline
$p\rightarrow \bar{\nu} + K^+, K^+ \rightarrow \pi^+ + \pi^0$ efficiency & 6.7 \% \\ 
atmospheric $\nu$ background & 6.7 events/Mton/year \\ \hline
\hline 
Vertex  resolution for 10 MeV electrons &  90 cm \\ \hline 
Angular resolution for 10 MeV electrons &  30$^\circ$ \\ \hline 
Energy  resolution for 10 MeV electrons &  20 \% \\ \hline 
\hline 

\end{tabular}
\end{center}
\label{tab:performance}
\end{table}
Relativistic charged particles, produced via neutrino interactions
or possible nucleon decays, radiate Cherenkov photons
while passing through the water \cite{Frank:1937fk}.
These photons' production angle compared to the charged particle's direction is
given by $\cos\theta_c = 1/n\beta$, and $\theta_c$ is
42~degrees in the case that the particle velocity $\beta \equiv v/c = 1$ and
the refractive index of the medium is $n = 1.33$ (i.e., water).
By detecting the spatial and timing distributions of the Cherenkov 
photons with an array of single photon capable sensors,
the parent charged particle's 
position, direction, and energy can be reconstructed.
Ring imaging water Cherenkov detectors are also able to discriminate charged muons ($\mu^\pm$) 
from electrons, positrons, and gammas ($e^\pm$ / $\gamma$), as the later group of particles 
induce electromagnetic showers in the medium which modify the resulting ring patterns. 
Figure \ref{fig:display} shows event displays of 
observed data from Super-Kamiokande's phase-I data-taking period (1996-2001). 
The fraction of mis-identification of muons (electrons)
as electrons (muons) is as low as $1\%$ as shown in 
Table \ref{tab:performance}.
Moreover, the high efficiency of tagging muon decay electrons 
provides additional information on the particle type 
with which to further purify muon or electron samples.
Thanks to their excellent particle identification capabilities, 
water Cherenkov detectors can determine 
neutrino flavor compositions in atmospheric and accelerator neutrinos
and thereby detect neutrino flavor transitions or oscillations.
Other critical features of water Cherenkov detectors are an 
excellent rejection efficiency 
for $\pi^0$ background interactions
as well as a good signal efficiency 
for $\nu_\mu \rightarrow \nu_e$ oscillation 
studies using the J-PARC neutrino beam.
These particle ID and tracking capabilities are also indispensable for identifying 
possible nucleon decay candidates while at the same time minimizing physics backgrounds 
to nucleon decay arising from atmospheric neutrino interactions.

Table \ref{tab:performance} presents a summary of Hyper-Kamiokande's high-energy 
performance relevant for nucleon decay searches as well as 
atmospheric and accelerator neutrino oscillation experiments
in which particle energies are above $\sim100$ MeV.
Detector performance for low-energy events  
such as solar neutrinos or supernova neutrinos are also shown.
Details of data analyses and the physics sensitivities of the Hyper-K 
detector are discussed in Sec. \ref{section:physics}.

%% file: experimental-setup/site.tex
\section{Experimental setup overview}
This section explains the baseline design of the Hyper-Kamiokande
detector, i.e. candidate site, cavern and water tank, 
water purification system, photosensors,
data acquisition system, and calibration system.  
Necessary technology for the detector
has been already established and the detector could be constructed
and fully operational in 10 years.

\subsection{Site, caverns, and tanks}
\label{section:site}.
\begin{figure}[htbp]
\includegraphics[scale=0.25]{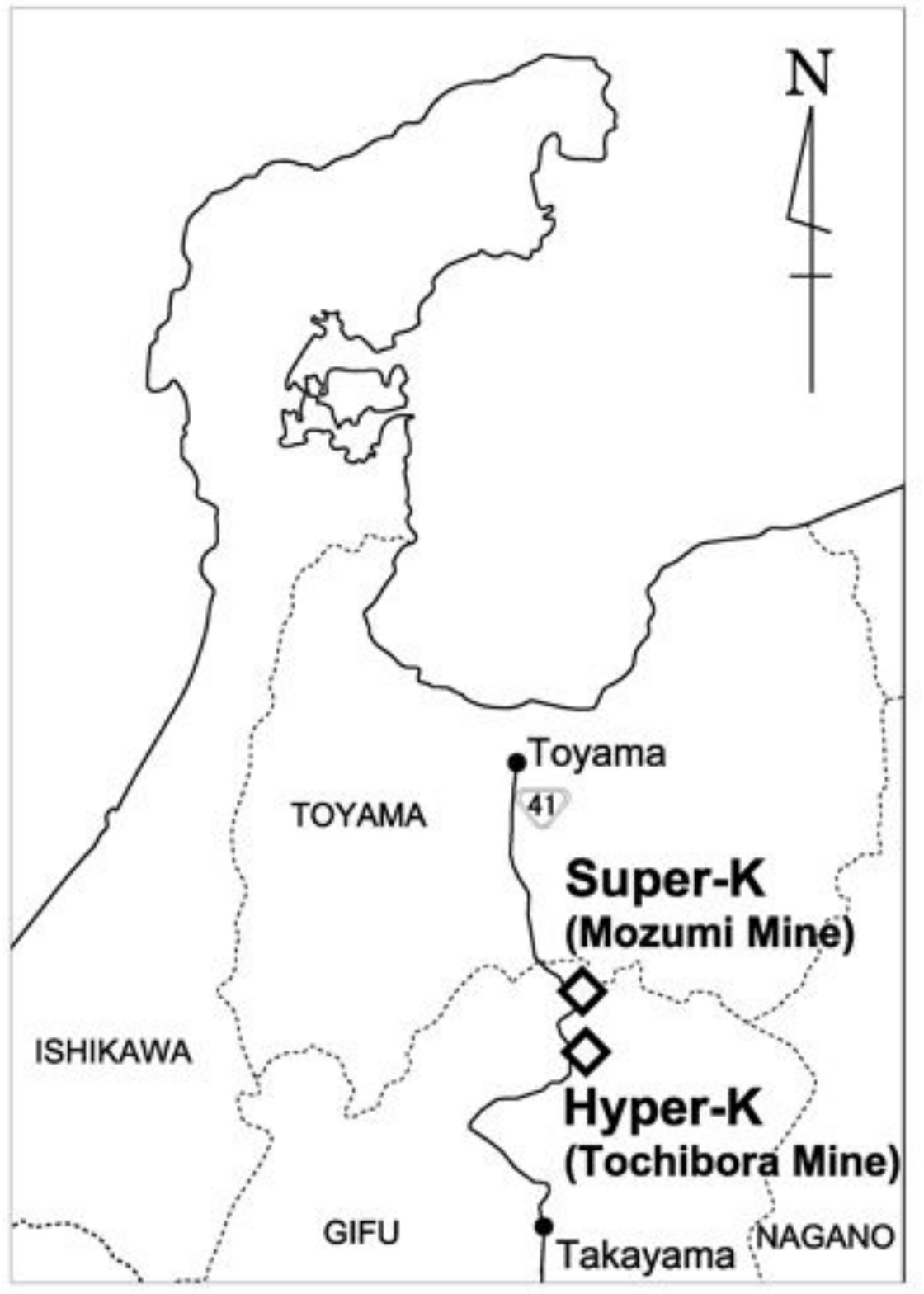}
\includegraphics[scale=0.25]{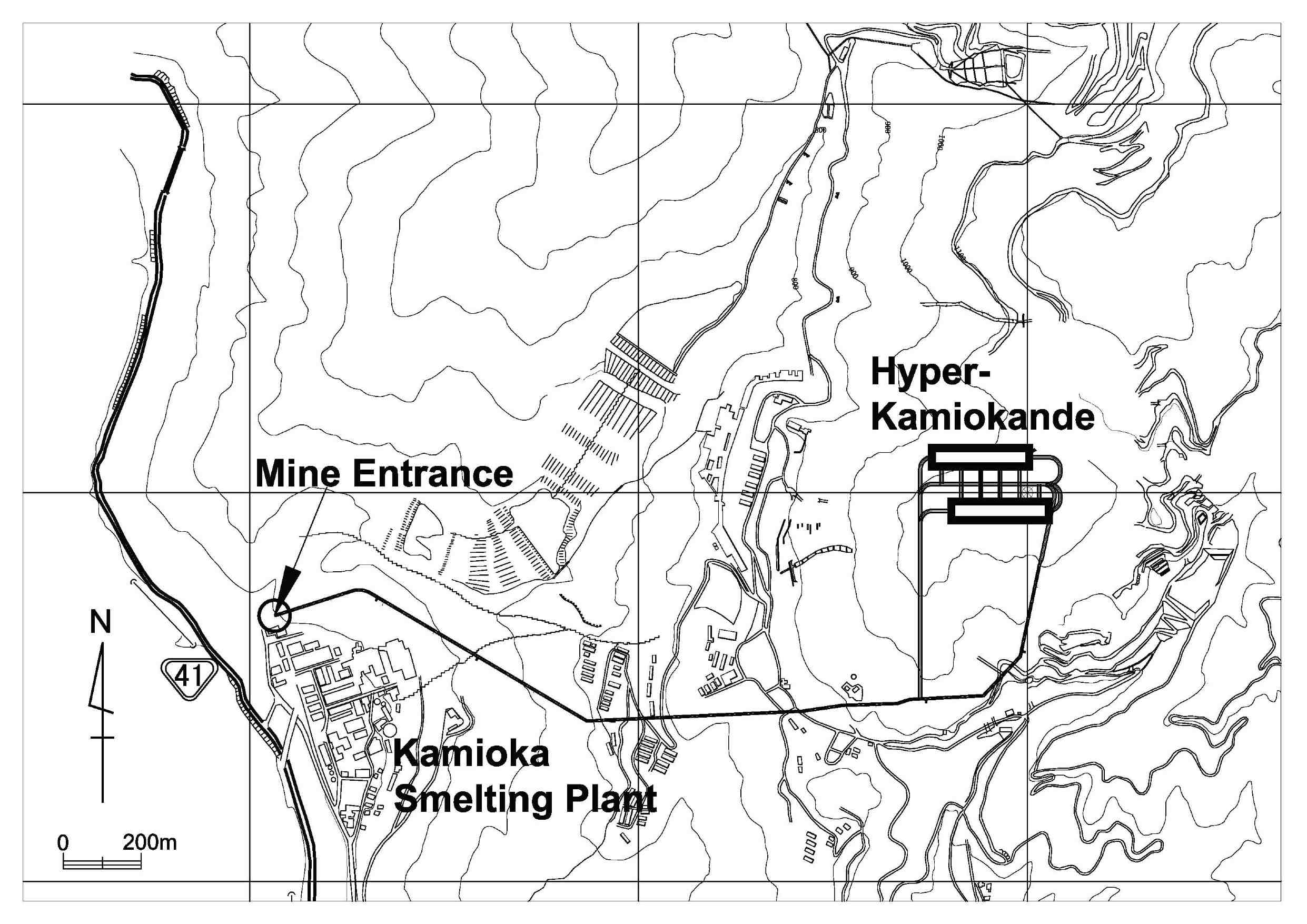}
\caption{The candidate site map.  The site is located 8 km south
of the Super-K site as shown in the left panel.  
The map of the Tochibora mine is shown in the right panel.\label{fig:map}}
\end{figure}

Hyper-Kamiokande 
is to be the third generation water Cherenkov detector in Kamioka 
designed for nucleon decay searches and neutrino studies.   
Its total volume of one million tons would be about 20 times larger than 
that of the world's largest currently operating underground 
detector, Super-Kamiokande. 
The Hyper-K detector candidate site,
located 8 km south of Super-K, 
is in the Tochibora mine
of the Kamioka Mining and Smelting Company,
near Kamioka town in Gifu prefecture, Japan, 
as shown in Fig. \ref{fig:map}.
The experiment site is accessible via a drive-in, 
$\sim$2.6 km long, (nominally) horizontal mine tunnel.
The detector will lie under the peak of Nijuugo-yama,
having 648 meters of rock
or 1,750 meters-water-equivalent (m.w.e.) overburden,
at geographic coordinates 
Lat.
$36^\circ21'08.928''$N,
Long.
$137^\circ18'49.688''$E, and an altitude of 508 m above sea level.
The rock wall in the existing tunnels
and sampled bore-hall cores are dominated by  
Hornblende Biotite Gneiss and Migmatite 
in the state of sound, intact rock mass.  This is desirable 
for constructing such unprecedented large underground cavities.
The site has a neighboring mountain, Maru-yama,  just 2.3 km away, whose
collapsed peak enables us to easily dispose of 
more than one million m$^3$ of waste rock from the detector cavern excavation.  
The site also has the benefit -- well-suited for a water Cherenkov experiment --
of abundant, naturally clean water located nearby.   
More than 13,000 m$^3$/day (i.e., one million tons per $\sim$80 days) will be available.
The Mozumi mine where the Super-K detector is located is another 
candidate site for which less geological information is available
at this moment.

\begin{figure}[tbp]
\includegraphics[scale=0.40]{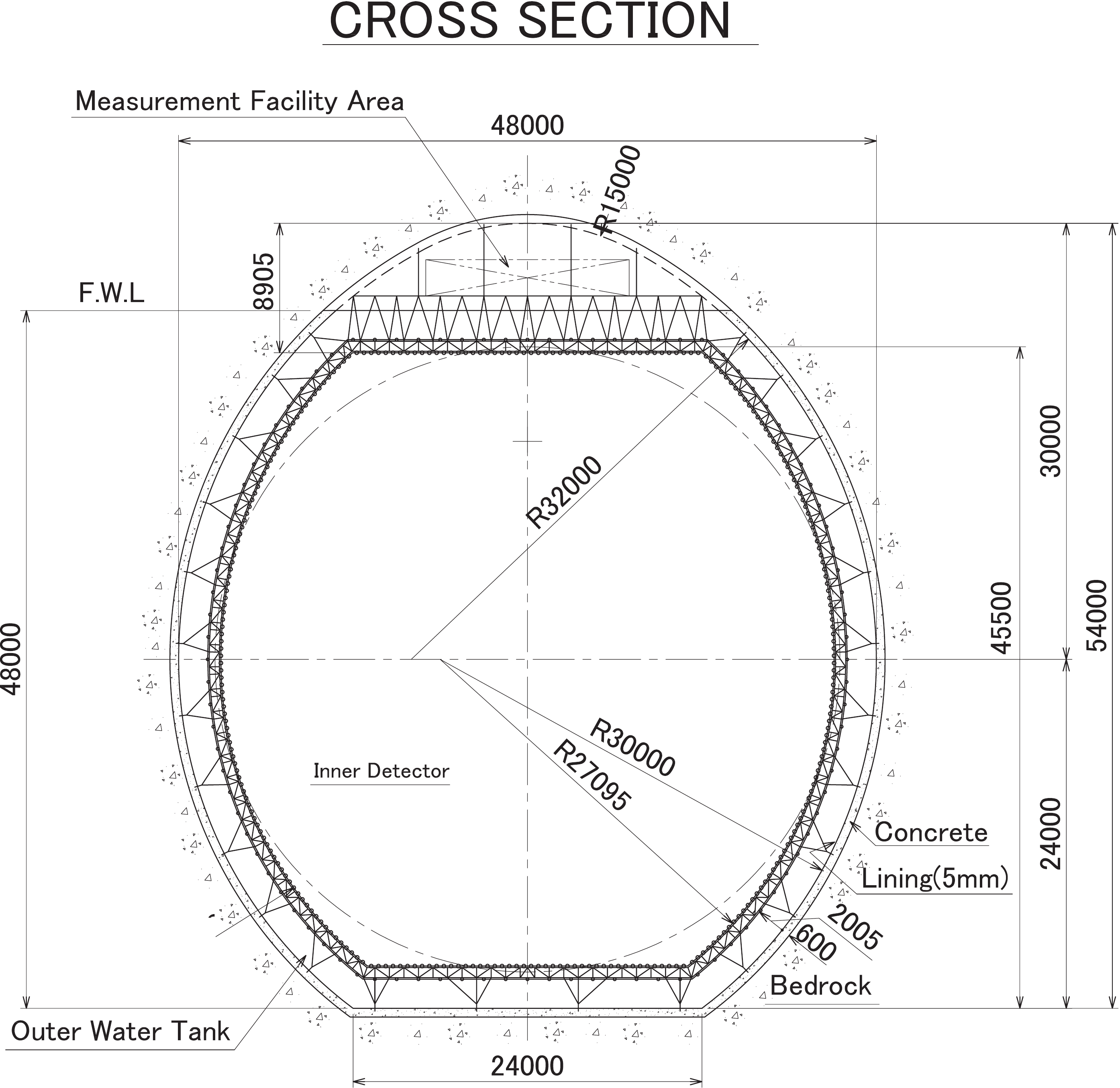}
\caption{Cross section view of the Hyper-Kamiokande detector.\label{fig:crosssection}}
\end{figure}
\begin{figure}[tbp]
\includegraphics[scale=0.50]{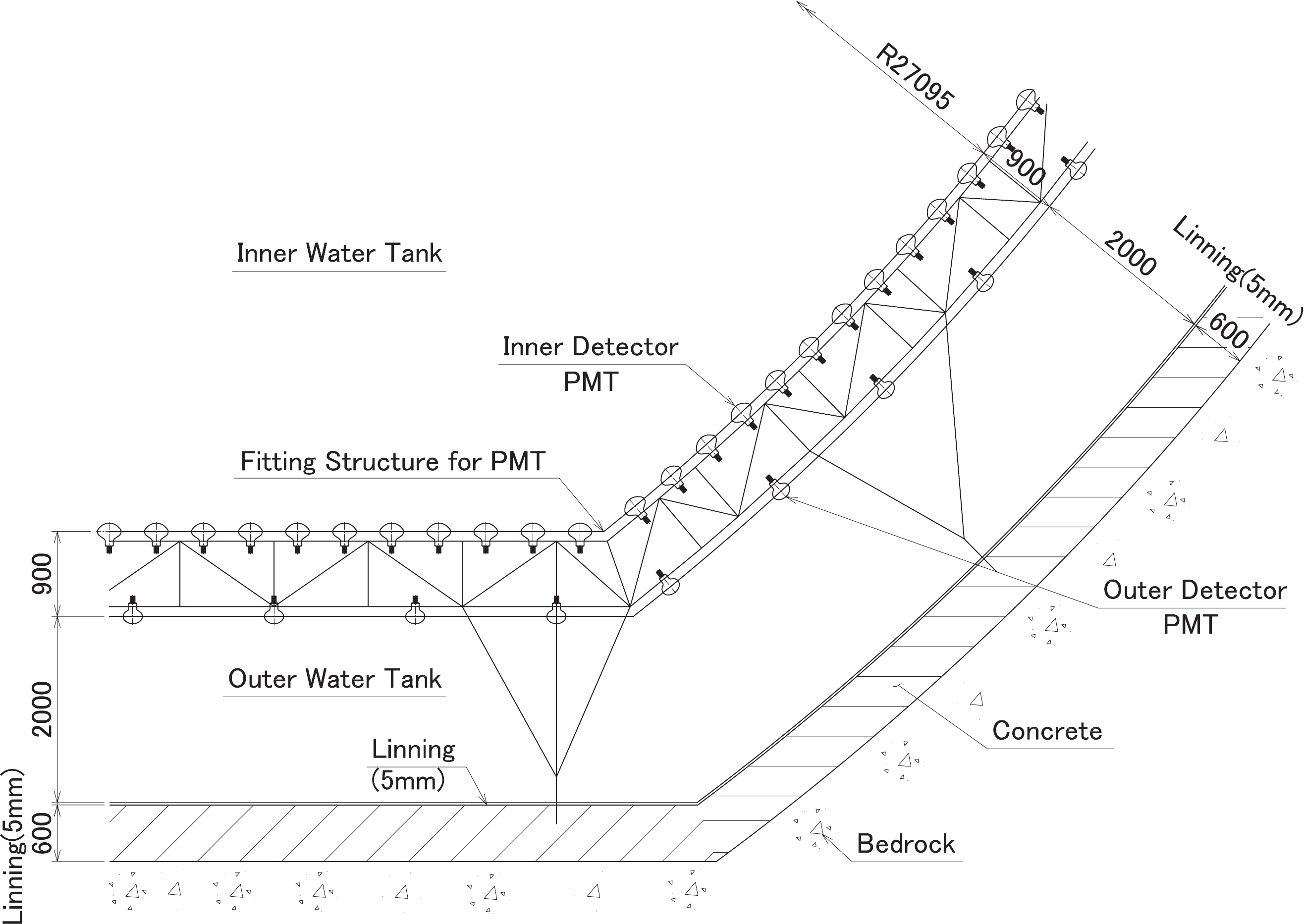}
\caption{Cross section view of the outer detector.\label{fig:crosssectionod}}
\end{figure}
\begin{figure}[htbp]
\includegraphics[scale=0.5]{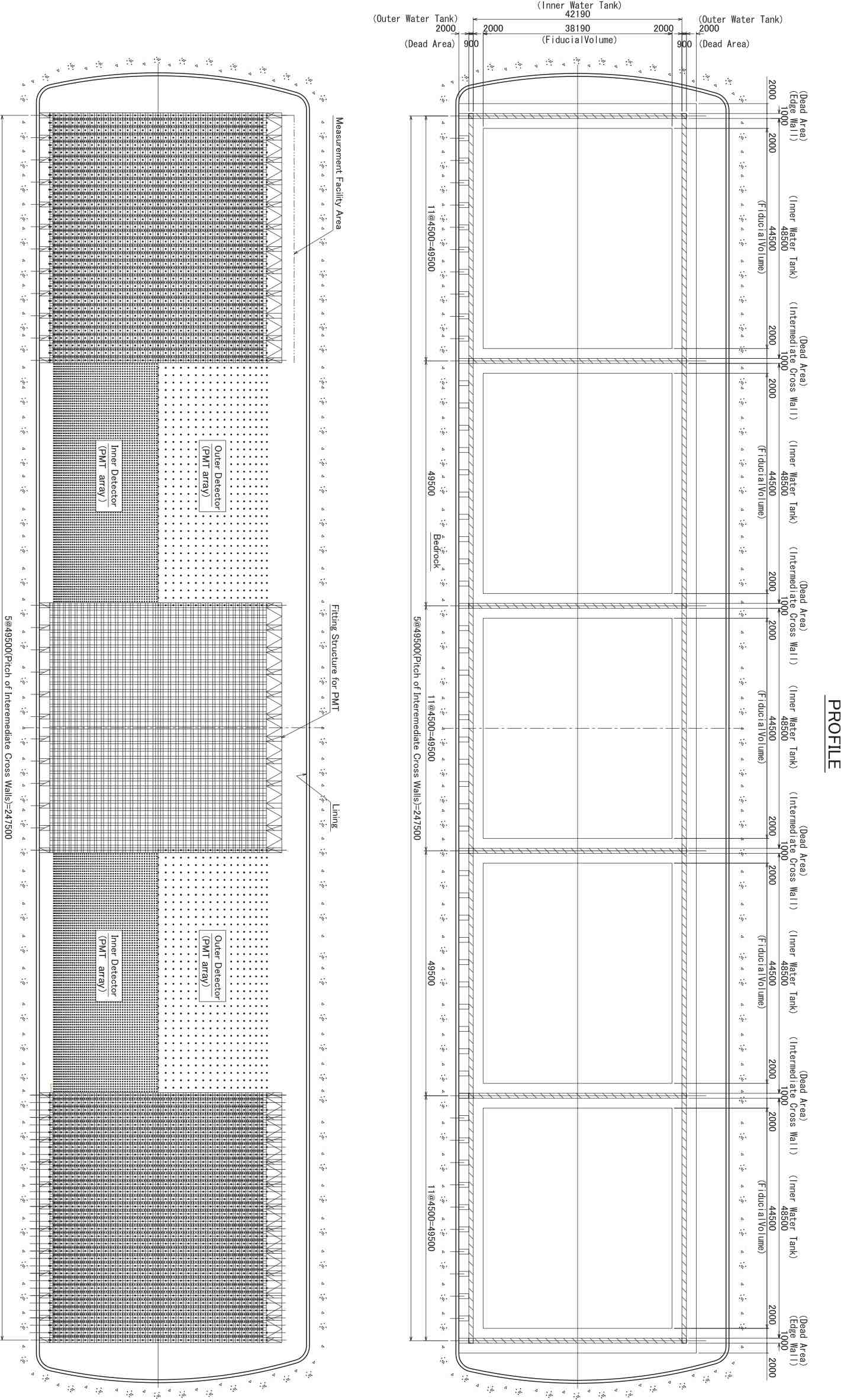}
\caption{
Profile of the Hyper-K detector.  
The left panel shows PMT arrays and the support structure for the inner
and outer detectors.
The right panel shows segmentation of the detector.
Each quasi-cylindrical tank 
lying horizontally is segmented by intermediate walls
into five sub-detectors.\label{fig:profile}}
\end{figure}
\begin{figure}[htbp]
\includegraphics[scale=0.55]{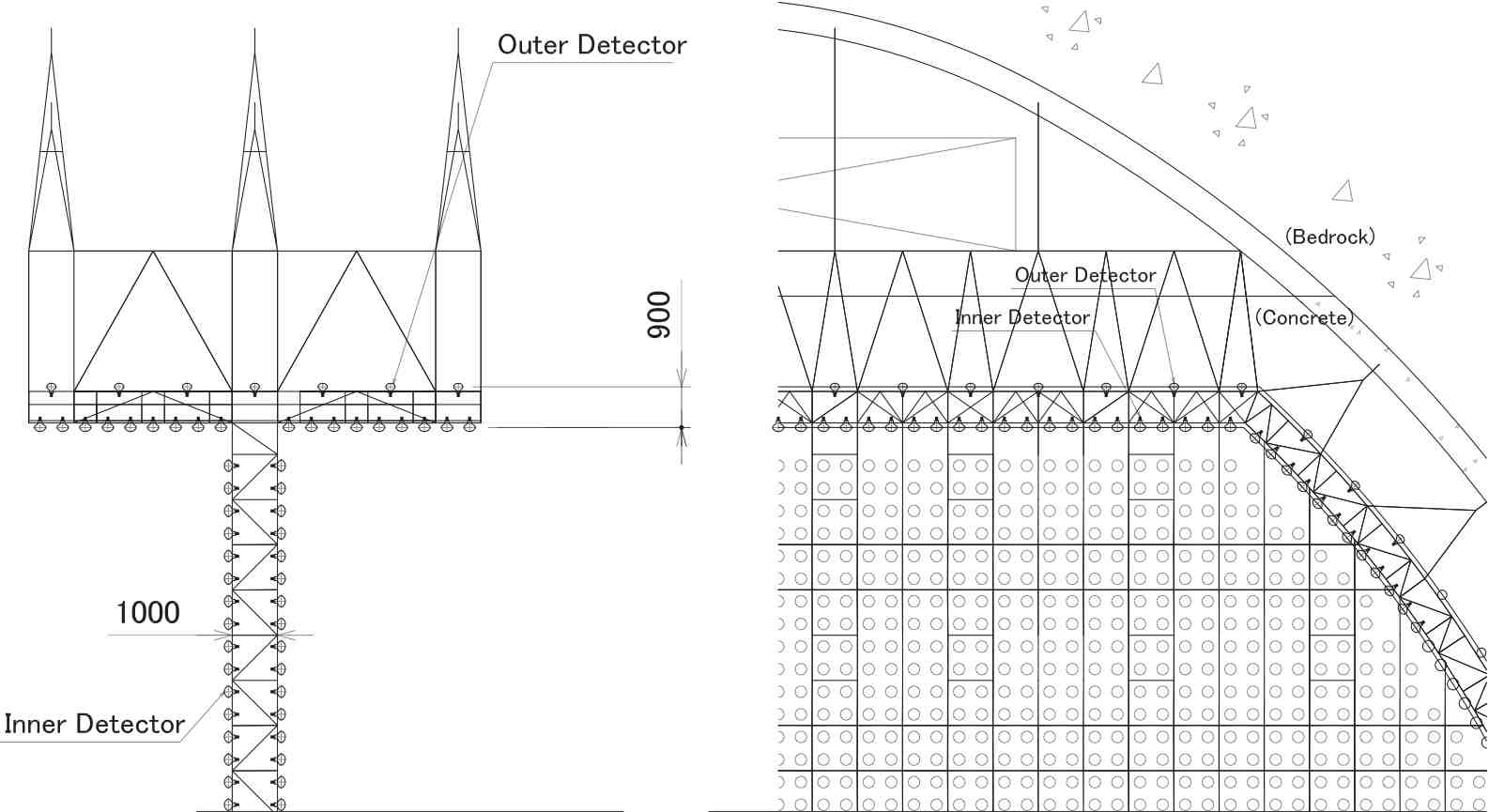}
\caption{
Magnified view of the top part of the segmentation wall.
\label{fig:segmentation}}
\end{figure}
In the base design, the Hyper-K detector is composed of 
two separated caverns 
as shown in Fig. \ref{fig:HK-schematic-view}, each 
having a egg-shape cross section 48 meters wide, 54 meters tall,   
and 250 meters long as shown 
in Fig. \ref{fig:crosssection}, \ref{fig:crosssectionod}, and \ref{fig:profile}.
These welded polyethylene tanks are filled up to a depth of 48 m with 
ultra-pure water: the total water mass equals 0.99 million tons.
The 5 mm thick polyethylene sheet which forms the water barrier 
has been used in waste disposal sites, 
and based upon that experience its expected lifetime is longer than 30 years.  
We expect a much longer lifetime than 30 years, 
as the underground experimental site is free from the sunshine which degrades 
plastics on the Earth's surface. 
This polyethylene lining sheet will be placed upon a concrete base to
achieve a strong, water tight structure as is shown in Fig. \ref{fig:crosssectionod}.
Detailed design of the structure of the concrete layer,
connection with the lining sheet, 
and the PMT support structure is yet to be made.
Each cavern will be optically separated by segmentation walls 
located every 49.5 m to form 5 (in total 10) compartments as 
shown in Fig. \ref{fig:segmentation}, such that 
event triggering and event reconstruction can be performed 
in each compartment separately and independently.
Because the compartment dimension of 50 m is comparable with that of Super-K (36 m)
and is shorter than the typical light attenuation length in water achieved by the Super-K water
filtration system, 
($>100$ m @ 400 nm),
we expect that detector performance of Hyper-K will be basically the same as
that of Super-K.
The water in each compartment is further optically 
separated into three regions.
The inner region has a barrel shape 42 m in height and in width, 
and 48.5 m in length, 
and is viewed by inward-facing array of 20-inch diameter photomultiplier tubes (PMTs).  
The entire array consists of 99,000 HAMAMATSU R3600 hemispherical PMTs,
uniformly surrounding the region and giving a photocathode coverage of 20\%.
The PMT type, size, and number density are subject to optimization.
An outer region completely surrounds the 5 (in total 10) inner regions
and is equipped with 25,000 8-inch diameter PMTs.  
This region is 2 m thick at top, bottom,
and barrel sides, except at both ends of each cavern, where 
the outer region is larger than 2 m due to rock engineering considerations. 
A primary function of the outer detector is to reject
entering cosmic-ray muon backgrounds and to help in identifying 
nucleon decays and neutrino interactions occurring in the inner detector.
The middle region or dead space is an uninstrumented, 0.9 m thick
shell between the inner and outer detector volumes 
and the stainless steel PMT support structure is located in this region.
Borders of both inner and the outer regions are lined with opaque sheets.
This dead space, along with the outer region, acts as a shield against
radioactivity from the surrounding rock.
The total volume of the inner region is 0.74 million tons
and the total fiducial volume is 10 times 0.056 = 0.56 million tons.
The fiducial volume is defined as the region formed by  
a virtual boundary located 2 m away
from the inner PMT plane (Fig. \ref{fig:profile}, right panel).
\begin{figure}[tbp]
\includegraphics[scale=0.65]{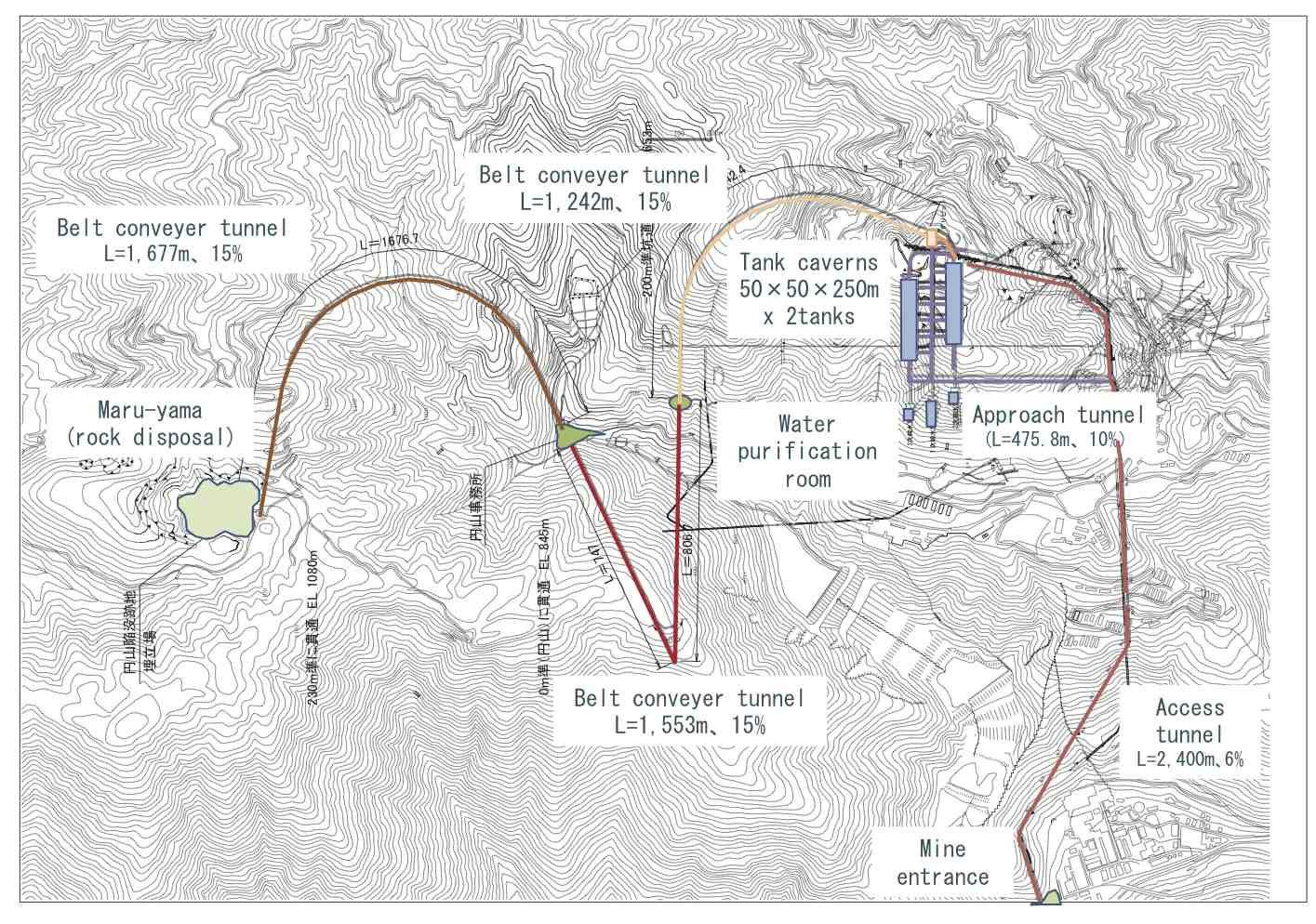}
\caption{Schematic view of the cavern construction.\label{fig:construction}}
\end{figure}
Figure \ref{fig:construction} shows a schematic view of the cavern construction.
After constructing access tunnels, approach
tunnels, and belt conveyor tunnels, construction of two water tank caverns
will start with the excavation of arches, 
and then excavation will proceed from top to bottom
for each layer utilizing the bench cut method.
The majority of waste rock is planned to 
be transported to the top of the adjacent mountain Maru-yama
via conveyors in the declined conveyor tunnels.

\begin{figure}[tbp]

\includegraphics[scale=0.52]{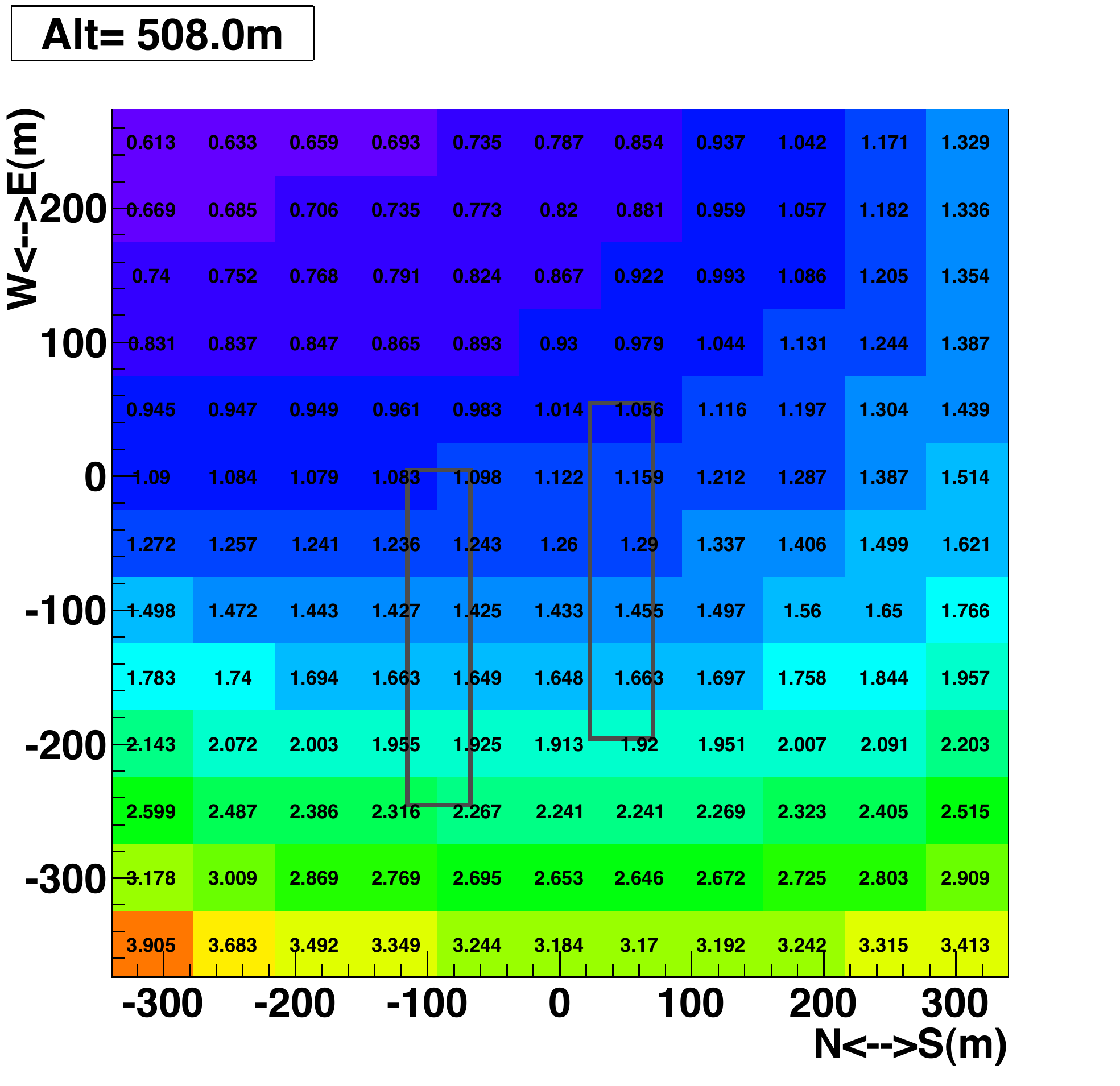}
\caption{Muon flux calculated by MC simulation around the Hyper-K candidate location
at Lat. $36^\circ21'08.928''$N, Long. $137^\circ18'49.688''$E, 
and altitude 508 m above sea level. Horizontal and vertical axes represent
distance from the candidate location in m in the North-South and West-East direction,
respectively. Text and color represents expected muon flux in units of $10^{-6}$sec$^{-1}$cm$^{-2}$. Two boxes represent the candidate locations of the two Hyper-K detector modules.
The calculation has been performed based on the predicted muon flux at the Earth's surface, 
attenuation by the rock around the detector. The depth in the mountain
is calculated using a 50 m mesh tomographic map. A rock density of 2.7g/cm$^2$ is assumed
in the attenuation calculation.
\label{fig:MuFlux}}
\end{figure}

The estimated cosmic-ray muon rate around the Hyper-K detector 
candidate site is shown in Fig.~\ref{fig:MuFlux}. The range of
estimated muon fluxes at the detector position
is 1.0 $\sim$ 2.3 $\times$ 10$^{-6}$ sec$^{-1}$cm$^{-2}$ which is
roughly 10 times higher than the flux at Super-K's location
(0.13 $\sim$ 0.14 $\times$ 10$^{-6}$ sec$^{-1}$cm$^{-2}$).
The expected deadtime due to these muons is less than 1\% and 
negligible for nucleon decay searches, atmospheric neutrino studies,
and long baseline experiments.
However, the radioactive spallation products produced by the higher 
cosmic-ray muon rate would increase background levels for low-energy 
physics targets such as solar or supernova neutrinos.
An estimation of the possible background increase by the higher 
cosmic-ray muon rate -- and its impact -- will be discussed in the Sec. \ref{section:solar}.

%% file: experimental-setup/water.tex
\subsection{Water purification system\label{section:water}}

Needless to say, water is the target material and signal-sensitive medium of the detector, and thus its quality directly affects the sensitivity. In order to realize such a huge Cherenkov detector, achieving good water transparency is the highest priority. In addition, as radon emanating from the photosensors and detector structure materials is the main background source for low energy neutrino studies, an efficient radon removal system is indispensable.

In Super-Kamiokande the water purification system has been continually modified and improved over the course of SK-I to SK-IV.  As a result, the transparency is now kept above 100 m and is very stable, and the radon concentration in the tank is held below 1 mBq/m$^3$.  Following this success, the Hyper-Kamiokande water system design will be based on the current Super-Kamiokande water system.

Naturally, ever-faster water circulation is generally more effective when trying to keep huge amounts of water clean and clear, but increasing costs limit this straightforward approach so a compromise between transparency and recirculation rate must be found.  In Super-Kamiokande, 50,000 tons of water is processed at the rate of 60 tons/hour in order to keep the water transparency (the attenuation length for 400 nm-500 nm photons) above 100 m, and 20 m$^3$/hour of radon free air is generated for use as a purge gas in degas modules, and as gas blankets for both buffer tanks and the Super-Kamiokande tank itself~\cite{Fukuda:2002uc}.
For the 0.99 million tons of water in Hyper-Kamiokande, these process speeds will need to be scaled-up to 1200 m$^3$/hour for water circulation and 400 m$^3$/hour for radon free air generation.

\begin{figure}
     \begin{center}
       \includegraphics[width=17cm]{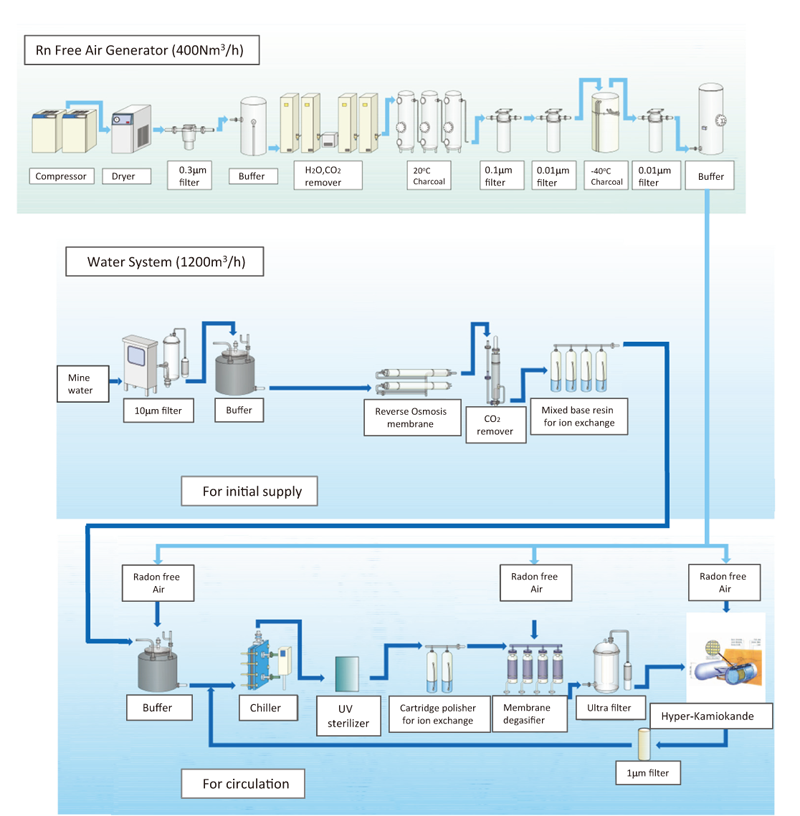}
       \caption{Current air \& water flow design of the Hyper-Kamiokande water system.}
       \label{water:water}
     \end{center}
\end{figure}

Figure \ref{water:water} shows the current design of the Hyper-Kamiokande air and water purification system. With these systems, the water quality in Hyper-Kamiokande is expected to be same as that in Super-Kamiokande.

%% file: experimental-setup/sensor.tex
\subsection{Photosensors\label{section:pmt}}

In addition to clean water, photosensors are the key component of a water Cherenkov detector. 
The intrinsic characteristics of such a detector impose a number of requirements on the crucial photodetectors.

Of course, a water Cherenkov detector uses the Cherenkov photons
to identify and to reconstruct the directions,
energies, and locations of the particles which produce them. The total number 
of Cherenkov photons produced has a linear correlation with the momentum 
of the particle. Therefore, the total number of Cherenkov photons 
of a low energy particle -- for example, a recoil electron from 
a solar neutrino elastic scattering interaction -- are few, and so  
the photosensors in the detector are required to have 
high efficiency and good resolution at the level of a single photon.
On the other hand, several hundred photons hit 
each PMT when high energy particles -- for example, the particles generated
by an inelastic atmospheric neutrino interaction -- pass through 
the detector. Therefore, the photosensors are also required 
to have wide dynamic range and good linearity.
Otherwise, it is not possible to achieve good particle
energy resolution.

The location of the interaction vertex is reconstructed
using photon arrival timing information at each PMT. Therefore,
good timing resolution of the photosensors is essential,  
and the jitter of the transit time is required 
to be small, i.e. less than 3 nanoseconds for a single photon.

\subsubsection{Photosensors in the baseline design}
\begin{figure}
     \begin{center}
       \includegraphics[width=7cm]{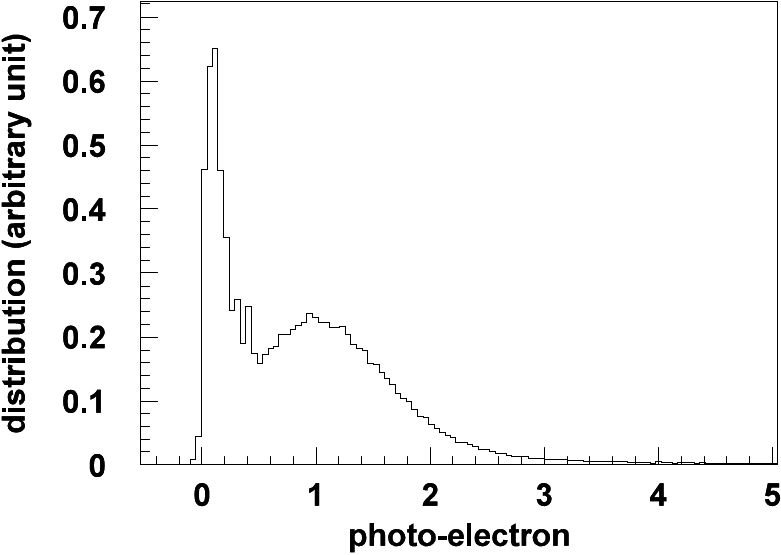}
       \caption{Single photoelectron distribution of a typical 20-inch PMT.}
       \label{pmt:pmt1pe}
     \end{center}
   \end{figure}

In the baseline design, we have selected the 20-inch PMT 
( R3600 ) used successfully in Super-Kamiokande as
the primary sensor candidate; this type of PMT
is already known to satisfy the requirements above. Moreover,
it been operated for more than 15 years in Super-Kamiokande and
thus not only the performance characteristics but also the long-term stability of 
this type of sensor are well understood.
The specifications of the 20-inch PMT is summarized in 
Table \ref{pmt:pmtspec}, while the single photoelectron distribution 
for this PMT is shown in Fig. \ref{pmt:pmt1pe}.

   \begin{table}
     \begin{center}
       \begin{tabular}{ll}
	 \hline \hline
	 Shape                        & Hemispherical \\
	 Photocathode area            & 50\,cm diameter \\
	 Window material              & Pyrex glass ($4\sim5$\,mm) \\
	 Photocathode material        & Bialkali (Sb-K-Cs) \\
	 Quantum efficiency           & 22\,\% at $\lambda=390$\,nm \\
	 Dynodes                      & 11\,stage Venetian blind type \\
	 Gain                         & 10$^7$ at $\sim2000$\,V \\
	 Dark current                 & 200\,nA at $10^7$ gain \\
	 Dark pulse rate              & 3\,kHz at $10^7$ gain \\
	 Cathode non-uniformity       & $<$ 10\,\% \\
	 Anode non-uniformity         & $<$ 40\,\% \\
	 Transit time                 & 90\,nsec at 10$^7$ gain \\
	 Transit time spread          & 2.2\,nsec (1\,$\sigma$) for 1 p.e. equivalent signals \\
	 Weight                       & 13\,kg \\
	 Pressure tolerance           & 6\,kg/cm$^2$\ water proof \\
	 \hline \hline
       \end{tabular}
       \caption{Specifications of 20-inch PMT.}
       \label{pmt:pmtspec}
     \end{center}
   \end{table}

Based on studies of the expected physics performance of Hyper-Kamiokande, 
it is necessary to cover 20\% of the detector wall with these PMTs. This means that in order 
to meet this physics-based requirement the total number of PMTs needed will be about 99,000.

In order to use R3600 in the Hyper-K detector, it is necessary
to equip each PMT with a protective case. Otherwise,
a collapse of one of the sensors could generate a shock wave
large enough to break the surrounding PMTs, potentially leading to a chain reaction
resulting in catastrophic damage to the detector.
In order to avoid this kind of severe accident, each PMT 
will be equipped with a protective, shock-absorbing case specifically designed to prevent  
a collapse of a single PMT from emitting such a fratricidal shock wave.

\begin{figure}
     \begin{center}
       \includegraphics[width=7cm]{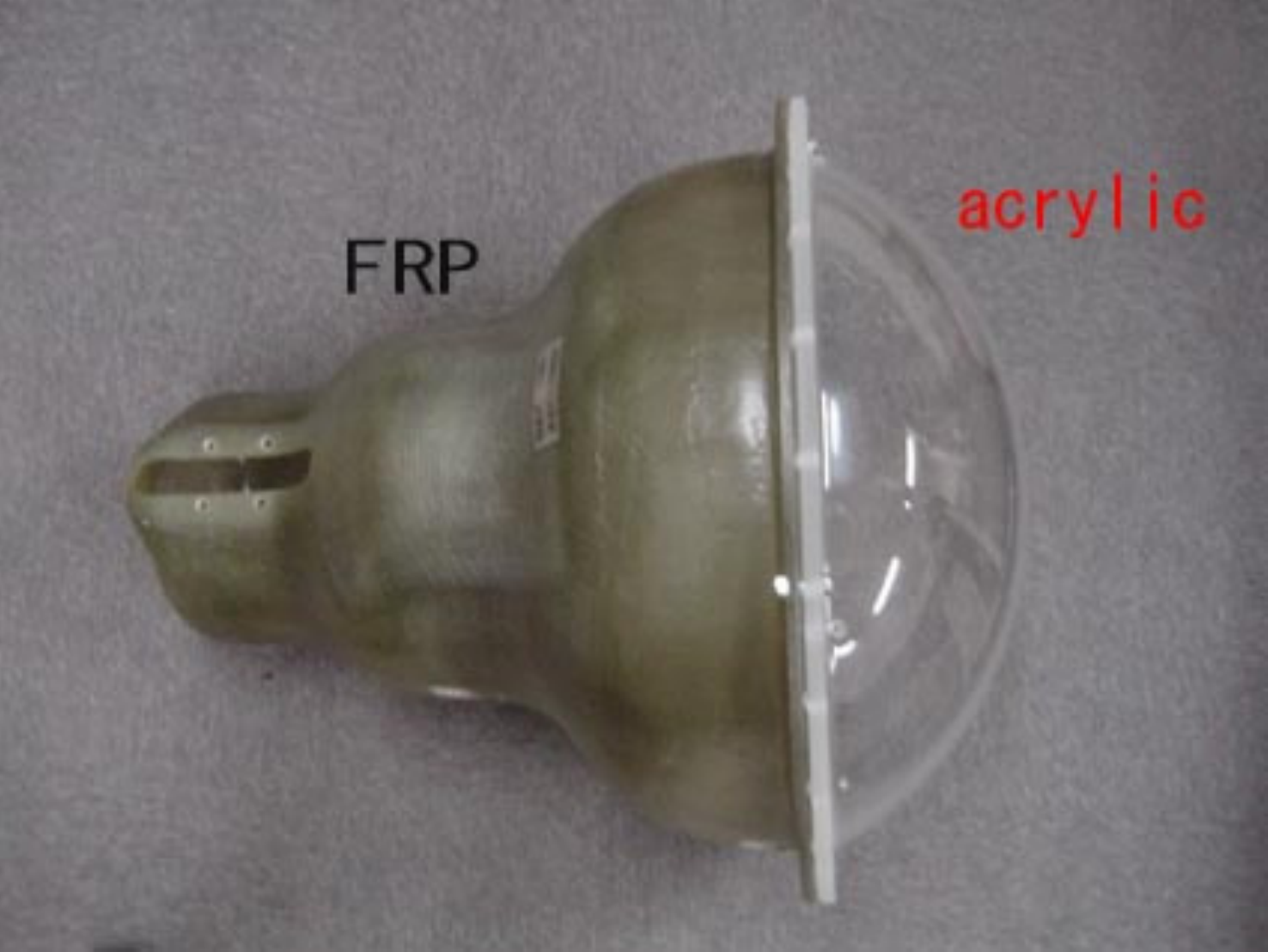}
       \caption{PMT case attached a 20-inch PMT.  These were added to all Super-K phototubes in 2002.}
       \label{fig:pmtcase}
     \end{center}
   \end{figure}

Currently, the PMTs in Super-K are all contained within 10 mm thick 
FRP cases in back and the 13 mm thick acrylic covers 
in front as shown in Fig. \ref{fig:pmtcase}.
This type of cover is known to satisfy the minimum requirements
of the Hyper-K detector.

\subsubsection{Possible alternative PMTs and other R\&D items}

If photosensors with higher efficiency are available, 
it would be possible to reduce the number or the size of 
the photosensors in Hyper-Kamiokande. 
There are several ongoing efforts to build high efficiency sensors. 
One approach is a small diameter PMT ( $\sim$ 10 inch ) 
with high quantum efficiency (QE).
Its QE is expected to be $\sim$ 40\%, which is almost double that of  
the R3600.
Another path would be to utilize a new type of sensor, called 
a Hybrid PMT. The Hybrid PMT is a sensor which utilize an avalanche 
photo diode (APD) instead of dynodes; it also
has higher single photon sensitivity and much higher timing
resolution.

Of course, it is necessary to increase the number of PMTs by 1.4 
if 10 inch PMTs with doubled QE are used. However, the relative smallness of
the sensor will be a great help in safe handling during the construction. 
Also, protective cases would presumably be less expensive compared
to the ones for R3600 for two reasons: smaller size means less case material, and a smaller PMT means considerably less stored energy to contain in case of breakage, so the cases could likely be thinner as well as smaller.

There have also been several attempts to improve the photon collection 
efficiency with special lens systems or mirrors attached to 
the existing sensors. It is well to study this kind of idea, but 
whether or not it can be made to work with this type of detector remains to be seen.
Often such optical approaches to increase light collection come with a price; for example, 
the loss of effective fiducial volume due to limited angles of acceptance, or decreased light  
arrival timing resolution.

The other important R\&D item is the design of the protective case. 
The current design used in Super-K is known to meet the requirements, 
but as a retrofitted design it was forced to satisfy several additional requirements 
only applicable to Super-Kamiokande.  In Super-K it was not possible
to change the mounting method of the PMTs, and the quantity 
of cases produced during each of two production runs 
was just 5,000 or so, not enough to benefit 
from economies of scale or automated processes. 
In the case of Hyper-Kamiokande it will be possible to design the frame mounts 
of the photosensors 
and the protective cases simultaneously, and it will be possible to 
simplify the design of the case as well as the mounting
methods. As for the choice of material used in the protective cases in Super-K,
there were restrictions due to the limited quantity required. 
Based on the Super-K experience, FRP is now known to contain non-negligible 
amounts radioactive substances. In Hyper-K we will produce at least 
20 times as many cases, and thus it should not be necessary to stick to FRP 
but instead to use a low-activity material like stainless steel or 
acrylic. The design of the PMT frame mounts and protective  
cases has been started in coordination with the design of
the detector structure.

\subsection{DAQ system \label{section:daq}}

The required specifications for the front-end electronics
are basically the same as the current Super-K modules, 
assuming that the Hyper-K PMT 
gain and the noise levels are also similar.
These specs are as summarized
in Table \ref{daq:frontend}.

   \begin{table}
     \begin{center}
       \begin{tabular}{ll}
	 \hline \hline
     Charge dynamic range         & $\sim$0.2 pC $\sim$ 2500 pC \\
	 Integral charge non-linearity& less than 1\% \\
	 Charge resolution            & 0.2 pC \\
	 Discriminator threshold (noise level) & less than 1mV \\
	 Timing resolution            & much better than 1ns @ 2pC \\
     Power consumption per channel& less than 1W/ch \\ 
     Data through rate            & 100kHz for 10 seconds \\
                                  & 10kHz as nominal dark rate \\
     Channel deadtime             & Smaller than $\sim$ 500ns \\
	 \hline \hline
       \end{tabular}
       \caption{Requirements for the front-end electronics.}
       \label{daq:frontend}
     \end{center}
   \end{table}

\begin{figure}
     \begin{center}
       \includegraphics[width=12cm]{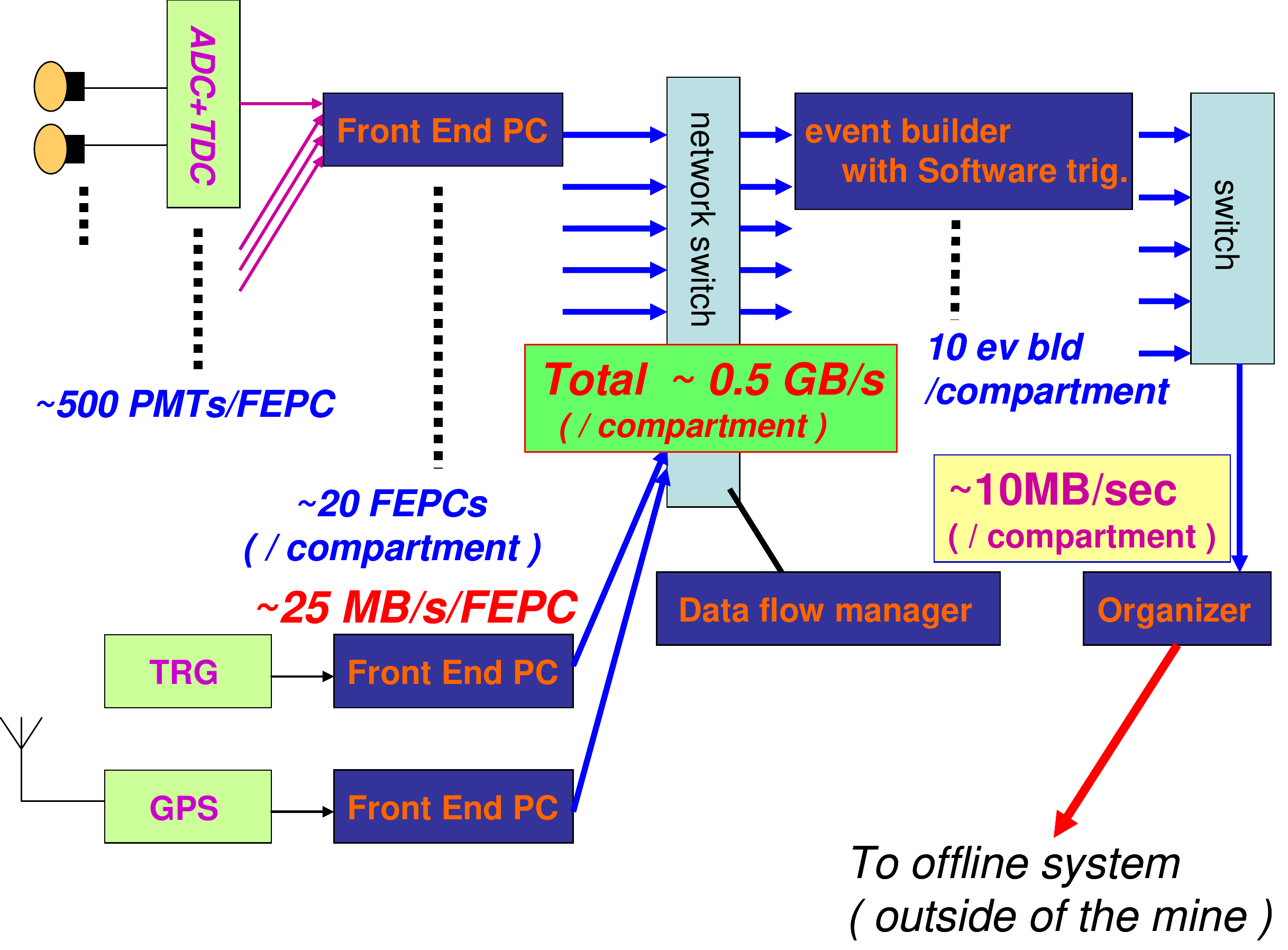}
       \caption{Schematic diagram of the data acquisition system.}
       \label{daq:daqschem}
     \end{center}
   \end{figure}

In Super-K we have developed a free-running, triggerless  (at the hardware level) DAQ 
system. The schematic diagram of this DAQ is shown 
in Fig. \ref{daq:daqschem}.

The front-end boards digitize any 
signals which exceed the discriminator threshold 
and send out the charge and timing information to 
the front-end computers. The front-end computers collect 
hit information from 30 electronics modules each, 
corresponding to $\sim$ 500 PMTs, sort the hit data 
in the order of timing and send it out to the merger 
computers.
The merger computers each collect all the hit information 
from a compartment and apply a software trigger to
remove noise hits. Then, the organizer computer
collects the event from the mergers, eliminates 
overlapping events, and send them to the offline computer
system.
This system reads out all the digitized hit information
from the PMTs and selects the events with software.
This system has been working without any problem in Super-K for
more than 2 years. 
Therefore, it is safe to say that there are no serious 
technical difficulties in preparing the DAQ system for 
this new detector with the currently available and well 
established tools at hand. 
The expected data rate from the entire Hyper-K detector is 
about 5~GB/s before applying the software trigger. 
After the first level software trigger, it will be 
reduced by 1/50 and another factor of five reduction could 
be achieved with Super-K-style intelligent vertex fitters applied
in the offline computer system. In the end, the data
rate written to disk is expected to be less than
100~MB/sec in total.

\subsubsection{R\&D items for the DAQ system}

A possible difficulty which could arise would involve physically 
running over 100,000 cables from the PMTs -- arranged as they must be 
throughout the immense tank -- into the DAQ system. 
The level of challenge will heavily depend on both the method selected to mount
the sensors to their supporting framework as well as how the cables are 
connected to the sensors, but it will surely not be a simple task and will contribute 
a significant amount of time to the total detector construction schedule. 
The great mass of cables inside the detector will also inevitably 
create some insensitive regions, particularly around the cable feed through 
leading out of the tank. Moreover, the length 
of the cables could be longer than 100~m, some 50\% longer than in Super-K. 
This may degrade the quality of the signal.

Therefore, it is worthwhile to consider putting 
the front-end electronics inside the detector itself. If it is possible
to place the electronics modules close to the photosensors
in the water then not only will the cable lengths be shorter, but also
the number of cables needing to pass through the detector surface will
be reduced.
We will need to consider the resulting temperature rise 
if we put electronics modules in the water. However, 
the current power consumption of the digitization module
and the high voltage supplies is a few watts per channel; this should  
be acceptable even if we use more than 100k sensors. 

There are still several R\&D items necessary to allow placing the electronics 
modules in the water. These include the fault-tolerance 
in the circuit, a water-proof case and its connectors 
or cable feed through, and a proper design of the heat 
exchanger from the circuit. The high voltage power supply
for the photosensor will also necessarily be placed in 
the same housing. There already exist possible 
solutions for each such item,  and thus what we have to do from an R\&D
standpoint is evaluate each implementation and select the proper 
ones considering both quality and cost. 

%% file: experimental-setup/calibration.tex
\subsection{Detector calibration\label{section:calibration}}

In Super-Kamiokande, various kind of calibrations have been performed.
A nitrogen+dye laser -- which emits pulses of light with a wavelength 
of 396~nanometers,  duration of $<4$~nanoseconds,  
and energy of 300 microjoules -- 
has been used to calibrate the timing response of PMTs as well as 
measure the charge linearity of PMTs + DAQ electronics. 
The laser light is feed to a diffuser ball in the Super-K tank
through an optical fiber with a large (200~$\mu$m) diameter core.

A nitrogen laser with 337~nm wavelength and laser diodes with
375~nm, 405~nm, 445~nm, and 473~nm wavelengths have been used
to measure water attenuation and scattering parameters.
Through optical fibers, these five laser light sources power eight light injectors which 
are mounted in various positions inside the Super-K tank. 
A computer-controlled optical switch directs the laser light to the various injectors, allowing the measurement of absorption and scattering parameters 
for light in the water with wavelength dependence taken into account. 

A xenon lamp which has stable( $\sim$5\% ) light intensity
has been used to calibrate PMT gain and to monitor
the position dependence of water quality.
UV light from the xenon lamp travels through an optical fiber with
very large (1000~$\mu$m) diameter core, and into a scintillator ball
sitting inside the Super-K tank.

An surplus medical grade electron linear accelerator (LINAC) system is 
permanently mounted near the top of the Super-K tank. It allows single 5$\sim$16 MeV 
electrons to be injected into various positions in the Super-K tank through an 
evacuated beam pipe and bending magnets which are positioned as needed to guide the beam.
At one electron every second this is likely the lowest flux particle accelerator in the world, 
but it has proven to be our most accurate low energy calibration source.  Through the use of a germanium counter the beam energy is known to within 20 keV.

A deuterium-tritium (DT) fusion generator which produces gamma and electrons via 
$^2$H + $^3$H $\rightarrow$ $^4$He + $n$, $^{16}$O($n$,$p$)$^{16}$N, 
$^{16}$N $\rightarrow \gamma$(6.1MeV), $e$(4.3MeV)  is able to be lowered 
into the Super-K tank when a gamma or electron source in this energy range is needed.
As it is much less labor intensive to operate than the LINAC, the DT generator is run about four
times as frequently -- a few times each year -- to ensure that the detector response is stable and its response uniform to low energy events.

A gamma ray source powered by neutrons emitted by 
from $^{252}$Cf produces $\sim$9~MeV gamma rays from nickel via
Ni$(n,\gamma)$Ni.  Like the DT generator, it is able to be lowered directly into the Super-K tank.
All of these low energy  gamma ray and electron sources are used to calibrate
vertex and energy determination, primarily for use in solar and supernova neutrino
analysis.

Cosmic ray muons which either pass all the way through or stop somewhere in the tank, as 
well as the decay electrons produced by these stopping muons are used to calibrate
energy reconstruction for more energetic events.  The cosmic ray muons are also used as a natural source of Cherenkov light for monitoring water transparency.

The laser and xenon light sources (and of course the cosmic ray muons) are 
used even during normal data taking by keeping the diffuser ball, 
scintillator ball, and light injectors in the tank and scheduling light emission 
times of each light source
at a low but steady rate($\sim$1~Hz in total) to minimize dead time of the observation.

The LINAC and radioactive calibration sources cannot be used during
normal data taking. Therefore, for calibration work with these
sources; for high rate data taking of laser and xenon; and to install,
uninstall, or move calibration sources in the tank, we need to stop
normal data taking, leading to observational dead time.  This is purposely kept
to a minimum, primarily to reduce the risk of missing a once-in-lifetime burst of neutrinos from a galactic supernova, but also to avoid wasting expensive accelerator beam time whenever a 
long baseline experiment is running. Calibrations can often be planned around scheduled 
accelerator downtime, but the risk of missing a supernova is obviously 
always present when the detector is turned off.

Almost all of these calibration sources will prove indispensable for Hyper-K as well 
to maintain the best performance of the new detector.
Since in the current baseline design of Hyper-K the two tanks are
separated into 10 individual compartments, it would require -- roughly speaking -- 
10 times more time and manpower to perform the same calibration work.
In addition, due to its egg-shaped cross section, the dome above the Hyper-K tank 
will be quite a bit narrower compared to the maximum tank width.  This may make calibrations 
near the wall of PMTs challenging, as will the bowed walls of the detector. 

Therefore, calibration methods should be carefully discussed during the 
detector design, and the detector should have dedicated, automated systems 
for placing various calibration sources at desired positions within the tank.
Due to the greater size and more complex geometry of Hyper-K, the 
positioning of calibration sources may need to be controlled by 
more advanced methods than the vertical wire drop lines used in Super-K.
Convenient methods like fixed rails, wire guides, or even submarine robots
installed during detector construction should be considered.
As for the large, immovable instruments like the LINAC, we may need to
consider installing a small network of simple beamlines to distribute 
particles to each compartment.

%% file: physics-cpv/cpv.tex
\section{Physics potential}\label{section:physics}

\subsection{Accelerator based neutrinos \label{sec:cp}}
\newcommand{\numu}{\ensuremath{\nu_{\mu}}}                   
\newcommand{\numubar}{\ensuremath{\overline{\nu}_{\mu}}}                   
\newcommand{\nue}{\ensuremath{\nu_{e}}}                   
\newcommand{\nuebar}{\ensuremath{\overline{\nu}_{e}}}                   
\newcommand{\enurec}{\ensuremath{E_\nu^\mathrm{rec}}}     
\newcommand{\deltacp}{\ensuremath{\delta}}

\subsubsection{$CP$ asymmetry measurement in a long baseline experiment}
If a finite value of $\theta_{13}$ is discovered by the ongoing and near-future accelerator and/or reactor neutrino experiments~\cite{Itow:2001ee,Ayres:2004js,Ardellier:2006mn,Ahn:2010vy,Guo:2007ug},
the next crucial step in neutrino physics will be the search for $CP$ asymmetry in the lepton sector.
A comparison of muon-type to electron-type transition probabilities between neutrinos and anti-neutrinos is 
one of the most promising methods to observe the lepton $CP$ asymmetry.
Recent indication of a nonzero, rather large value of $\theta_{13}$~\cite{Abe:2011sj} makes this exciting possibility more realistic with near-future experiments such as Hyper-Kamiokande.

In the framework of the standard three flavor mixing, 
the oscillation probability is written using the parameters of the MNS matrix (see Sec.~\ref{section:intro-neutrino}),
to the first order of the matter effect, as~\cite{Richter:2000pu}:
\begin{eqnarray}
P(\numu \to \nue) & = & 4 C_{13}^2S_{13}^2S_{23}^2 \cdot \sin^2\Delta_{31}  \nonumber \\
& & +8 C_{13}^2S_{12}S_{13}S_{23} (C_{12}C_{23}\cos\deltacp - S_{12}S_{13}S_{23})\cdot \cos\Delta_{32} \cdot \sin\Delta_{31}\cdot \sin\Delta_{21} \nonumber \\
& & -8 C_{13}^2C_{12}C_{23}S_{12}S_{13}S_{23}\sin\deltacp \cdot \sin\Delta_{32} \cdot \sin\Delta_{31}\cdot \sin\Delta_{21} \nonumber \\
& & +4S_{12}^2C_{13}^2(C_{12}^2C_{23}^2 + S_{12}^2S_{23}^2S_{13}^2-2C_{12}C_{23}S_{12}S_{23}S_{13}\cos\deltacp)\cdot \sin^2\Delta_{21} \nonumber \\
& & -8C_{13}^2S_{13}^2S_{23}^2\cdot \frac{aL}{4E_\nu} (1-2S_{13}^2)\cdot \cos\Delta_{32}\cdot \sin\Delta_{31} \nonumber \\
& & +8 C_{13}^2S_{13}^2S_{23}^2 \frac{a}{\Delta m^2_{31}}(1-2S_{13}^2)\cdot\sin^2\Delta_{31}, \label{Eq:cpv-oscprob}
\end{eqnarray}
\noindent where $C_{ij}$, $S_{ij}$, $\Delta_{ij}$ are $\cos\theta_{ij}$, $\sin\theta_{ij}$, $\Delta m^2_{ij}\, L/4E_\nu$, respectively, 
and $a \mathrm{[eV^2]}= 7.56\times 10^{-5} \times \rho \mathrm{[g/cm^3]} \times E_\nu[\mathrm{GeV}] $.
The parameter $\deltacp$ is the complex phase that violates $CP$ symmetry.
The corresponding probability for $\numubar \to \nuebar$ transition is obtained by replacing $\deltacp \rightarrow -\deltacp$
and $a \rightarrow -a$.
The third term, containing $\sin\deltacp$, is the $CP$ violating term which flips the sign between $\nu$ and $\bar{\nu}$ and thus introduces $CP$ asymmetry if $\sin\deltacp$ is non-zero.
The last two terms are due to the \textit{matter effect}; caused by coherent forward scattering in matter, they produce a fake (i.e., not $CP$-related) asymmetry between 
neutrinos and anti-neutrinos.
As seen from the definition of $a$, the amount of asymmetry due to the matter effect is proportional to the neutrino energy at a fixed value of $L/E_\nu$.

\begin{figure}[tbp]
\includegraphics[width=0.45\textwidth]{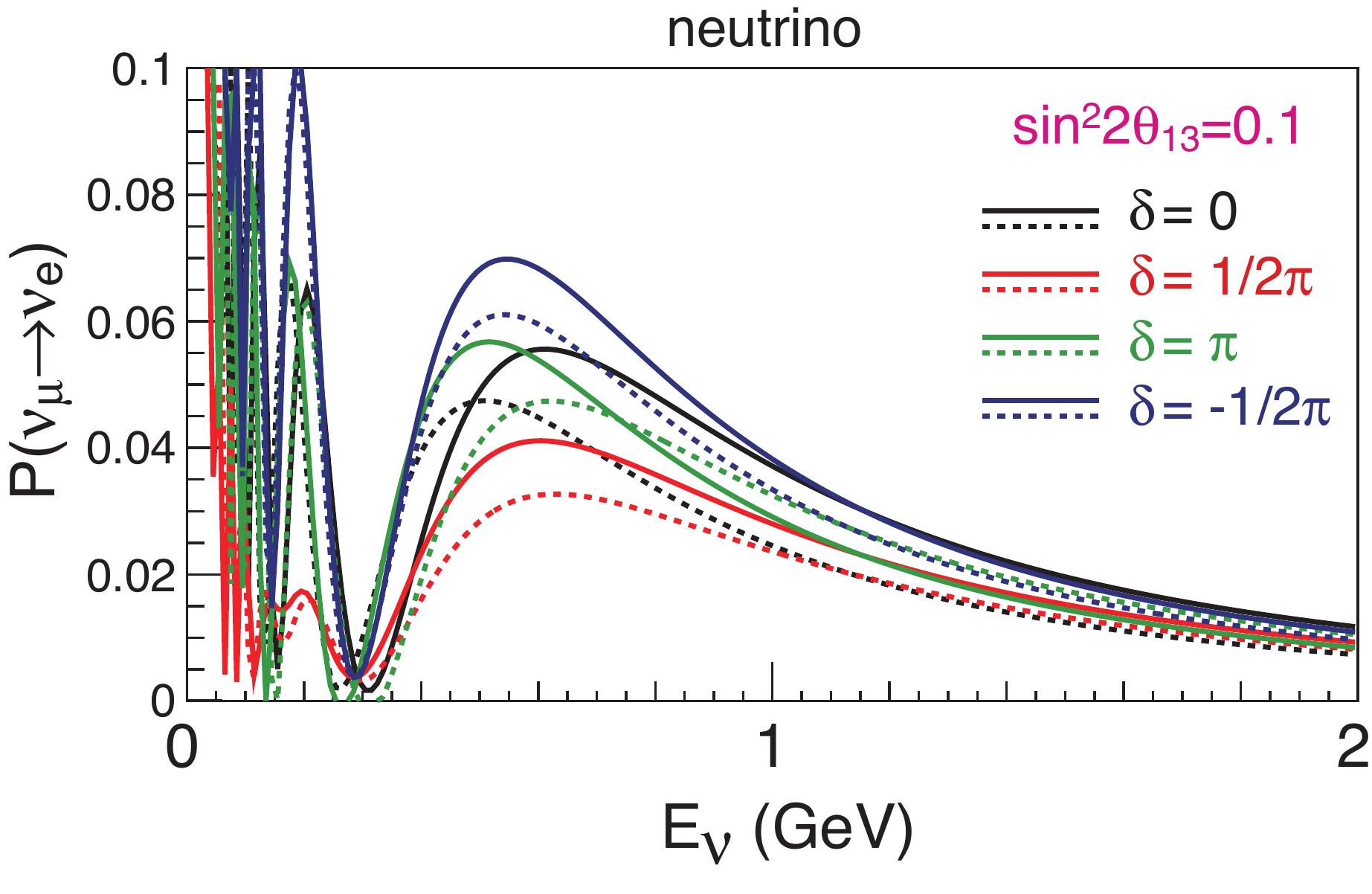}
\includegraphics[width=0.45\textwidth]{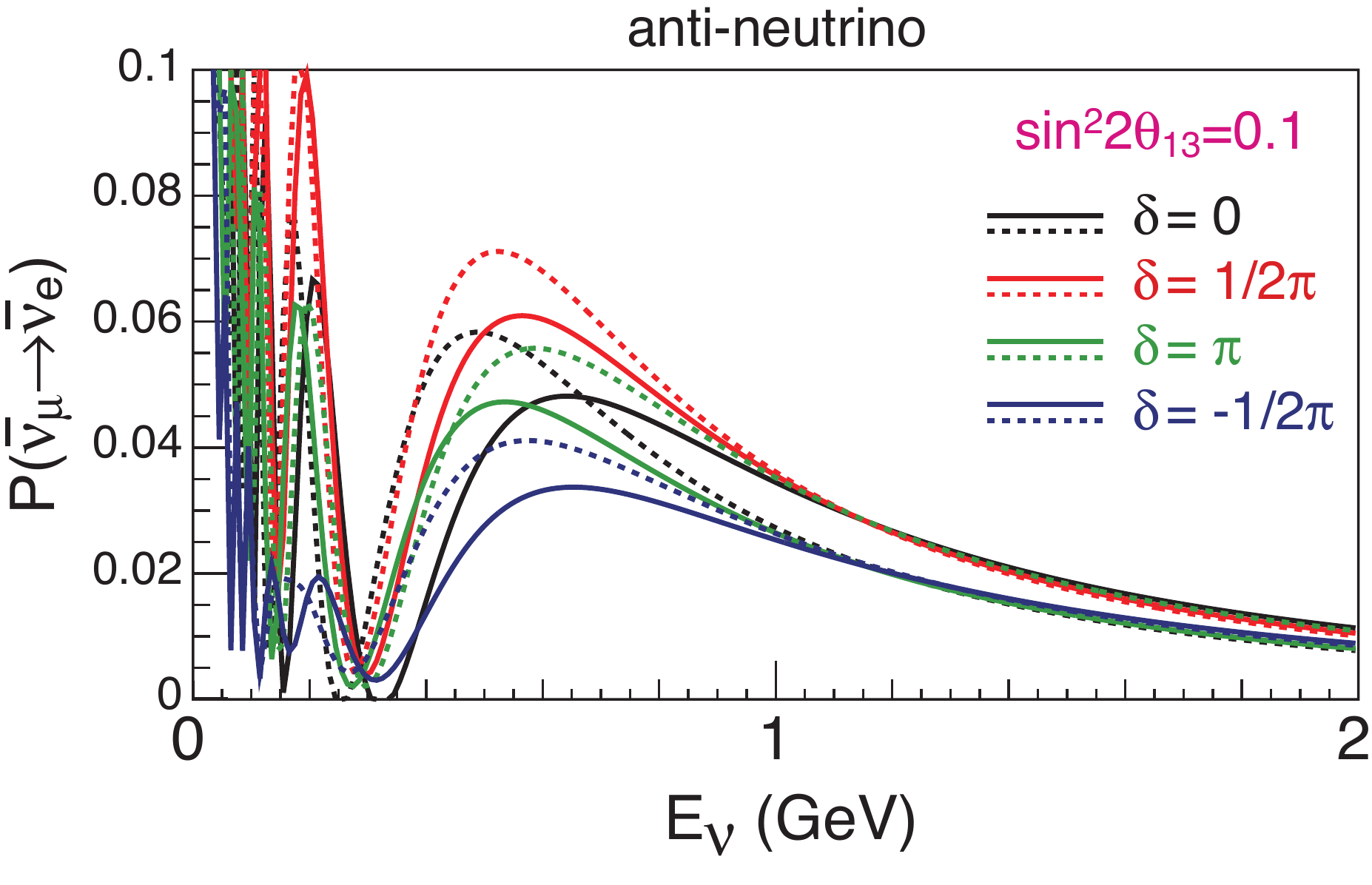}
\caption{Oscillation probabilities as a function of the neutrino energy for $\numu \to \nue$ (left) and $\numubar \to \nuebar$ (right) transitions with L=295~km and $\sin^22\theta_{13}=0.1$. 
Black, red, green, and blue lines correspond to $\deltacp = 0, \frac{1}{2}\pi, \pi$, and $-\frac{1}{2}\pi$, respectively.
Other parameters are listed in Table~\ref{tab:cp-params}.
Solid (dashed) line represents the case for a normal (inverted) mass hierarchy.
\label{fig:cp-oscpob}}
\end{figure}

\begin{figure}[tbp]
\includegraphics[width=0.45\textwidth]{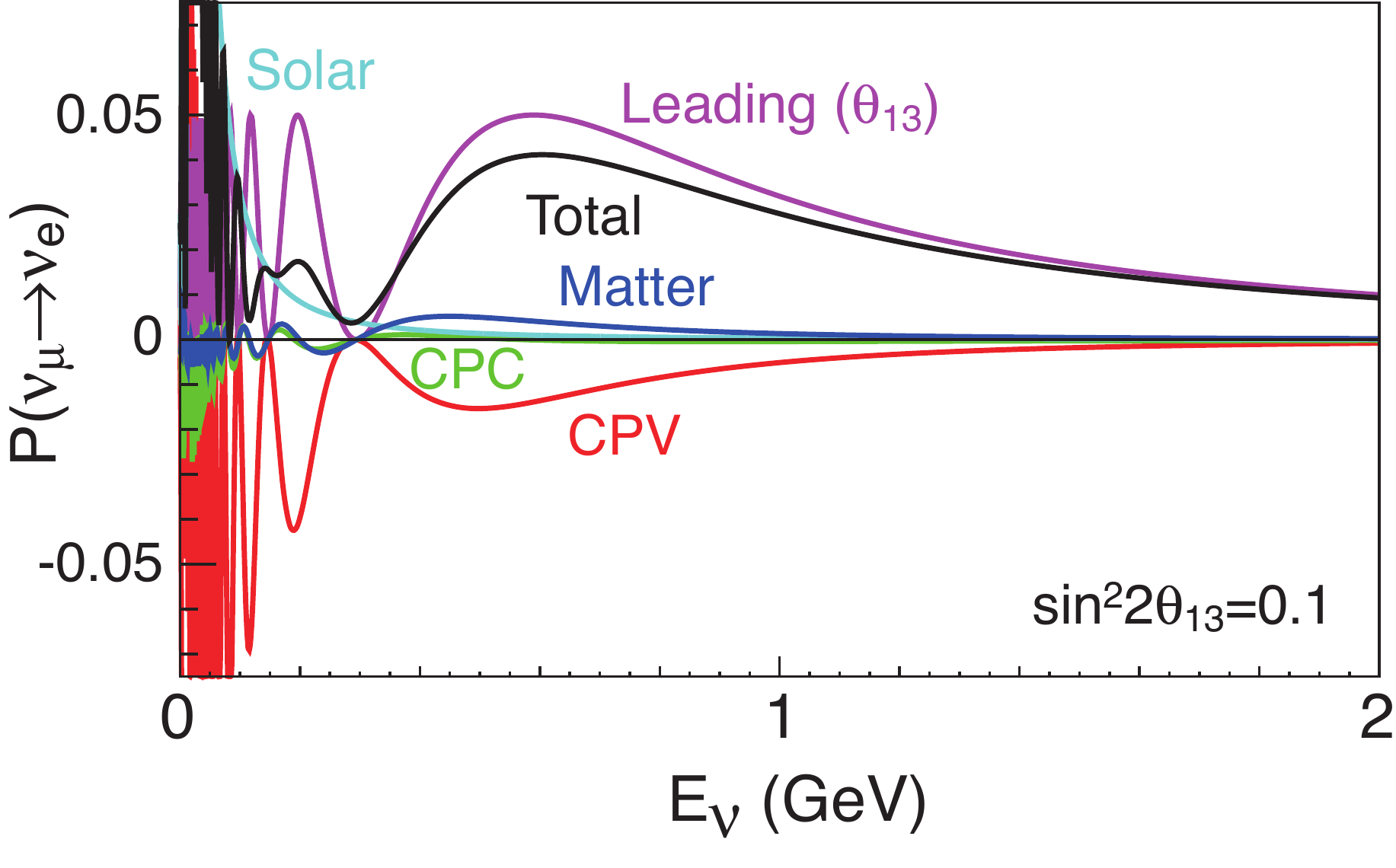}
\includegraphics[width=0.45\textwidth]{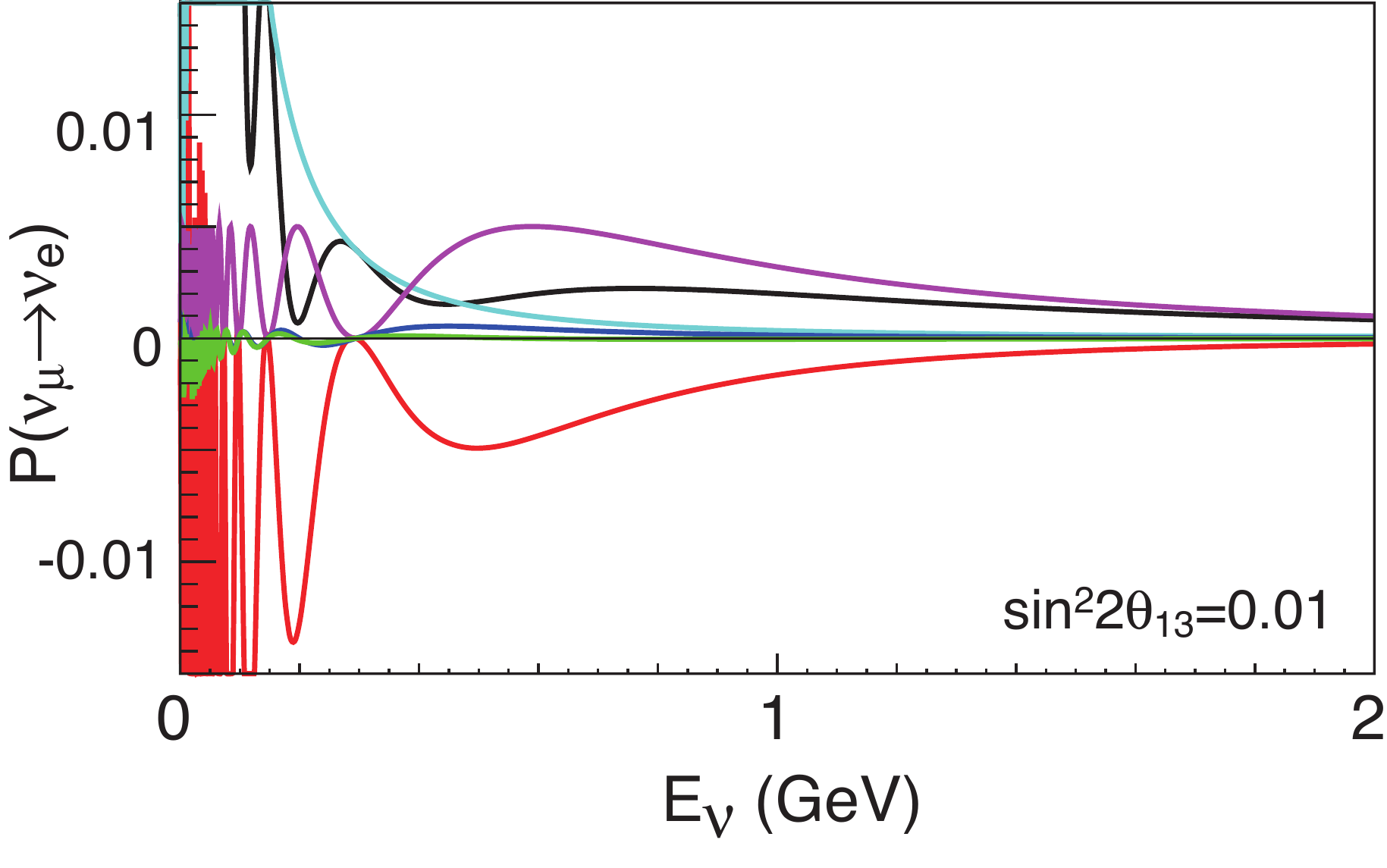}
\caption{Oscillation probability of $\numu \to \nue$  as a function of the neutrino energy with a baseline of 295~km. Left: $\sin^22\theta_{13}=0.1$, right: $\sin^22\theta_{13}=0.01$. 
$\deltacp = \frac{1}{2}\pi$ and normal hierarchy is assumed.
Contribution from each term of the oscillation probability formula is shown separately.
\label{fig:cp-oscpob-bd}}
\end{figure}

\begin{table}[htdp]
\caption{Parameters other than $\theta_{13}$ and $\deltacp$ assumed in this section.
}
\begin{center}
\begin{tabular}{cc} \hline \hline
Name & Value \\ \hline \hline
L & 295~km \\ \hline
$\Delta m^2_{21}$ & 7.6$\times 10^{-5}$~eV$^2$ \\ \hline
$|\Delta m^2_{32}|$ & 2.4$\times 10^{-3}$~eV$^2$ \\ \hline
$\sin^2\theta_{12}$ & 0.31 \\ \hline
$\sin^2\theta_{23}$ & 0.5 \\ \hline
Density of the earth ($\rho$) & 2.6~g/cm$^3$\\ \hline \hline
\end{tabular}
\end{center}
\label{tab:cp-params}
\end{table}%

Figure~\ref{fig:cp-oscpob} shows the $\numu \to \nue$ and $\numubar \to \nuebar$ oscillation probabilities as a function of the true neutrino energy for a baseline of 295~km.
The parameters other than $\theta_{13}$ and $\deltacp$ assumed in this section are summarized in Table~\ref{tab:cp-params}.
The value of $\sin^2\theta_{23}$ is set to the maximal mixing, as suggested by the current best fit values of atmospheric and accelerator-based experiments ~\cite{Ashie:2005ik,Ashie:2004mr,Hosaka:2006zd,Ahn:2006zza,Adamson:2011ig}.
The density of earth is based on \cite{Hagiwara:2011kw}.
Other parameters are based on the global fit~\cite{Fogli:2011qn}.
The cases for $\deltacp = 0, \frac{1}{2}\pi, \pi$, and $-\frac{1}{2}\pi$, are overlaid in Fig.~\ref{fig:cp-oscpob}.
Also shown are the case of normal mass hierarchy ($\Delta m^2_{32}>0$) with solid lines and inverted mass hierarchy ($\Delta m^2_{32}<0$) with dashed lines.
The oscillation probabilities depend on the value of $\deltacp$, and by comparing the neutrinos and anti-neutrinos,
one can see the effect of $CP$ violation.
There are sets of different mass hierarchy and values of $\deltacp$ which give similar oscillation probabilities.
This is known as the degeneracy due to unknown mass hierarchy and may introduce a fake solution 
if we do not know the true mass hierarchy.
The effect of this degeneracy on the measurement of $CP$ asymmetry will be discussed later.

Figure~\ref{fig:cp-oscpob-bd} shows the contribution from each term of the $\numu \to \nue$ oscillation probability formula, Eq.(\ref{Eq:cpv-oscprob}).
For $\sin^22\theta_{13}=0.1 (0.01)$ and $\deltacp = \pi/2$ with normal hierarchy, the contribution from the leading term, the $CP$ violating ($\sin\delta$) term, and the matter term to the $\numu \to \nue$ oscillation probability at 0.6~GeV neutrino energy are 0.05, $-0.014$, and 0.004 (0.005, $-0.004$, and 0.0004), respectively.
The fraction of the contribution from the $CP$ violating term to the total oscillation probability depends on the value of $\theta_{13}$.
For a relatively large value of $\theta_{13}$, the first term of Eq.~(\ref{Eq:cpv-oscprob}) --- which is proportional to $\sin^2\theta_{13}$ --- is dominant.
Because the $CP$ violating term is proportional to $\sin\theta_{13}$,
 for smaller $\theta_{13}$ the asymmetry ${\cal A}= \{ P(\numu \to \nue) - P(\numubar \to \nuebar) \} / \{P(\numu \to \nue) + P(\numubar \to \nuebar)\}$ becomes larger.
However, the number of $\nu_\mu \to \nu_e$ signal events decreases as $\theta_{13}$ gets smaller, resulting in larger statistical uncertainty.
As a result, the sensitivity to the $CP$ asymmetry has only modest dependence on $\theta_{13}$,
if it is not too small to observe the signal.

Due to the relatively short baseline and thus lower neutrino energy at the oscillation maximum, 
the contribution of the matter effect is smaller for the J-PARC to Hyper-Kamiokande experiment
compared to other proposed experiments like LBNE in the United States~\cite{LBNE}.
Because the matter effect terms have $\sin^22\theta_{13}$ dependence, 
its significance relative to the $\sin\delta$ term ($\propto \sin\theta_{13}$) increases for a larger value of $\theta_{13}$.
For $\sin^22\theta_{13} \sim 0.1$, the matter effect has the same order of contribution to 
the oscillation probability as the $CP$ violating term even with a baseline of 295~km,
and thus the J-PARC to Hyper-Kamiokande experiment would have sensitivity to the mass hierarchy.

\subsubsection{J-PARC to Hyper-Kamiokande long baseline experiment}
\begin{figure}[tbp]
\includegraphics[width=0.75\textwidth]{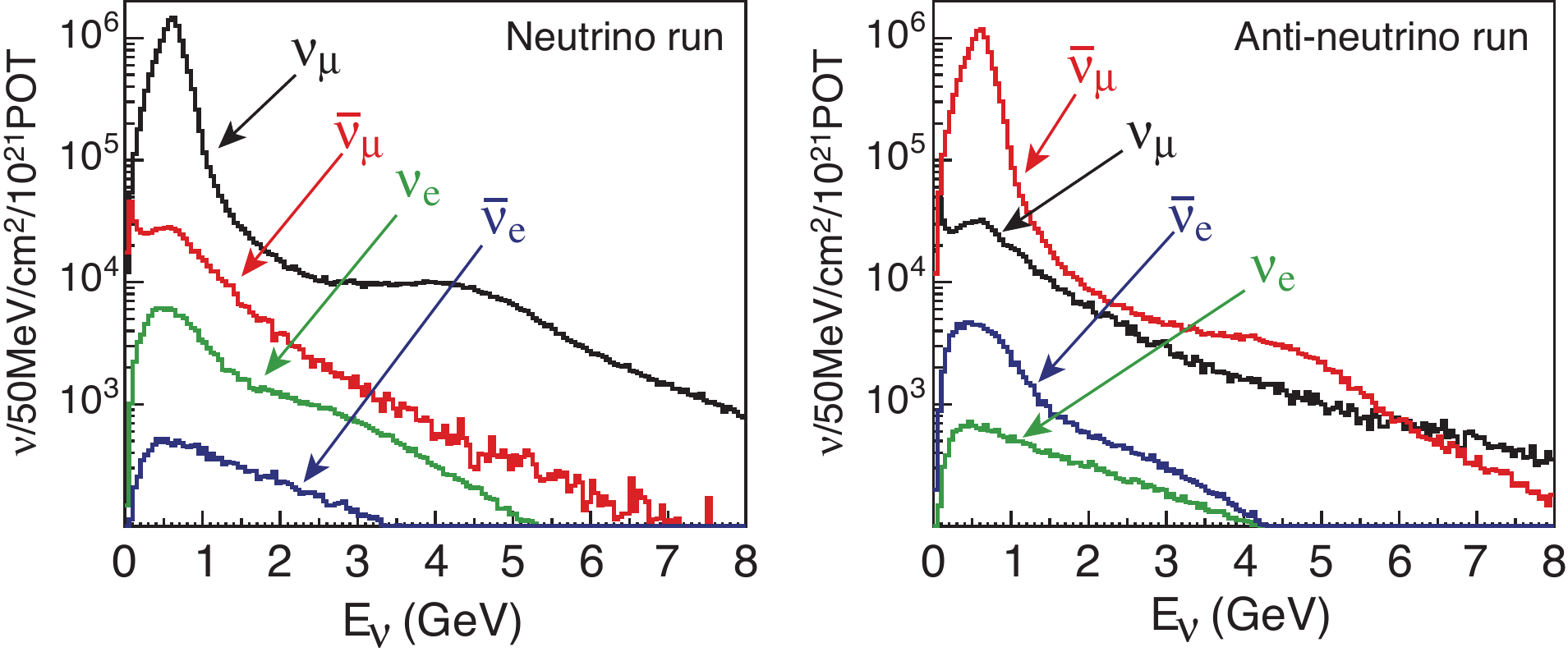}
\caption{Expected neutrino flux at Hyper-K. Left: neutrino mode, right: anti-neutrino mode. $\numu, \numubar, \nue, \nuebar$ components are shown separately.\label{fig:cp-nuflux} }
\end{figure}

The J-PARC neutrino beamline, which currently provides a neutrino beam to the T2K experiment,
is designed to realize a common off-axis angle to Super-K and the candidate Hyper-K site in the Tochibora mine~\cite{Nakamura:2003hk}.
In this study, the off-axis angle is set to 2.5$^\circ$, the same as the current T2K configuration.
A beam power of 1.66~MW is assumed as the nominal case based on the KEK roadmap~\cite{KEKRoadmap}, while feasibility with lower beam intensity is also explored.
The neutrino beam Monte Carlo simulation is used to estimate the neutrino flux and energy spectrum.
In the beam simulation, the hadron production cross section is tuned based on the measurements of pion and kaon production cross sections by the NA61 collaboration~\cite{Abgrall:2011ae, NA61:kaon}.
A proton beam energy of 30~GeV and a magnetic horn current of 320~kA are assumed.
Figure~\ref{fig:cp-nuflux} shows the expected neutrino flux at Hyper-K for neutrino and anti-neutrino mode running.
Thanks to the off-axis method, the spectrum has a narrow peak around the energy where oscillation probability is the maximum, with small high energy tail.
Contamination of $\nue(\nuebar)$ in the beam is well below 1\% at the peak for both cases.

Interactions of neutrinos in the Hyper-K detector is simulated with the NEUT program library~\cite{hayato:neut,Mitsuka:2007zz,Mitsuka:2008zz}, which is used in both the Super-K and T2K experiments.
The response of the detector is simulated using the Super-K full Monte Carlo simulation 
based on the GEANT3 package~\cite{Brun:1994zzo}.
The simulation is based on the SK-IV configuration with the upgraded electronics and DAQ system, 
while the number of PMTs is reduced to about half to simulate 20\% photocathode coverage of Hyper-K.
Events are reconstructed with the Super-K reconstruction software.
As described in Sec.~\ref{section:site}, each Hyper-K tank is divided into five optically separated compartments, 
each one of which has about twice the volume of Super-K.
Thus, the Super-K full simulation gives a realistic estimate of the Hyper-K performance.

The electron (anti-)neutrinos from $\numu \to \nue$ oscillation are identified via their charged current interaction.
The charged current quasielastic (CCQE) interactions, $\nue + n \to e^- + p$ and $\nuebar + p \to e^+ + n$,
have the largest cross section for $E_\nu<1$~GeV.
Because the momentum of recoil protons are typically below Cherenkov threshold and neutrons do not emit Cherenkov light,
they are identified as events with single electron-like ring in Hyper-K.
Thanks to the two-body kinematics of the CCQE interactions, we can reconstruct the incident neutrino energy using only the energy and scattering angle of the outgoing electron(positron).
Thus, we select events with a single electron-like ring as $\nu_e$ candidate signal events.
The background is mainly intrinsic $\nu_e$ contamination in the original neutrino beam,
and neutral current interactions with $\pi^0$ misidentified as an electron due to mis-reconstruction of one $\gamma$ from $\pi^0 \to \gamma \gamma$.

The criteria to select $\nu_e$ candidate events are based on those developed for and established with the Super-K and T2K experiments:
\begin{enumerate}
\item The event is fully contained (FC) inside the inner detector.
\item Reconstructed vertex is inside the fiducial volume (FV).
\item Visible energy ($E_\mathrm{vis}$) is greater than 100 MeV.
\item Number of reconstructed rings is one.
\item The reconstructed ring is identified as electron-like ($e$-like).
\item There is no decay electron associated to the event.
\item The invariant mass ($M_\mathrm{inv}$) of the reconstructed ring and a second ring force-found with a special $\pi^0$ fitter is less than 100~MeV/$c^2$. This selection is imposed in order to reduce the background from mis-reconstructed $\pi^0$.
\item The reconstructed energy ($E_\nu^\mathrm{rec}$) is less than 2~GeV.
\end{enumerate}
Assuming a charged current quasielastic interaction, the neutrino energy  ($E_\nu ^{\rm rec}$) is reconstructed from the electron energy ($E_e$) and the angle between the neutrino beam and the electron direction ($\theta_e$) as
\begin{eqnarray}
E_\nu ^{\rm rec}=\frac {2(m_n-V) E_e +m_p^2 - (m_n-V)^2 - m_e^2} {2(m_n-V-E_e+p_e\cos\theta_e)},
\label{eq:Enurec}
\end{eqnarray}
where $m_n, m_p, m_e$ are the mass of neutron, proton, and electron, respectively, $p_e$ is the electron momentum, and $V$ is the nuclear potential energy (27~MeV).
Here, rather than selecting a narrow signal region by cutting on the reconstructed energy,
the reconstructed neutrino energy spectrum over a wider range is used to obtain maximum information on signal and background.
Thus, the selection criterion regarding reconstructed neutrino energy is much looser than the past studies.
If we require 0.1~GeV$<E_\nu ^{\rm rec}<1.25$~GeV, the rejection efficiency for `NC1$\pi^0$' events,
defined as neutral current events having no particle with momentum above the Cherenkov threshold other than a single $\pi^0$, is 95\%
while the efficiency for $\numu \to \nue$ CC signal events is 64\%.
If necessary, a tighter selection can be applied by requiring $M_\mathrm{inv}$ 
to be farther away from the $\pi^0$ mass and
an additional requirement on the difference of likelihoods between the double and single ring assumptions ($\Delta {\cal L}$) as described in~\cite{Itow:2001ee}.
When $M_\mathrm{inv}<70$~MeV/$c^2$,  $\Delta {\cal L}<20$ and 0.1~GeV$<E_\nu ^{\rm rec}<1.25$~GeV are required, the NC1$\pi^0$ background rejection efficiency is 97.6\% and the signal efficency is 50\%.

The number of events after each selection step is shown in Tables~\ref{Tab:cp-selection-nu} and \ref{Tab:cp-selection-anti}  for each signal and background component.
Unless otherwise stated, in this section running times of 1.5 years for neutrino mode and 3.5 years for anti-neutrino mode (five years in total) are assumed, with \textit{one year} of running time corresponding to $10^7$~sec.
For the signal, $\sin^22\theta_{13}=0.1$ and $\delta=0$ are assumed in these tables.

\begin{table}[tbp]
\caption{The number of events after each selection step for neutrino mode. $\sin^22\theta_{13}=0.1$, $\deltacp=0$ is assumed.}
\begin{center}
\begin{tabular}{c||c|c|c|c|c||c|c} 
$\nu$ mode (1.5 years)				&$\numu$ CC	&$\numubar$ CC	&$\nue$ CC	&$\nuebar$ CC	&	 NC		&$\numu \to \nue$ CC	& $\numubar \to \nuebar$ CC		\\ \hline \hline
Interaction in FV 					&	64843  	&	2859  		&	4548  	&	368 		&	69898 	&	6024 	&	81  	\\
FCFV \& E$_\mathrm{vis}$$>$100~MeV	&	46106 	&	2030  		&	3742  	&	297 		&	17437 	&	5858   	&	79 	\\
1-ring       						&	24506  	&	1436 		& 	2054  	&	180 		&	4201 	&	4972   	&	66 	\\
$e$-like       						&	711 	&	21			&	1993  	&	175 		&	3211 	&	4893   	&	65 	\\
No decay-e   						&	145 	&	4  			&	1639  	&	165 		&	2795 	&	4435   	&	64 	\\
$M_\mathrm{inv}<$100~MeV/$c^2$  	&	45	 	&	1  			&	1173  	&	100 		&	811 	&	3966   	&	54 	\\
$E_\nu^\mathrm{rec}<2$~GeV   		&	38 		&	1 			&	917 	&	57 			&	718 	&	3940   	&	51 	\\ \hline
Efficiency(\%)						&	0.06 	&	0.03 		& 	20.2  	& 	15.4  		&	1.03 	&	65.4  	&	62.6 	\\ 
\end{tabular}
\end{center}
\label{Tab:cp-selection-nu}
\end{table}

\begin{table}[tbp]
\caption{The number of events after each selection step for anti-neutrino mode. $\sin^22\theta_{13}=0.1$, $\deltacp=0$ is assumed.}
\begin{center}
\begin{tabular}{c||c|c|c|c|c||c|c} 
$\overline{\nu}$ mode (3.5 years)	&$\numu$ CC	&$\numubar$ CC	&$\nue$ CC	&$\nuebar$ CC	&	 NC		&$\numu \to \nue$ CC	& $\numubar \to \nuebar$ CC		\\ \hline \hline
Interaction in FV    				&	34446  	&	30121  		&	2905 	&	2335 		&	58569   &	817  	&	2983 	\\
FCFV \& E$_\mathrm{vis}$$>$100~MeV	&	24981 	&	20960  		&	2470 	&	1824 		&	17677 	&	798  	&	2907 	\\
1-ring       						&	11830 	&	15925  		&	1168  	&	1225 		&	4433 	&	604 	&	2577 	\\
$e$-like       						&	401 	&	210  		&	1131 	&	1197 		&	3412 	&	596 	&	2539 	\\
No decay-e   						&	88 		&	30  		&	893 	&	1148 		&	3000 	&	520 	&	2526 	\\
$M_\mathrm{inv}<$100~MeV/$c^2$     	&	20 		&	11  		&	564 	&	803 		&	834 	&	439 	&	2187 	\\
$E_\nu^\mathrm{rec}<2$~GeV 			&	15 		&	10 			&	374  	&	598 		&	750 	&	421 	&	2168 	\\ \hline
Efficiency(\%)						&	0.04 	&	0.03 		&	12.86 	&	25.63 		&	1.28 	&	51.6 	&	72.7  \\  

\end{tabular}
\end{center}
\label{Tab:cp-selection-anti}
\end{table}

\begin{figure}[tbp]
\includegraphics[width=0.8\textwidth]{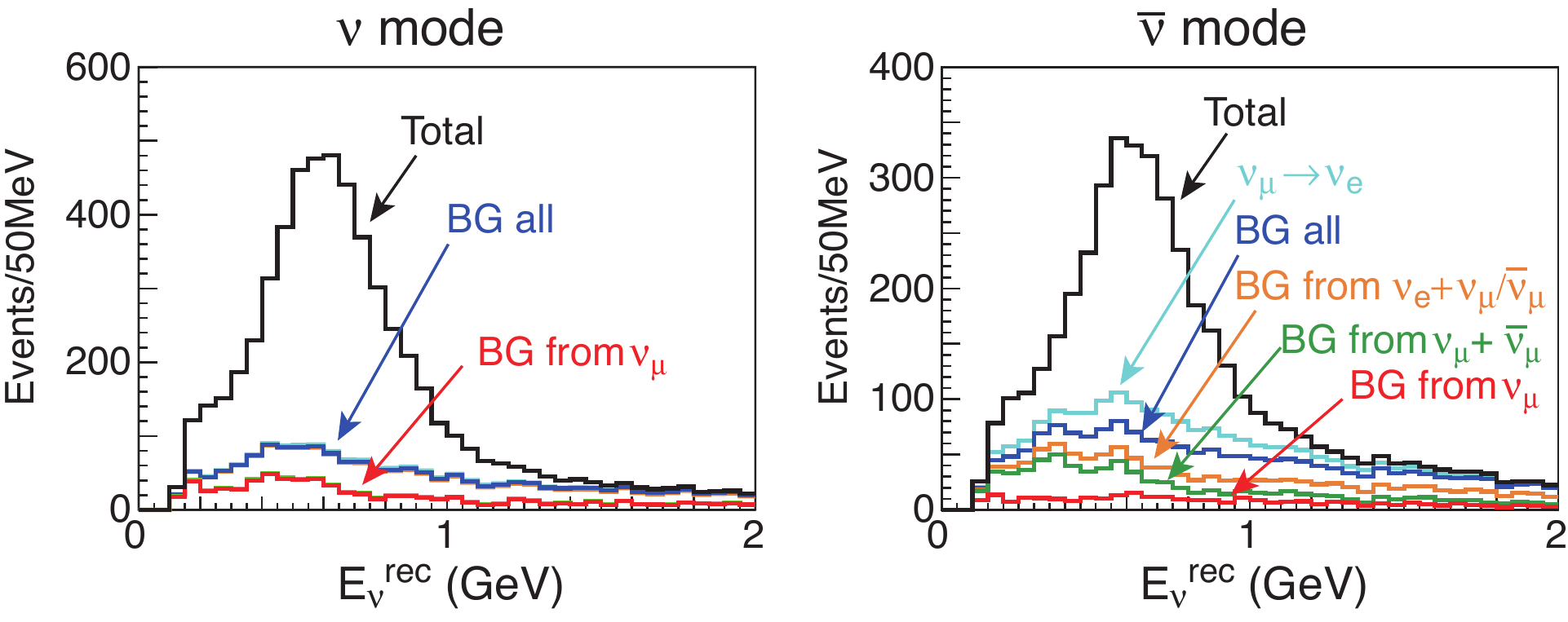}
\caption{Reconstructed neutrino energy distribution of the $\nue$ candidate events. $\sin^22\theta_{13}=0.1$, $\delta=0$ and normal hierarchy.
\label{fig:cp-enurec}}
\end{figure}

\begin{figure}[tbp]
\includegraphics[width=0.75\textwidth]{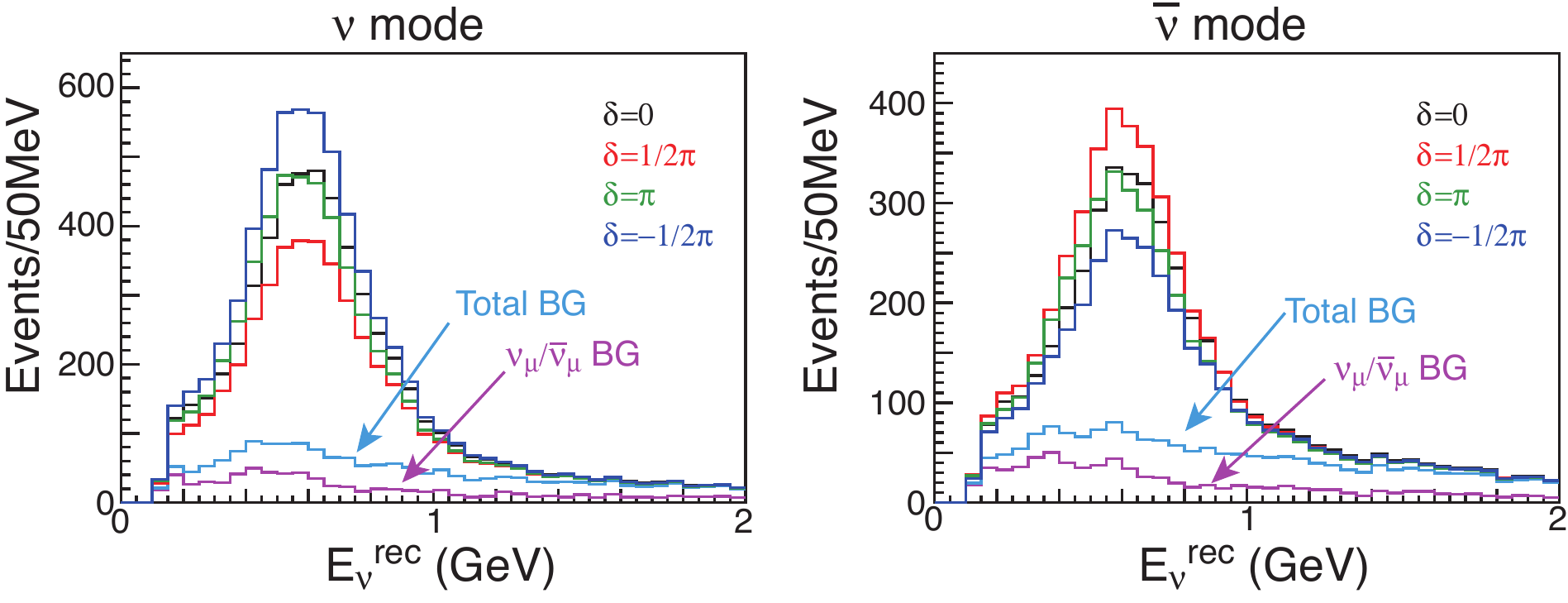}
\includegraphics[width=0.75\textwidth]{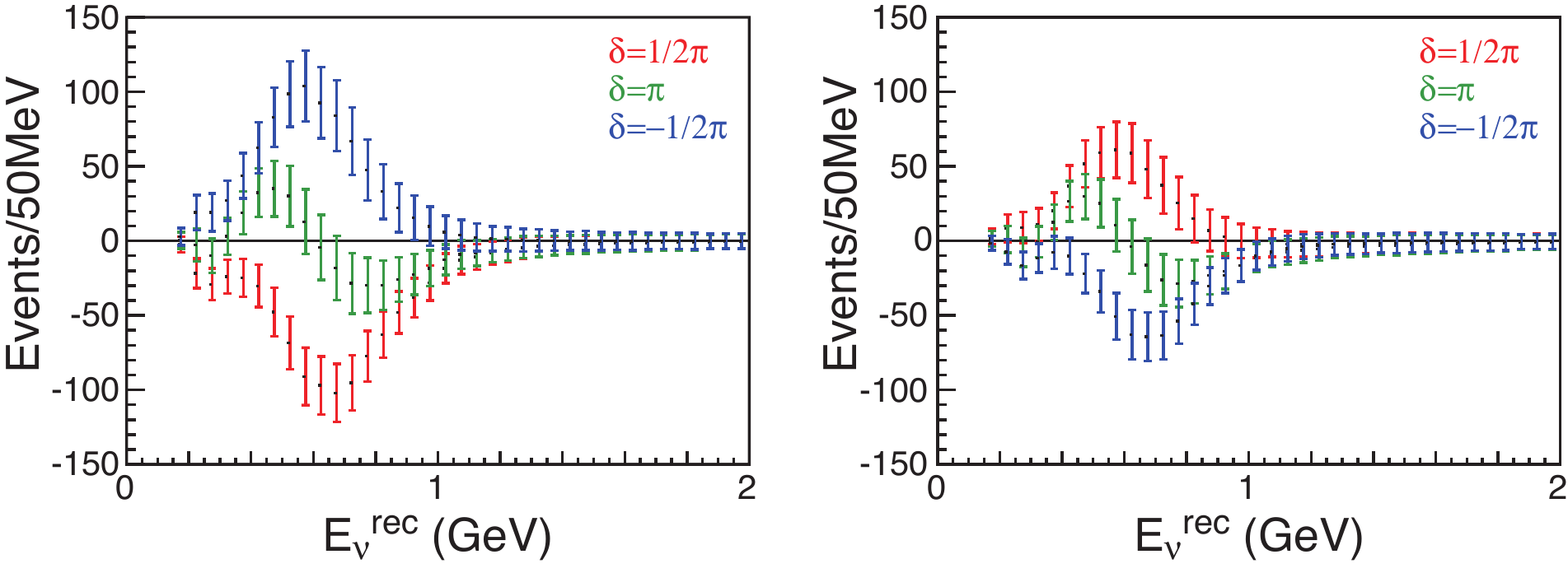}
\caption{Top: Reconstructed neutrino energy distribution for several values of $\deltacp$.
$\sin^22\theta_{13}=0.1$ and normal hierarchy is assumed. 
Bottom: Difference of the reconstructed neutrino energy distribution from the case with $\deltacp=0$.
The error bars represent the statistical uncertainties of each bin for  1.5 (3.5) years of running in  neutrino (anti-neutrino) mode.
\label{fig:cp-enurecdiff}}
\end{figure}

Figure~\ref{fig:cp-enurec} shows the reconstructed neutrino energy distribution after all the selections.
In the neutrino mode, the dominant background components are 
intrinsic $\nue$ contamination in the beam and mis-identified neutral current $\pi^0$ production events originating from $\numu$.
In the anti-neutrino mode, in addition to $\nuebar$ and $\numubar$, $\nue$ and $\numu$ components have non-negligible contributions due to larger fluxes and cross-sections compared to their counterparts in the neutrino mode.

The reconstructed neutrino energy distributions for several values of $\deltacp$,
with $\sin^22\theta_{13}=0.1$ and the normal mass hierarchy,
are shown in the top plots of Fig.~\ref{fig:cp-enurecdiff}.
The effect of $\delta$ is clearly seen using the reconstructed neutrino energy.
The bottom plots show the difference of reconstructed energy spectrum from $\delta=0$ for the cases $\delta = \frac{1}{2}\pi, \pi$ and $-\frac{1}{2}\pi$.
The error bars correspond to the statistical uncertainty for the total five year experiment.
By using not only the total number of events but also the reconstructed energy distribution,
the sensitivity to $\delta$ can be improved, 
and one can discriminate all the values of $\delta$, including the difference between $\delta = 0$ and $\pi$.

One has to note that, although the study described here is done within the standard MNS framework 
that attributes the source of $CP$ violation to one parameter $\delta$,
the comparison of neutrino and anti-neutrino oscillation probabilities gives a direct measurement of the $CP$ asymmetry,
without assuming an underlying model.

\subsubsection{Sensitivity to the $CP$ asymmetry in the neutrino oscillation}
A $\chi^2$ analysis based on the reconstructed neutrino energy distribution has been performed to study the sensitivity of 
the `J-PARC to Hyper-K' experiment to the $CP$ asymmetry in the neutrino oscillation.

\subparagraph{Analysis method}
A binned $\chi^2$ is constructed from the $E_\nu ^{\rm rec}$ distribution, with 50~MeV bin width for the energy range of 0--2~GeV.
As the systematic uncertainty, uncertainties in the normalizations of signal, background originating from $\numu$ and $\numubar$, those from $\nue$ and $\nuebar$, and the relative normalization between neutrino and anti-neutrino are taken into account.
The $\chi^2$ is defined as
\begin{eqnarray*}
\chi^2 &=& \sum_{\nu, \overline{\nu}} \sum_{i} 
\left[ N_{}^i - \left\{ 1\pm \frac{1}{2} f_{\nu/\overline{\nu}}\right\} \cdot
\left( (1+f_\mathrm{sig})\cdot n_\mathrm{sig}^i + (1+f_{\numu})\cdot n_{\numu}^i + (1+f_{\nue})\cdot n_{\nue}^i\right) 
\right]^2 / N_{}^i  \\
 & & + \frac{{f_\mathrm{sig}}^2}{{\sigma_\mathrm{sig}}^2} + \frac{{f_{\numu}}^2}{{\sigma_{\numu}}^2}
 + \frac{{{f_{\nue}}}^2}{{\sigma_{\nue}}^2} + \frac{{f_{\overline{\nu}/\nu}}^2}{{\sigma_{\nu/\overline{\nu}}}^2},
\end{eqnarray*}
where the index $i$ runs over bins of reconstructed neutrino energy, and $+$ and $-$ are applied for neutrino and anti-neutrino mode, respectively.

$N^i$ is the number of expected events for the $i$-th \enurec\ bin for oscillation parameters
$(\theta_{13}^\mathrm{true},$ $\delta^\mathrm{true},$ $\mathrm{sign}(\Delta m^2_{32})^\mathrm{true})$.
$n_\mathrm{sig}^i$, $n_{\numu}^i$, and $n_{\nue}^i$ are the expected number of events for the appearance signal, the background originating from $\numu/\numubar$, and the background from intrinsic $\nue/\nuebar$, respectively,
with a set of oscillation parameter tested $(\theta_{13}^\mathrm{test},$ $\delta^\mathrm{test},$ $\mathrm{sign}(\Delta m^2_{32})^\mathrm{test})$.
The systematic parameters $f_\mathrm{sig}$, $f_{\numu}$, $f_{\nue}$, $f_{\overline{\nu}/\nu}$ represent uncertainties of the signal, the background from $\numu$, those from $\nue$, and relative normalization of neutrino and anti-neutrino, respectively.
Those systematic parameters are assumed to be energy independent and just the overall scale for each component is changed.
$\sigma_X$ is the assumed size of uncertainty for corresponding systematic parameter $f_X$.

In the recent search for $\nue$ appearance by T2K~\cite{Abe:2011sj},
the total systematic uncertainty of the number of expected events at the far detector (Super-K) is 17.6\% for $\sin^22\theta_{13}=0.1$,
with contributions from neutrino flux uncertainty (8.5\%), neutrino interaction cross section (10.5\%), the near detector efficiency (5.6\%), and the far detector related systematics (9.4\%).

The neutrino flux uncertainty is dominated by the hadron production uncertainty.
The kaon production~\cite{NA61:kaon} and long target data~\cite{NA61:long} from the NA61 experiment will significantly reduce the flux uncertainty.
Furthermore, systematic uncertainties from the neutrino beam and interaction cross section will be reduced 
by using energy dependent extrapolation from near detector measurements to far detector expectations.

The neutrino interaction cross section  and near detector systematics
will be reduced with more data from the near detector.
The T2K off-axis near detector consists of several subdetectors inside a 0.2~T magnetic field.
In \cite{Abe:2011sj}, inclusive $\numu$ charged current measurement in the near detector is used to normalize 
the expected event rate at the far detector, while the cross section uncertainty is estimated based on external measurements and interaction models.
The uncertainty will be reduced by precise measurements of cross section 
with detailed analysis of the near detector data, which is in progress.

For measurement of antineutrinos, background from the wrong sign component ($\nu$ contamination in $\bar{\nu}$ beam) 
introduces additional uncertainty if there is no charge sign selection capability.
Thus, the magnetic field of the near detector will be a powerful tool to reduce uncertainty of anti-neutrino cross sections.
Furthermore, the near detector may be upgraded for the J-PARC to Hyper-K experiment, based upon the experience with T2K.

For the systematic uncertainties with the far detector, 
most of them are estimated by using atmospheric neutrinos as a control sample.
For example, the uncertainty of the $\pi^0$ background rejection efficiency is estimated 
with the $\pi^0$ topological control sample made by combining one data electron and one simulated gamma event; 
its uncertainty is limited by the number of atmospheric neutrino events available.
Similarly, ring counting and particle identification uncertainties are limited by statistics of the 
$\nue$-enriched atmospheric neutrino sample.
Another source of uncertainty is the limited knowledge of the neutrino interaction cross section.
With more than an order of magnitude larger statistics available with Hyper-K,
the beam neutrino events as well as the atmospheric neutrino events can be used to study systematics. 
Together with improved understanding of the neutrino interaction, uncertainties associated with the far detector will be reduced.

Based on the experience from T2K analysis and prospects for future improvements described above,
$\sigma_X$ is set to 5\% for all four systematic parameters.

For each set of ($\theta_{13}^\mathrm{test}$, $\deltacp^\mathrm{test}$, $\mathrm{sign}(\Delta m^2_{32})^\mathrm{test}$), 
the $\chi^2$ is minimized by changing the systematic parameters, $f_X$.
The $\chi^2$ is then compared to the value at the true oscillation parameters, and the difference $\Delta \chi^2 \equiv (\chi^2\mathrm{(test)} - \chi^2\mathrm{(true)}) $ is used to evaluate the significance of the measurement.
When allowed regions are drawn on a $(\sin^22\theta_{13})$-$\delta$ plane, 
the 68.3\% ($1\sigma$), 95.5\%($2\sigma$), and 99.7\% ($3\sigma$) CL
allowed regions are defined as the regions of parameters where $\Delta \chi^2 < 2.30, 6.18,$ and 11.83, respectively.
For measurements where a single parameter is concerned, e.g. the uncertainty of $\delta$, mass hierarchy determination, 
and exclusion of $\sin\delta = 0$, $\Delta \chi^2$ values corresponding to a single parameter are used~\cite{Nakamura:2010zzi}.

\subparagraph{Sensitivity if the mass hierarchy is known}
\begin{figure}[tbp]
\includegraphics[width=0.5\textwidth]{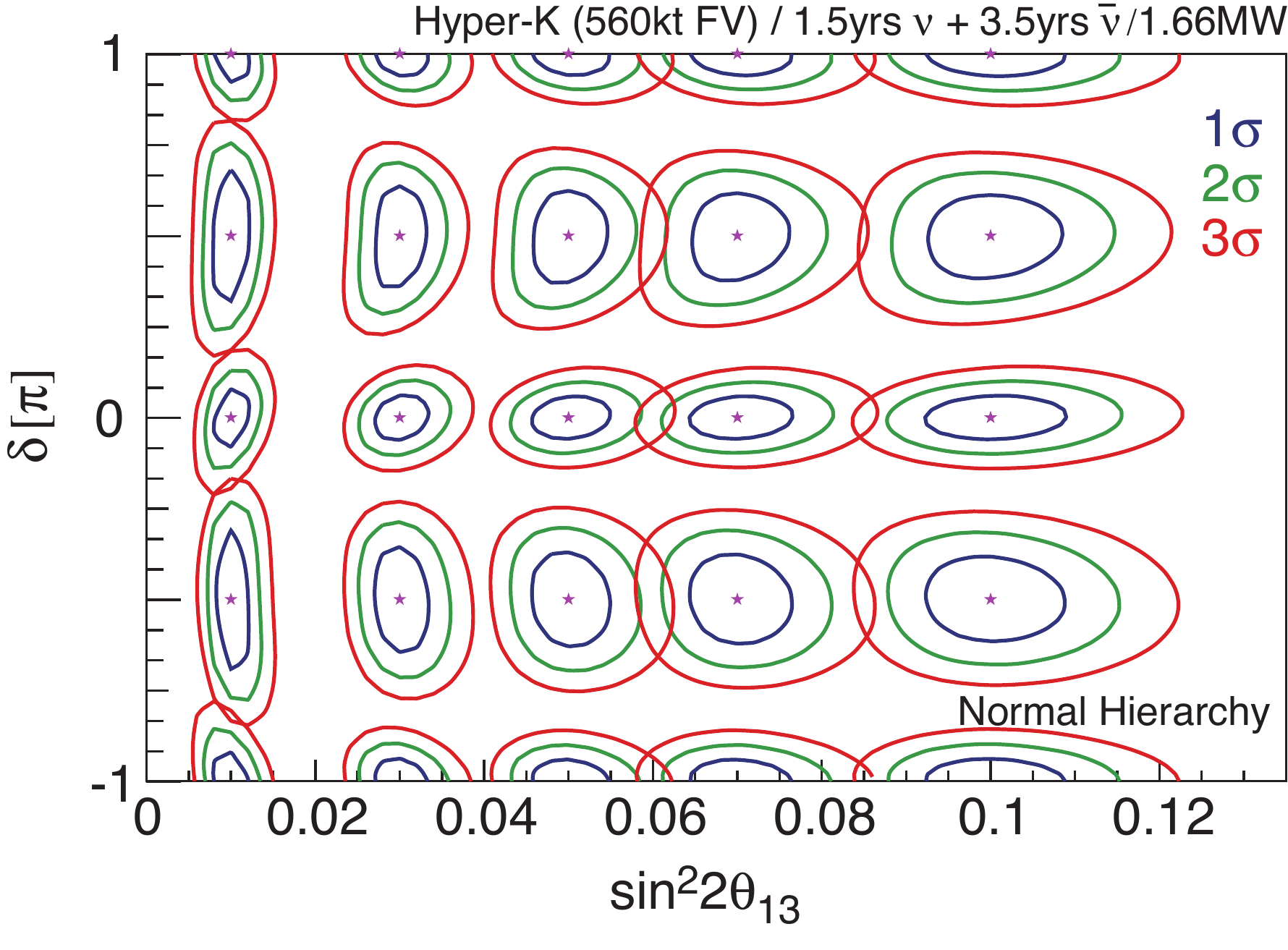}
\caption{Allowed regions for the parameter sets of $\sin^22\theta_{13} = (0.01, 0.03, 0.05, 0.07, 0.10)$ and $\deltacp = (-\frac{1}{2} \pi, 0, \frac{1}{2}\pi, \pi)$ overlaid together.
Blue, green, and red lines represent 1, 2, 3 $\sigma$ allowed regions, respectively.
Stars indicate the true parameters.
It is assumed that the mass hierarchy is known to be the normal hierarchy.
\label{fig:cp-allow1}}
\end{figure}

\begin{figure}[tbp]
\includegraphics[width=0.5\textwidth]{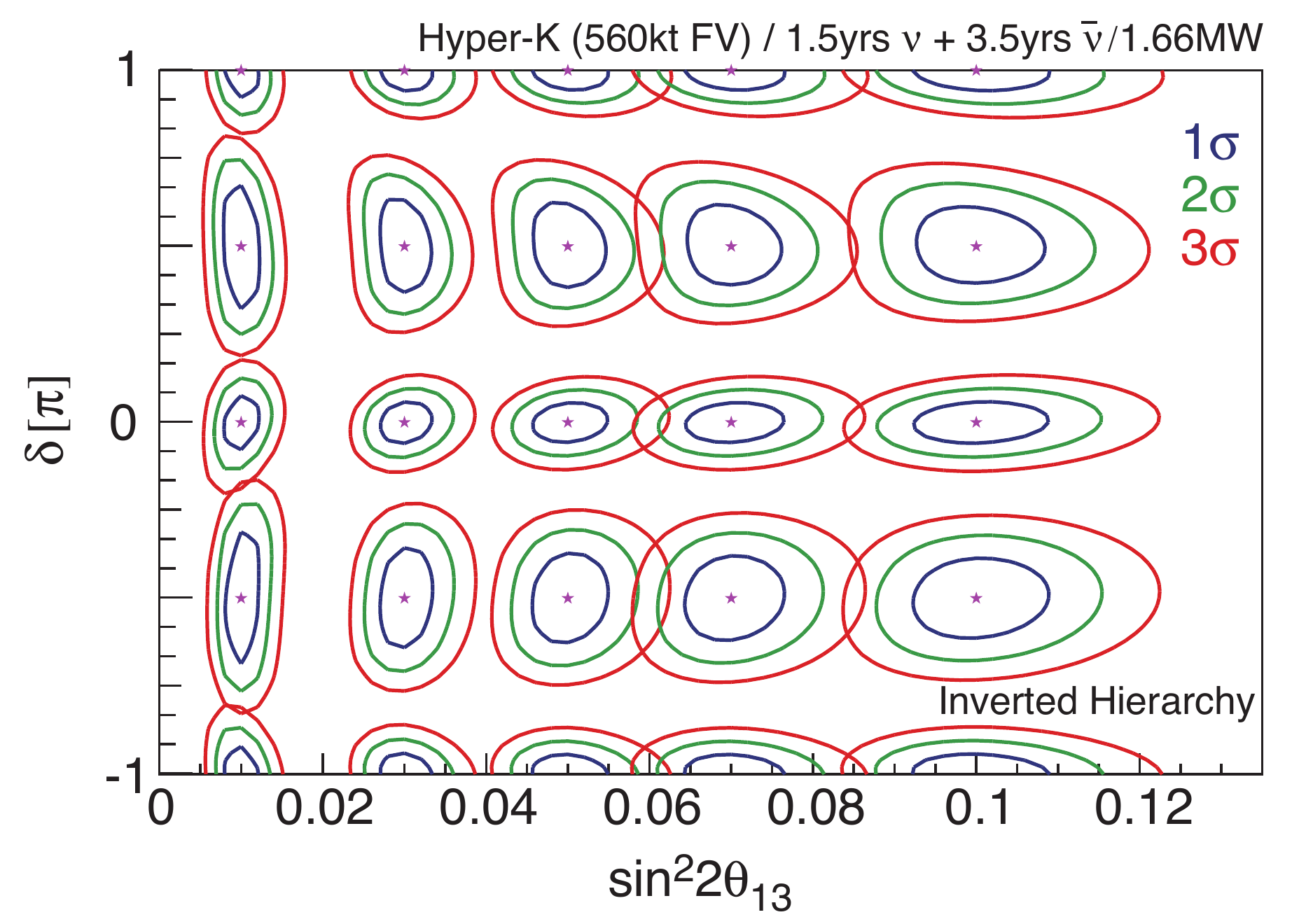}
\caption{Allowed regions for inverted hierarchy. See caption of Fig.~\ref{fig:cp-allow1}.
\label{fig:cp-allow2}}
\end{figure}

\begin{figure}[tbp]
\includegraphics[width=0.5\textwidth]{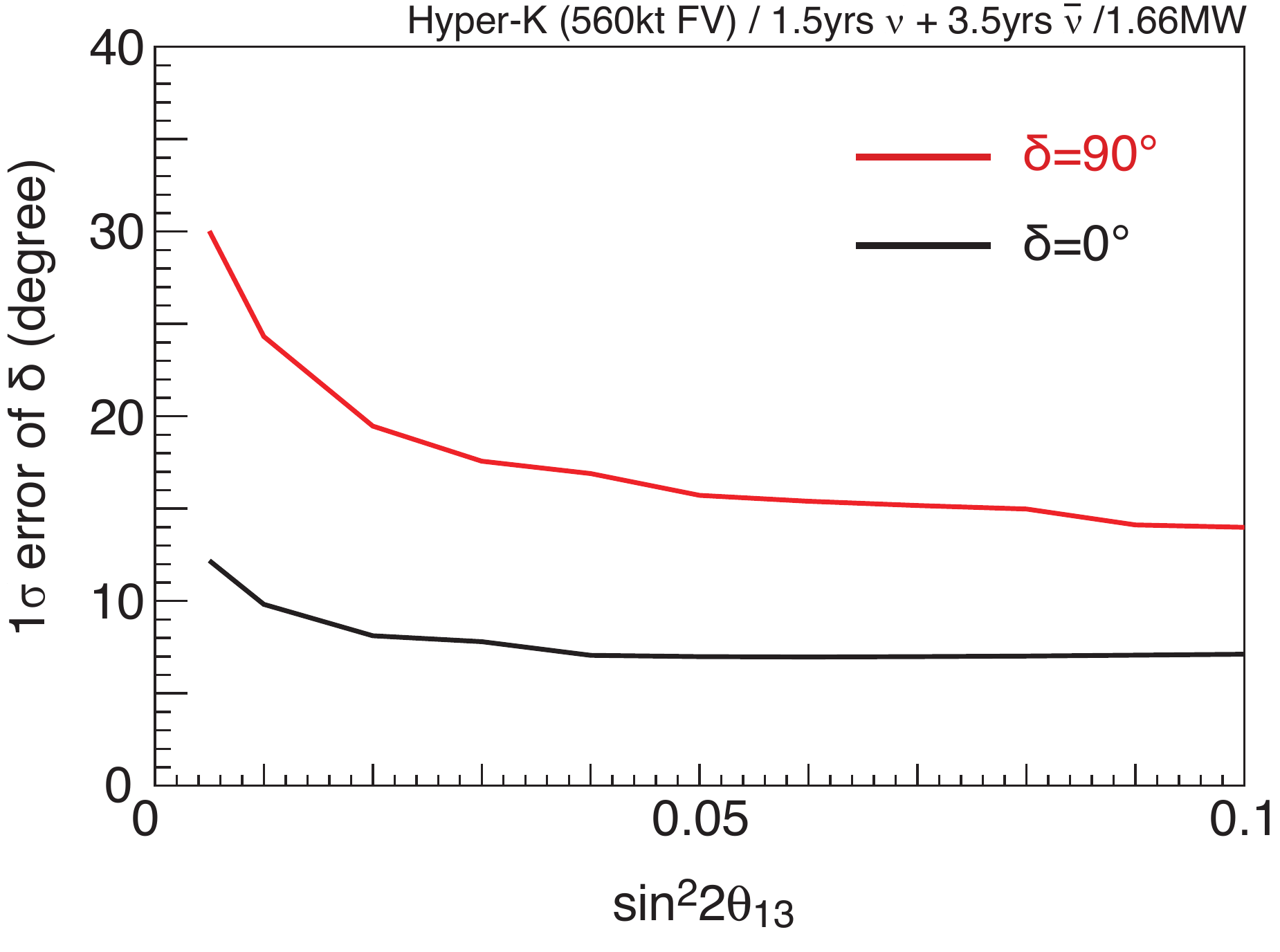}
\caption{1$\sigma$ error of $\deltacp$ as a function of $\sin^22\theta_{13}$ for the normal hierarchy case.
\label{fig:cp-delta-error}}
\end{figure}

\begin{figure}[tbp]
\includegraphics[width=0.48\textwidth]{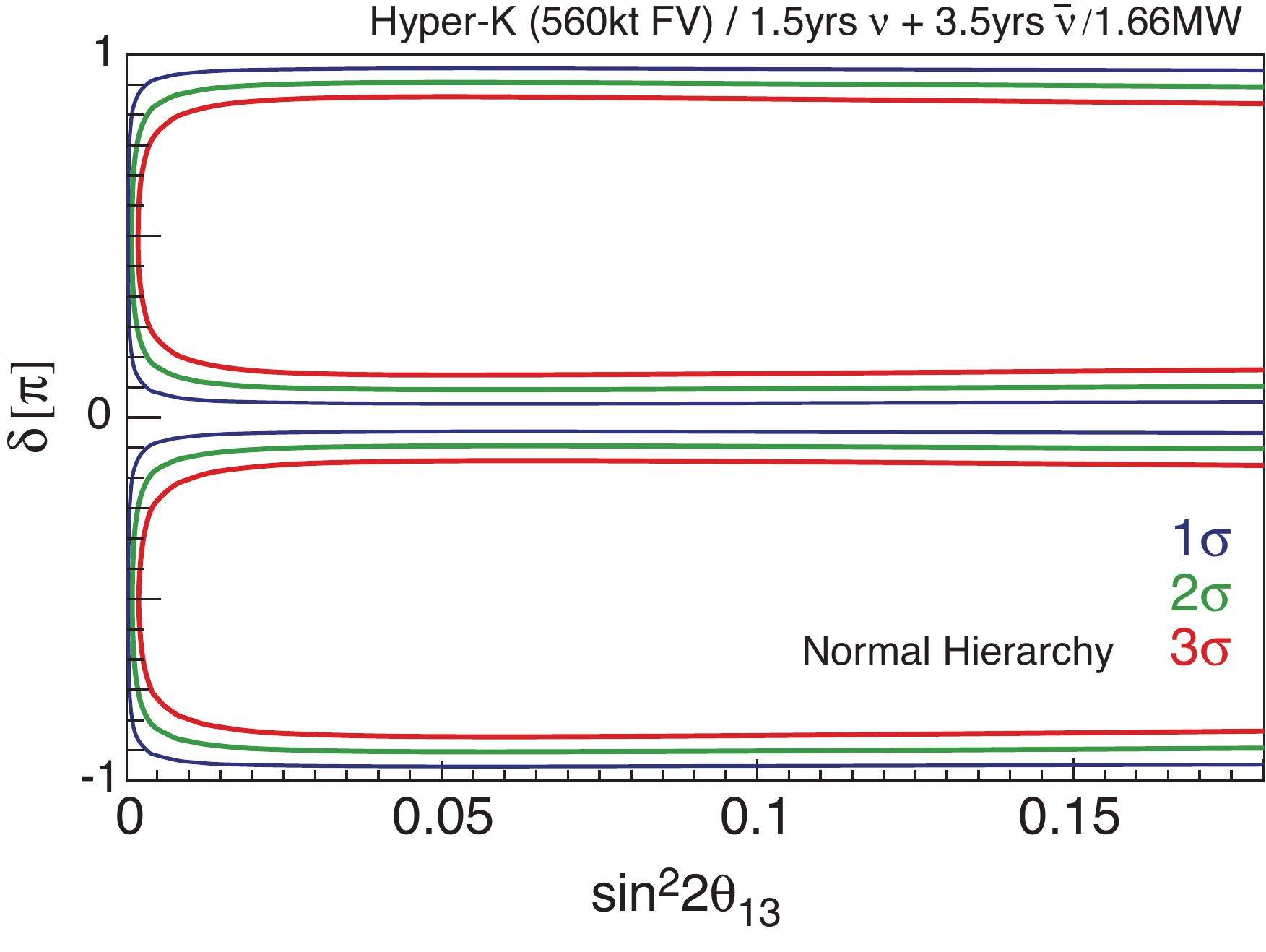}
\includegraphics[width=0.48\textwidth]{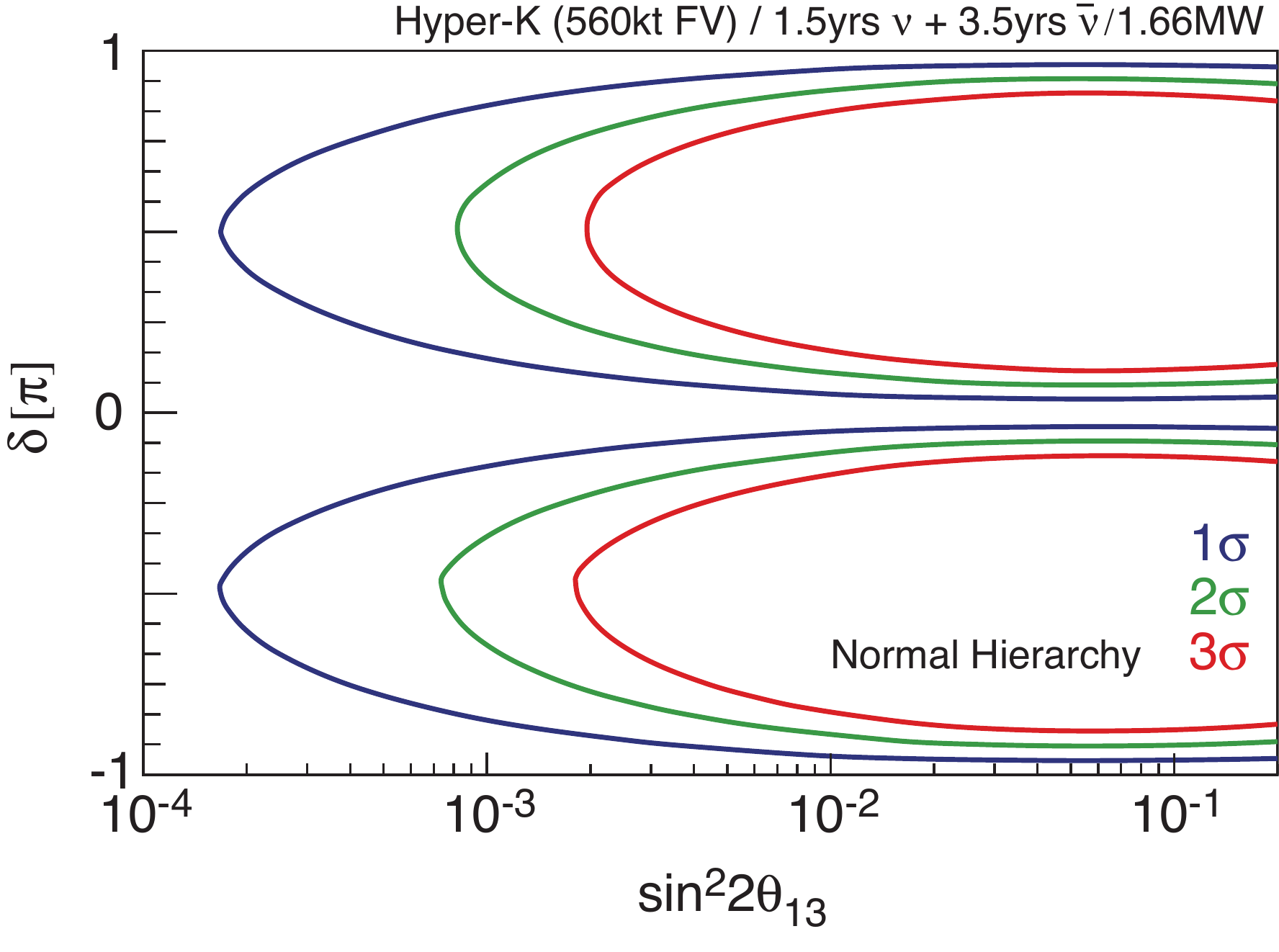}
\caption{Sensitivity to CP violation. Blue, green, and red lines correspond to 1, 2, and 3 $\sigma$ exclusion of $\sin\delta=0$, respectively.
Left: horizontal axis is linear scale, right: log scale.
\label{fig:cp-sens1}
}
\end{figure}

Let us first consider  the case where the mass hierarchy is known by other measurements.
If the mass hierarchy is known to be the normal hierarchy, $\mathrm{sign}(\Delta m^2_{32})$ can be fixed to be positive.
The allowed regions for the true parameter sets of combinations of $\sin^22\theta_{13} = (0.01, 0.03, 0.05, 0.07, 0.1)$ 
and $\deltacp = (0, \frac{1}{2}\pi, \pi, -\frac{1}{2} \pi)$, for the normal hierarchy, are shown in Fig.~\ref{fig:cp-allow1}.
The same plot for the inverted hierarchy case is shown in Fig.~\ref{fig:cp-allow2}.
For both cases, $\deltacp$ and $\sin^22\theta_{13}$ can be well determined for the region of $\sin^22\theta_{13}$ indicated by the T2K result of June 2011~\cite{Abe:2011sj}, $\sin^22\theta_{13}>0.03 (0.04)$ (90\% CL) for normal (inverted) hierarchy. 

Figure~\ref{fig:cp-delta-error} shows the 1$\sigma$ error of $\deltacp$ as a function of $\sin^22\theta_{13}$ for the normal hierarchy.
The size of the error is almost the same for the inverted hierarchy.
For $\sin^22\theta_{13}>0.03$, the value of $\delta$ can be determined to better than 8$^\circ$ for $\deltacp=0^\circ$, and better than 18$^\circ$ for $\deltacp=90^\circ$.

Figure~\ref{fig:cp-sens1} shows the regions where $\sin\deltacp=0$ is excluded, 
i.e.\ $CP$ is found to be violated in the lepton sector, with  1$\sigma$, 2$\sigma$, and 3$\sigma$ significance, in the case of normal hierarchy and the mass hierarchy is known.
The J-PARC Hyper-K experiment will have a sensitivity to the $CP$ asymmetry down to $\sin^22\theta_{13}\sim 0.003$ with 3 $\sigma$ significance.

\begin{figure}[tbp]
\includegraphics[width=0.75\textwidth]{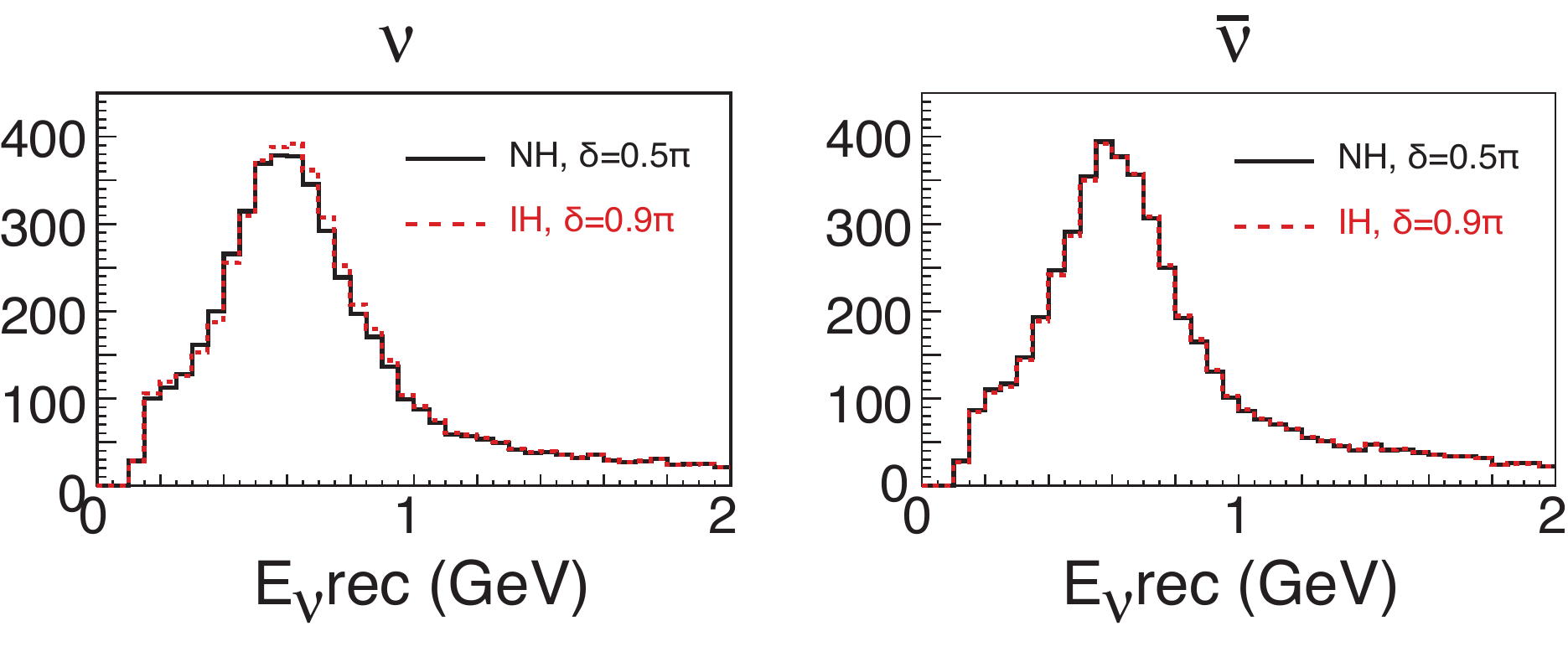}
\caption{ Reconstructed neutrino energy distributions for 
($\sin^22\theta_{13}, \deltacp$) = (0.10, 0.5$\pi$) with normal mass hierarchy (black solid) and 
($\sin^22\theta_{13}, \deltacp$) = (0.10, 0.9$\pi$) with inverted mass hierarchy (red dashed).
\label{fig:cp-degeneracy-enurec}}
\end{figure}

\begin{figure}[tbp]
\includegraphics[width=0.5\textwidth]{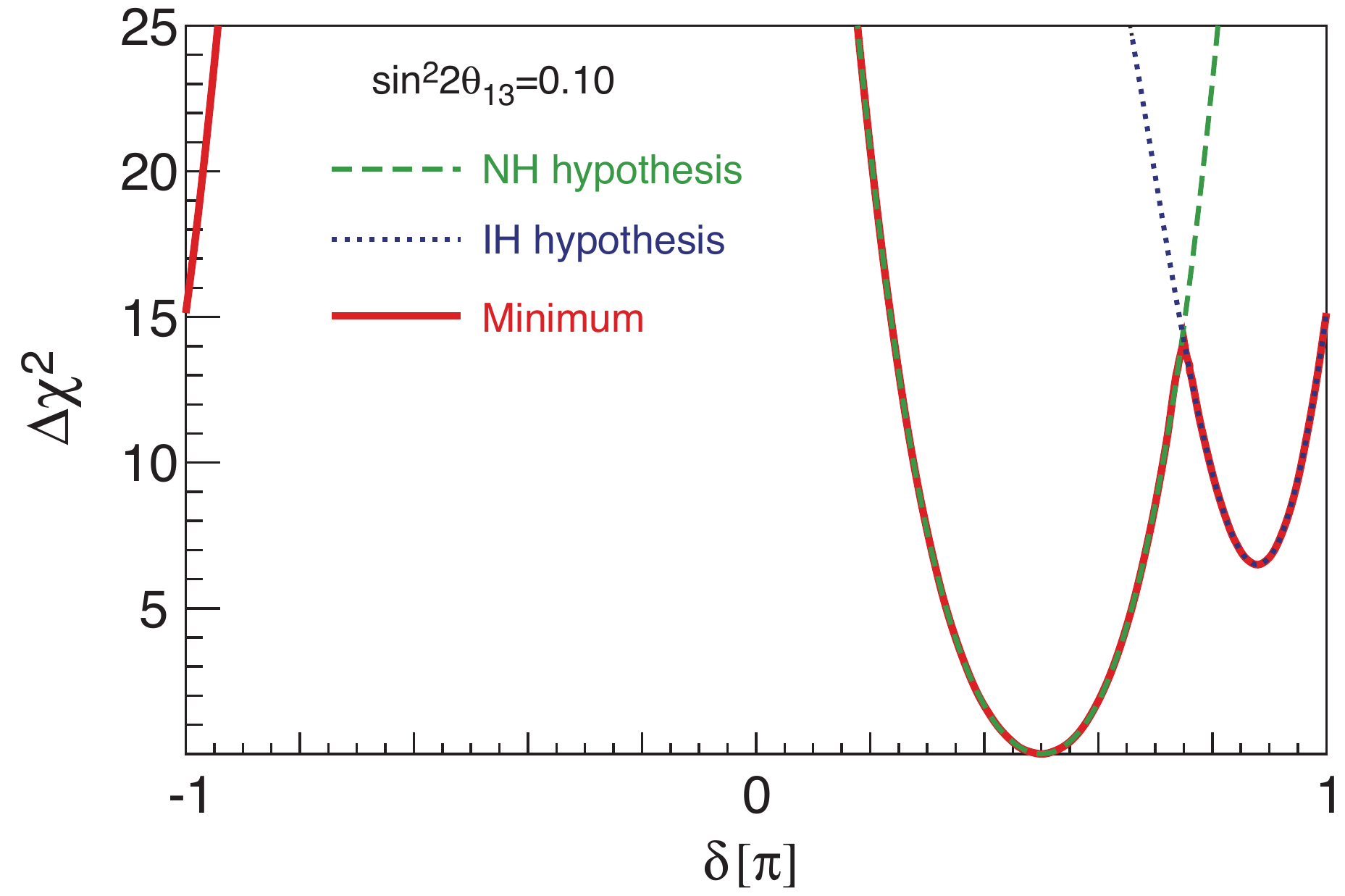}
\caption{ The $\Delta \chi^2$ as a function of $\deltacp$, for the case the mass hierarchy is unknown. $\sin^22\theta_{13}$ is fixed to 0.1.
The true parameters are ($\sin^22\theta_{13}, \deltacp$) = (0.10, 0.5$\pi$) and normal hierarchy.
In this case, $\chi^2$ is calculated with both normal and inverted hierarchy and the smaller is taken as the minimum value.
Due to the parameter degeneracy, there is a local minimum with inverted hierarchy hypothesis.
\label{fig:cp-chi2_1}}
\end{figure}

\subparagraph{Effect of unknown mass hierarchy}
If the mass hierarchy is unknown, there might be a fake solution due to the parameter degeneracy~\cite{Minakata:2001qm}.
Figure~\ref{fig:cp-degeneracy-enurec} shows the reconstructed neutrino energy distributions for 
($\sin^22\theta_{13}, \deltacp$) = (0.10, 0.5$\pi$) with normal mass hierarchy, and 
($\sin^22\theta_{13}, \deltacp$) = (0.10, 0.9$\pi$) with inverted mass hierarchy.
Those two parameters give similar oscillation probabilities, hence similar energy distributions.

Figure~\ref{fig:cp-chi2_1} shows $\Delta \chi^2$ as a function of $\deltacp$, for the case the mass hierarchy is unknown. 
In this plot, $\sin^22\theta_{13}$ is fixed to 0.1.
The true parameters are ($\sin^22\theta_{13}, \deltacp$) = (0.10, 0.5$\pi$) and normal hierarchy.
In the case the mass hierarchy is unknown, $\chi^2$ is calculated with both normal and inverted hierarchy hypotheses and the smaller is taken to calculate $\Delta \chi^2$.
Due to the parameter degeneracy, there is a local minimum with the inverted hierarchy hypothesis  in addition to the minimum around the true value with the normal hierarchy hypothesis.
If this fake solution is consistent with null $CP$ asymmetry even though the true solution violates the $CP$ symmetry, then the sensitivity to the $CP$ violation will be lost for that parameter set.

\begin{figure}[tbp]
\includegraphics[width=0.5\textwidth]{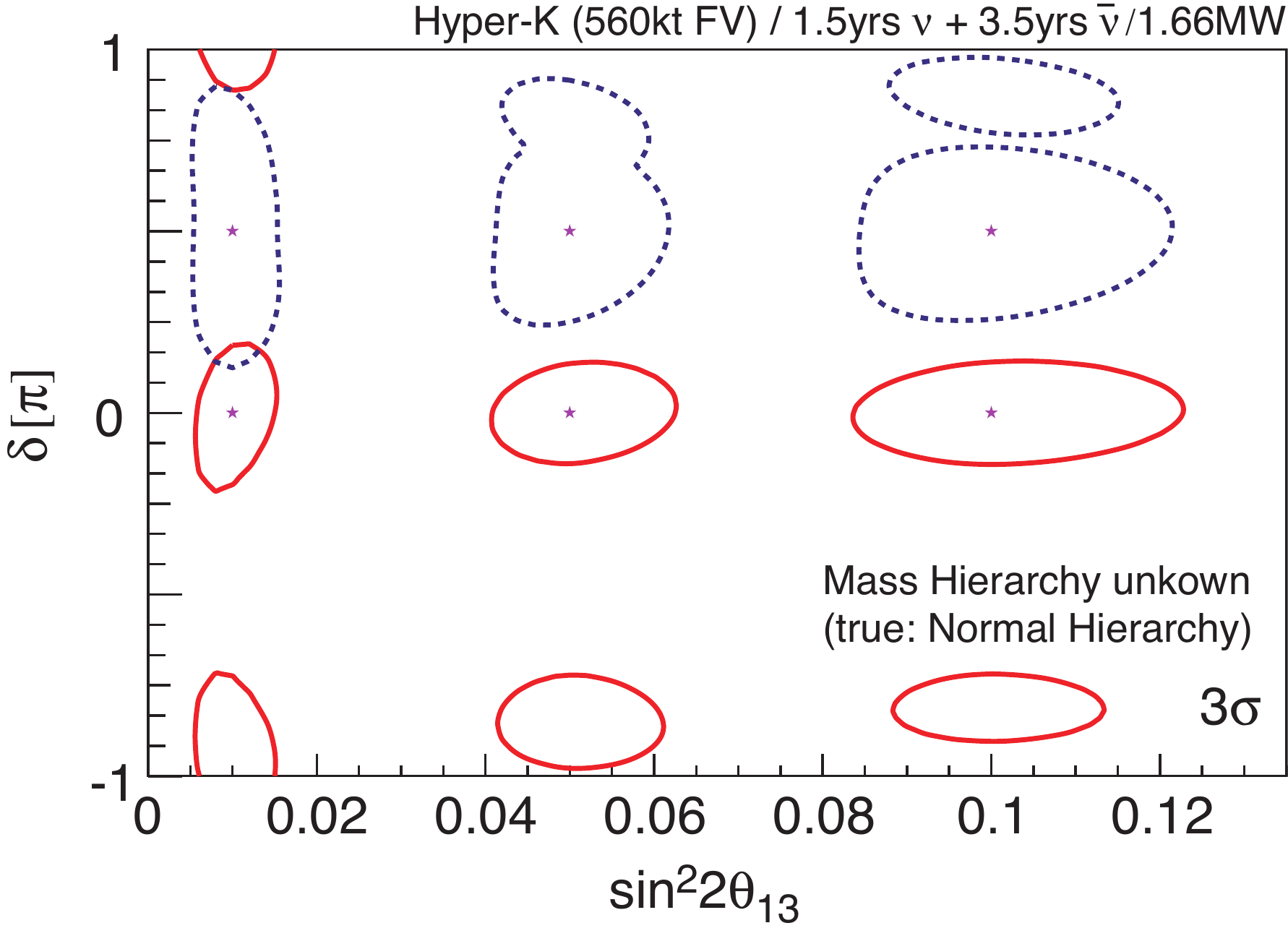}
\caption{$3\sigma$ allowed regions  for the case the mass hierarchy is unknown. 
The true mass hierarchy is the normal hierarchy.
Solid (dashed) line shows the contour for true $\deltacp=0 (\frac{1}{2}\pi)$.
Stars indicate the true parameter values.
Plots with $\sin^22\theta_{13}$ of 0.01, 0.05, and 0.1 are overlaid.
In addition to the region around the true values, there are fake solutions around $\delta=\pi(-\pi)$ for true $\delta=0(\frac{1}{2}\pi)$ due to degeneracy.
\label{fig:cp-allow3} }
\end{figure}

Figure~\ref{fig:cp-allow3} shows the 3$\sigma$ allowed regions for the case the true mass hierarchy is normal but not determined prior to this experiment.
Solid red (dashed blue) line shows the contour for true $\deltacp=0\, (\frac{1}{2}\pi)$.
Stars indicate the true parameter values.
Plots for $\sin^22\theta_{13}$ of 0.01, 0.05, and 0.1 are shown together.
The areas indicated by dashed lines around $\sin^22\theta_{13}=0.1$ correspond to the case shown in Fig.~\ref{fig:cp-chi2_1}.
In addition to the region around the true values, there is a fake solution due to unknown mass hierarchy.

\begin{figure}[tbp]
\includegraphics[width=0.5\textwidth]{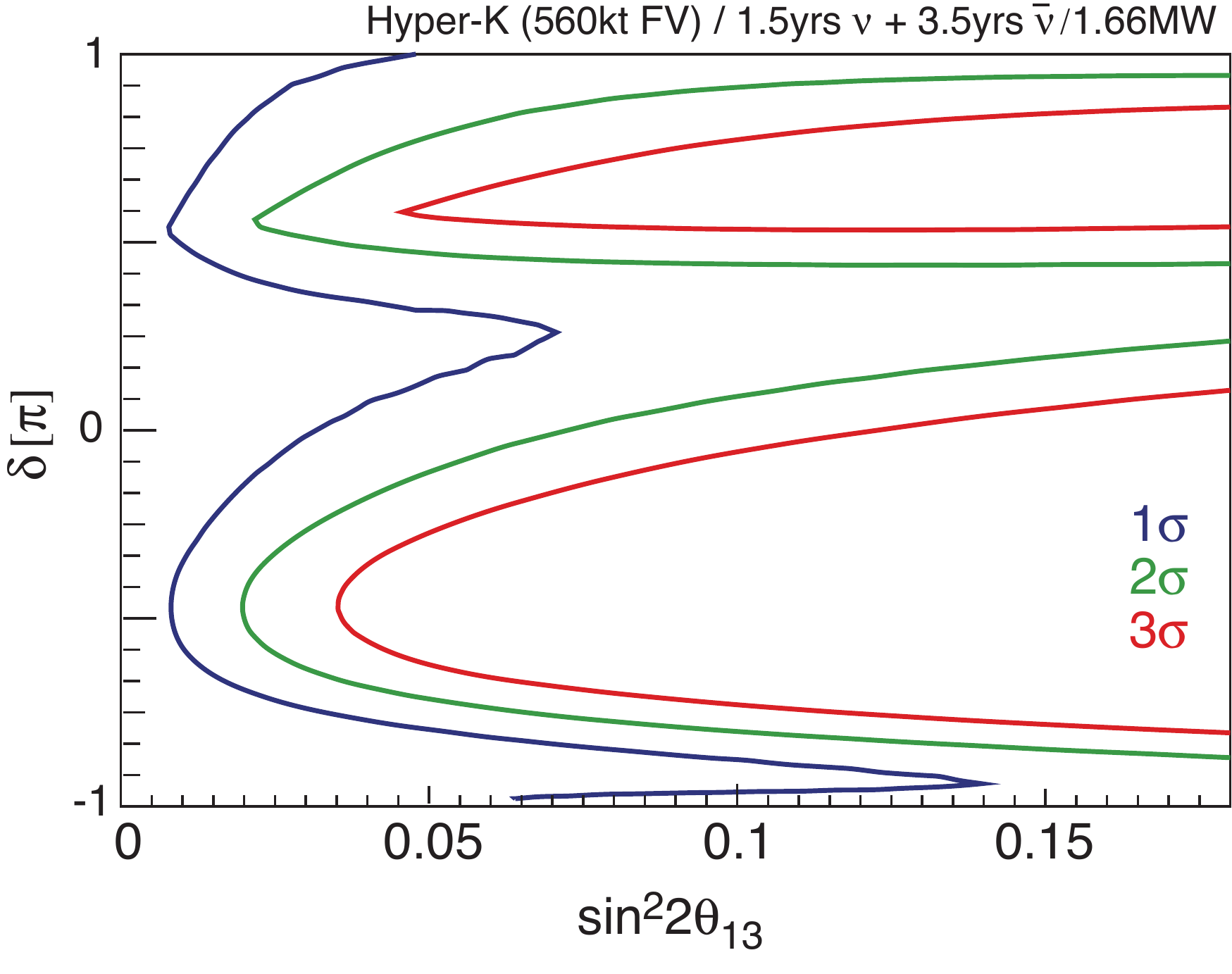}
\caption{Sensitivity to the mass hierarchy.
Blue, green, and red lines correspond to 1, 2, and 3 $\sigma$, respectively.
The true mass hierarchy is the normal hierarchy.
\label{fig:mh-sensitivity}}
\end{figure}

If $\sin^22\theta_{13}$ is as large as $\sim 0.1$,  the NO$\nu$A experiment in 
United States~\cite{Ayres:2004js} together with the T2K experiment will have sensitivity to the mass hierarchy. 
In addition, for a large value of $\sin^22\theta_{13}$, Hyper-K itself has considerable sensitivity to the mass hierarchy.
Figure~\ref{fig:mh-sensitivity} shows the sensitivity to the mass hierarchy from the 
J-PARC to Hyper-K experiment alone.
For each set of $(\theta_{13}^\mathrm{true},$ $\delta^\mathrm{true}$) with normal mass hierarchy
(${\Delta m^2_{32}}^\mathrm{true}>0$), 
the inverted hierarchy hypothesis is tested by calculating $\Delta \chi^2$ for various sets of
$(\theta_{13}^\mathrm{test}$, $\delta^\mathrm{test})$ with ${\Delta m^2_{32}}^\mathrm{test}<0$.
The J-PARC to Hyper-K experiment has sensitivity to the mass hierarchy for $\sin^22\theta_{13} > 0.05$.
For $\sin^22\theta_{13} = 0.1$, the mass hierarchy can be determined with more than 3$\sigma$ significance for 46\% of the $\deltacp$ parameter space.
Also, the atmospheric neutrino observation will have a sensitivity to the mass hierarchy as described in Sec.~\ref{section:atmnu}.
If $\sin^22\theta_{13}$ is too small for those experiments to determine the mass hierarchy, 
the effect of mass hierarchy becomes smaller, as it becomes relatively insignificant compared to the genuine $CP$ asymmetry. 

\begin{figure}[tbp]
\includegraphics[width=0.48\textwidth]{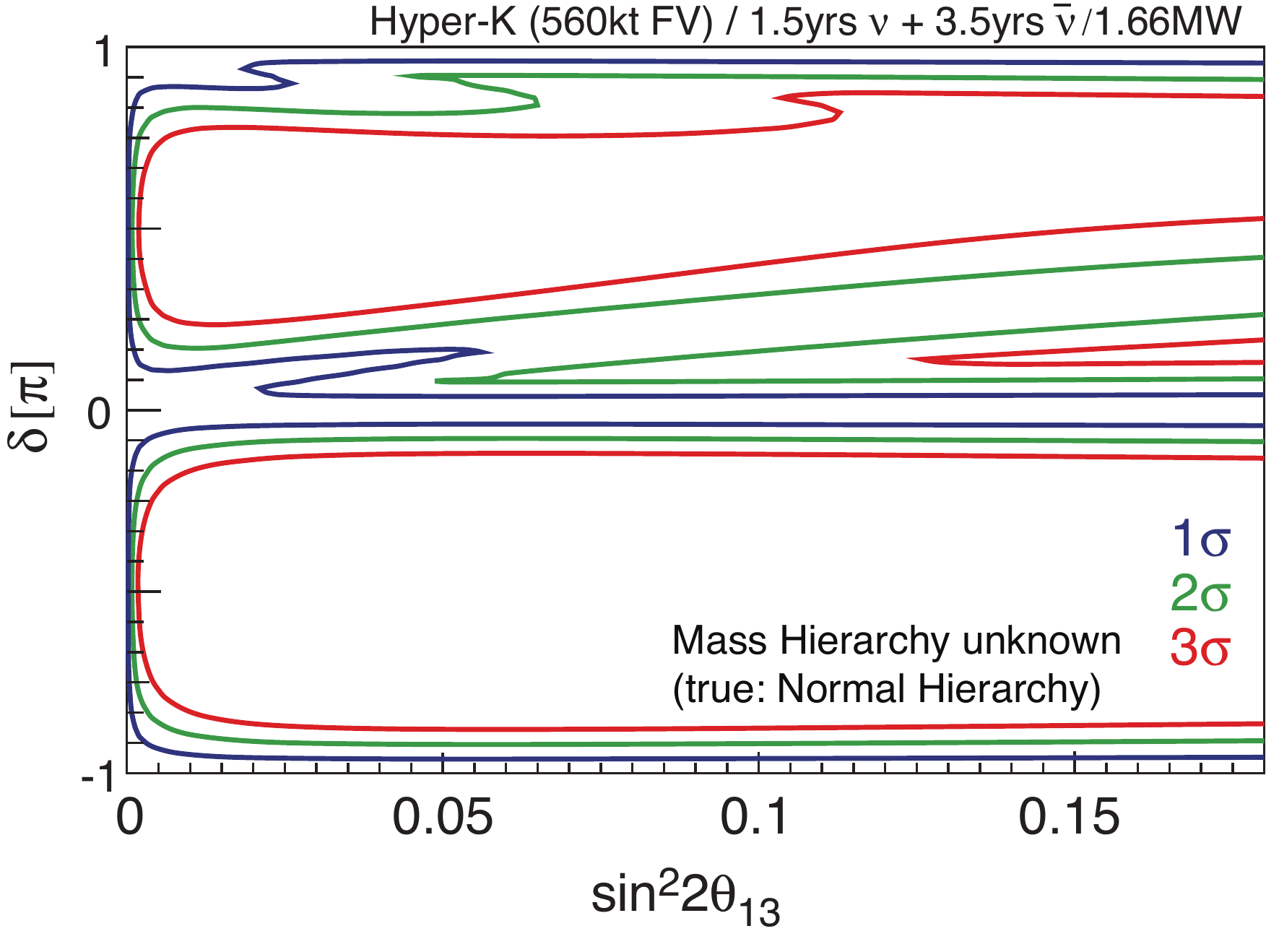}
\includegraphics[width=0.48\textwidth]{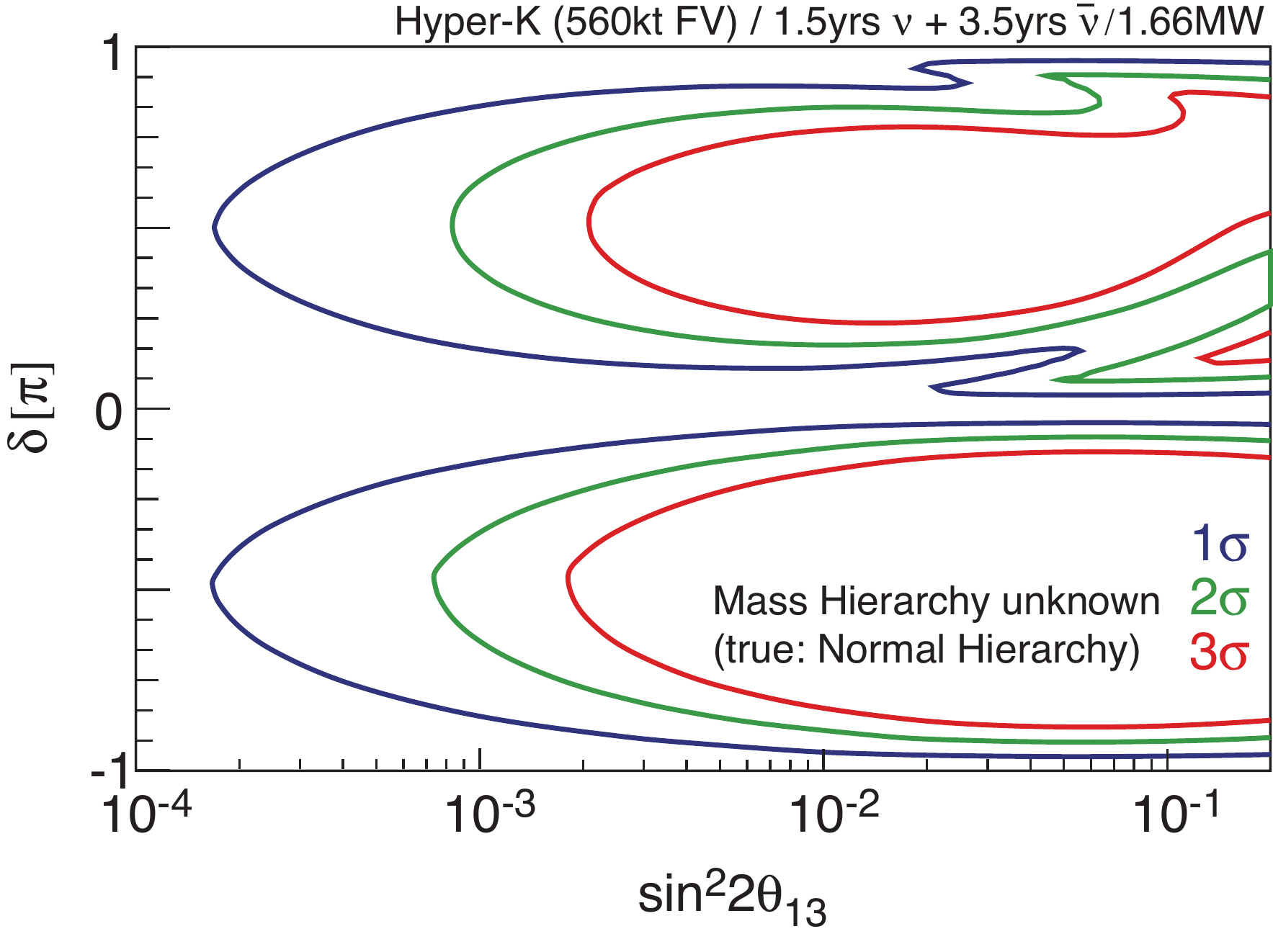}
\caption{Sensitivity to CP violation for the case the mass hierarchy is not known.
Blue, green, and red lines correspond to 1, 2, and 3 $\sigma$ exclusion of $\sin\delta=0$, respectively.
The true mass hierarchy is the normal hierarchy. \label{fig:cp-sens2}}
\end{figure}

If we assume that the mass hierarchy is not known, the discovery potential for $CP$ violation is as shown in Figure~\ref{fig:cp-sens2}.
For the parameter $0<\deltacp<\pi$, the exclusion region becomes smaller compared to the case where mass hierarchy is known (Fig.~\ref{fig:cp-sens1}) because of the fake solution coming from unknown mass hierarchy.
However, there are sets of parameters for which $CP$ asymmetry can be observed, and for $-\pi<\deltacp<0$, there is little effect even if the mass hierarchy is not measured.
This is because the mass hierarchy can be determined and the degeneracy is resolved for these parameter sets as shown in Fig.~\ref{fig:mh-sensitivity}.
For the case in which the true mass hierarchy is inverted, a similar argument holds with $-\pi<\deltacp<0$ and $0<\deltacp<\pi$ inverted.

In reality, the mass hierarchy can be determined for wider range of parameters by combining other experiments and atmospheric neutrino observations with Hyper-K.
In conclusion, the knowledge of the neutrino mass hierarchy will have only limited impact on the discovery potential of leptonic $CP$ violation by a J-PARC to Hyper-K long baseline program as discussed here.

\subparagraph{Sensitivity vs running time}

\begin{figure}[tbp]
\includegraphics[width=0.5\textwidth]{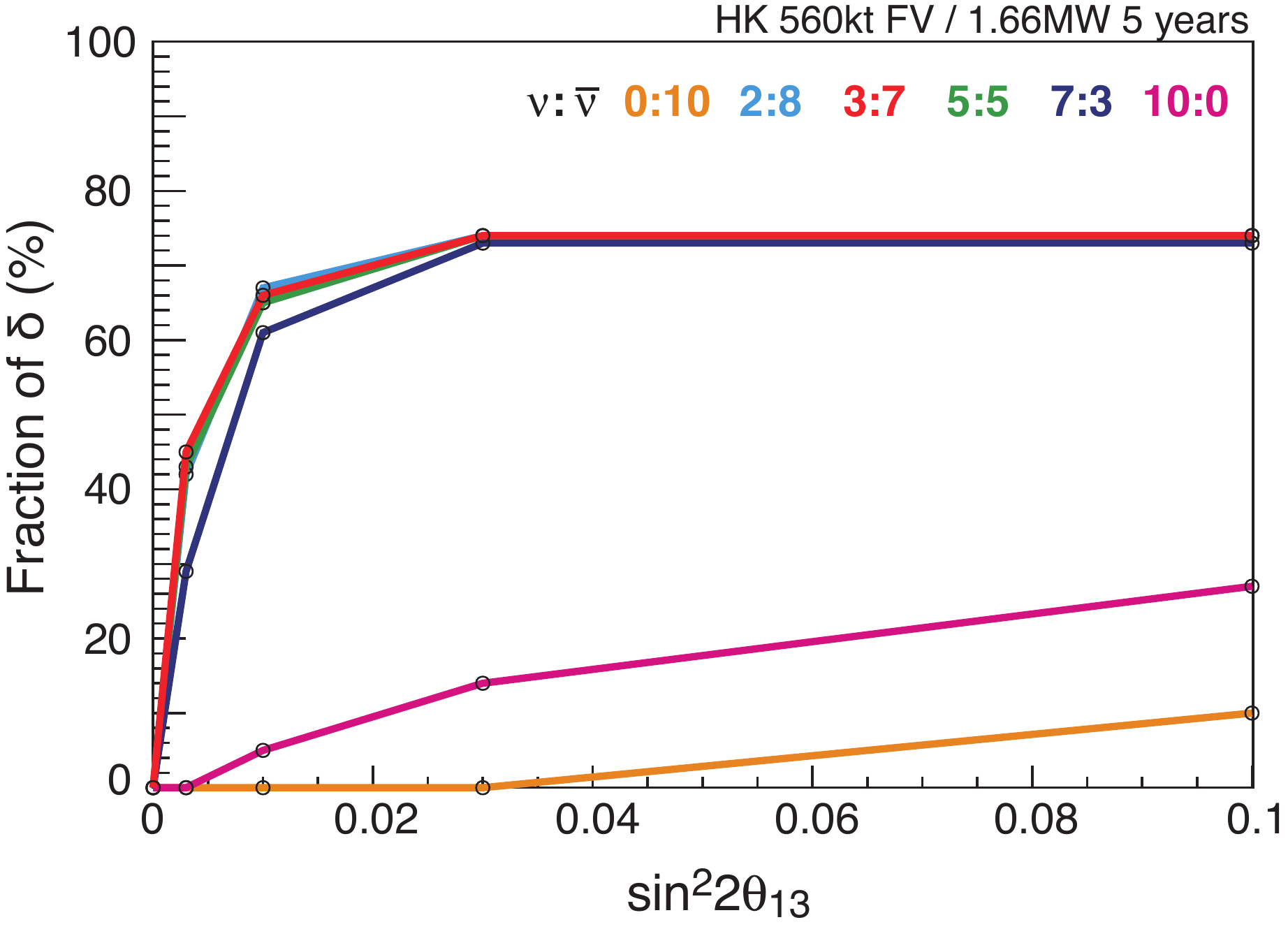}
\caption{Fraction of $\deltacp$ for which $\sin\deltacp= 0$ is excluded with 3 $\sigma$ significance as a function of true $\sin^22\theta_{13}$,
with different ratio of $\nu$ and $\bar{\nu}$ running time, while  total running time is fixed to five years with 1.66~MW.
Normal hierarchy is assumed.
\label{fig:cp-sens-ratio}}
\end{figure}

Figure~\ref{fig:cp-sens-ratio} shows the fraction of $\deltacp$ for which $\sin\deltacp= 0$ is excluded with 3~$\sigma$, 
with different ratio of $\nu$ and $\bar{\nu}$ running time, while the total running time and the beam power is fixed to five years with 1.66~MW.
The case with $\nu : \bar{\nu} = 3 : 7$ has the best sensitivity, although the difference is marginal around this value.

\begin{figure}[tbp]
\includegraphics[width=0.45\textwidth]{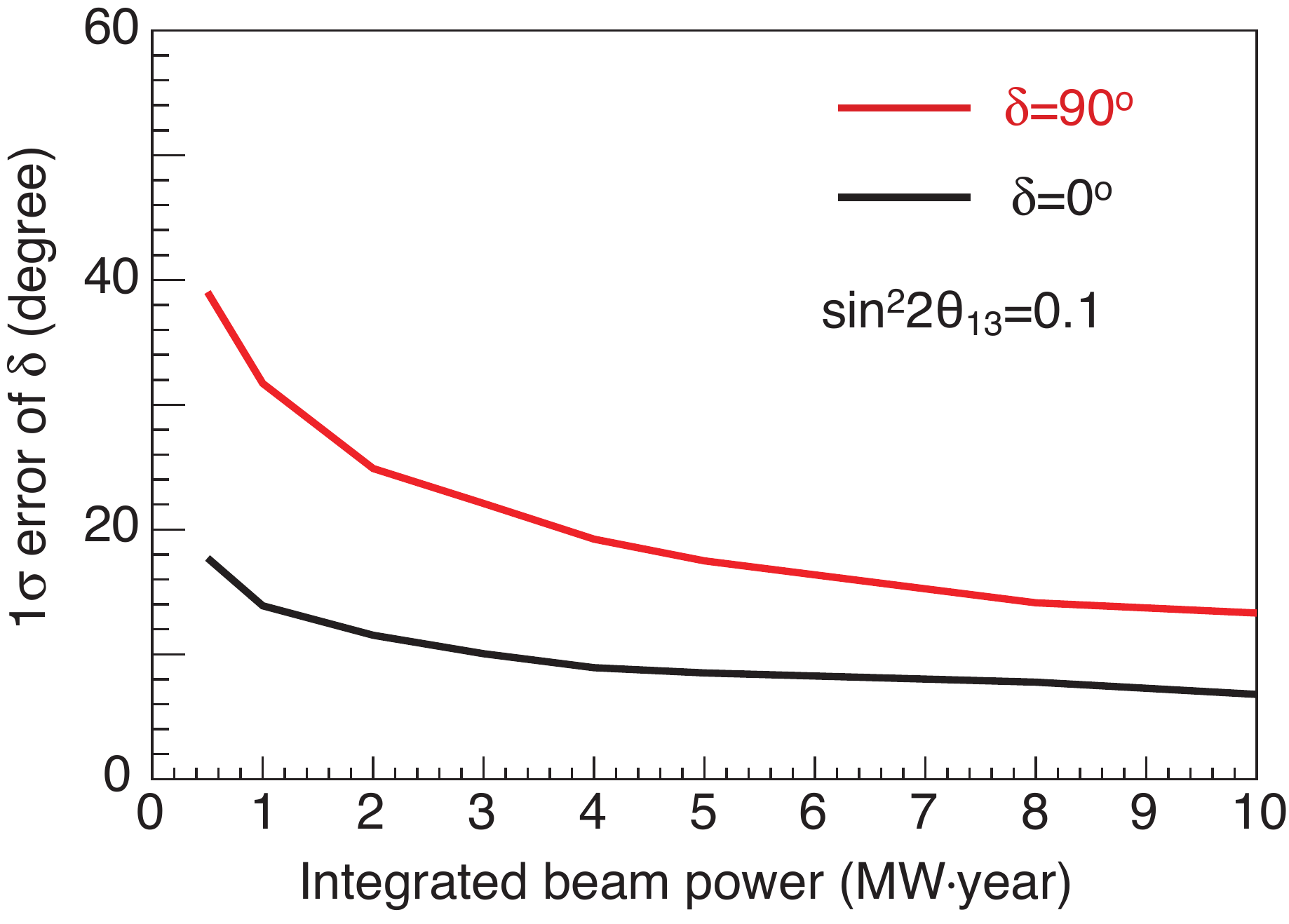}
\includegraphics[width=0.45\textwidth]{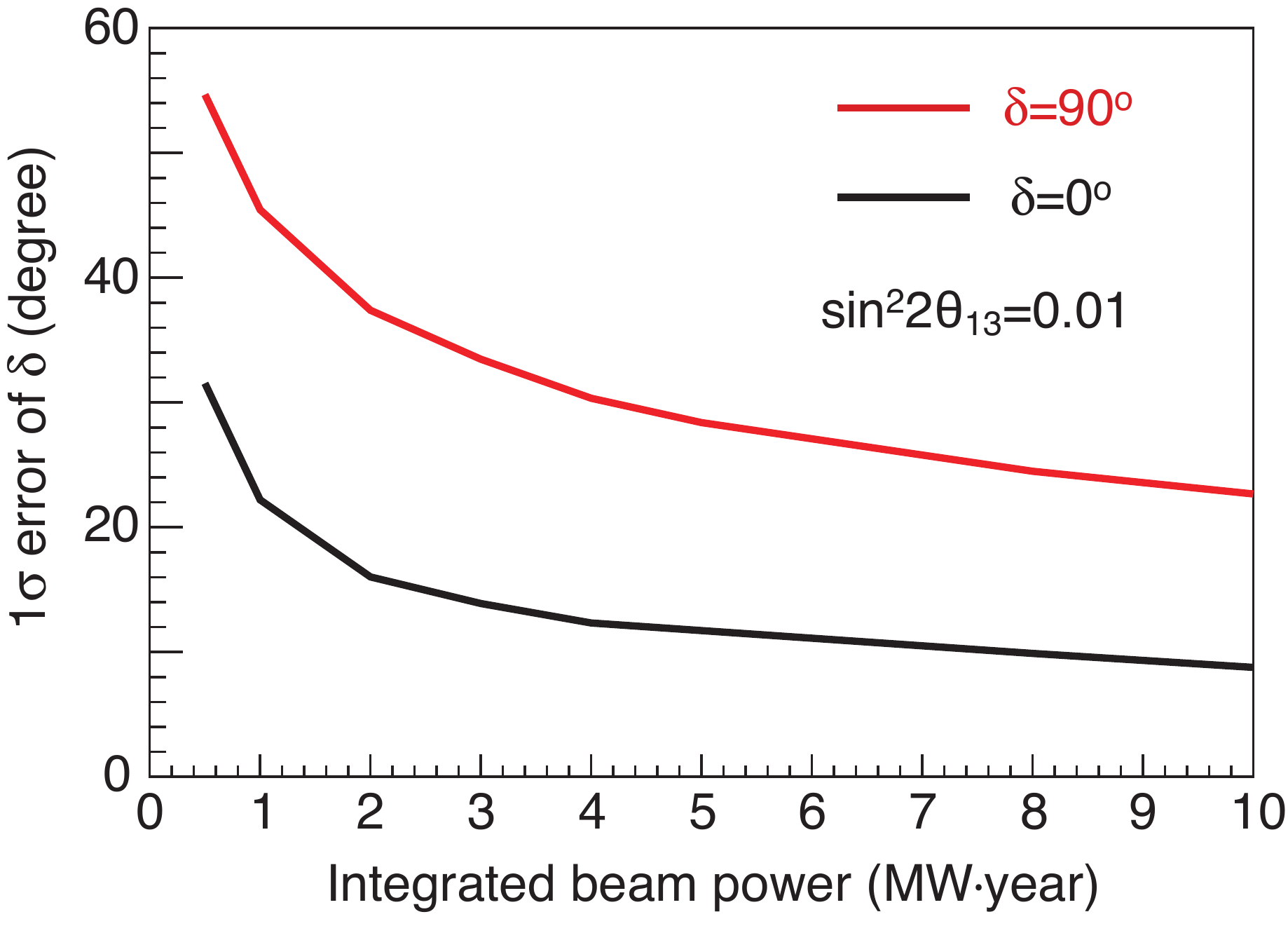}
\caption{1$\sigma$ uncertainty of $\deltacp$ as a function of integrated beam power. 
The ratio of $\nu$ and $\bar{\nu}$ running time is fixed to 7:3.
left: $\sin^22\theta_{13} =0.1$, right: $\sin^22\theta_{13} =0.01$.
\label{fig:cp-delta-error-time}}
\end{figure}

Figure~\ref{fig:cp-delta-error-time} shows the 1$\sigma$ uncertainty of $\deltacp$ as a function of the integrated beam power for $\sin^22\theta_{13} =0.1$ and 0.01.
The mass hierarchy is assumed to be known.
The ratio of $\nu$ and $\bar{\nu}$ running time is fixed to 7:3.

\begin{figure}[tbp]
\includegraphics[width=0.5\textwidth]{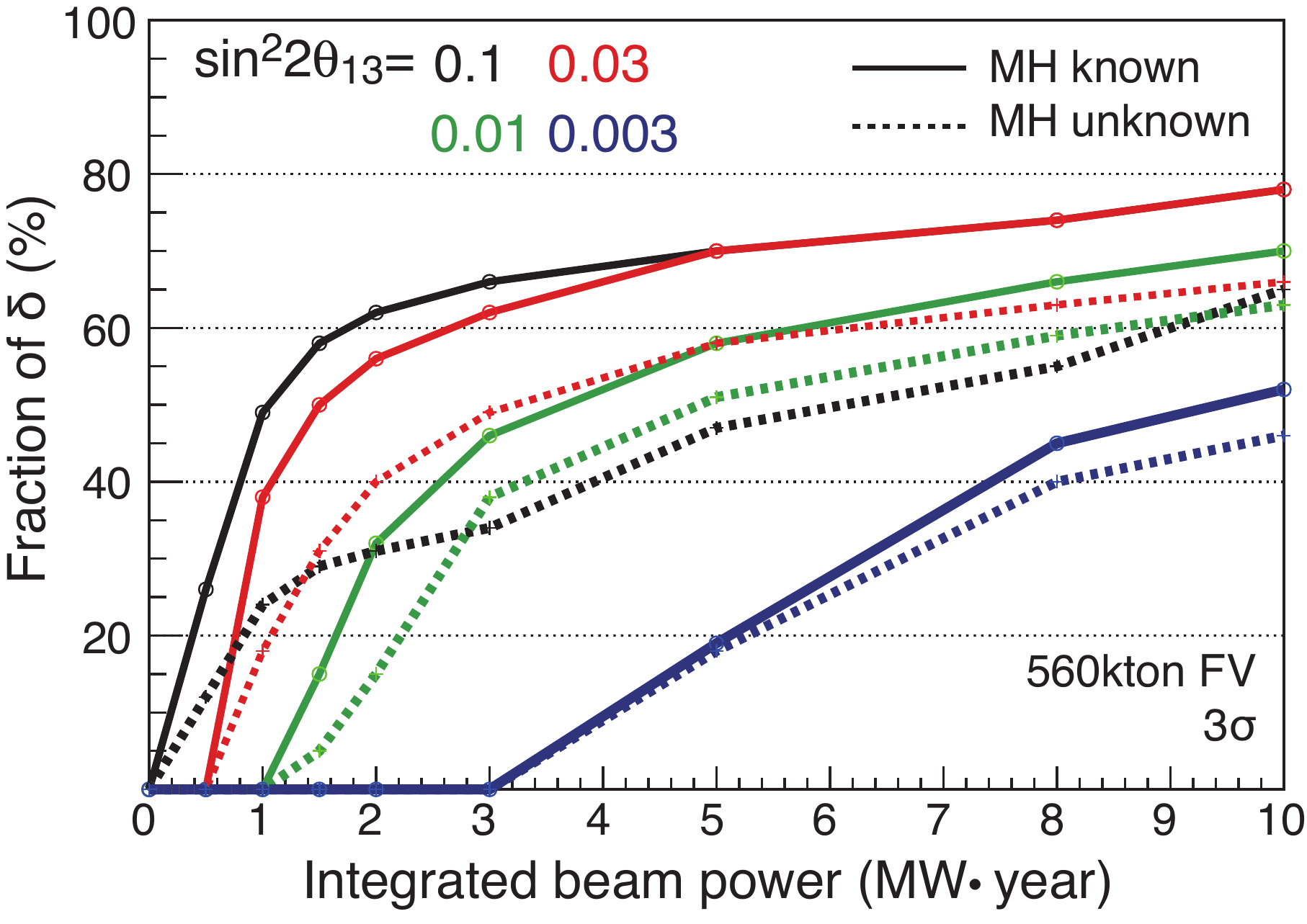}
\caption{Fraction of $\deltacp$ for which $\sin\deltacp= 0$ can be excluded with 3 $\sigma$ as a function of the integrated beam power.
The ratio of neutrino and anti-neutrino mode is fixed to 3:7. 
\label{fig:cp-sens-time}}
\end{figure}
Figure~\ref{fig:cp-sens-time} shows the sensitivity to $CP$ violation as a function of the integrated beam power.
The vertical axis shows the fraction of $\deltacp$ for which $\sin\deltacp= 0$ is excluded with 3$\sigma$ significance.
The ratio of neutrino and anti-neutrino mode is fixed to 3:7.
Solid and dashed lines correspond to the case the mass hierarchy is known and unknown, respectively.
The true mass hierarchy is normal in both cases.
Although the sensitivity becomes worse if the mass hierarchy is unknown, $CP$ asymmetry can be accessed for more than half of $\deltacp$ parameter region if $\sin^22\theta_{13} >0.01$.

\begin{figure}[tbp]
\includegraphics[width=0.5\textwidth]{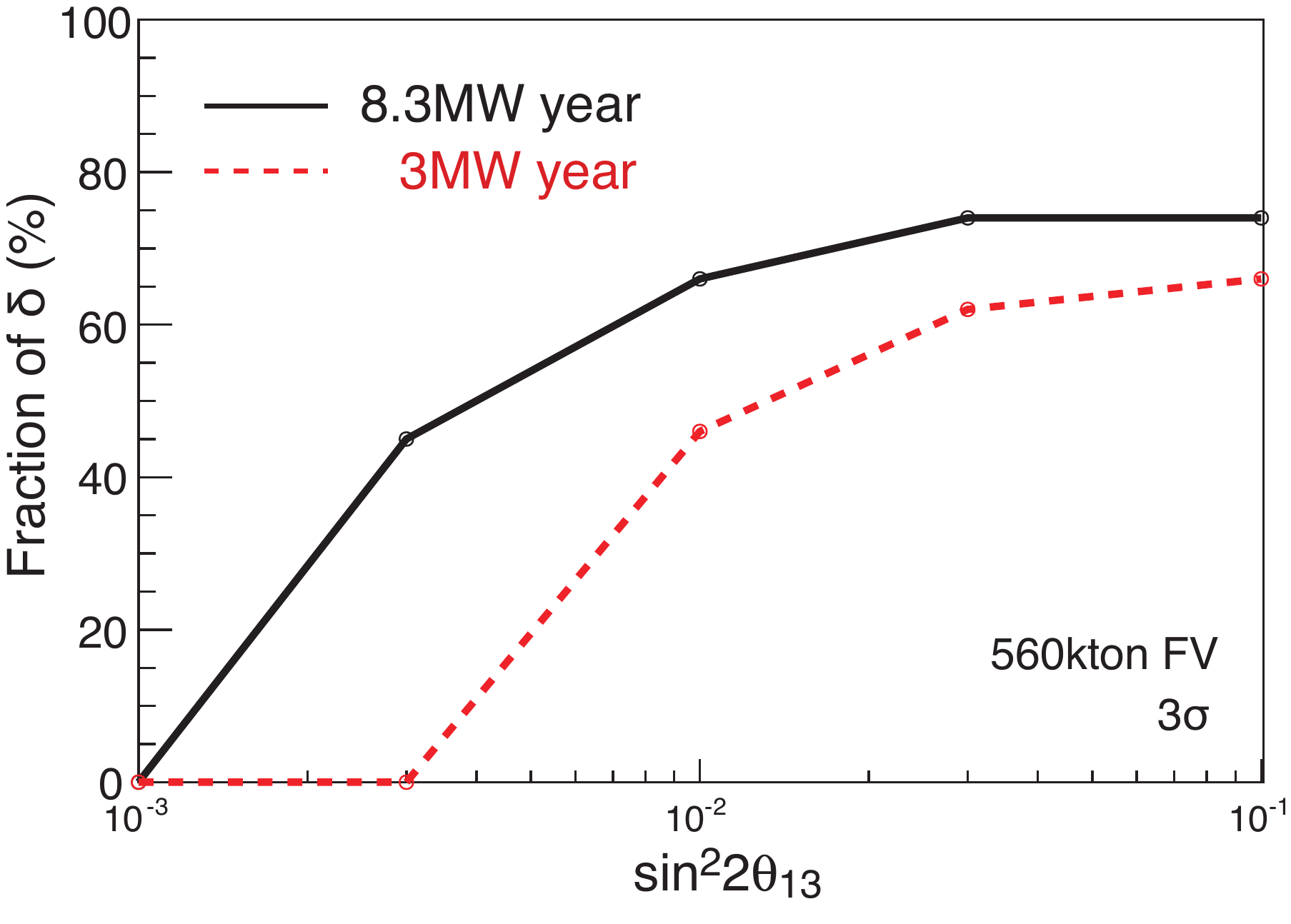}
\caption{Fraction of $\deltacp$ for which $\sin\deltacp= 0$ is excluded with three sigma as a function of the true value of 
$\sin^22\theta_{13}$. Red and Black lines represent 3~MW$\cdot$year and 8.3~MW$\cdot$year, respectively.
\label{fig:cpv-dfrac-vs-theta}}
\end{figure}

Figure~\ref{fig:cpv-dfrac-vs-theta} shows the fraction of $\deltacp$ for which $\sin\deltacp= 0$ is excluded with  3$\sigma$ as a function of the true value of $\sin^22\theta_{13}$, for 3~MW$\times 10^7$s and 8.3~MW$\times 10^7$s of integrated beam power. 
The mass hierarchy is assumed to be normal hierarchy and measured prior to this experiment.

The fraction of $\deltacp$ for which $CP$ asymmetry can be discovered with more than 3$\sigma$ under several assumptions are summarized in Table~\ref{Tab:CPsummary}.

\begin{table}[htdp]
\caption{Fraction of $\deltacp$ (in \%) for which $CP$ asymmetry can be discovered with more than 3$\sigma$ under several assumptions. Integrated beam power is 8.3~MW$\cdot$year.}
\begin{center}
\begin{tabular}{cc||c|c}   
& & \multicolumn{2}{c}{mass hierarchy} \\ 
& & known & unknown \\ \hline \hline
 & 0.1 & 74 &  55 \\
$\sin^22\theta_{13}$& 0.03 & 74 &  63  \\
& 0.01 & 66 &  59  \\ 
\end{tabular}
\end{center}
\label{Tab:CPsummary}
\end{table}%

\subsubsection{Summary}
The sensitivity to leptonic $CP$ asymmetry of a long baseline experiment using 
a neutrino beam directed from J-PARC to the Hyper-Kamiokande detector has been studied.
The running time is assumed to be five years in total, with 1.66~MW of beam power.
Based on a full simulation of beamline and detector, it is found that 
$\deltacp$ and $\sin^22\theta_{13}$ can be well determined for the values of $\sin^22\theta_{13}$ indicated by the recent T2K result.

If the mass hierarchy is known,
for $\sin^22\theta_{13}>0.03$ the value of $\delta$ can be determined to better than 18$^\circ$ for all values of $\deltacp$
and
$CP$ violation in the lepton sector can be observed with 3$\sigma$ significance for 74\% of the possible values of $\deltacp$.
If we assume that the mass hierarchy is not known, the sensitivity to $CP$ violation is reduced due to degeneracy.
Even for this case, $CP$ violation can be observed with 3$\sigma$ significance for  55\% of $\deltacp$ parameter space if $\sin^22\theta_{13}=0.1$, and 63\% if $\sin^22\theta_{13}=0.03$ with five years of experimental operations.

For $\sin^22\theta_{13} > 0.05$, it is also possible to determine the mass hierarchy for some of $\deltacp$ with the J-PARC to Hyper-K experiment alone.
For $\sin^22\theta_{13} = 0.1$, the mass hierarchy can be determined with more than 3$\sigma$ significance for 46\% of the $\deltacp$ parameter space.

\clearpage

%% file: physics-atmnu/atmnu.tex
\subsection{Atmospheric neutrinos}\label{section:atmnu}

\subsubsection{Goals of the atmospheric neutrino study}

Atmospheric neutrinos are a guaranteed neutrino source 
in the Hyper-Kamiokande experiment.
The indication by T2K \cite{Abe:2011sj}
that $\theta_{13}$ is potentially large
is encouraging for future atmospheric neutrino studies, 
as there will be a good chance
to extract information on neutrino properties via
the three flavor oscillation effect.
Assuming $\sin^22\theta_{13} > 0.04$ 
as the global fit result suggests \cite{Fogli:2011qn}, 
the targets of the atmospheric neutrino studies in Hyper-K would be:
\begin{itemize}
\item  mass hierarchy determination, namely to select $\Delta m^2_{32}>0$ or $\Delta m^2_{32}<0$ with more than 3$\sigma$ significance provided 
$\sin^2\theta_{23}>0.4$.
\item  to solve $\sin^2\theta_{23}$ octant degeneracy, namely to discriminate
$\sin^2\theta_{23}<0.5$ (first octant) from $\sin^2\theta_{23}>0.5$ 
(second octant), 
when the mixing is not maximal as $\sin^22\theta_{23}<0.99$.
\item  to obtain complementary information on $CP$ phase $\delta$.
\end{itemize}
To extract the expected three flavor oscillation effects,
we will study atmospheric electron neutrino flux variations 
as well as muon neutrino flux variations.
Expected sensitivities for all of these topics are discussed in this section.
\begin{figure}[thb]
  \begin{center}
    \includegraphics[scale=1.0]{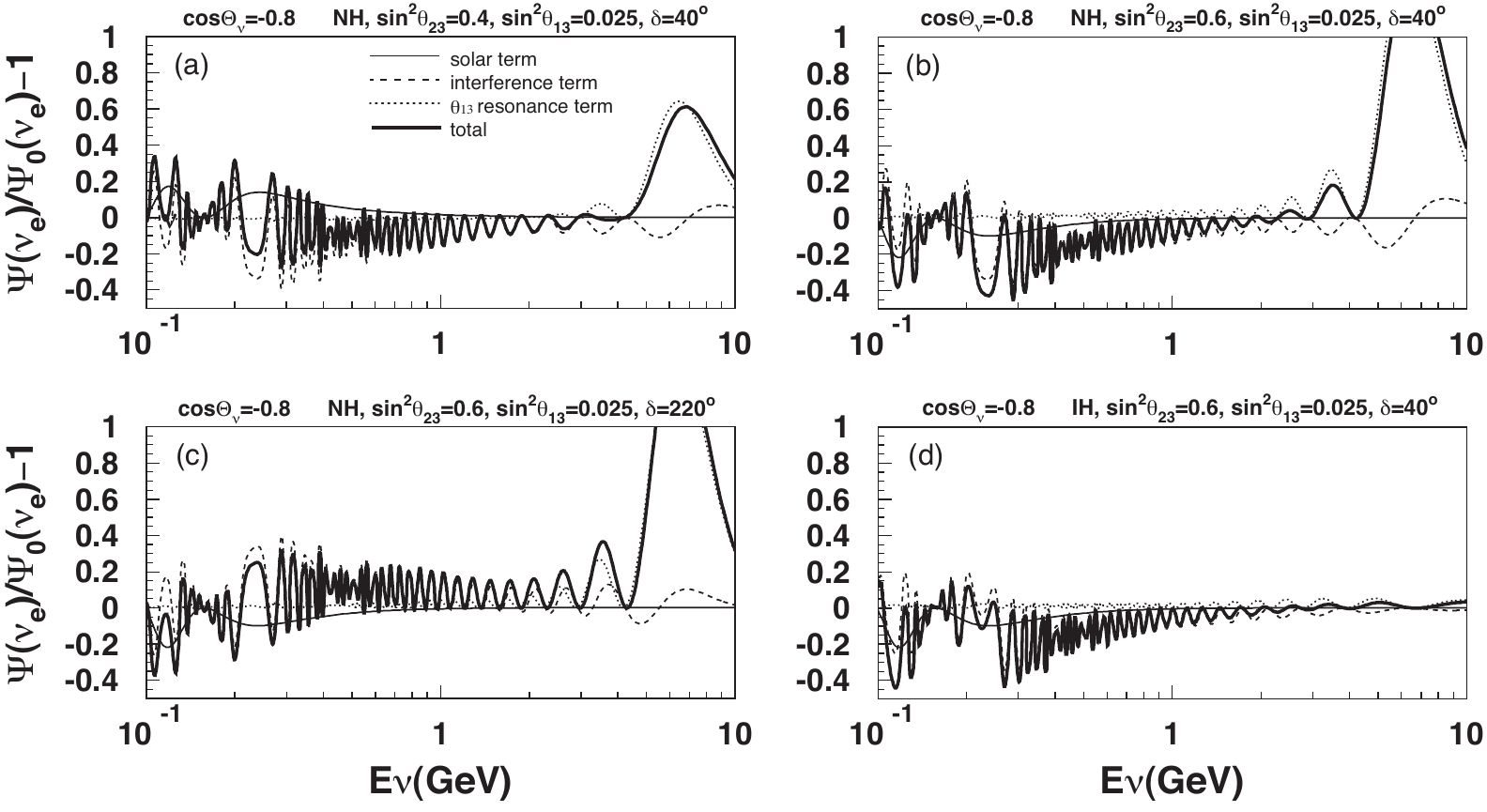}
  \end{center}
  \caption{Oscillated \(\nu_e\) flux relative to the non-oscillated flux
           as a function of neutrino energy
           for the upward-going neutrinos 
           with zenith angle $\cos\Theta_\nu=-0.8$.
           $\bar \nu_e$ is not included in the plots.
           Thin solid lines, dashed
           lines, and dotted lines correspond to the solar
           term, the interference term, and the 
           \(\theta_{13}\) resonance term, 
           respectively
           (see Eq. \ref{eqn:nue-oscillation}).  Thick solid lines
           are total fluxes.
           Parameters are set as
           \((\sin^2\theta_{12},\sin^2\theta_{13},\sin^2\theta_{23},\delta,
           \Delta m^2_{21},\Delta m^2_{32})\)
           \(=(0.31, 0.025, 0.6, 40^\circ,\)
           \(7.6\times 10^{-5}\mbox{eV}^2, +2.4\times 10^{-3}\mbox{eV}^2)\)
           unless otherwise noted.
           The $\theta_{23}$ octant effect can be seen by comparing  
           (a) (\(\sin^2\theta_{23}=0.4\)) and 
           (b) (\(\sin^2\theta_{23}=0.6\)).
           $\delta$ value is changed to $220^\circ$ in (c) to be compared
           with $40^\circ$ in (b).  
           The mass hierarchy is inverted only in (d) so
           \(\theta_{13}\) resonance (MSW) effect disappears in this plot.
           For the inverted hierarchy the MSW
           effect should appear in the $\bar \nu_e$ flux, 
           which is not shown in the plot.
  \label{fig:nue-example}
           }
\end{figure}

Oscillation probabilities of atmospheric neutrinos
in the three flavor neutrino mixing scheme have been 
discussed by many authors,
and the oscillation 
effect in electron neutrino flux is analytically calculated 
\cite{Peres:2003wd} as: 
\begin{eqnarray}
 \frac{\Phi(\nu_e)}{\Phi_0(\nu_e)} -1 & \approx & P_2\cdot(r\cdot \cos^2\theta_{23}-1) \nonumber \\
&& -r\cdot \sin\tilde{\theta}_{13}\cdot \cos^2\tilde{\theta}_{13}\cdot \sin{2\theta_{23}}\cdot(\cos{\delta}\cdot R_2-\sin{\delta}\cdot I_2) \nonumber \\
&& +2\sin^2\tilde{\theta}_{13}\cdot(r\cdot \sin^2\theta_{23}-1)
\label{eqn:nue-oscillation}
\end{eqnarray}
where we call the first, second, and third terms the 
``solar term'', ``interference term'', and ``$\theta_{13}$ resonance term'',
respectively.
\(P_2\) is the two neutrino transition probability of \(\nu_e
\rightarrow \nu_{\mu,\tau}\) 
which is driven by the solar neutrino
mass difference \(\Delta m^2_{21}\).
$R_2$ and $I_2$ represent oscillation amplitudes for $CP$ even and odd
terms.
For anti-neutrinos, the probabilities $P_2, R_2, I_2$ are obtained
by replacing the matter potential $V \rightarrow -V$, and the sign of
the $\delta$ (see \cite{Peres:2003wd} for details).
\(r\) is the \(\nu_\mu/\nu_e\) flux
ratio as a function of neutrino energy; $r \approx 2$ at sub-GeV energies, 
starts deviating from 2 at 1 GeV, and reaches to $\sim 3$ at 10 GeV.
The $\tilde{\theta}_{13}$ is an 
effective mixing angle in the Earth; 
\(\sin^2\tilde{\theta}_{13}\) could become large at \(5\sim10\) GeV neutrino
energy due to the matter potential
\cite{Wolfenstein:1977ue,Mikheyev:1985zz,Mikheyev:1986zz}.
This MSW resonance happens with neutrinos in the case of
normal mass hierarchy ($\Delta m^2_{32} > 0$), and with anti-neutrinos
in the case of inverted mass hierarchy ($\Delta m^2_{32} < 0$).

In order to demonstrate the behavior of these three terms,
Fig. \ref{fig:nue-example} shows how the \(\nu_e\) flux changes 
as a function of neutrino energy based on a numerical
calculation of oscillation probabilities,
in which the matter density profile in the Earth
is taken into account  
\cite{Barger:1980tf,Wendell:2010md}.
We adopted an Earth model constructed by the median density in
each of the dominant regions of the preliminary reference Earth model (PREM)
\cite{PREM}: 
inner core $(   0  \leq r < 1220 \mbox{km})$ 13.0 $\mbox{g/cm}^{3}$,
outer core $(1220  \leq r < 3480 \mbox{km})$ 11.3 $\mbox{g/cm}^{3}$,
mantle     $(3480  \leq r < 5701 \mbox{km})$  5.0 $\mbox{g/cm}^{3}$, and
the crust  $(5701  \leq r < 6371 \mbox{km})$  3.3 $\mbox{g/cm}^{3}$.
In Fig. \ref{fig:nue-example}
dotted lines correspond to the \(\theta_{13}\) resonance term 
(the third term in Eq. \ref{eqn:nue-oscillation}),   
which could make a significant contribution in the \(5\sim10\) GeV region 
if \(\sin^2\theta_{13}\) is a few percent
\cite{Abe:2011sj,Fogli:2011qn}.
The resonance effect in $\nu_e$ (not $\bar \nu_e$) can be seen in 
Fig. \ref{fig:nue-example}(b) which assumes normal mass hierarchy,
but resonance does not occur in Fig. \ref{fig:nue-example}(d) 
which assumes inverted hierarchy.
For the case of $\bar \nu_e$, the situation is reversed.
By using statistically enhanced $\nu_e$ and $\bar \nu_e$ samples 
in observed data,
this resonance term would enable us 
to reveal the neutrino mass hierarchy.
In addition, this term (as well as the solar term)  
has discrimination
power for the  \(\theta_{23}\) octant because this term is proportional to
\((r\cdot \sin^2\theta_{23}-1)\).
The effect of $\theta_{23}$ can be seen by comparing
Fig. \ref{fig:nue-example}(a) and (b).
Dashed lines in Fig. \ref{fig:nue-example} correspond to the
interference term (second term in Eq. \ref{eqn:nue-oscillation}).
The $CP$ phase effect would be dominant  
in the region from a few 100 MeV to a few GeV neutrino energy
as shown in Fig. \ref{fig:nue-example}(b) ($\delta=40^\circ$)
and (c) ($\delta=220^\circ$).
This interference term is proportional to $\sin2\theta_{23}$
(the first power) but the ambiguity of the parameter
is as small as a few \% and therefore $CP$ sensitivity
does not much depend on the value of $\theta_{23}$.

\subsubsection{Analysis method}
Atmospheric neutrino interactions
are simulated by the Monte Carlo method
using a flux calculation \cite{Honda:2006qj}
and interaction models \cite{hayato:neut,Mitsuka:2007zz,Mitsuka:2008zz}.  
The propagation of secondary particles and Cherenkov photons
in detector water and event
reconstructions are performed by using the Super-K detector simulator 
and standard reconstruction tools \cite{Ashie:2005ik}.  
We expect that -- by design -- the detector performance of Hyper-K will be basically the same
as that of Super-K.   In addition to using similar photodetectors and electronics, and having similar quality water, the length scale of each segmented Hyper-K subdetector compartment is 50 m, which is only 1.5 times larger than that of Super-K.
We have generated Monte Carlo (MC) events with
11 Megaton$\cdot$years statistics in the fiducial volume
which correspond to a 500 year exposure of Super-K
or 25 years of Hyper-K.
Oscillation probabilities are calculated by taking into account
the full parameters in the standard 3 flavor neutrino scheme;
\(\theta_{12}\), \(\theta_{13}\), \(\theta_{23}\), \(\delta\),
and two squared mass differences of \(\Delta m^2_{21}\) and \(\Delta m^2_{32}\).
\begin{table}[htdp]
\caption{Oscillation parameters are set to be these values 
         in this section, unless otherwise noted.}
\begin{center}
\begin{tabular}{cc} \hline \hline
$\Delta m^2_{21}$ & 7.6$\times 10^{-5}$~eV$^2$ \\ \hline
$\Delta m^2_{32}$ & 2.4$\times 10^{-3}$~eV$^2$ \\ \hline
$\sin^2\theta_{12}$ & 0.31 \\ \hline
$\sin^2\theta_{13}$ & 0.025 \\ \hline
$\sin^2\theta_{23}$ & 0.5 \\ \hline
$\delta$ & $40^\circ$ \\ \hline\hline
\end{tabular}
\end{center}
\label{tab:atmnu-params}
\end{table}%
Oscillation parameters, if they are not otherwise specified,
are set to the values shown in Table~\ref{tab:atmnu-params}.
The $\theta_{23}$ value is set to be 
the maximal mixing suggested by atmospheric and
accelerator neutrino experiments
\cite{Ashie:2005ik,Ashie:2004mr,Hosaka:2006zd,Ahn:2006zza,Adamson:2011ig},
and other parameters are taken from a recent global fit result \cite{Fogli:2011qn}
except for the random choice of the $\delta$ value.
In the mass hierarchy determination study, 
we test both signs of $\Delta m^2_{32}$.

\begin{table}[htdp]
\caption{Expected number of $\nu_e$-like and $\bar{\nu}_e$-like events
         in 10 Hyper-K years
         for each interaction component.}
\begin{center}
\begin{tabular}{lccccc} \hline \hline
& ~~~CC $\nu_{e}$~~~ & ~~~CC $\bar \nu_{e}$~~~ & CC $\nu_{\mu} + \bar \nu_{\mu}$ & ~~~~NC~~~~ & ~~~Total~~~ \\
\hline
$\nu_e$-like sample & 15247 & 2831 & 3731 & 4792 & 26601 \\
~~~~~~- single-ring & 6356 & 1086 & 1682 & 1740 & 10864 \\ 
~~~~~~- multi-ring & 8891 & 1745 & 2049 & 3052 & 15737 \\ 
Percentage (\%) & 57.3 & 10.6 & 14.0 & 18.0 & 100.0 \\ 
\\
$\bar \nu_e$-like sample & 28309 & 17255 & 1232 & 4559 & 51355 \\
~~~~~~- single-ring & 20470 & 13401 & 444 & 2496 & 36811 \\ 
~~~~~~- multi-ring & 7839 & 3854 & 788 & 2063 & 14544 \\ 
Percentage (\%) & 55.1 & 33.6 & 2.4 & 8.9 & 100.0 \\ 
 \hline\hline
\end{tabular}
\end{center}
\label{tab:nueandnuebar}
\end{table}%
The MC events are divided into three classes: fully contained (FC),
partially contained (PC), and upward-going muons (UP$\mu$).
The FC events are further divided into several sub-samples based 
on the Super-K analyses by using reconstructed
variables such as visible energy, the number of Cherenkov rings,
particle type ($e$-like or $\mu$-like), the number of muon decay
electrons, and so on \cite{Wendell:2010md}.
In addition, the multi-GeV single-ring and multi-ring $e$-like events 
are regrouped into $\nu_e$-like and $\bar{\nu}_e$-like samples
in the following manner.
In the charged current (CC) non quasi-elastic (QE) interaction components
among the single-ring sample, 
$\pi^+ (\pi^-)$ is expected to be more copiously produced in 
$\nu_e (\bar \nu_e)$ interactions because the secondary charged lepton
is $e^- (e^+)$.
The $\pi^{+}$ decays 
into $\mu^{+}$ which in turn produces a delayed signal of a decay electron, 
while in the case of $\pi^{-}$ 
it is often absorbed in water before decaying into $\mu^-$,
so no decay electron would be produced. 
Therefore, events with more than one decay electron are classified
as $\nu_{e}$-like,
while events having no decay electron are classified 
as $\bar \nu_{e}$-like.
As for the multi-GeV multi-ring $e$-like sample, the events are also divided 
into $\nu_{e}$-like and $\bar \nu_{e}$-like samples 
via a likelihood method. 
CC $\nu_{e}$ interactions 
tend to have a larger Feynman $y$ distribution than CC $\bar \nu_{e}$, 
therefore CC $\nu_{e}$ 
is expected to have larger transverse momentum, more rings,
and more muon decay electrons. 
Hence these three observables
are used in the construction of a likelihood function, 
and multi-ring events are divided into $\nu_{e}$-like and $\bar \nu_{e}$-like
sub-samples by applying the likelihood cut.
Table \ref{tab:nueandnuebar} shows the expected number of 
non-oscillated $\nu_e$ and 
$\bar \nu_e$-like events in 10 years exposure 
(for the $\nu_e$-like sample: CC $\nu_e$ 57.3$\%$, CC $\bar \nu_e$ 10.6$\%$, 
while for the $\bar \nu_e$-like sample: CC $\nu_e$ 55.1$\%$, CC $\bar \nu_e$ 33.6$\%$).
Due to the small fraction of wrong sign electrons  
(CC $\bar \nu_e \sim 10\%$)
in the $\nu_e$-like sample, we expect 
a significant enhancement of the CC $\nu_e$ component 
(CC $\nu_e \sim 60\%$) in the case of the normal hierarchy,
while less excess is to be expected in the case of the inverted hierarchy.  
The $\bar \nu_e$-like sample will provide independent information
on the mass hierarchy with its higher CC $\bar \nu_e$ fraction.

These MC events are used as fake data as well as expectation.  
To make fake data, the MC is weighted by livetime
and oscillation probabilities.
The $\chi^2$ is defined by comparing each fake data set with
the expectation in terms of Poisson statistics: 
\begin{eqnarray}
\chi^2 & \equiv & \sum_n
\left[
2
\left(
N^n_{\rm MC}(1+\sum_i f^n_i \cdot \epsilon_i)-N^n_{\rm DT}
\right)
+2N^n_{\rm DT}\ln
\left(
\frac{N^n_{\rm DT}}{N^n_{\rm MC}(1+\sum_i f^n_i \cdot \epsilon_i)}
\right)
\right] \nonumber \\
& + & \sum_i
\left(
\frac{\epsilon_i}{\sigma_i}
\right)^2
\label{eqn:chi2}
\end{eqnarray}
where
\begin{eqnarray*}
n & : & \mbox{counter for event type, momentum, and zenith angle} \\
N^n_{\rm MC} & : & \mbox{the number of MC events in the \(n\)-th bin} \\
N^n_{\rm DT} & : & \mbox{the number of fake data events in the \(n\)-th bin} \\
i & : & \mbox{counter for uncertainties} \\
\epsilon_i & : & \mbox{the list of systematic uncertainties (nuisance parameters)} \\
f^n_i & : & \mbox{error coefficients for $i$-th uncertainty and $n$-th event bin} \\
\sigma_i & : & \mbox{estimated size of systematic uncertainties} 
\end{eqnarray*}
The definition of \(\chi^2\) and data binning are
same as the latest Super-K three flavor neutrino oscillation analyses
\cite{Wendell:2010md}.
Moreover, 
systematic uncertainties (\(\sigma_i\)) 
based on \cite{Wendell:2010md} which covers
uncertainties of neutrino flux, interactions, detector response,
and event reconstructions 
are incorporated in this analysis.
Because the fake data are made from MC events, \(\chi^2\) minimum
is always obtained as \(\chi^2_{{\rm min}} = 0\) 
with all \(\epsilon_i=0\)
at the true oscillation parameters.
Then sensitivities of each oscillation parameters are evaluated by using
\(\Delta\chi^2 \equiv \chi^2_{\rm min}(\mbox{test point}) -
\chi^2_{\rm min}(\mbox{true point}) 
= \chi^2_{\rm min}(\mbox{test point}) \).

\begin{figure}[htb]
  \begin{center}
    \includegraphics[scale=0.35]{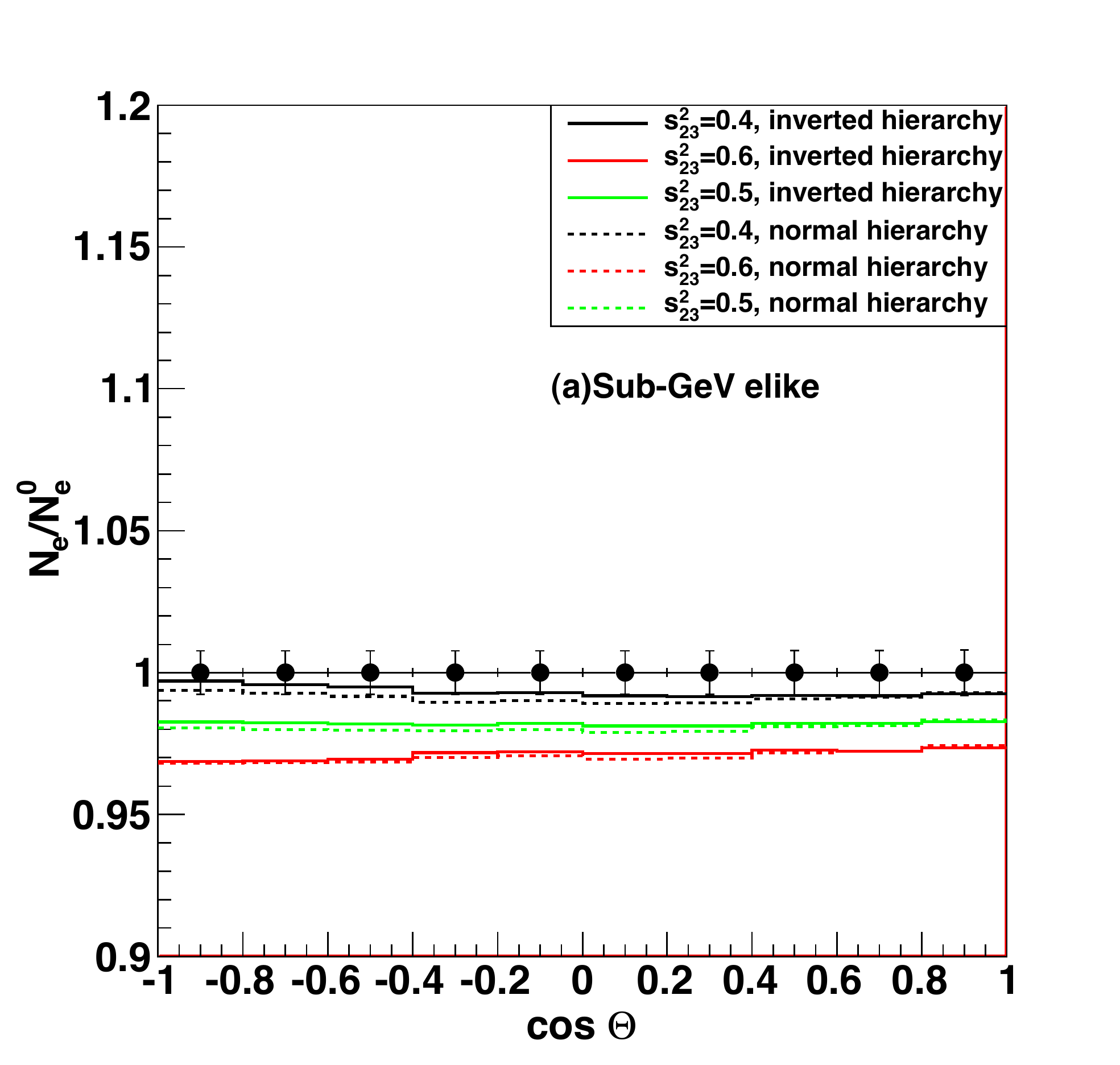}\\
    \includegraphics[scale=0.35]{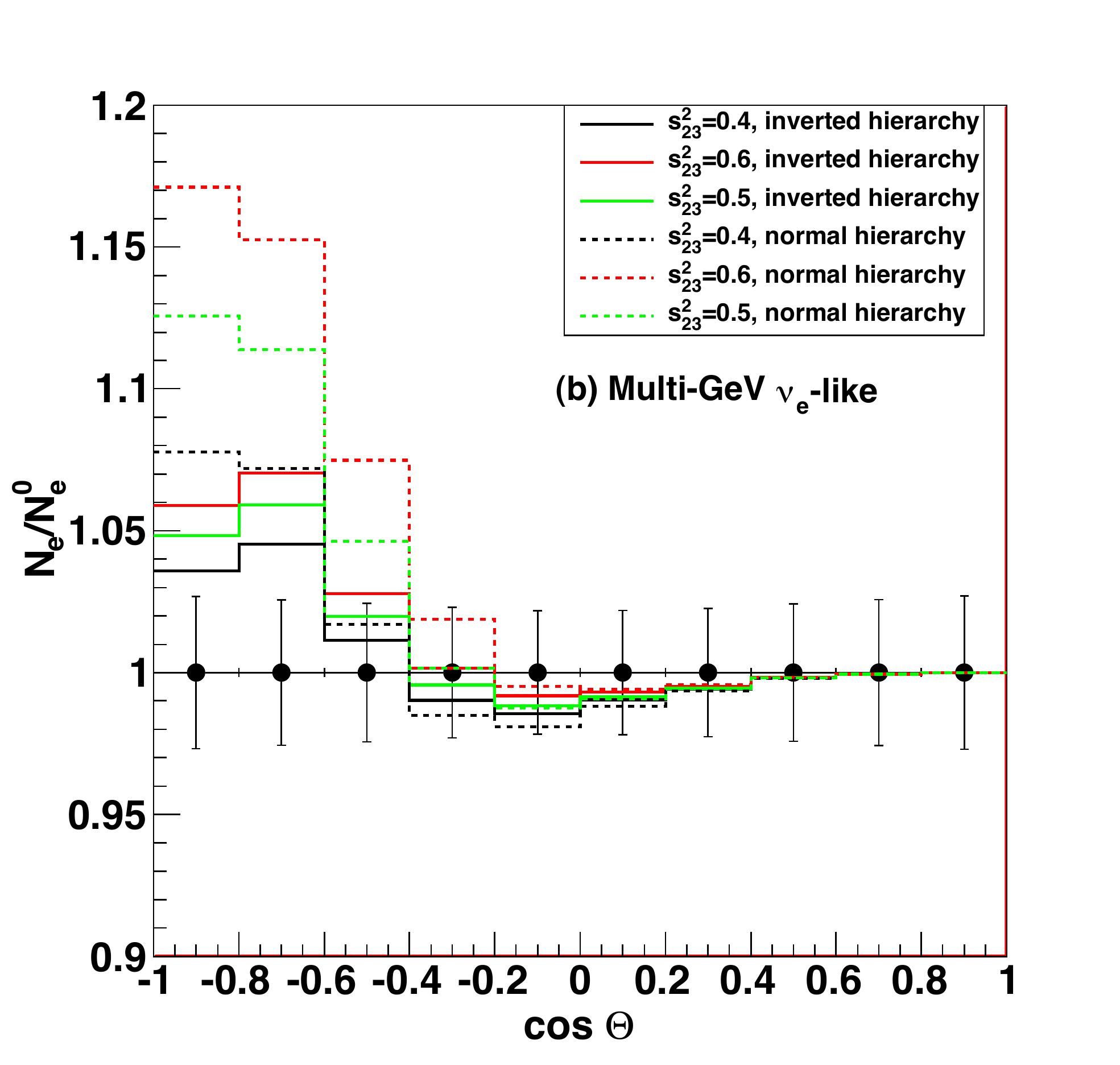}
    \includegraphics[scale=0.35]{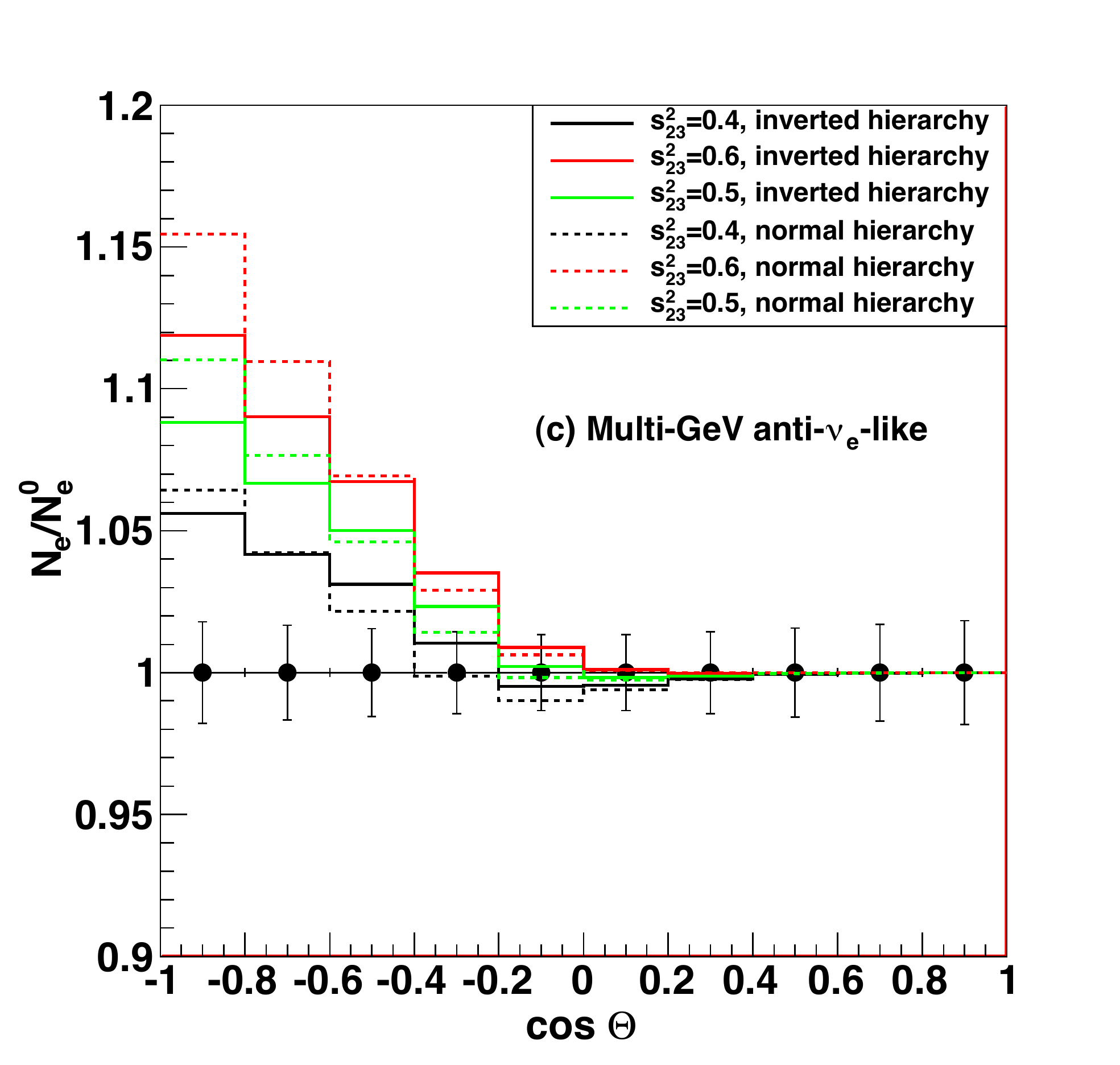}
  \end{center}
  \caption{Expected event rate changes
           in (a) sub-GeV single-ring \(e\)-like, 
           (b) multi-GeV $\nu_{e}$-like, 
           and (c) multi-GeV $\bar \nu_{e}$-like event samples.
           The vertical axis shows the ratio of oscillated $e$-like event
           rate to the non-oscillated one.
           Mass hierarchy is normal for dashed lines and inverted 
           for solid lines.
           Colors show $\sin^2\theta_{23}$ values as 0.4 (black),
           0.5 (green), and 0.6 (red).
           Points with error bars represent null oscillation expectations
           with expected statistical errors for 5.6 Megaton$\cdot$years 
           exposure or 10 years of Hyper-K.
           }
  \label{fig:hierarchy-zenith}
\end{figure}
\begin{figure}[htbp]
  \begin{center}
    \includegraphics[scale=0.35]{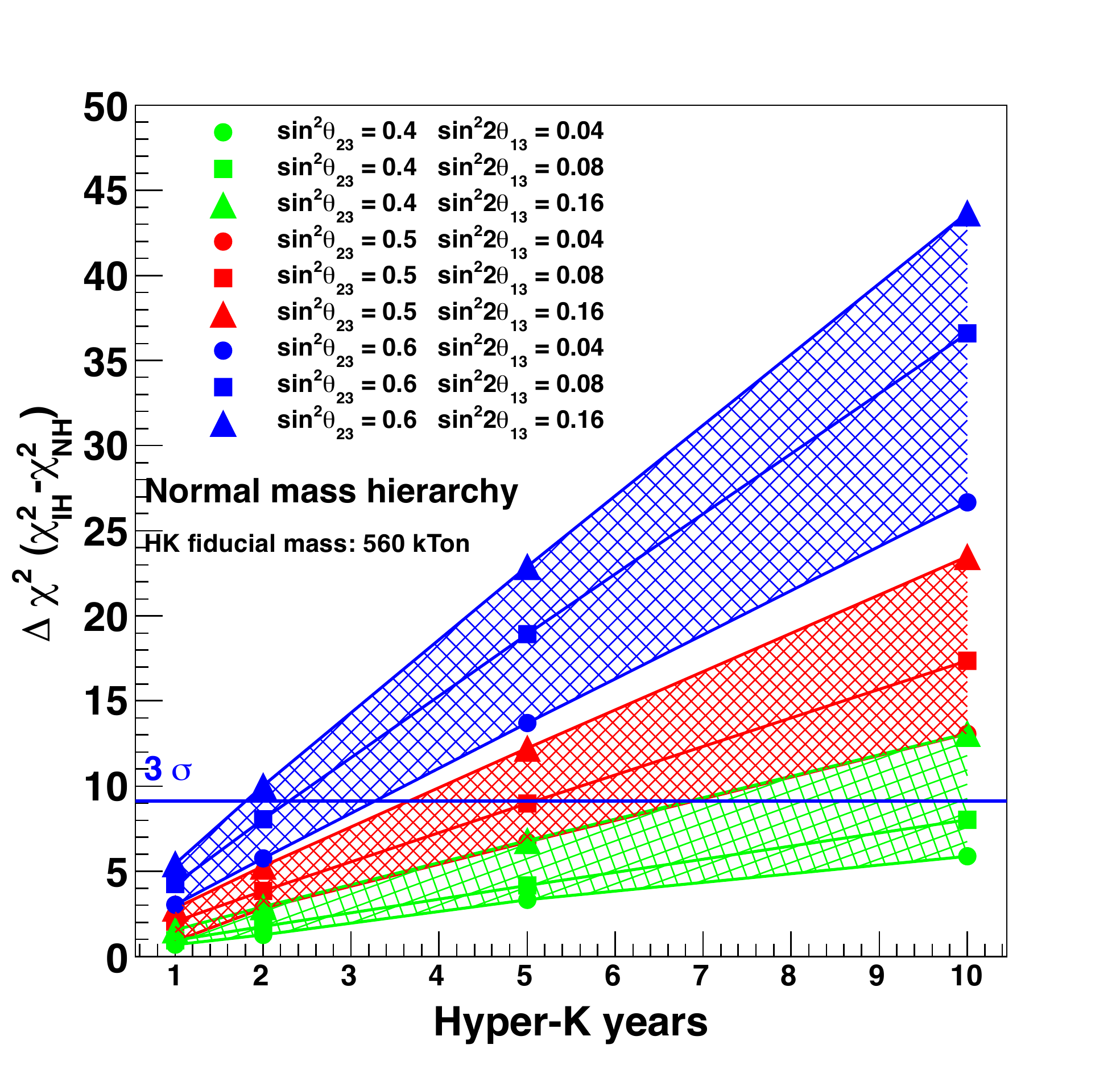}
    \includegraphics[scale=0.35]{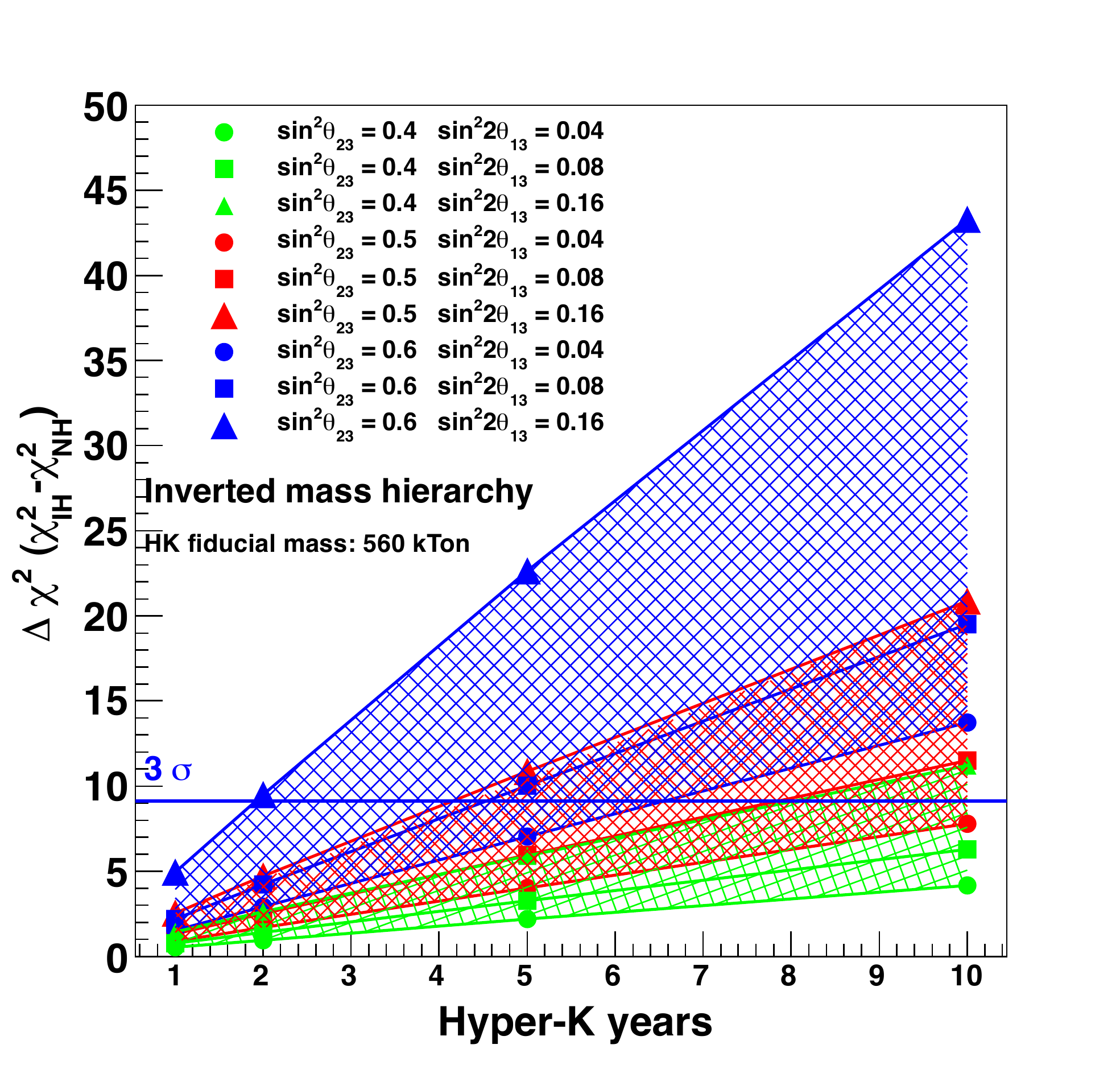}
  \end{center}
  \caption{Expected significance for mass hierarchy determination
           as a function of Hyper-K livetime in years.
           All of $\theta_{23}$, $\theta_{13}$, and $\delta$ 
           are assumed to be unknown and allowed to vary without any
           constraints.
           }
  \label{fig:hierarchy-time-thetafree}
\end{figure}
\begin{figure}[htbp]
  \begin{center}
    \includegraphics[scale=0.35]{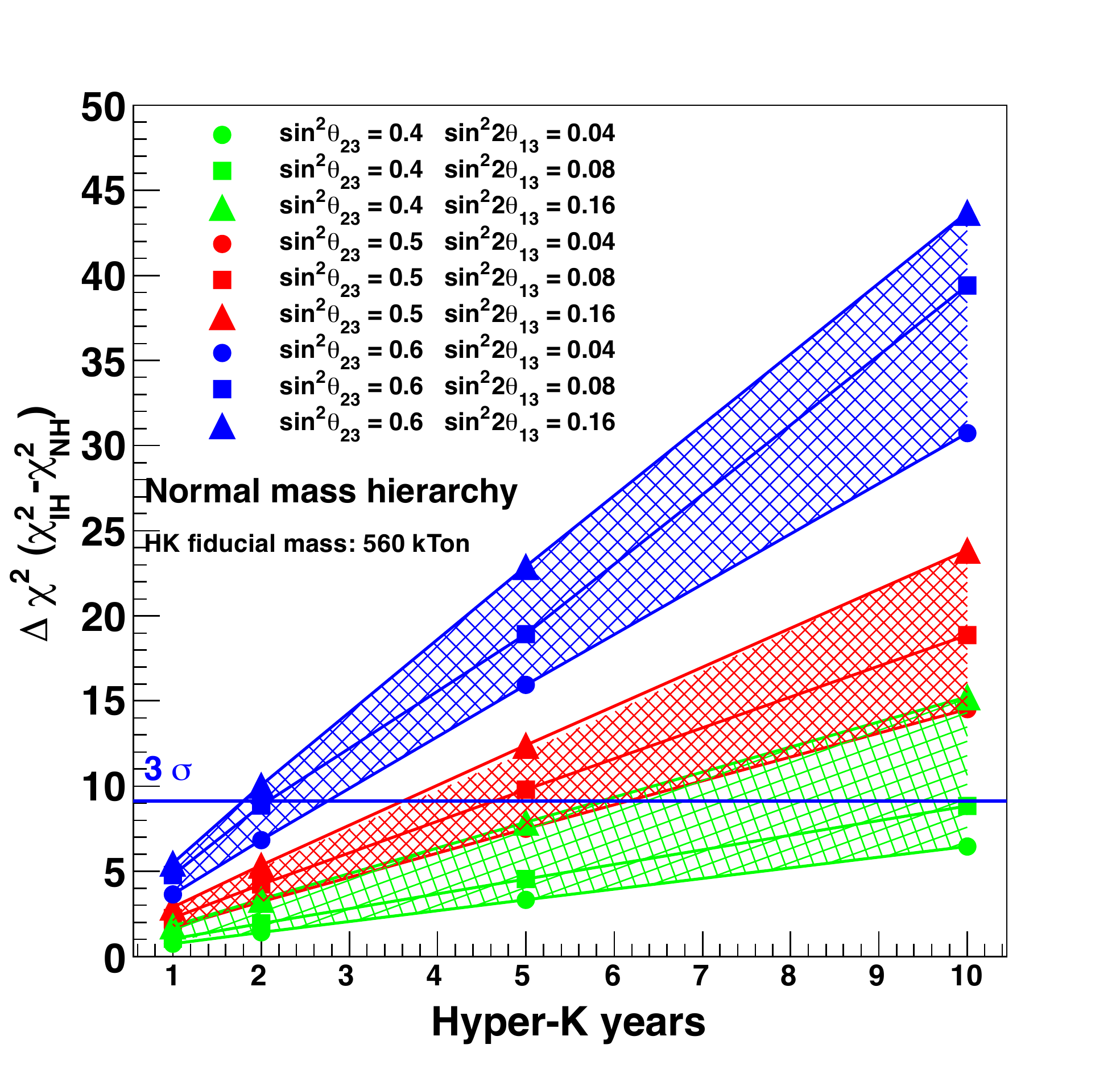}
    \includegraphics[scale=0.35]{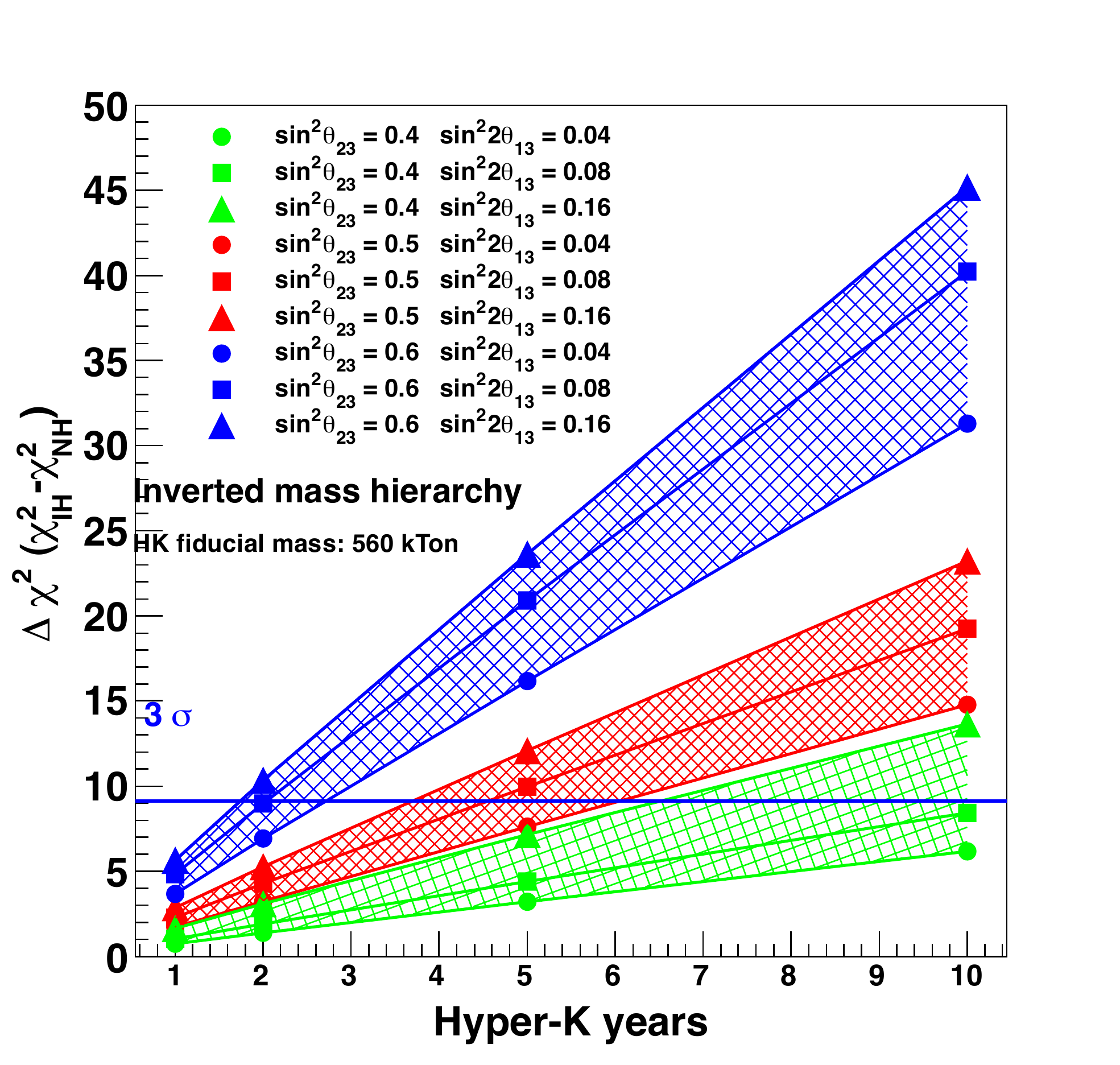}
  \end{center}
  \caption{Expected significance for mass hierarchy determination
           as a function of Hyper-K livetime in years.
           Both $\theta_{23}$ and $\theta_{13}$ are assumed to be well known
           and are fixed.
           }
  \label{fig:hierarchy-time}
\end{figure}
\begin{figure}[htbp]
  \begin{center}
    \includegraphics[scale=0.35]{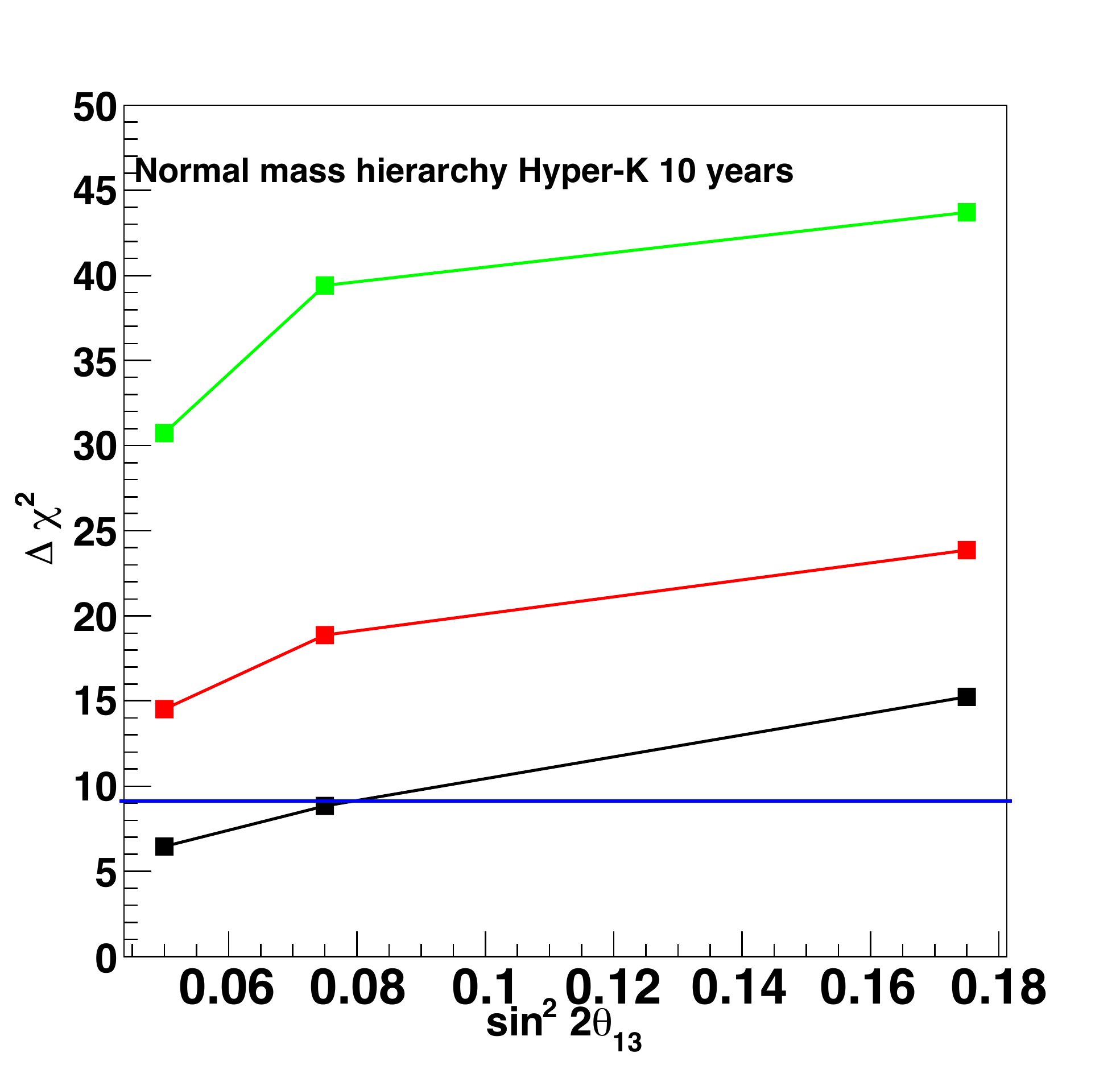}
    \includegraphics[scale=0.34]{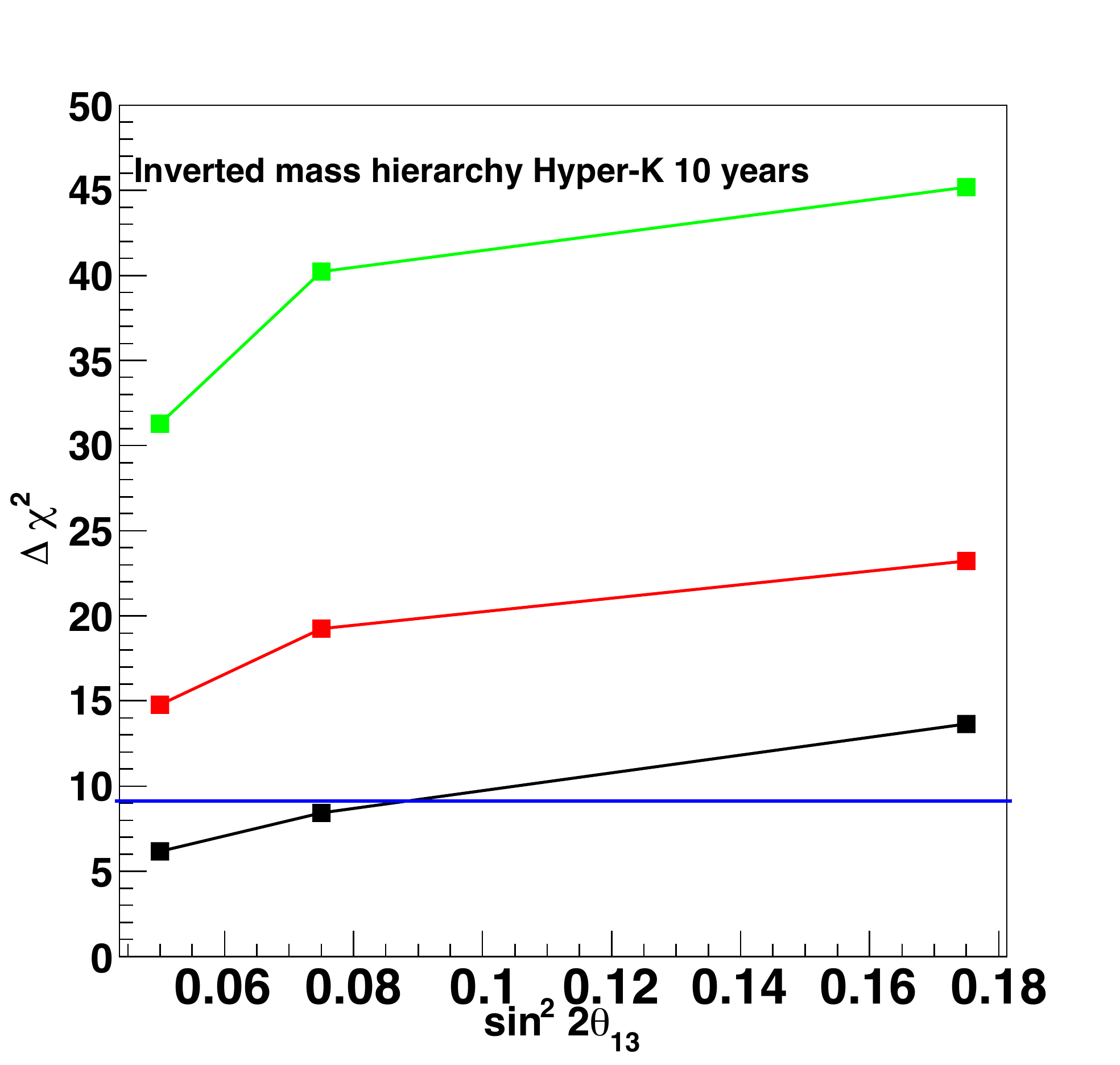}
  \end{center}
  \caption{Expected significance for the mass hierarchy determination.
           In the left panel, normal mass hierarchy is the case and
           $\chi^2$ for the wrong assumption; 
           $\Delta\chi^2 \equiv 
           \chi^2_{\rm min}{\rm (inverted)} - \chi^2_{\rm min}{\rm (normal)}$
           is shown for various true values of $\sin^22\theta_{13}$.
           The right panel is for the inverted hierarchy case.
           Each colors show the case of $\sin^2\theta_{23} =$
           0.4 (black), 0.5 (red), and 0.6 (green), 
           and the blue horizontal lines
           show $3\sigma (\Delta\chi^2=9.2)$.
           }
  \label{fig:hierarchy}
\end{figure}
\subsubsection{Neutrino mass hierarchy}
In Fig.~\ref{fig:hierarchy-zenith}, 
the expected zenith angle distributions of $e$-like events
are shown separately for sub-GeV $e$-like, multi-GeV $\nu_e$-like,
and multi-GeV $\bar \nu_e$-like sub-samples
as the ratio against the non-oscillated case.
We expect a sizable difference between normal (dashed lines)
and inverted (solid lines) hierarchy both in the $\nu_e$-like and
$\bar \nu_e$-like samples.  The difference becomes larger for
larger $\sin^2\theta_{23}$ because the resonance term 
is proportional to \((r\cdot \sin^2\theta_{23}-1)\) 
in Eq. \ref{eqn:nue-oscillation}.
Sensitivity for determining the neutrino mass hierarchy is studied
and shown in Fig. \ref{fig:hierarchy-time-thetafree}.
The $\Delta\chi^2$ for the wrong mass hierarchy assumption,
-- interpreted as significance of the mass hierarchy determination -- 
is shown as a function of livetime in years.
In the calculation of the significance, 
fitting parameters 
$\theta_{13}$, $\theta_{23}$, and $\delta$ are assumed to be unknown
and allowed to freely vary when obtaining the $\chi^2$ minimum.
The expected significance is larger for larger $\sin^2\theta_{23}$
and $\sin^2\theta_{13}$ as is expected.
With 10 years exposure,
the significance is more than $3\sigma$
for most of
the 
parameter sets
suggested by the global fit results \cite{Fogli:2011qn}. 
Figure \ref{fig:hierarchy-time} also shows the significance
but in this case both $\theta_{13}$ and $\theta_{23}$ are assumed to be well known
and are fixed.
The effect of the assumption that both mixing angles are known or not
can be seen most readily 
in the inverted hierarchy case, but even there the difference is rather limited.
In the case of 
$(\sin^22\theta_{13}, \sin^2\theta_{23}) = (0.08, 0.5)$ for example,
required exposure time to confirm inverted hierarchy with $3\sigma$ CL
would be
between 5 years ($\theta_{13}$ and $\theta_{23}$ are well known)
and 8 years ($\theta_{13}$ and $\theta_{23}$ are unknown).
On the other hand, we need $\sim 5$ years to confirm normal hierarchy
no matter what the state of knowledge regarding $\theta_{13}$ and $\theta_{23}$ is.
In conclusion
we need between $2\sim 10$ years of data to reach $3\sigma$ significance
for the parameter space
suggested by the global fit results
except for the case that $\sin^2\theta_{23}$
is smaller than 0.4.
Figure \ref{fig:hierarchy} summarizes the 10 year significance
for each $\sin^2\theta_{23}$ and $\sin^2\theta_{13}$ parameter set.

\subsubsection{Octant of \(\theta_{23}\)}
\begin{figure}[htbp]
  \begin{center}
    \includegraphics[scale=0.35]{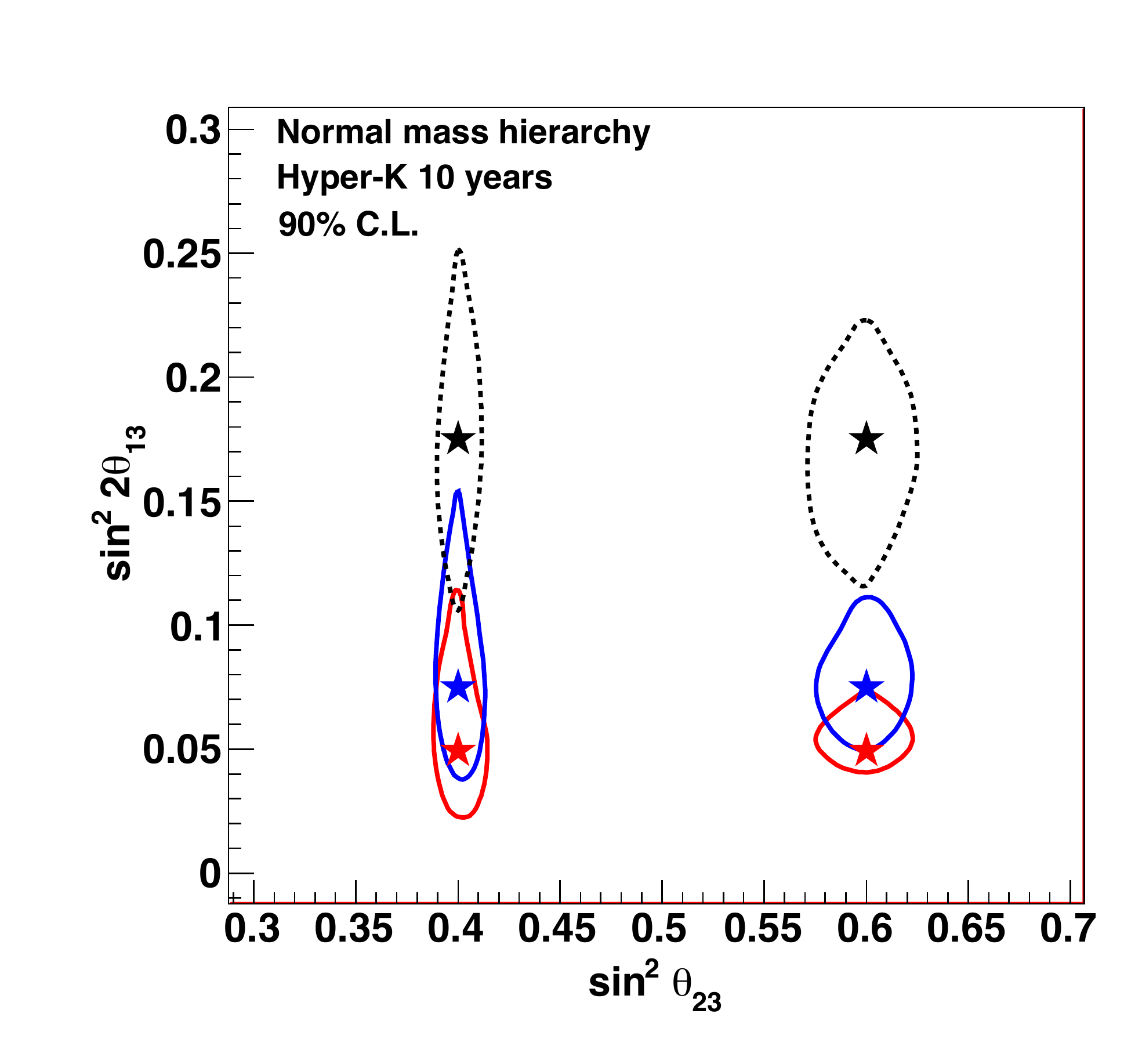}
    \includegraphics[scale=0.35]{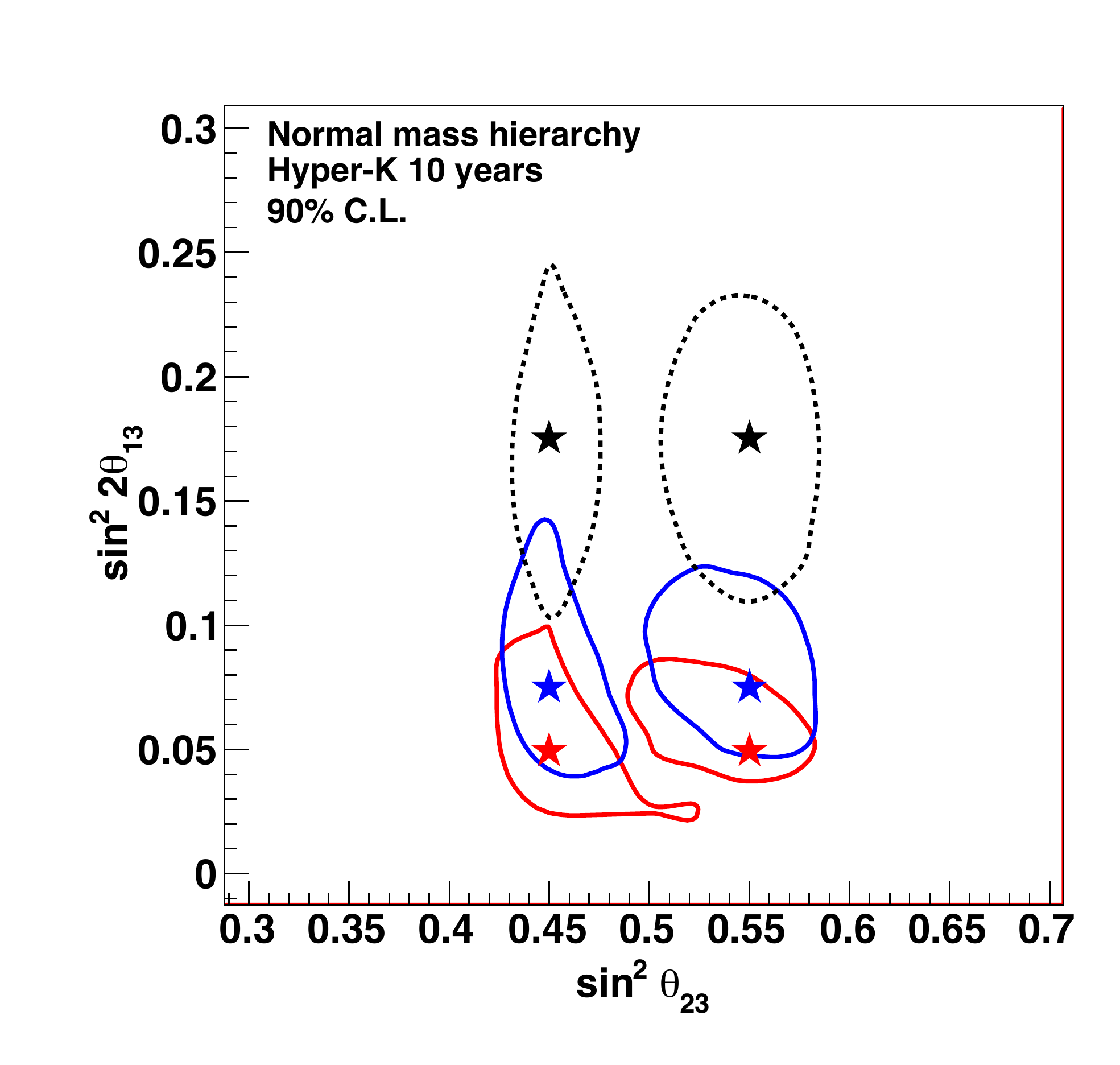}
  \end{center}
  \caption{Expected sensitivities for \(\sin^2\theta_{23}\) 
           and \(\sin^22\theta_{13}\) at 90\% CL 
           with a livetime of 10 years in Hyper-K.  Stars in the contours
           represent the assumed true mixing angles.
           The left panel shows the case of $\sin^22\theta_{23} = 0.96$
           (\(\sin^2\theta_{23}=0.4\) or 0.6).
           The right panel shows the case of $\sin^22\theta_{23} = 0.99$
           (\(\sin^2\theta_{23}=0.45\) or 0.55).           
           Normal mass hierarchy is assumed.  
           }
  \label{fig:s23-octant-contour}
\end{figure}
Figure \ref{fig:hierarchy-zenith} already showed how the event rate changes
for different \(\sin^2\theta_{23}\) values.  
We expect event rate change and zenith angle distortions 
in multi-GeV $\nu_e$ and $\bar \nu_e$-like samples
as well as small rate changes in sub-GeV electrons.

Figure \ref{fig:s23-octant-contour} shows the expected sensitivities
for \(\sin^2\theta_{23}\) and \(\sin^2\theta_{13}\) 
with an exposure of 10 Hyper-K years.
Normal mass hierarchy is assumed in the figure.
We can expect to discriminate between \(\sin^2\theta_{23}=0.4\) and
0.6, which corresponds to \(\sin^22\theta_{23}=0.96\).
In the case of \(\sin^2\theta_{23}=0.45\) or 0.55, 
which corresponds to \(\sin^22\theta_{23}=0.99\),
discrimination is marginal but 
could be achieved.
\begin{figure}[htbp]
  \begin{center}
    \includegraphics[scale=0.35]{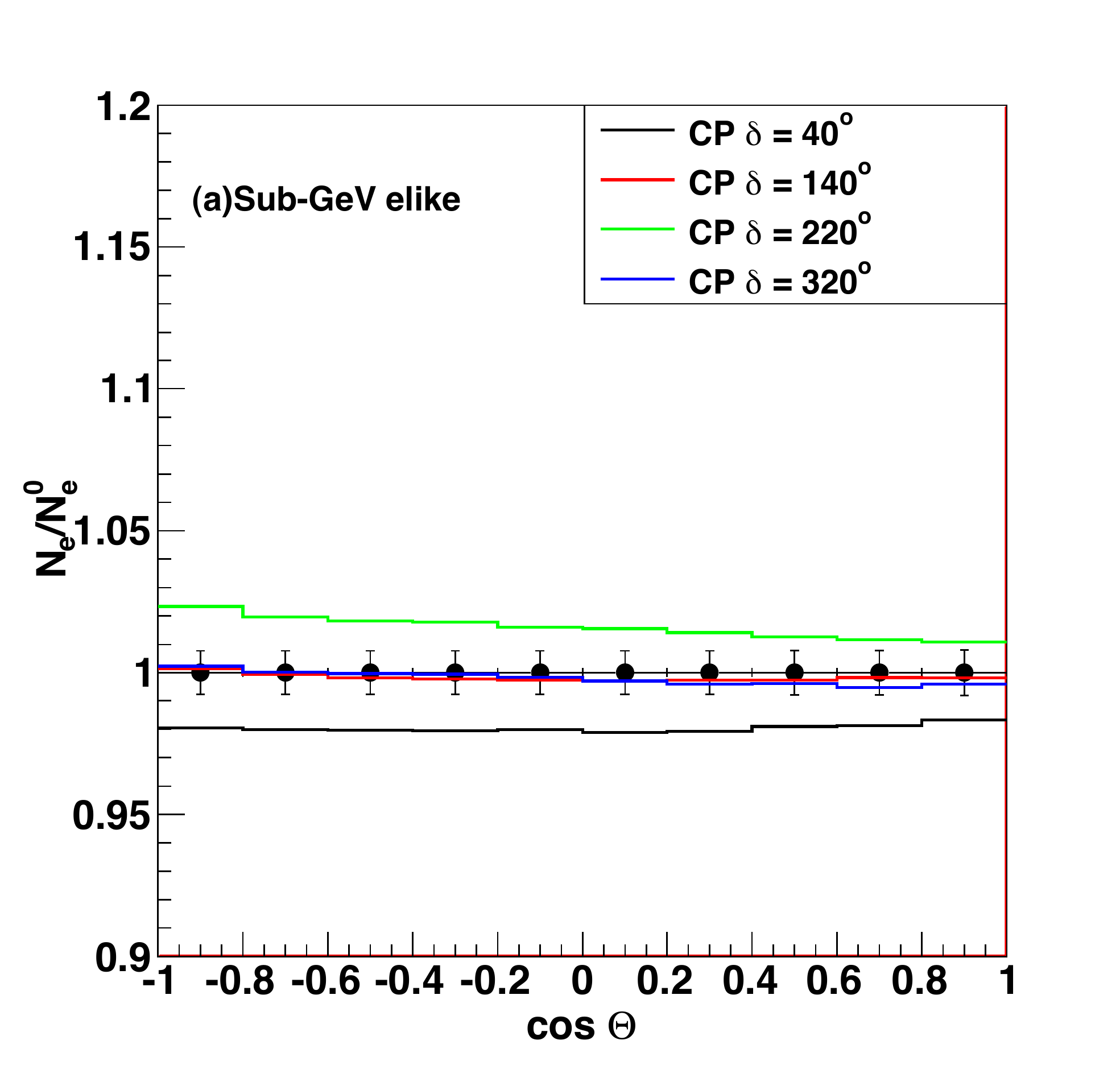}\\
    \includegraphics[scale=0.35]{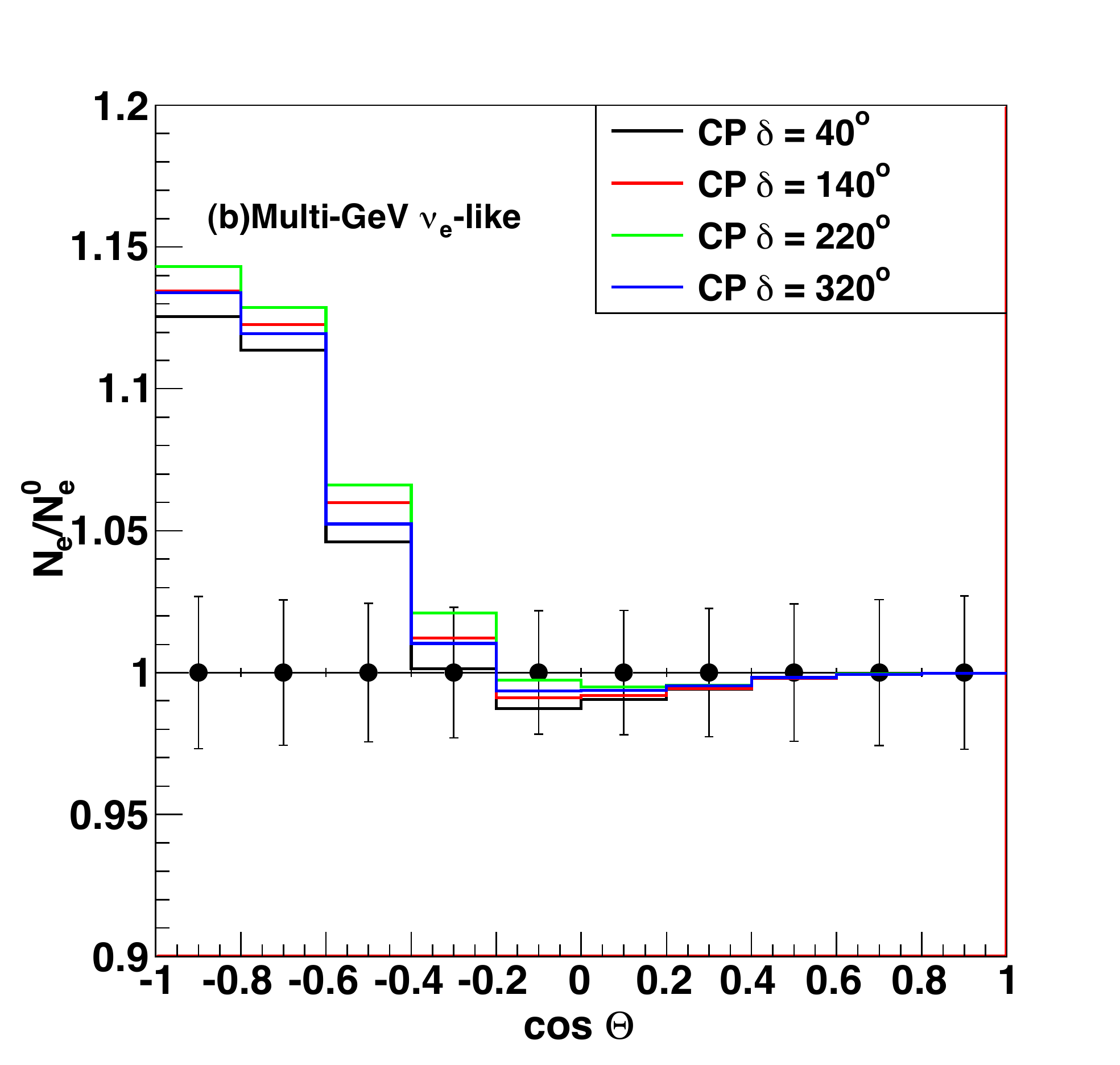}
    \includegraphics[scale=0.34]{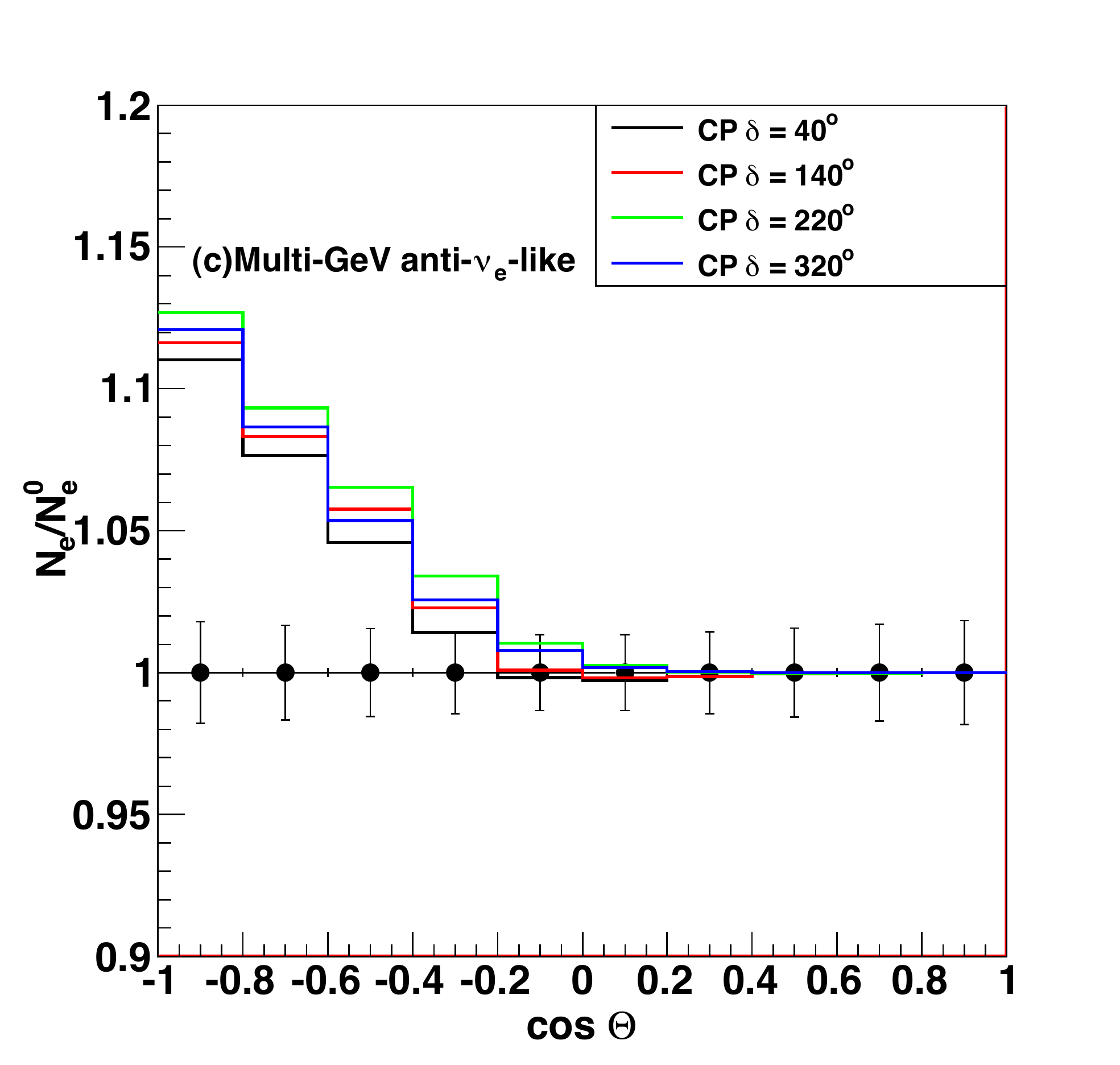}
  \end{center}
  \caption{
           Expected event rate changes
           in sub-GeV single-ring \(e\)-like, 
           multi-GeV $\nu_{e}$-like, 
           and multi-GeV $\bar \nu_{e}$-like event samples.
           The vertical axis shows the ratio of oscillated $e$-like event
           rate to the non-oscillated one.
           True $CP$ phase \(\delta\) is varied as
           \(40^\circ\) (black), 
           \(140^\circ\) (red), 
           \(220^\circ\) (green), 
           and \(320^\circ\) (blue).
           Normal mass hierarchy is assumed.  
           Points with error bars represent null oscillation expectation
           with expected statistical errors for 10 years of Hyper-K.
           }
  \label{fig:cp-distribution}
\end{figure}
\begin{figure}[htbp]
  \begin{center}
    \includegraphics[scale=0.2]{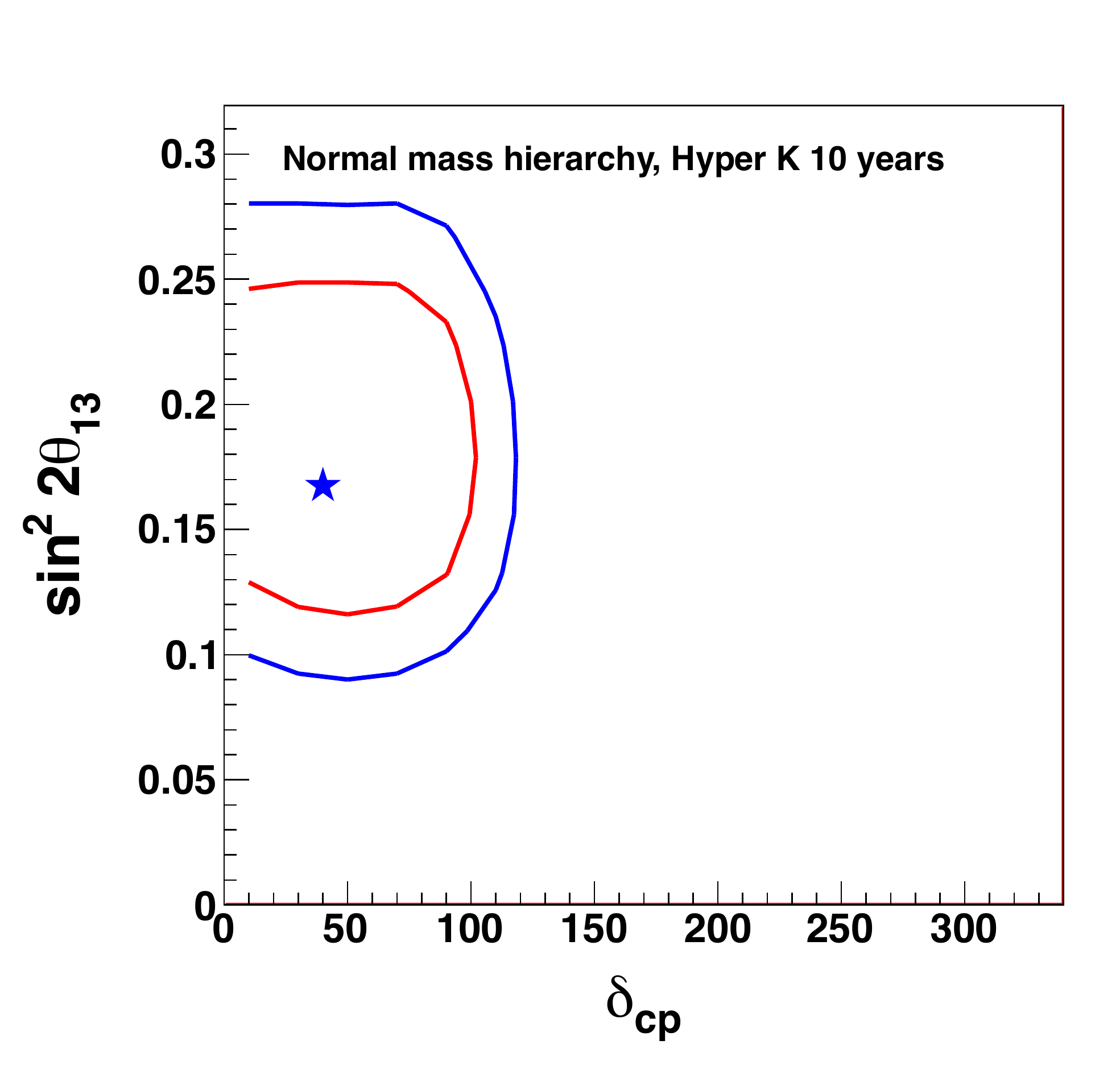}
    \includegraphics[scale=0.2]{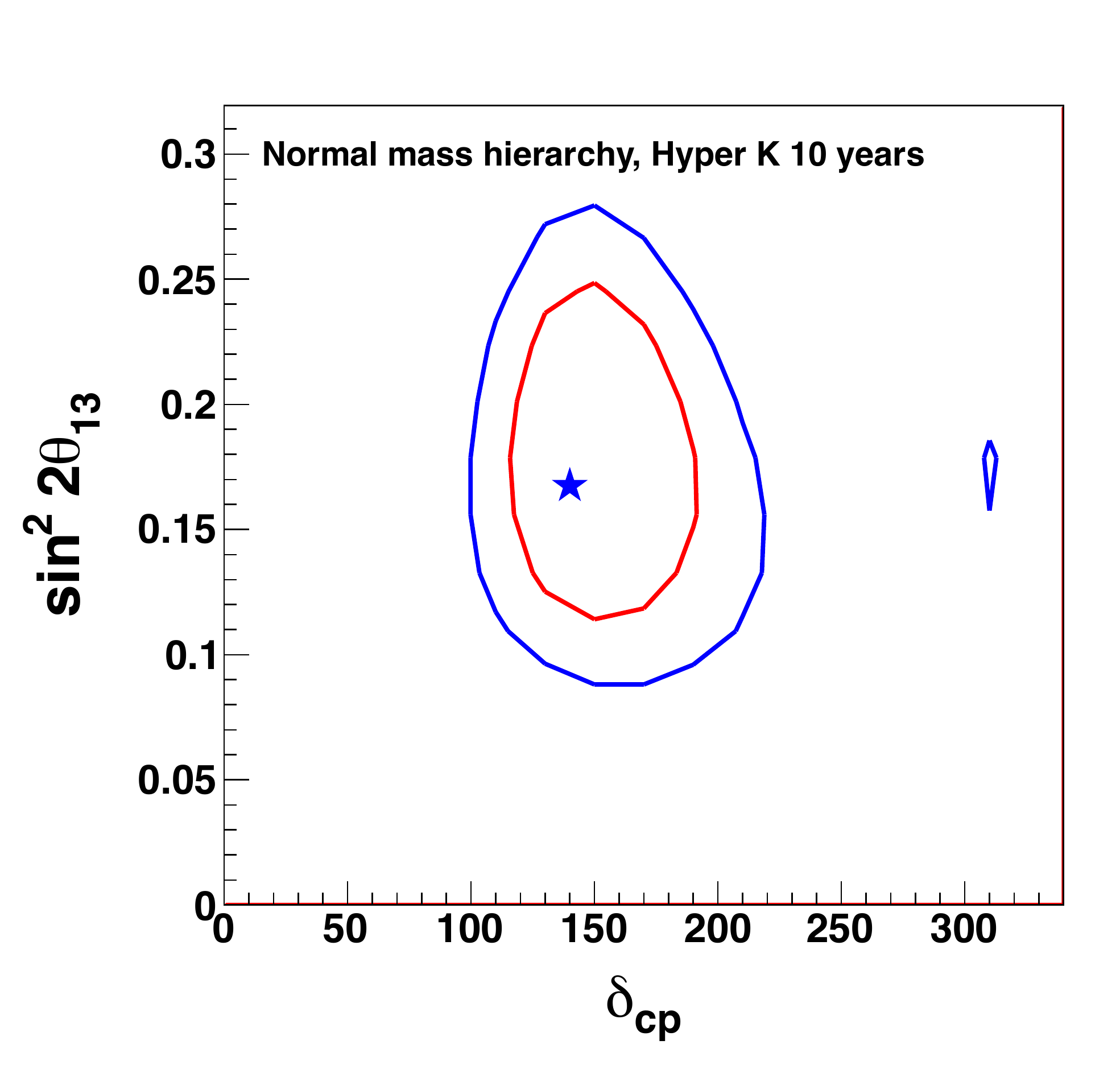}
    \includegraphics[scale=0.2]{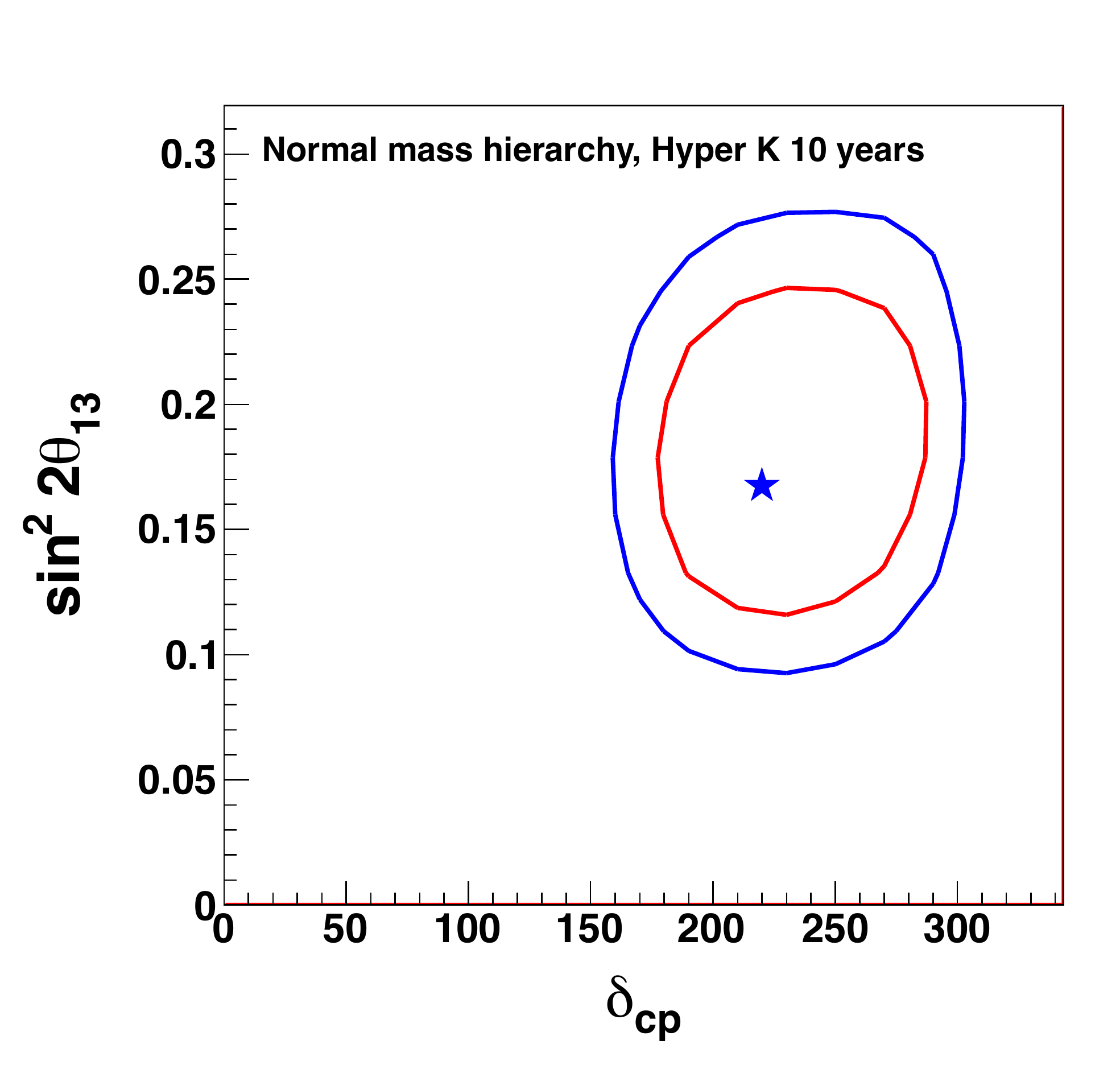}
    \includegraphics[scale=0.2]{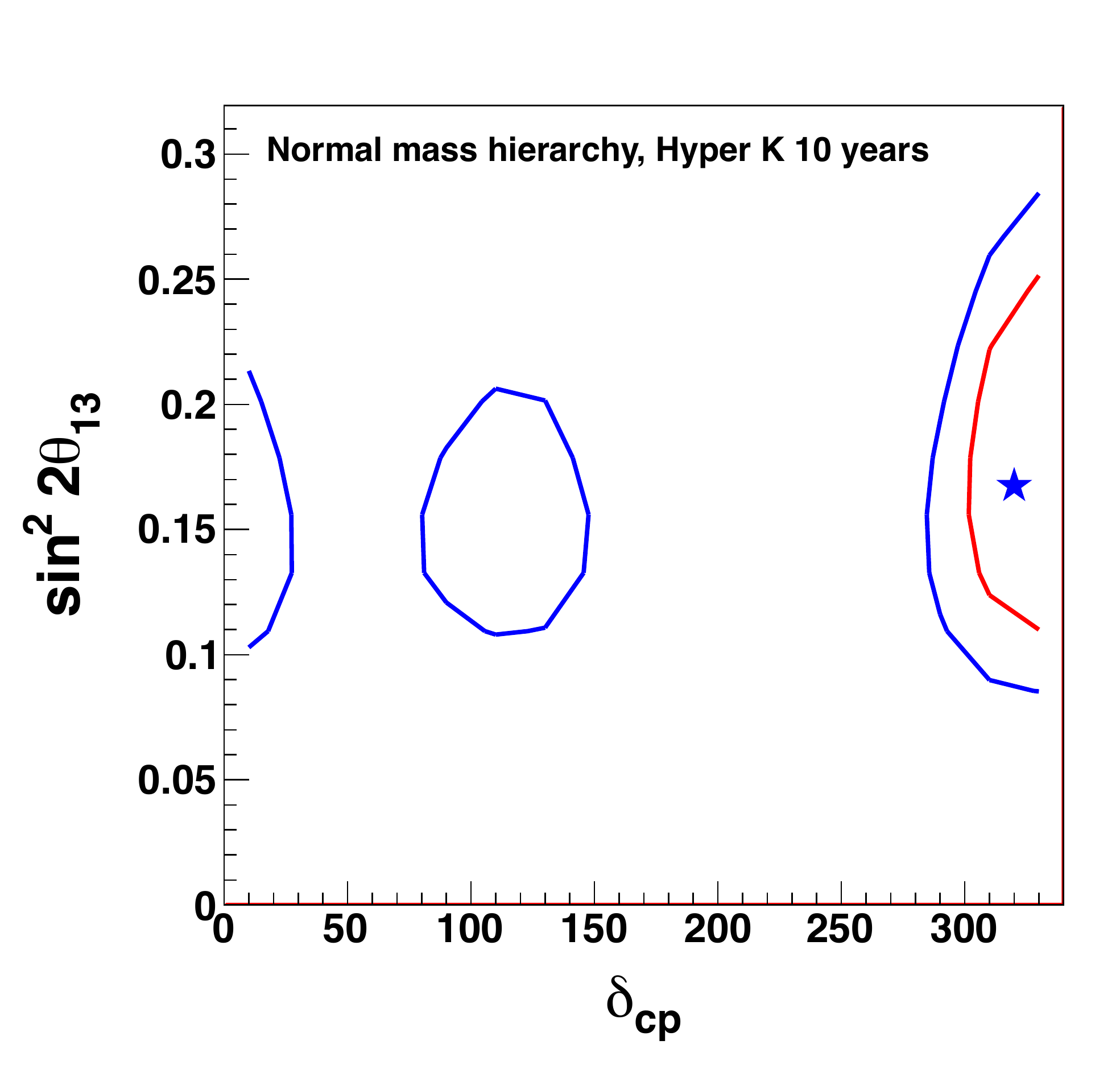}
    \includegraphics[scale=0.2]{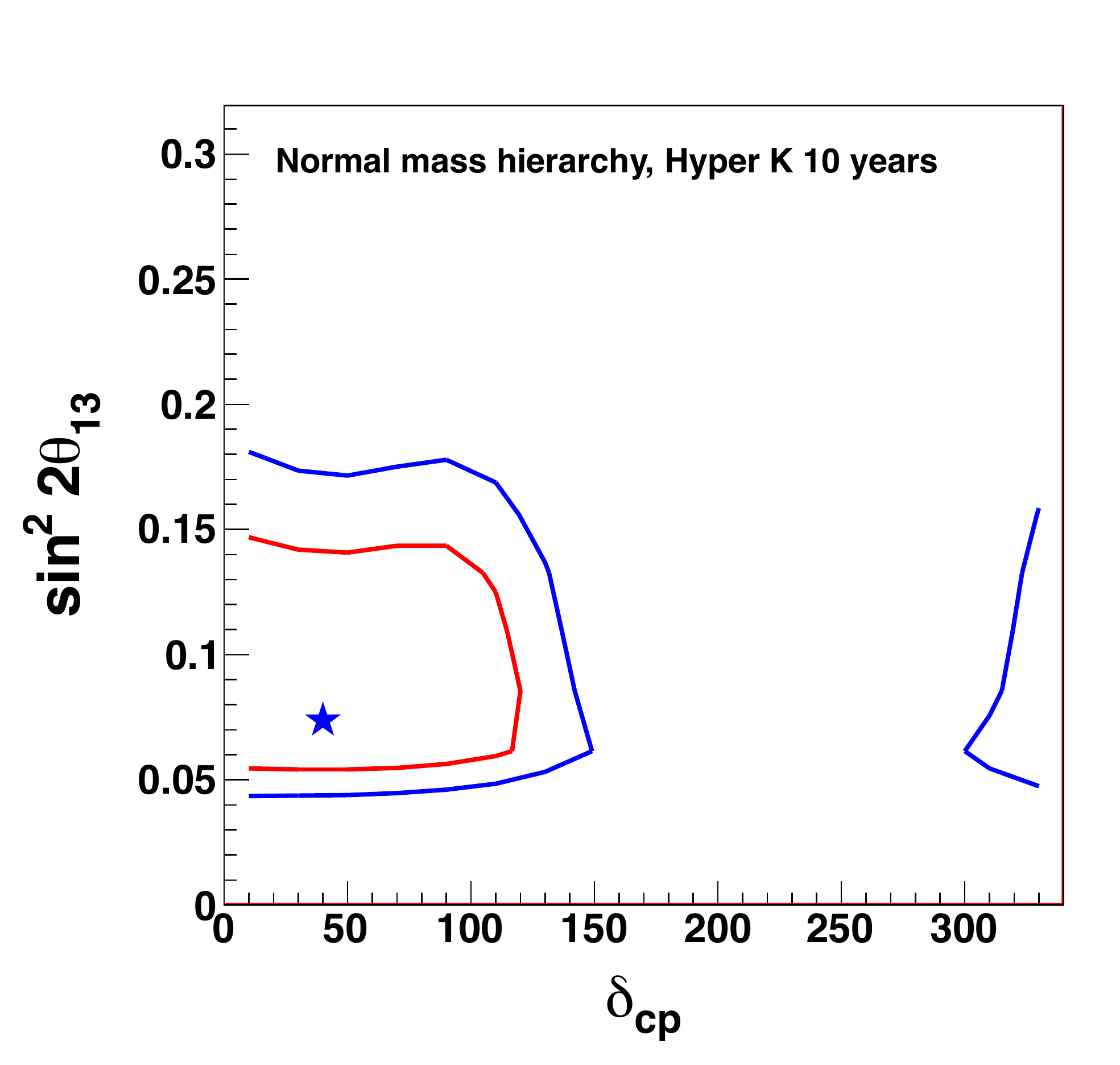}
    \includegraphics[scale=0.2]{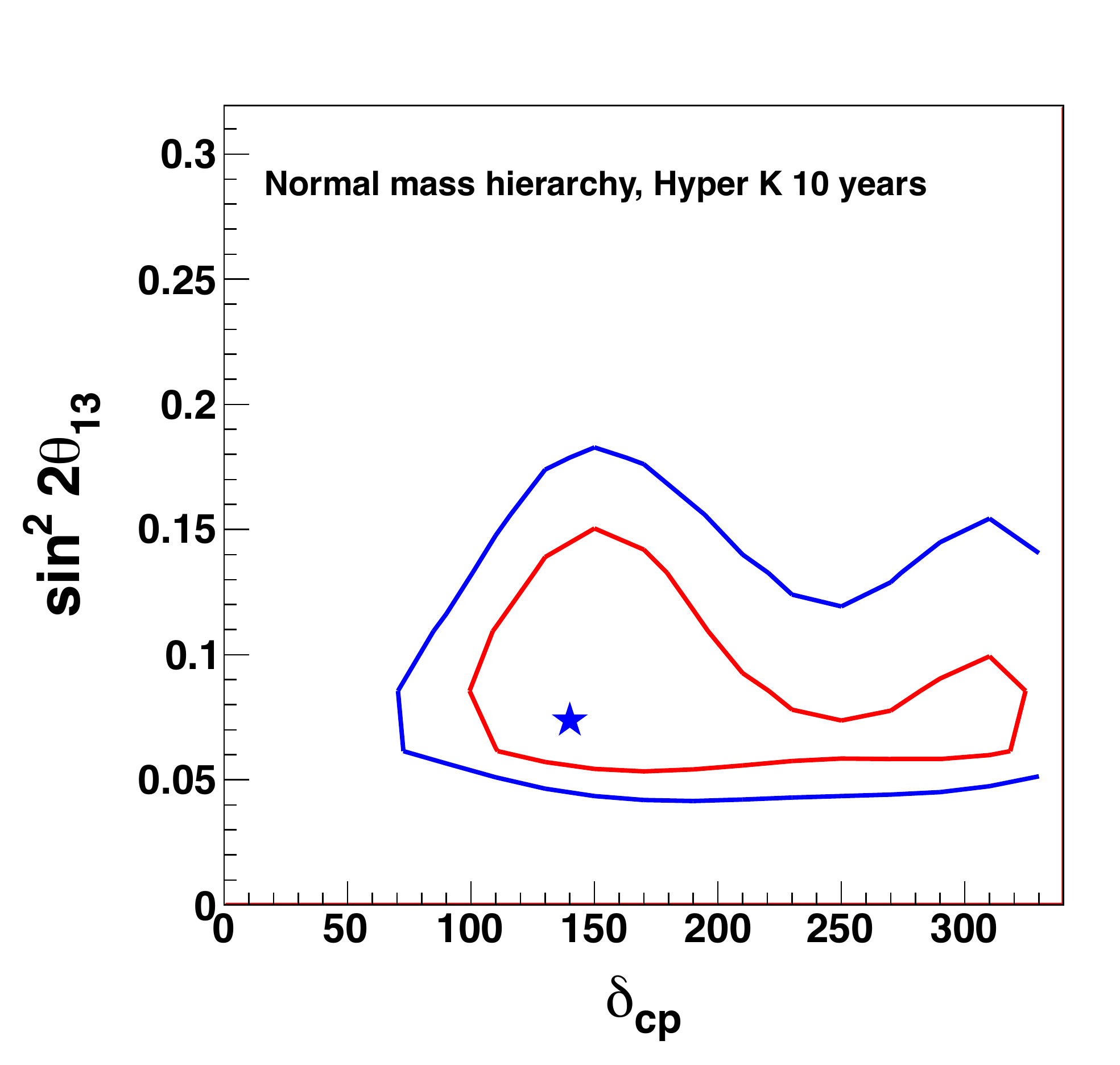}
    \includegraphics[scale=0.2]{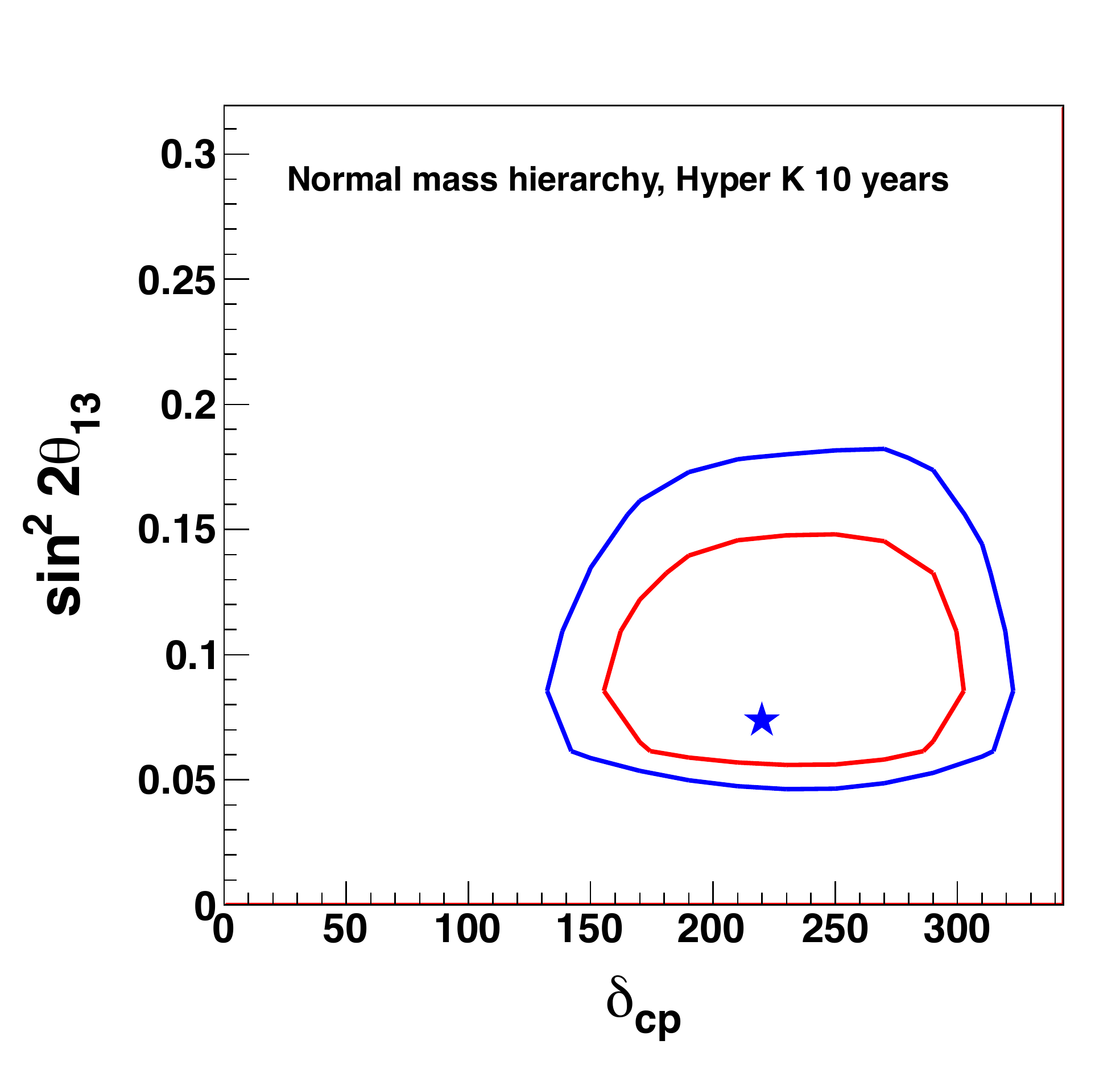}
    \includegraphics[scale=0.2]{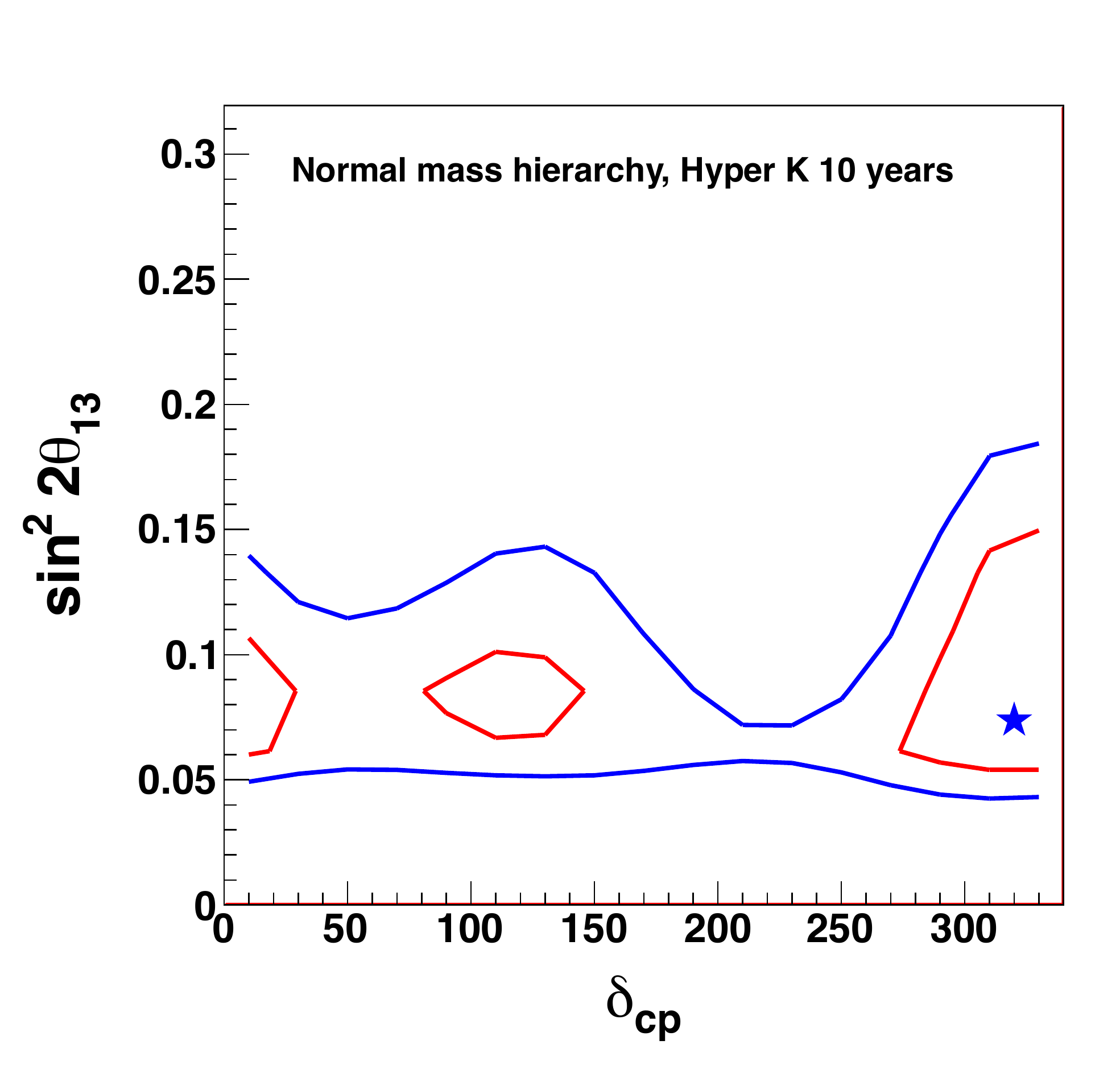}
  \end{center}
  \caption{ Expected sensitivities for \(\delta\) and \(\sin^22\theta_{13}\)
           at 90\% CL (red) and 99\% CL (blue)
           with a livetime of 10 Hyper-K years.  Stars in the contours
           represent the true parameter set of $\theta_{13}$ and $\delta$.
           Normal mass hierarchy is assumed.  
           }
  \label{fig:cp-contour}
\end{figure}

\subsubsection{$CP$ phase}
The effect of $CP$ phase \(\delta\) expressed by the second term in
Eq. \ref{eqn:nue-oscillation} can appear in the neutrino energy
region from 100 MeV to a few GeV
as shown in Fig. \ref{fig:nue-example}.
Figure \ref{fig:cp-distribution} shows that the event rate changes
will appear both in sub-GeV and multi-GeV energy samples as expected.

Figure \ref{fig:cp-contour}
shows the expected sensitivity for the 
$CP$ phase evaluated by \(\Delta\chi^2\).  
Normal mass hierarchy is assumed in this analysis.
These sensitivities do not depend on \(\theta_{23}\) very much and so 
\(\sin^2\theta_{23}\) is fixed to be 0.5 in the figure.
In the case of
\(\sin^22\theta_{13}=0.16\), discrimination of the $CP$ phase for 
\(\delta=40^\circ\), \(140^\circ\), \(220^\circ\),
and \(320^\circ\) is expected to be possible at the 90\% CL
and even \(99\%\) CL discrimination is possible for 
\(\delta=40^\circ\) and \(220^\circ\).
In the case of \(\sin^22\theta_{13}=0.08\) the accuracy of 
$CP$ phase measurements gets worse, 
but constraints could still be obtained in some cases.
Although the sensitive parameter space and $\delta$ precision are limited,
there is a good chance to obtain precious information
on $CP$ $\delta$ and to provide supplemental information
for the $CP$ asymmetry study conducted by the J-PARC to Hyper-K neutrino oscillation
experiment discussed in Sec. \ref{sec:cp}.

\subsubsection{Summary}
Sensitivity to several neutrino oscillation parameters by using
high statistic atmospheric neutrino data in Hyper-Kamiokande has been studied
assuming $\sin^22\theta_{13} > 0.04$. 
We have applied the full MC simulation and latest systematic uncertainties
used in the current Super-Kamiokande analyses.
It has been found that the expected significance for the mass hierarchy determination
is more than $3\sigma$ 
provided $\sin^2\theta_{23}>0.4$.
We expect to be able to discriminate between \(\sin^2\theta_{23}<0.5\) (first octant)
and $>0.5$ (second octant)
if \(\sin^22\theta_{23}\) is less than 0.99.
In addition, there is a good chance to obtain supplemental information
on $CP$ $\delta$.

%% file: physics-pdecay/pdecay.tex
\subsection{Nucleon decays}\label{section:pdecay}

Large water Cherenkov detectors have very good sensitivities for nucleon decays; indeed, the technology was originally developed and the first generation of such detectors was constructed in the 1980's primarily to search for proton decays predicted by the then-popular Grand Unified Theory (GUT), Minimal $SU$(5).  That model was eventually ruled out through non-observation of decays, leading to new models with longer lifetime predictions based on the experimentally demonstrated rarity of these decays.   The search for these predicted decays continues: for more than a decade Super-Kamiokande has had the world's best limits, generally by an order of magnitude or more,  on most of the current theoretically favored decay modes.  
As described in Sec.~\ref{section:intro-ndecay}, 
among many possible nucleon decay modes $p \rightarrow e^{+} \pi^{0}$ and $p \rightarrow \overline{\nu} K^{+}$ have been the subjects of the most intense interest, and they will be discussed in this section. 

The sensitivity of the Hyper-Kamiokande experiment for nucleon decays has been studied with a MC simulation based on the Super-Kamiokande analysis.  An estimate of the atmospheric neutrino background is necessarily included in the study.

\subsubsection{Sensitivity study for the $p \rightarrow e^{+} \pi^{0}$ mode}

For the $p \rightarrow e^{+} \pi^{0}$ mode where a $\pi^{0}$ decays primarily to two $\gamma$s, all of the final state particles are visible in a water Cherenkov detector. 
Signal candidates are selected with the following criteria: 
(A-1) the event is fully contained in the detector (FC event), and the vertex is reconstructed in the fiducial volume of 0.56~Megatons, 
(A-2) all rings are $e$-like, and the number of rings is two or three (two ring events could occur due to  an overlap of rings),
(A-3) there is no Michel decay electron (eliminate the events with an invisible muon), 
(A-4) for three ring events only, the invariant mass of two $e$-like rings is reconstructed 
between 85 to 185 MeV/$c^{2}$ to identify the $\pi^0$, 
(A-5) the reconstructed total momentum is less than 250 MeV/$c$ and the total invariant mass is between
800 to 1050 MeV/$c^{2}$. 

Water molecules consist of two free protons and eight bound protons in an oxygen nucleus.
In the MC simulation, the binding energy, Fermi motion, and interactions in nucleus are taken into account.

A significant but unavoidable inefficiency of the $p \rightarrow e^{+} \pi^{0}$ signal is caused by missing the $\pi^0$ which interacts in the nucleus (absorption and scattering).
Although the $p \rightarrow e^{+} \pi^{0}$ efficiency of a free proton is as high as 87\%, the average efficiency of both free and bound protons is lower because the $\pi^0$ from the bound  proton decay often suffers these interactions inside the nucleus.  
As a result, the overall proton decay efficiency of $p \rightarrow e^{+} \pi^{0}$ is estimated to be 
45\%.

The main background source of the proton decay search is the atmospheric neutrino events, which can occasionally produce an electron and a $\pi^{0}$ in the final state.
With the selection criteria of (A-1)-(A-5), the remaining background events are estimated to be 1.6~events/Megaton$\cdot$year from the atmospheric neutrino MC simulation.
This result of the MC simulation has been experimentally confirmed by the K2K experiment~\cite{mine}. K2K's one kiloton water Cherenkov near detector accumulated muon neutrino beam data corresponding to a 10~Megaton$\cdot$year exposure of atmospheric neutrinos. Using a  calibrated and well-understood neutrino flux, the corresponding atmospheric neutrino-induced background rate of false $p \rightarrow \mu^{+} \pi^{0}$  events was estimated based on the observed beam-induced events in the detector.  This rate was then extrapolated in turn to the atmospheric neutrino-induced background rate of $p \rightarrow e^{+} \pi^{0}$,  $1.63^{+0.42}_{-0.33}({\rm stat})^{+0.45}_{-0.51}({\rm syst})$ events/Megaton$\cdot$year.

\begin{figure}[htbp]
  \begin{center}
    \includegraphics[height=22pc]{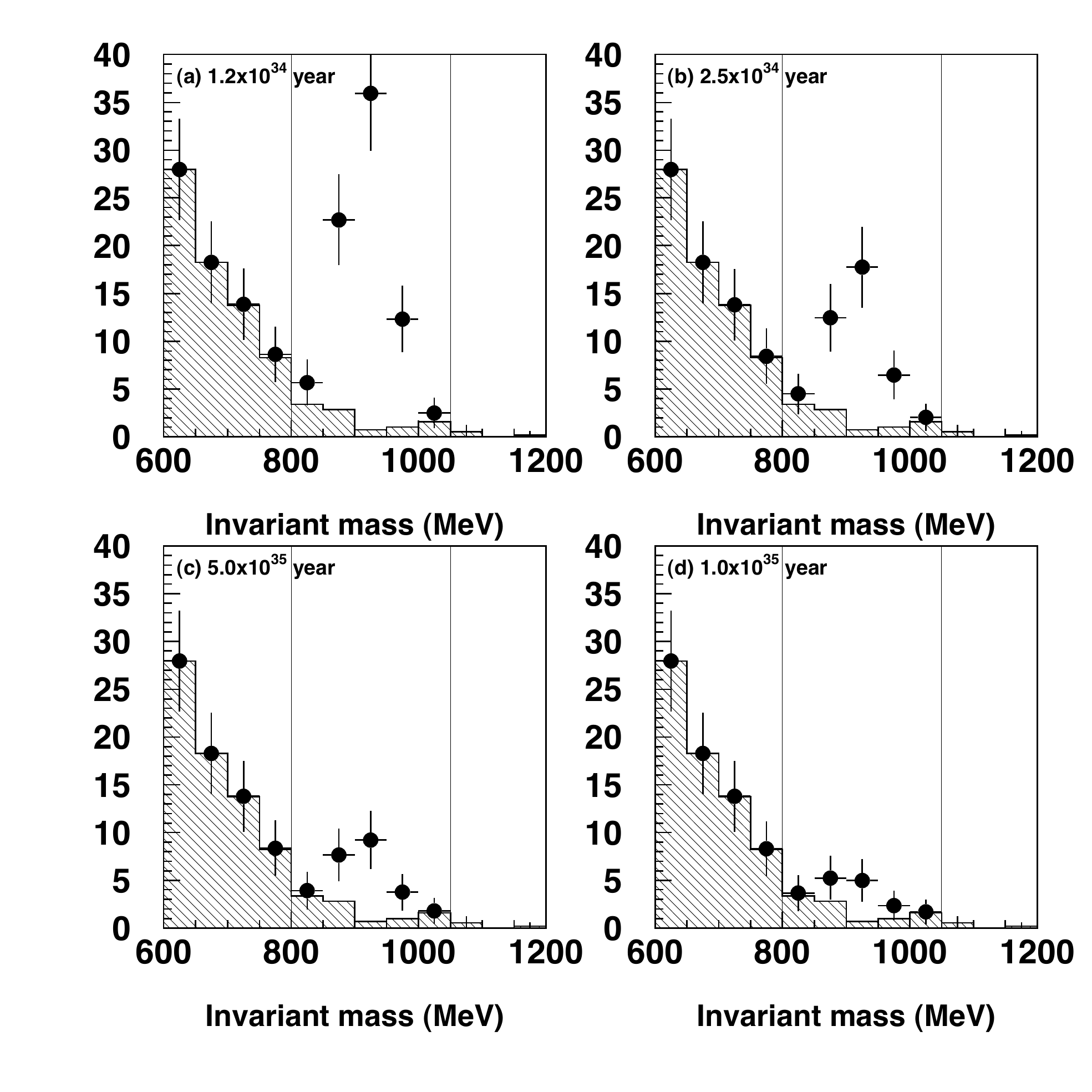}
  \end{center}
\caption {Invariant mass distributions of MC events after all cuts except for the mass cut with a 5.6~Megaton$\cdot$year exposure. The proton lifetime is assumed to be: (a) 1.2$\times 10^{34}$, (b) 2.5$\times 10^{34}$, (c) 5.0$\times 10^{34}$, and (d) 1.0$\times 10^{35}$ years. Dots show the combined signal plus background events and the hatched histograms are atmospheric neutrino background events.}
  \label{fig:exp_epi0}
\end{figure}

\begin{figure}[htbp]
  \begin{center}
    \includegraphics[height=19pc]{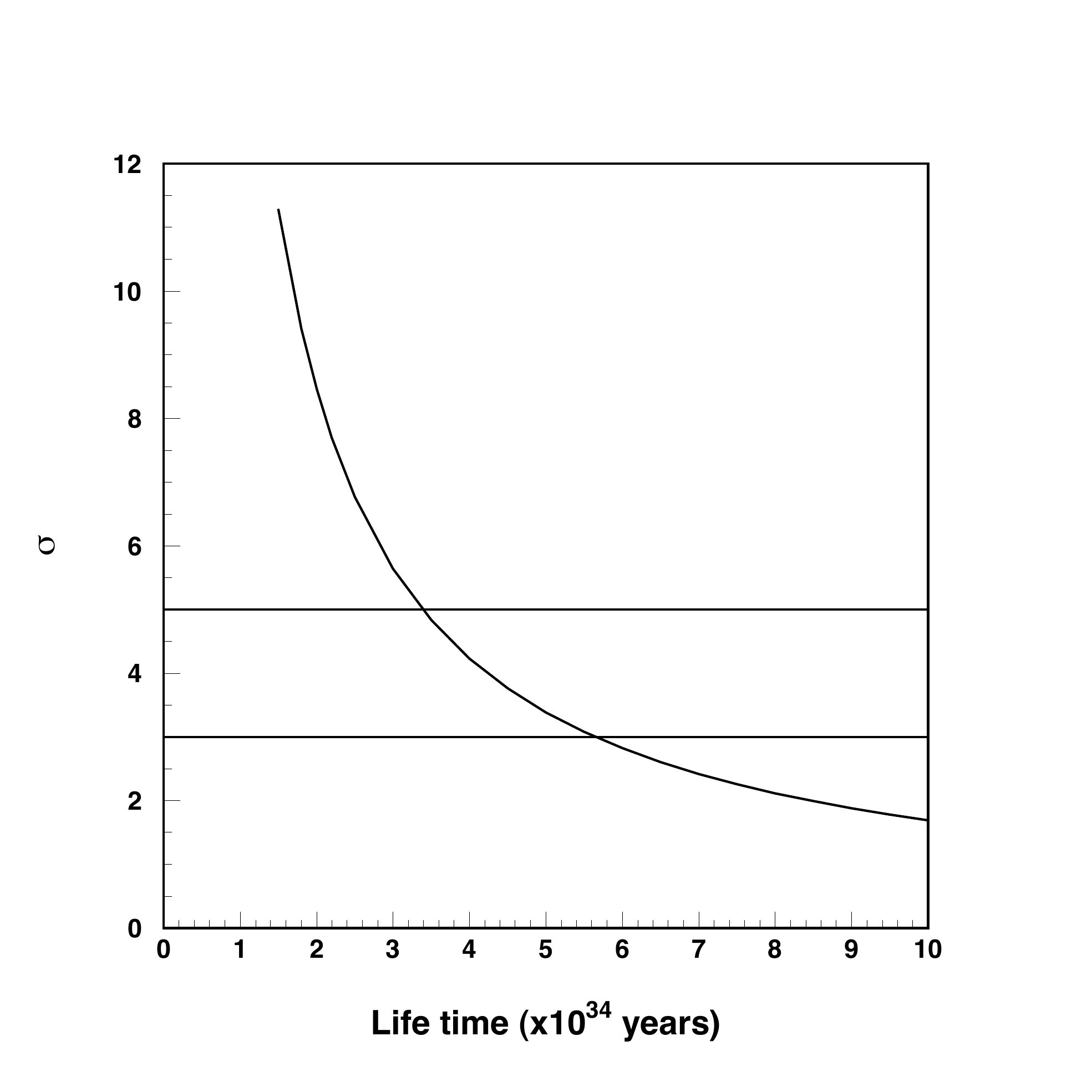}
  \end{center}
\caption {The significance of the $p \rightarrow e^{+} \pi^{0}$ signal derived from counting the events in the signal region (Fig.~\ref{fig:exp_epi0}) following a 5.6~Megaton$\cdot$year exposure, as a function of the true proton lifetime. The upper line indicates 5~$\sigma$ significance and the lower line indicates 3~$\sigma$.}
  \label{fig:dis-epi0}
\end{figure}

Fig.~\ref{fig:exp_epi0} shows the total invariant mass distributions of the proton decay signal events and the atmospheric neutrino background events after all the cuts except the mass cut.
The running time of Hyper-Kamiokande is assumed to be 10 years which corresponds to a 5.6~Megaton$\cdot$ year exposure. The proton lifetimes in these plots are taken to be: (a) 1.2$\times 10^{34}$ years
 (current limit), (b) 2.5$\times 10^{34}$ years, (c) 5.0$\times 10^{34}$ years, and (d) 
1.0$\times 10^{35}$ years. The number of atmospheric neutrino background events is estimated to be 9.0 events. 
Fig.~\ref{fig:dis-epi0} shows the significance of the signal events
as a function of the proton lifetime by counting the events in the signal region. 
In this figure, the systematic uncertainty of the background is assumed to be 44\%, the same as Super-K.
We could separate a signal from the background events at a  5~$\sigma$
significance for a proton lifetime of 3.4$\times 10^{34}$ years, and at 3~$\sigma$ with 
5.7$\times 10^{34}$ years. 

\begin{figure}[htbp]
  \begin{center}
    \includegraphics[height=19pc]{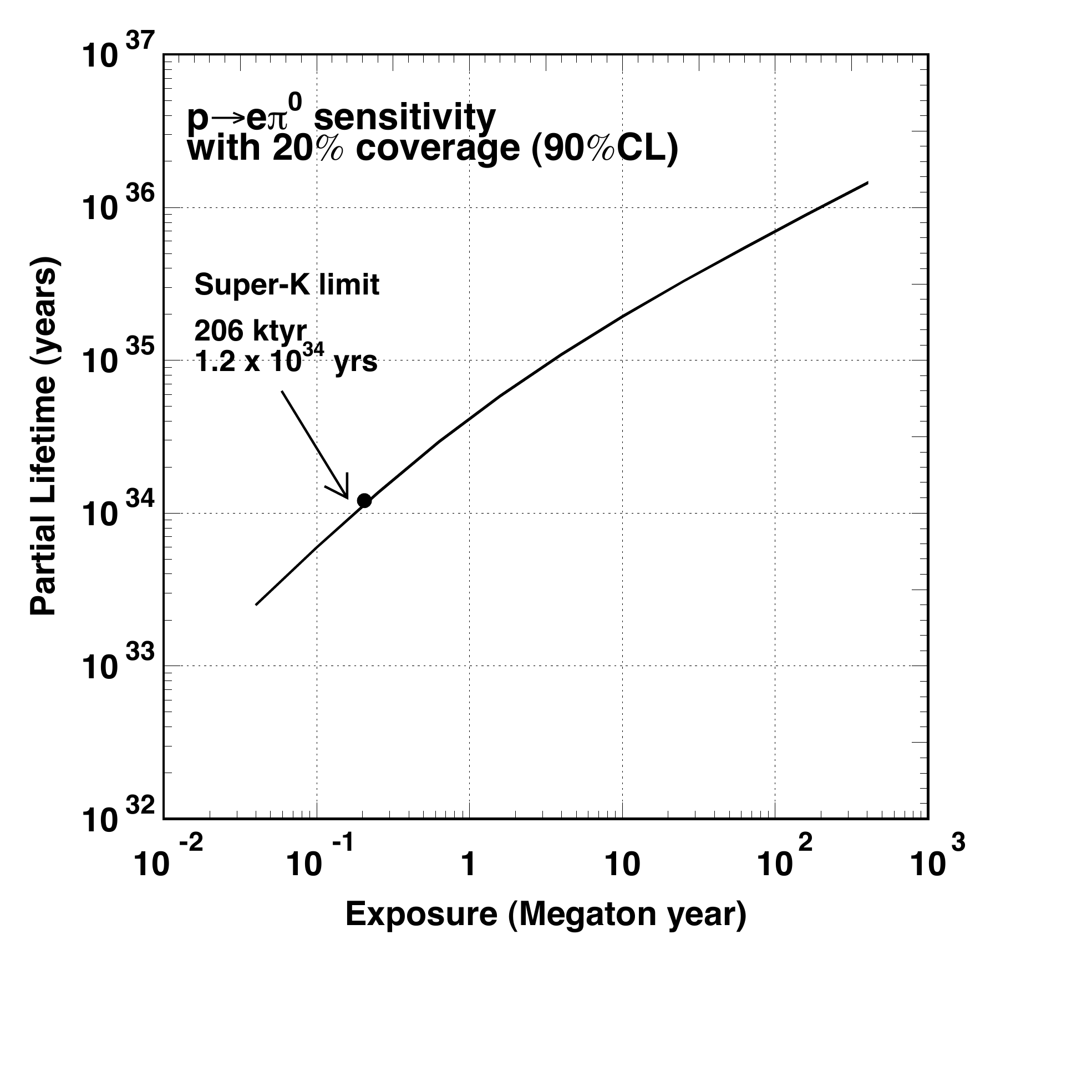}
  \end{center}
\caption {Sensitivity of the proton decay search for the $p \rightarrow e^{+} \pi^{0}$ mode as a function of detector exposure.}
  \label{fig:sens_epi0}
\end{figure}

Fig.~\ref{fig:sens_epi0} shows the sensitivity for proton decay with a 90$\%$ CL as a function of the detector exposure. We could reach 1.0$\times 10^{35}$ years partial lifetime with a 4~Megaton$\cdot$year exposure, which corresponds to 8 years running Hyper-Kamiokande; by contrast, it will  take Super-Kamiokande 178 years to reach this level.

\subsubsection{Sensitivity study for the $p \rightarrow \overline{\nu} K^{+}$ mode}
For the $p \rightarrow \overline{\nu} K^{+}$ mode, $K^{+}$ itself is not visible in a water Cherenkov detector due to having a low, sub-Cherenkov threshold,  momentum. 
However, $K^{+}$ can be identified by the decay products of $K^+ \to \mu^{+}+\nu$  (64\% branching fraction) and $K^+ \to \pi^{+}+\pi^{0}$  (21\% branching fraction).
The muons and pions from the $K^+$ decays have monochromatic momenta due to being produced via two-body decays. 
Furthermore, when a proton in an oxygen nucleus decays, the proton hole is filled by de-excitation of another proton, resulting in  $\gamma$ ray emission. The probability of a 6~MeV $\gamma$ ray being emitted is about 40\%. 
This 6~MeV $\gamma$ is a characteristic signal used to identify a proton decay and to reduce the atmospheric neutrino background. 
There are three established methods for the $p \rightarrow \overline{\nu} K^{+}$ mode search~\cite{Kobayashi:2005pe}: 
(1) look for single muon events with a de-excitation $\gamma$ ray just before the time of the muon, since the $\gamma$ ray is emitted at the time of $K^+$ production; 
(2) search for an excess of muon events with a momentum of 236~MeV/$c$ 
in the momentum distribution; and 
(3) search for  $\pi^{0}$ events with a momentum of 205~MeV/$c$. 

\begin{figure}[htbp]
  \begin{center}
    \includegraphics[height=19pc]{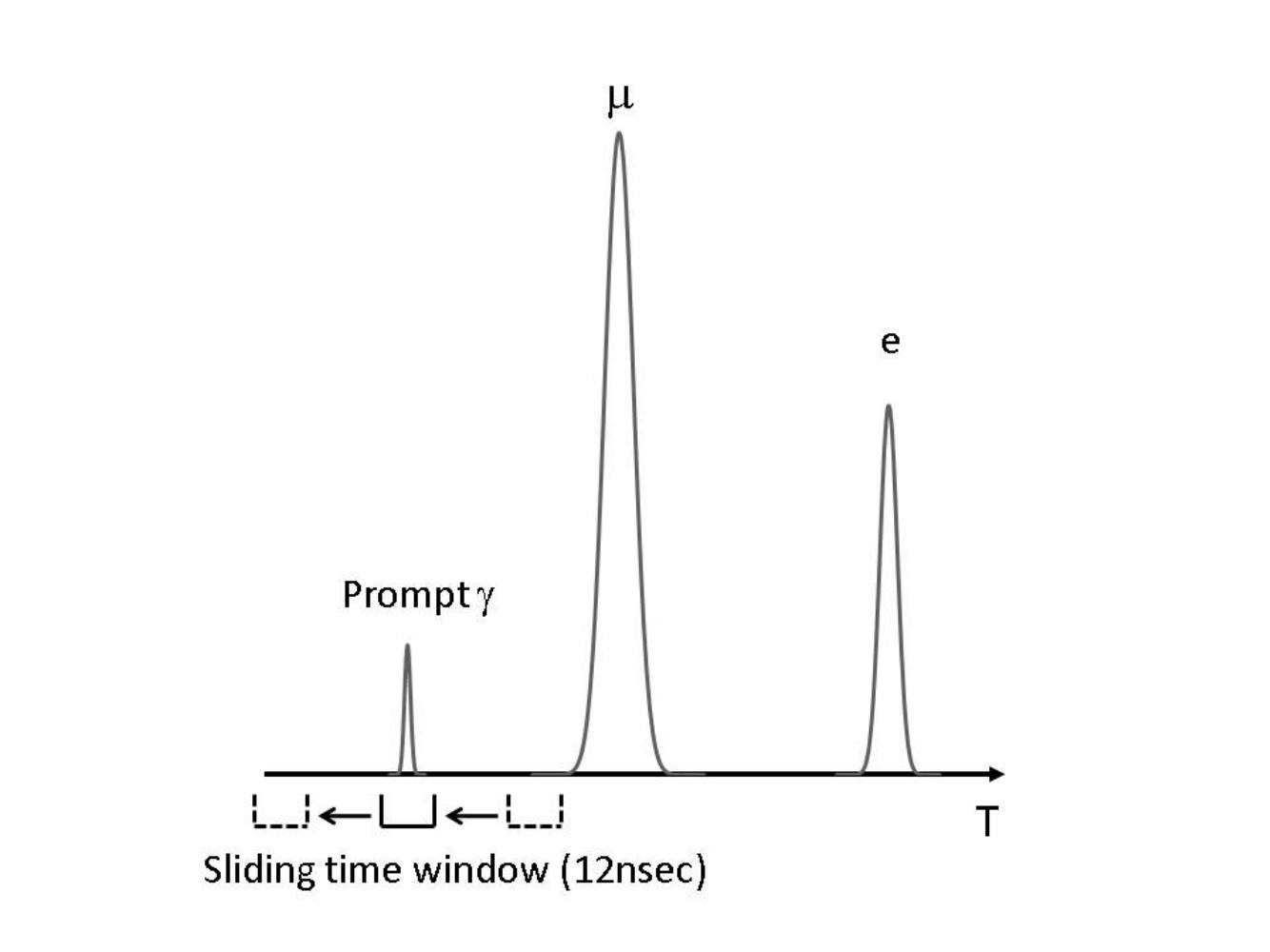}
  \end{center}
\caption {Schematic view of the prompt $\gamma$ hit search in the $p \rightarrow \overline{\nu} K^{+} (+ \gamma^*), K^{+} \to \mu^{+}+\nu$ mode.}
  \label{fig:prompt-gamma}
\end{figure}

In method (1), the $p \rightarrow \overline{\nu} K^{+} (+ \gamma^*), K^{+} \to \mu^{+}+\nu$ candidate events are selected with the following criteria: 
(B-1) a fully contained (FC) event with one ring, 
(B-2) the ring is $\mu$-like, 
(B-3) there is a Michel decay electron, 
(B-4) the muon momentum is between 215 and 260~MeV/$c$, 
(B-5) the distance of the vertices between the muon and the Michel electron is less than 200~cm, 
(B-6) the time difference between the $\gamma$ and the muon is less than 75 ns ($\sim 6\tau_{K^+}$),  and 
(B-7) the number of PMTs hit by the $\gamma$ is between 4 and 30 to (select 6~MeV energy). 
The prompt $\gamma$ hits are searched for by sliding a 12~ns wide timing window just before the time of the muon signal as shown in Fig.~\ref{fig:prompt-gamma}.

In method (2), the previous selection criteria are relaxed; the events are not to satisfy (B-5)-(B-7), but rather only (B-1)-(B-4).  An excess of muon signals in the momentum distribution is then searched for by fitting the data with the 
proton decay signal expectation over the atmospheric neutrino background events.
In method (3), the $\pi^0$ events with a momentum of 205~MeV/$c$ are selected.
Although the $\pi^+$ does not make a clear Cherenkov ring due to its low momentum, hit activity caused by $\pi^+$ in the opposite direction of the $\pi^0$ is used to identify the $K^{+} \rightarrow \pi^{+}\pi^{0}$ signal.
The following selection criteria are used:
(C-1) FC events with two rings (from $\pi^0 \to \gamma \gamma$), 
(C-2) both rings are $e$-like, 
(C-3) one Michel decay electron from the muon produced by $\pi^+ \to \mu^+$, 
(C-4) the invariant mass of two rings is between 85 and 185 MeV/$c^{2}$ ($\pi^0$ mass region), 
(C-5) the reconstructed $\pi^{0}$ momentum is between 175 and 250 MeV/$c$,
(C-6) the visible energy opposite the $\pi^{0}$ (140 - 180 degrees) is between 7 and 17~MeV, and 
(C-7) the visible energy 90 - 140 degrees from $\pi^{0}$ direction is less than 12~MeV. 

\begin{figure}[htbp]
  \begin{center}
    \includegraphics[height=22pc]{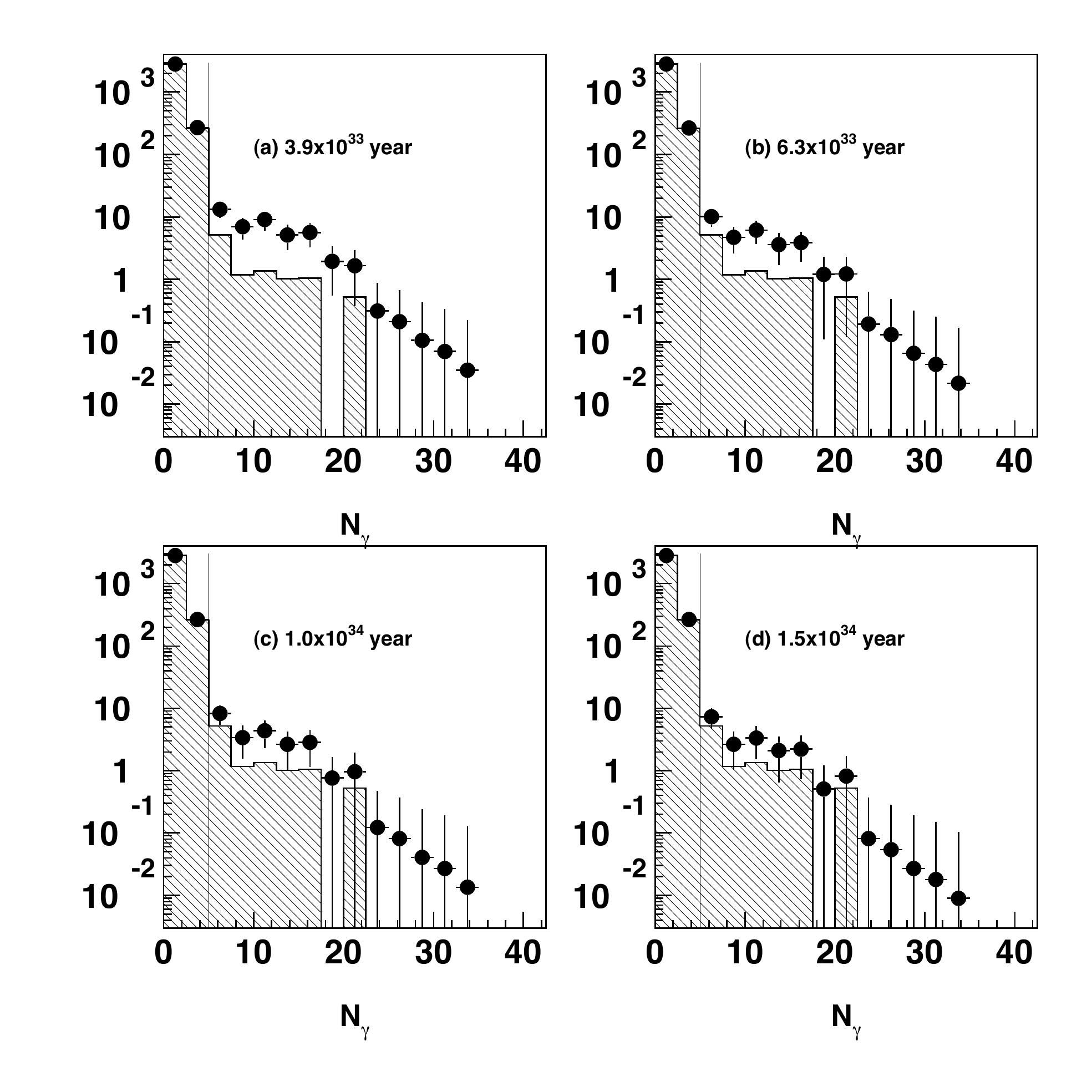}
  \end{center}
\caption {Number of hits distributions in the prompt $\gamma$ hits search for MC events passing cuts (B-1)-(B-6) for a 5.6~Megaton$\cdot$year exposure. The proton lifetime is assumed to be: (a) 3.9$\times 10^{33}$, (b) 6.3$\times 10^{33}$, (c) 1.0$\times 10^{34}$, and (d) 1.5$\times 10^{34}$ years. Dots show the signal plus background events and the hatched histograms are the atmospheric neutrino background events.}
  \label{fig:exp_nuk}
\end{figure}

\begin{figure}[htbp]
  \begin{center}
    \includegraphics[height=18pc]{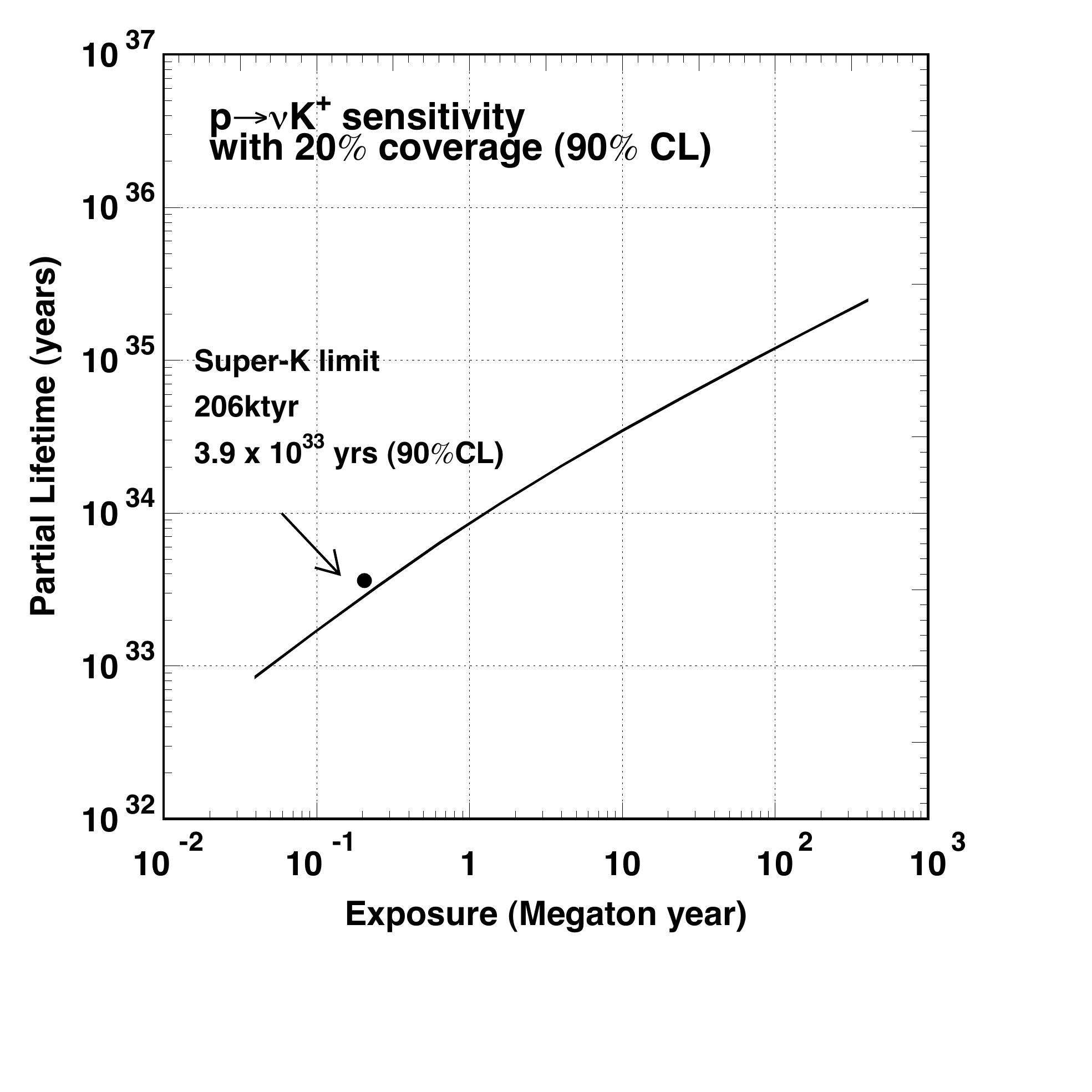}
  \end{center}
\caption {Sensitivity plot for the $p \rightarrow \overline{\nu} K^{+}$ mode as a function of detector exposure.}
  \label{fig:sens_nuk}
\end{figure}

The detection efficiencies are calculated to be 7.1\% for method (1), 43\% for method (2), and 6.7\% for method (3).
The background rates from atmospheric neutrinos are 1.6, 1940, and 6.7 events/Megaton$\cdot$year for the methods (1), (2), and (3), respectively.  
Fig.~\ref{fig:exp_nuk} shows the number of hits distributions in the prompt $\gamma$ hits search with all other cuts (B-1)-(B-6) of method (1) applied. The running time of Hyper-Kamiokande is assumed to be 10 years which corresponds to a 5.6~Megaton$\cdot$
year exposure. 
The proton lifetime is taken to be: (a) 3.9$\times 10^{33}$ years (current limit), (b) 6.3$\times 10^{33}$ years, 
(c) 1.0$\times 10^{34}$ years, and (d) 1.5$\times 10^{34}$ years.
The number of atmospheric neutrino background events is estimated to be 9.0 events.
With an assumed background uncertainty of 33\% (the same as Super-K), proton decay signals from the $p \rightarrow \overline{\nu} K^{+} (+ \gamma^*), K^{+} \to \mu^{+}+\nu$ mode could be separated from the background events with a 5~$\sigma$ significance for a  proton lifetime of 6.3$\times 10^{33}$ years as shown in 
Fig.~\ref{fig:exp_nuk}(b), and with 3~$\sigma$ for  1.0$\times 10^{34}$ years as shown in  
Fig.~\ref{fig:exp_nuk}(c). 
Fig.~\ref{fig:sens_nuk} shows the 90\% CL sensitivity curve  for the  $p \rightarrow \overline{\nu} K^{+}$ mode, by combining all three methods,  as a function of the detector exposure. 
In this case, a 1.2~Megaton$\cdot$year exposure is needed to reach a lifetime 
sensitivity of $10^{34}$ years. That corresponds to two years running Hyper-Kamiokande, or
53 years of livetime for Super-Kamiokande. 

\subsubsection{Sensitivity study for other nucleon decay modes}

Although the $p \rightarrow e^{+} \pi^{0}$ mode is predicted to be the dominant decay mode in many GUT models, 
the branching ratios of other nucleon decay modes are not small.
Table~\ref{tab:branch} shows several proton decay modes with the branching ratios predicted by several GUT models.
The ratio of neutron to proton lifetimes is also shown in the table.
Since we do not know which model (and there are many others) is correct, in order to discover  proton decay -- as well as to examine and experimentally constrain the various GUT models --  searching for a variety of nucleon decay modes is clearly important.

\begin{table}[htb]
\caption{Branching ratios for proton decay and the
ratio of neutron to proton lifetimes predicted by the SU(5) and SO(10) models.
.\label{tab:branch}}
\vspace{0.4cm}
\begin{center}
\begin{tabular}{l|rrrrr}
\hline
        & \multicolumn{5}{c}{Br.($\%$)} \\ 
\hline
        & \multicolumn{4}{c}{SU(5)}& SO(10)  \\ 
\hline
References& ~\cite{mach} & ~\cite{gav} & ~\cite{dono} & ~\cite{bucc} & ~\cite{bucc}   \\ 
\hline
$p \rightarrow e^{+} \pi^{0}$ & 33   & 37  & 9  & 35 & 30 \\ 
\hline
$p \rightarrow e^{+} \eta^{0}$ & 12  & 7 & 3 & 15 & 13 \\
\hline 
$p \rightarrow e^{+} \rho^{0}$ & 17    & 2   & 21   & 2 & 2   \\ 
\hline
$p \rightarrow e^{+} \omega^{0}$ & 22  & 18 & 56 & 17 & 14 \\
\hline 
Others & 17   & 35   & 11   & 31 & 31  \\ 
\hline
\hline
$\tau_{p}/\tau_{n}$ & 0.8 &1.0 & 1.3 &  & \\
\hline  
\end{tabular}
\end{center}
\end{table}

The Hyper-Kamiokande sensitivities for other nucleon decay modes are estimated based on the 
efficiencies and background rates of SK-II~\cite{:2009gd}. Table~\ref{tab:other_mode}
shows the 90~$\%$ CL sensitivities with a 5.6~Megaton$\cdot$year exposure 
(10 years running of Hyper-Kamiokande). 
The current lifetime limits are also shown in the table for reference.
In all cases, we could explore an order of magnitude longer lifetime regions.

The decay modes in Table~\ref{tab:other_mode} all conserve the baryon number minus the lepton number, $(B-L)$. 
Recently another $(B+L)$ conserving mode, $n \rightarrow e^{-} K^{+}$, was also given attention and searched for by Super-Kamiokande.
In $n \rightarrow e^{-} K^{+}$, the $K^{+}$ stops in the water and decays into $\mu^{+} + \nu$.
The final state particles observed in $n \rightarrow e^{-} K^{+},K^+ \to \mu^+ \nu$ are $e^{-}$ and  $\mu^{+}$. Both $e^{-}$ and $\mu^{+}$ have monochromatic momenta as a result of originating from two-body decays.
Furthermore, the timing of $\mu^{+}$ events are delayed with respect to the $e^{-}$ events because of the $K^{+}$ lifetime.  In SK-II, the estimated efficiencies and the background rate are
8.4\% and 1.1~events/Megaton$\cdot$year, respectively. From those numbers, the sensitivity
to the  $n \rightarrow e^{-} K^{+}$ mode with a 5.6~Megaton$\cdot$year exposure is estimated to be
2.2$\times 10^{34}$ years.

The possibility of $n \overline{n}$ oscillation is another interesting phenomenon; it violates the baryon number $(B)$ by 
$|\Delta B|$ = 2.  These $n \overline{n}$ oscillations have been searched for in Super-Kamiokande with a 0.09~Megaton$\cdot$year worth of data~\cite{jang}. Further improvement of the $n \overline{n}$ oscillation search is expected in Hyper-Kamiokande.

\begin{table}[htp]
\caption{Summary of the sensitivities from a 5.6~Megaton$\cdot$year exposure compared with  current lifetime limits. The current limits for $p \rightarrow e^{+} \pi^{0}$, $p \rightarrow \mu^{+} \pi^{0}$, and $p \rightarrow \overline{\nu} K^{+}$ are the result of analyzing a 0.22~Megaton$\cdot$year data sample, while the other modes are from a 0.14~Megaton$\cdot$year exposure.
\label{tab:other_mode}}
\vspace{0.4cm}
\begin{center}
\begin{tabular}{|l|l|l|}
\hline
Mode        & Sensitivity (90$\%$ CL)  & Current limit \\ 
\hline
\hline
$p \rightarrow e^{+} \pi^{0}$ & 13$\times 10^{34}$ years   & 1.3$\times 10^{34}$ years \\ 
\hline
$p \rightarrow \mu^{+} \pi^{0}$ & 9.0$\times 10^{34}$  & 1.1$\times 10^{34}$  \\ 
\hline
$p \rightarrow e^{+} \eta^{0}$& 5.0$\times 10^{34}$  & 0.42$\times 10^{34}$ \\ 
\hline
$p \rightarrow \mu^{+} \eta^{0}$& 3.0$\times 10^{34}$  & 0.13$\times 10^{34}$ \\ 
\hline
$p \rightarrow e^{+} \rho^{0}$& 1.0$\times 10^{34}$  & 0.07$\times 10^{34}$ \\ 
\hline
$p \rightarrow \mu^{+} \rho^{0}$& 0.37$\times 10^{34}$  & 0.02$\times 10^{34}$ \\ 
\hline
$p \rightarrow e^{+} \omega^{0}$& 0.84$\times 10^{34}$  & 0.03$\times 10^{34}$ \\ 
\hline
$p \rightarrow \mu^{+} \omega^{0}$& 0.88$\times 10^{34}$  & 0.08$\times 10^{34}$ \\ 
\hline
$n \rightarrow e^{+} \pi^{-}$ & 3.8$\times 10^{34}$   & 0.20$\times 10^{34}$ \\ 
\hline
$n \rightarrow \mu^{+} \pi^{-}$ & 2.9$\times 10^{34}$   & 0.10$\times 10^{34}$ \\
\hline
$p \rightarrow \overline{\nu} K^{+}$ & 2.5$\times 10^{34}$   & 0.40$\times 10^{34}$  \\ 
\hline
\end{tabular}
\end{center}
\end{table} 

\subsubsection{Summary}

We have studied the sensitivity of various nucleon decay searches in Hyper-Kamiokande. 
Table~\ref{tab:pdk-summary} shows the summary of the study for the highlighted modes, 
$p \rightarrow e^{+} \pi^{0}$ and $p \rightarrow \overline{\nu} K^{+}$. If the
proton lifetime is shorter than 5.7$\times 10^{34}$ years for the $p \rightarrow e^{+} \pi^{0}$ mode, or shorter than 1.0$\times 10^{34}$ years for $p \rightarrow \overline{\nu} K^{+}$, we could discover a signal over the atmospheric neutrino background events with a 3$\sigma$ significance by collecting a  5.6~Megaton$\cdot$year exposure. 

\begin{table}[htb]
\caption{Summary of the sensitivity study for a 5.6~Megaton$\cdot$year exposure 
for the $p \rightarrow e^{+} \pi^{0}$ and $p \rightarrow \overline{\nu} K^{+}$ modes. 
For $p \rightarrow \overline{\nu} K^{+}$,  method (1) $\mu$ + 6MeV$\gamma$ is labeled "Meth.1", method (2) ($\mu$) as "Meth.2", and method (3) $\pi^{+}\pi^{0}$ as "Meth.3".
\label{tab:pdk-summary}}
\vspace{0.4cm}
\begin{center}
\begin{tabular}{l|r|r|r|r}
\hline
        & $p \rightarrow e^{+} \pi^{0}$ & 
        \multicolumn{3}{c}{$p \rightarrow \overline{\nu} K^{+}$} \\ 
\hline
       &  & Meth.1 & Meth.2 & Meth.3 \\
\hline
\hline
Efficiency ($\%$)& 45   & 7.1  & 43  & 6.7  \\ 
\hline
Background (/Mton$\cdot$yr) & 1.6  & 1.6 & 1940 & 6.7 \\
\hline 
90$\%$ Sensitivity ($\times 10^{34}$ yrs)
                   & 13  & \multicolumn{3}{c}{2.5} \\
\hline
3$\sigma$ Discovery potential ($\times 10^{34}$ yrs)
                   & 5.7  & \multicolumn{3}{c}{1.0} \\
\hline 
\end{tabular}
\end{center}
\end{table}

\begin{figure}[htbp]
  \begin{center}
    \includegraphics[height=24pc]{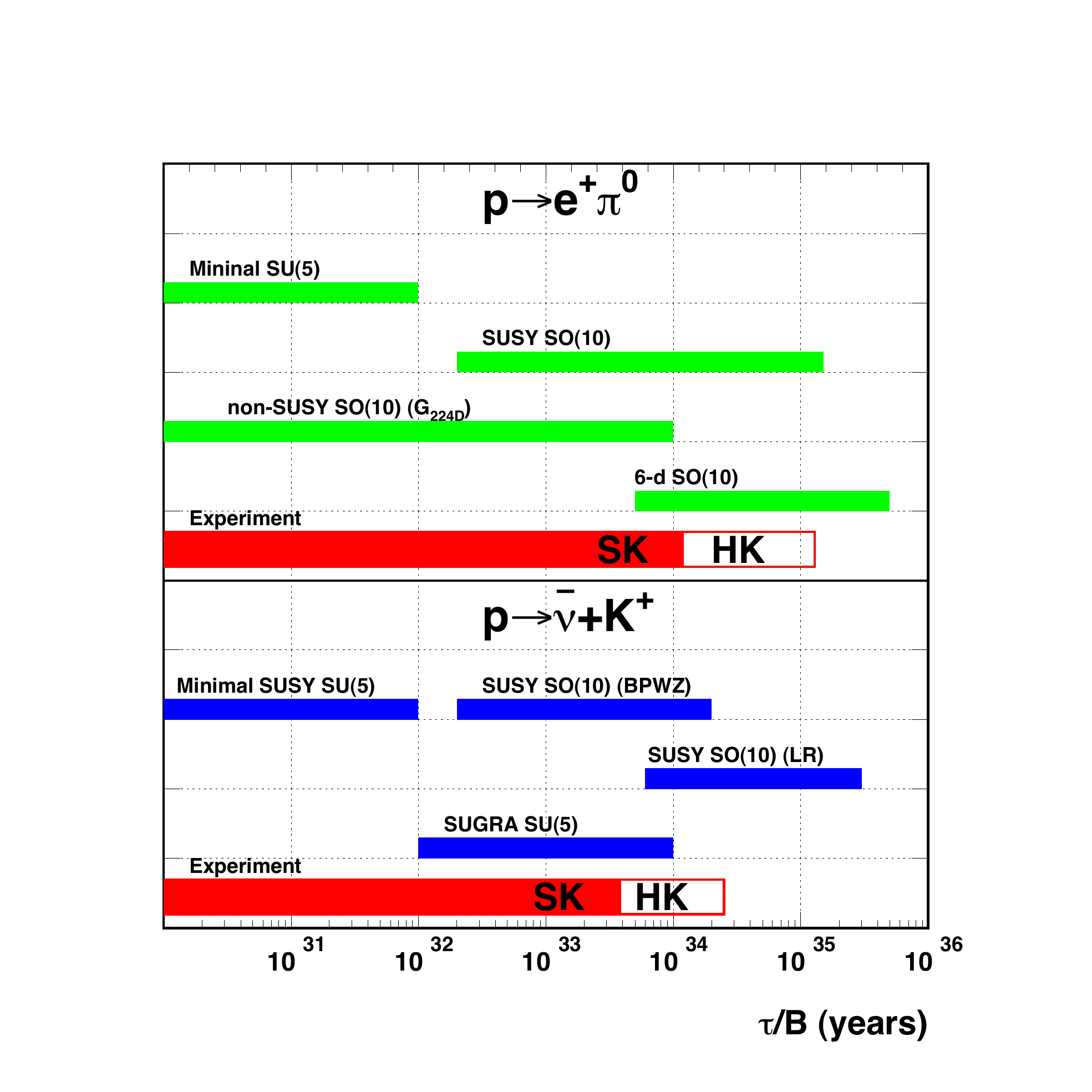}
  \end{center}
\caption {Proton lifetime predictions of several GUT models, the current experimental 
limits (90$\%$ CL) by Super-K, and the sensitivities of Hyper-Kamiokande with a 5.6~Megaton$\cdot$year exposure. Hyper-Kamiokande can cover most of the predicted range of the leading GUT models.}
  \label{fig:lifetime}
\end{figure}

\begin{figure}[htbp]
  \begin{center}
    \includegraphics[height=15pc]{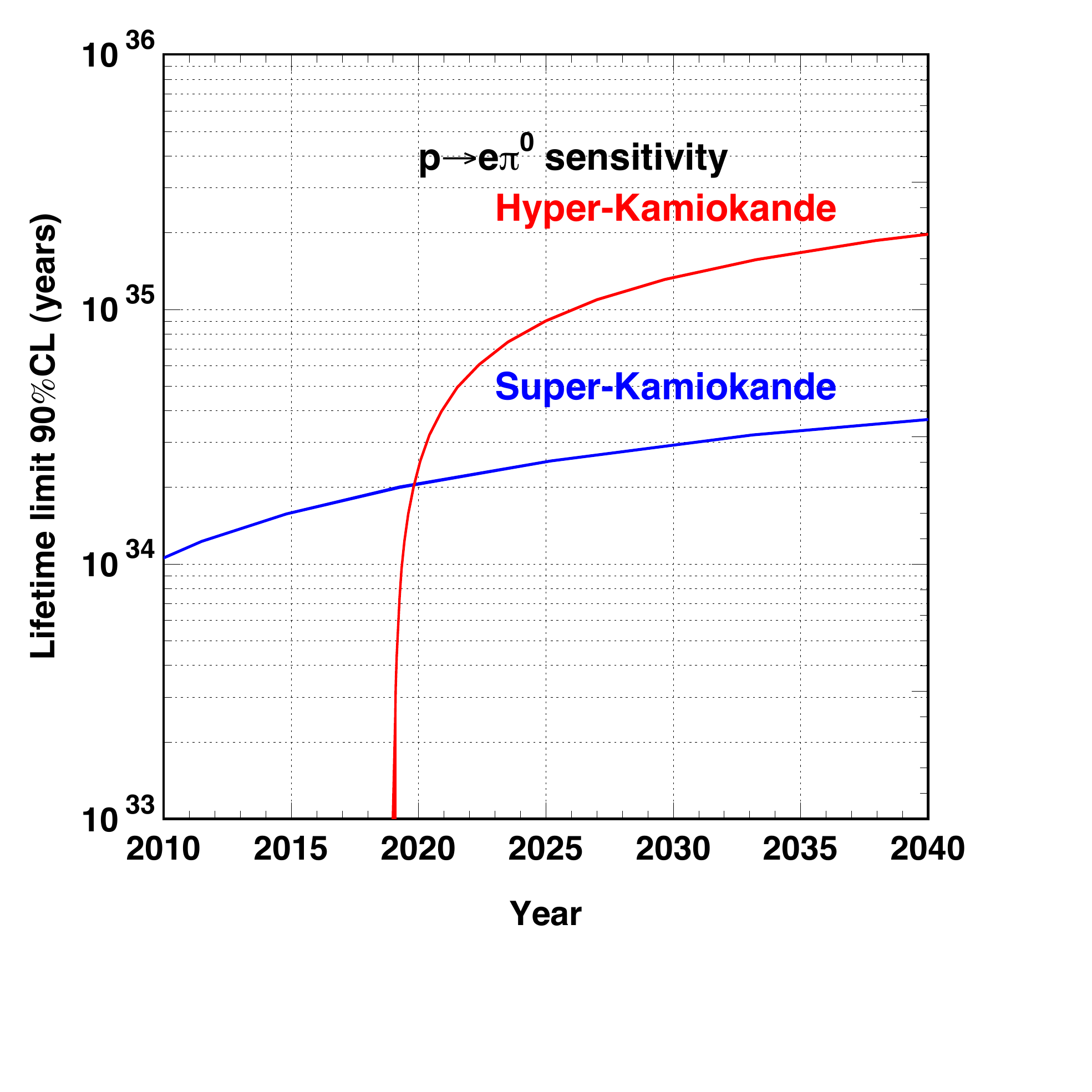}
    \includegraphics[height=15pc]{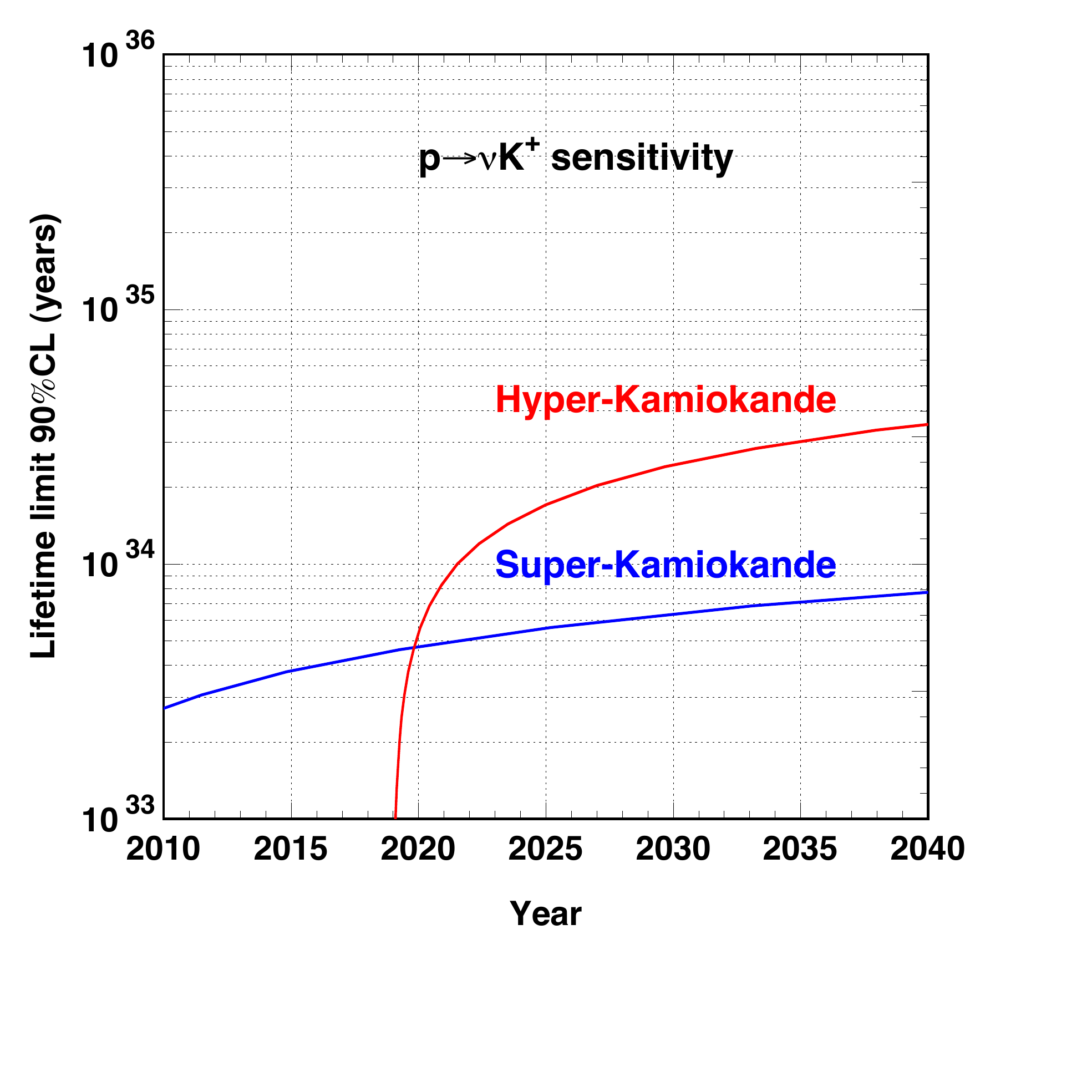}
  \end{center}
\caption {The proton decay search sensitivity as a function of year. The left plot is for the $p \rightarrow e^{+} \pi^{0}$ mode and the right is for the $p \rightarrow \overline{\nu} K^{+}$ mode. Hyper-Kamiokande is assumed to start from 2019; its results will overtake the Super-Kamiokande limits within one year.}
  \label{fig:sk-hk}
\end{figure}

Fig.~\ref{fig:lifetime} compares the proton lifetime predictions made by several leading theoretical models with the current experimental limits from Super-Kamiokande and the sensitivity of a 10 year Hyper-Kamiokande run.
Although Super-Kamiokande has been running well and gives us the most stringent limits for various nucleon decay searches, we have not observed any evidence of nucleon decay yet. 
The sensitivity of Hyper-Kamiokande can cover most of the predicted range of the major GUTs models.
Fig.~\ref{fig:sk-hk} shows the sensitivity of Super-Kamiokande and 
Hyper-Kamiokande as a function of the year. Hyper-Kamiokande is assumed to start data-taking from 2019. The nucleon decay results of Hyper-Kamiokande can overtake Super-Kamiokande within a single year of running.  There is no question that  to explore an order of magnitude longer lifetime regions, a larger detector is absolutely necessary.  As seen in this sensitivity study, Hyper-Kamiokande will open up a new decade, in more ways than one, in the search for nucleon decay. 

\clearpage

%% file: physics-solarnu/solarnu.tex
\clearpage
\subsection{Solar neutrinos}
\label{section:solar}

The effective fiducial volume for the solar neutrino analysis in Hyper-Kamiokande 
will be increased by a factor of 27 compared to Super-Kamiokande.
In this section, high statistics measurements of solar neutrinos
in Hyper-K are discussed.

In solar neutrino oscillations, regeneration of the electron
neutrinos through the Mikheyev-Smirnov-Wolfenstein (MSW) matter effect
\cite{Wolfenstein:1977ue,Mikheyev:1985zz,Mikheyev:1986zz} 
in the Earth is expected. 
Regeneration of the solar electron neutrinos in the Earth would constitute 
concrete evidence of the MSW matter effect, and so it is important to experimentally observe 
this  phenomenon.
However, the matter effect has not been directly confirmed yet, 
since the sensitivities of the current solar neutrino experiments 
are not sufficient.
According to the MSW model, the observed solar neutrino event rate 
in water Cherenkov detectors in the nighttime is expected to be higher 
-- by about a few percent in the current solar
neutrino oscillation parameter region --
than that in the daytime. 
We would like to measure this difference in Hyper-Kamiokande. 

Hyper-K also could be used for variability analyses of the Sun.
For example, the $^8$B solar neutrino flux highly depends on 
the Sun's present core temperature.
Unlike multiple scattered, random-walking photons or 
slow-moving helioseismic waves, free streaming 
solar neutrinos are the only available messengers with which to 
precisely investigate ongoing conditions in the core region of the Sun.
Hyper-K, with its unprecedented statistical power,  could measure the solar 
neutrino flux over short time periods. Therefore, short time variability of the 
temperature in the solar core could be monitored by the solar neutrinos in Hyper-K.

In order to achieve these precision measurements, 
background event levels must be sufficiently small.
Here, we have estimated the basic performance of Hyper-K for 
low energy events assuming some typical background levels.
In this study, the current analysis tools and the detector simulation 
for the low energy analysis \cite{Abe:2011xx} in Super-K were used.
The dark rate of the PMTs and the water transparency were assumed to be similar to
those in the current Super-K detector.
A brief summary of the low energy event reconstruction performance in Hyper-K 
is listed in Table~\ref{tab:performance}. 

The analysis threshold of the total energy of the recoil electrons 
in Hyper-K will be 7.0~MeV or lower, since a 7.0~MeV threshold was 
previously achieved in the SK-II solar neutrino analysis~\cite{sk2-solar}.
The current analysis tools will work all the way down to 4.5~MeV in Hyper-K with a 
vertex resolution of 3.0~m.  Not surprisingly, higher energy events will be reconstructed 
with even better vertex resolution.

\subsubsection{Background estimation}
\label{section:solar_bg}
The major background sources for the $^8$B solar neutrino measurements
are the radioactive daughter isotopes of Rn-222 in water and 
the radioactive spallation products created by cosmic-ray muons. 
Rn-222 will be reduced to a similar (or lower) level as that currently in the Super-K detector, 
since Hyper-K will employ a similar water purification system and design improvements 
may well occur over the next several years. 
However, the spallation products will be increased in Hyper-K.
In the current design, the cosmic-ray muon rate is expected to be 
increased by a factor of about 10 in equal volumes, as discussed in  
Sec.~\ref{section:site}. 
The spallation products will not simply be increased by the same factor.
This is because high energy cosmic-ray muons tend to produce the spallation
products, while the average energy of the cosmic-ray muons at the shallower Hyper-K site is 
expected to be lower than that at the deeper Super-K site; greater overburden means 
less muons, but it also means those that do get through are more energetic.
We have estimated the average energies of the cosmic-ray muons to be  
$\sim 560$~GeV at the Super-K site and $\sim 300$~GeV at the Hyper-K site.
Considering that the spallation production cross section is proportional 
to the 0.7-th power of the cosmic-ray muons' energy,  
the density of spallation products will be increased by a factor of  6--7.
We found the remaining spallation products will be increased by another 
factor of 3 (at most) with the current analysis tools.  This is due to decreasing efficiency
for separating spallation products from signal events 
with increasing the cosmic-ray muon rate. 
So, the density of the remaining spallation products will be increased 
by a factor of 20 at most in Hyper-K.
However, this could (and most likely will) be reduced by ongoing 
improvements of the analysis tools.

In Super-K, angular information is used to extract the solar neutrino
signal events \cite{full-solar}. 
We have estimated the possible effect of the background level in the
signal extraction after considering angular information.

In this study, we used 9.0--9.5~MeV Super-K-I data as a reference.
The extracted solar neutrino signal events and background events 
in this energy region over the entire run period (0.09~Megaton$\cdot$years) were
1350 events and 7700 events, respectively.
So, the Signal-to-Noise (S/N) ratio is 18\%. 
We made artificial data samples with reduced S/N ratios, then 
applied the signal extraction.
As a result, we found the expected statistical error is almost the square root 
of 2 $\sim$ 15 times the number of signal events for 1 $\sim$ 20 times the Super-K-I
background level, respectively.
Table~\ref{tab:sol-bgtest} shows a summary of the expected statistical
errors in a Super-K-I type detector with increased backgrounds, as well as that of Hyper-K.
\begin{table}[htb]
 \caption{Expected statistical uncertainties for 10000 signal events with increased background 
 levels. The Super-K-I solar neutrino data sample between 9.0--9.5~MeV was used as a reference 
 The 3rd column is the Hyper-K factor relative to Super-K given the same observation time.
  To estimate the 3rd column, the same detector resolution
  and 0.56~Mton fiducial volume are assumed in Hyper-K. }
\begin{center}
\begin{tabular}{cccc}
\hline \hline
Background level & Stat. err. in SK &  & Stat. err. in HK\\
\hline
SK-I BG $\times 20$ & 3.6\% &  & $\times 1/2.0$ \\ 
SK-I BG $\times 10$ & 2.7\% &  & $\times 1/2.7$ \\ 
SK-I BG $\times 7$  & 2.4\% &  & $\times 1/3.1$ \\ 
SK-I BG $\times 5$  & 2.1\% &  & $\times 1/3.5$ \\ 
SK-I BG             & 1.4\% &  & $\times 1/5.2$ \\ 
\hline \hline
\end{tabular}
\end{center}
\label{tab:sol-bgtest}
\end{table}
Once the angular distribution is used to extract the solar signal, 
the statistical error on this signal would be reduced by a factor of 2.0 in Hyper-K, 
even though the background level is increased by a factor of 20, 
for the same observation time assuming both detectors have identical resolution.

In summary, Hyper-K will provide higher statistical measurements of
solar neutrinos than Super-K, even though there will be more spallation
backgrounds.

\subsubsection{Oscillation study}

In solar neutrino oscillations, a difference in the solar neutrino 
event rates during the daytime and the nighttime is expected from 
the MSW effect in the Earth. 
This is called the day/night asymmetry;  it has not yet been observed. 
In Hyper-K, a precise measurement of the day/night asymmetry will be
performed using higher statistics than those available in Super-K.

\begin{figure}[htb]
 \begin{center}
  \includegraphics[height=9.5cm]{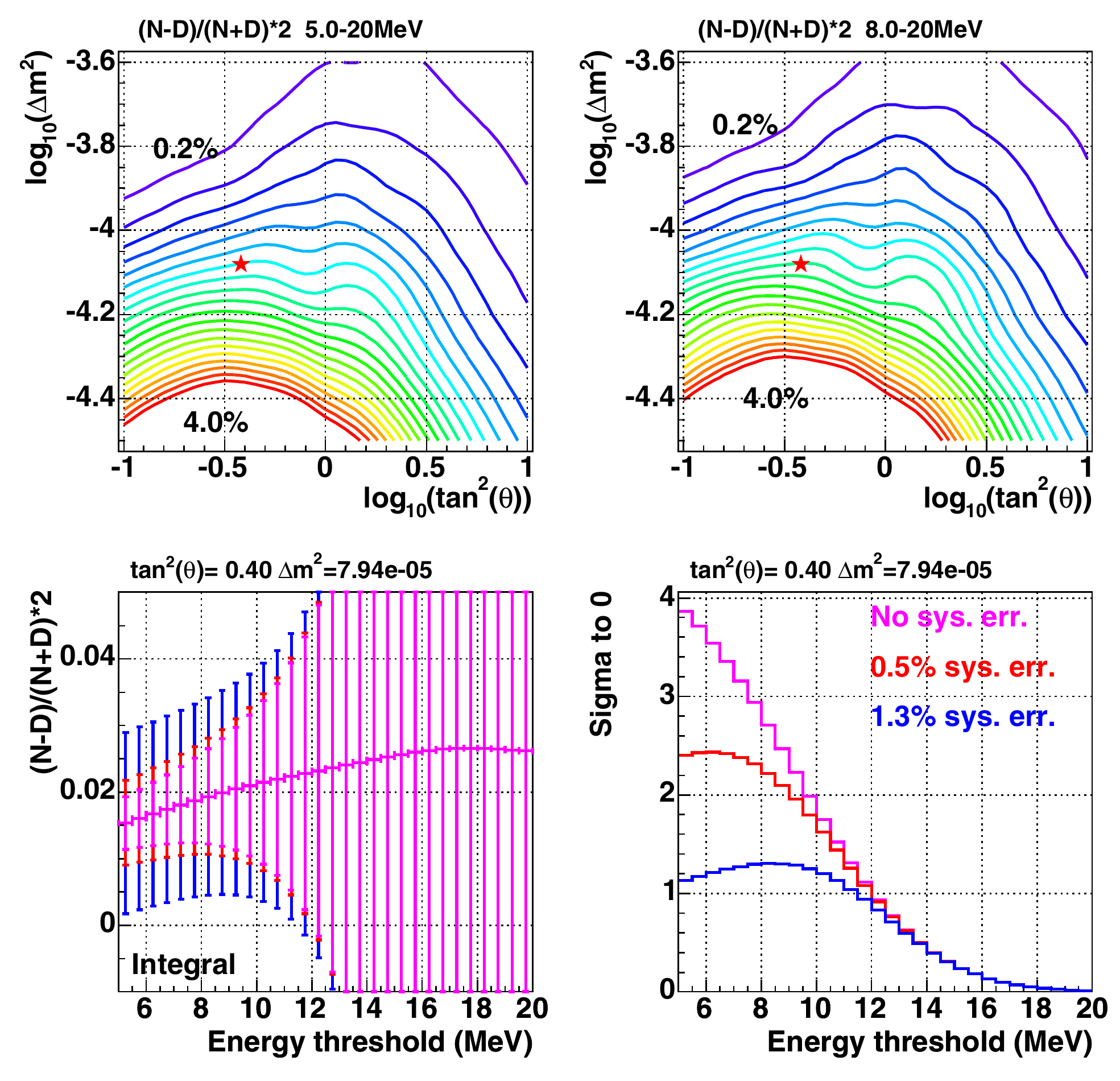} \\
 \end{center}
 \caption{Expected day/night asymmetry in a megaton water Cherenkov
  detector. 40\% photo-coverage, 0.5~Megaton$\cdot$years daytime data 
  and 0.5~Megaton$\cdot$years nighttime data are assumed. 
  The effect of background events and reduction efficiencies are not considered.   
  Upper left: expected day/night asymmetry in the 5.0--20~MeV electron 
  total energy region. 
  Upper right: expected day/night asymmetry in the 8.0--20~MeV region.
  Lower left: expected day/night asymmetry with uncertainties as a
  function of the lower energy threshold at 
 $(\tan^2 \theta_{12}, \Delta \rm{m}^2_{21}) = (0.40, 7.9 \times 10^{-5} {\rm eV}^2)$. 
 The upper energy threshold is 20~MeV.
 The meaning of the different colors are defined in the lower right plot.
 Lower right: expected day/night significance as a function of the
 energy threshold.}
 \label{fig:sol-dn1}
\end{figure}

The upper plots in Fig.~\ref{fig:sol-dn1} show the expected day/night
asymmetries with different lower energy thresholds. 
The expected day/night asymmetry is at about the 1\% level around the current
solar global oscillation parameters. 
In order to observe the day/night asymmetry in Hyper-K, 
we must reduce the up-down systematic uncertainty below that level.

\begin{figure}[htb]
 \begin{center}
  \includegraphics[height=6.5cm]{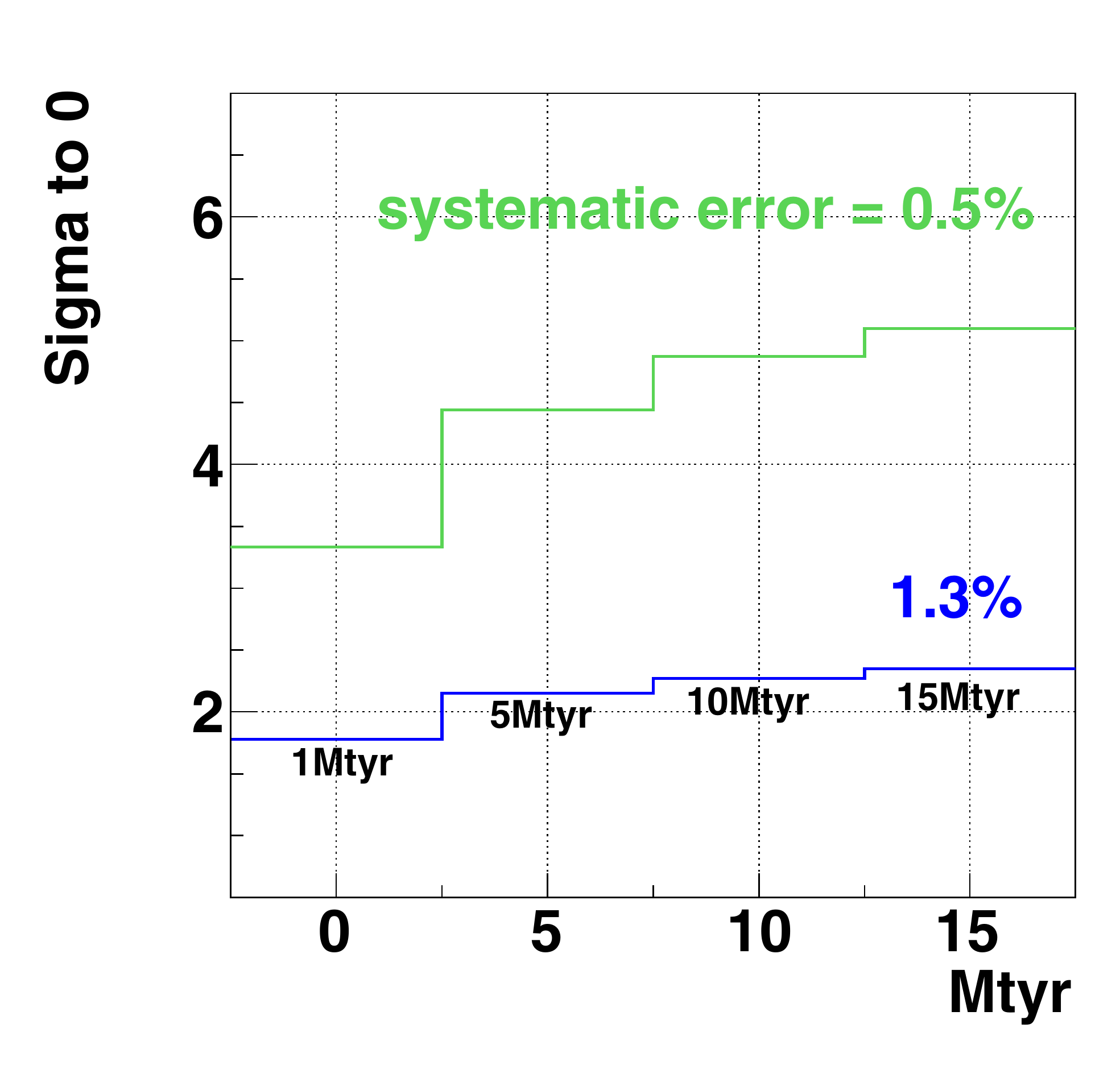} \\
 \end{center}
 \caption{Expected day/night significance as a function of the
 observation time near the solar global oscillation best-fit parameters.
 The Super-K-I value of S/N is assumed. 
 The total electron energy region is 5.0--20~MeV.}
 \label{fig:sol-dn2}
\end{figure}

The expected day/night asymmetry in the high energy region is larger 
than that in the low energy region, as shown in the lower left plot in
Fig.~\ref{fig:sol-dn1}.
So, high statistics data in this higher energy region would be desirable. 
We have studied two typical values of systematic uncertainty, where the one at 
1.3\% corresponds to the up-down systematic uncertainty of Super-K.
From the lower right plot in Fig.~\ref{fig:sol-dn1},  
the most sensitive lower energy threshold would be 6~MeV and 8~MeV for the
0.5\% and 1.3\% up-down systematic uncertainties, respectively.
Figure~\ref{fig:sol-dn2} shows expected day/night significance 
as a function of the observation time.
Since the expected day/night asymmetry is small, it will be important to
reduce the systematic uncertainties in order to observe the day/night
asymmetry with high precision.
We believe that this should be possible, especially if we design and prepare 
the necessary calibration devices during detector construction.

\subsubsection{Time variation study}

Solar neutrinos could be used as a direct probe of the nuclear 
reactions taking place in the solar core.
In particular, the $^8$B solar neutrino flux has a remarkable $T^{18}$ dependence 
according to Standard Solar Model (SSM) \cite{bahcall-textbook}. 
Here, $T$ is the solar core temperature, and with such a high-order dependence
it is possible that even modest changes in the solar core temperature 
could be amplified into something detectable via measurements of the $^8$B solar neutrino flux.

Assuming the statistical uncertainties estimated in 
Sec.~\ref{section:solar_bg} can be used for Hyper-K,
the expected uncertainty on the solar core temperature  
when the background level is increased by a factor of 20 
would be the following:
\[
 \frac{\sigma_T}{T} 
  = \frac{1}{18} \frac{\sigma_N}{N} 
  = \frac{\sqrt{15 \cdot N}}{18 \cdot N}
\]
Here $N$, $\sigma_{\rm T}$, and $\sigma_{\rm N}$ are the number of 
observed $^8$B solar neutrinos, error in $T$, and error in $N$,
respectively. 
The expected number of observed $^8$B solar neutrinos 
in Hyper-K is 200 events per day above 7.0~MeV, as shown 
in Table~\ref{tab:targets}. 
When $N$ is $200$, $\sigma_T/T$ will be $0.015$.
Therefore, the solar core temperature could be monitored 
within a few percent accuracy day by day.  Naturally, by 
integrating over longer periods, more 
subtle temperature changes - potentially down to the 0.1\% level - could be monitored.

\subsubsection{Summary}

In this section, rough estimates of potential solar neutrino measurements 
are reported.
The solar neutrino analysis is sensitive to the detector resolutions 
and background levels.
We have estimated expected sensitivities based on the 
current Super-K analysis tools.

As a result of its shallower site,  the increase of the background level 
in Hyper-K will be up to a factor of 20 as compared to Super-K.  
However -- due to its much greater size --  
the statistical uncertainties on solar neutrino measurements would actually be 
reduced by a factor of at least two in Hyper-K as compared to Super-K on an equal time 
basis, assuming similar detector resolutions.

The day/night asymmetry of the solar neutrino flux -- concrete evidence of the matter effect on oscillations -- could be discovered and then precisely 
measured in Hyper-K, given that the detector up-down 
response is understood to better than about 1\%.  Good calibration tools 
will be a must for this physics. 

Hyper-K will provide short time and high precision variability analyses of  the solar core activity. 
The solar core temperature could be monitored within a few percent
accuracy day by day, and to a tenth of a percent over the period of several months.

%% file: physics-supernova/supernova.tex
\subsection{Astrophysics}\label{section:astro}

In this section, potential signals arising from several astrophysical sources are considered, and
the prospects for their detection and study in Hyper-Kamiokande are briefly discussed.

\subsubsection{Supernova burst neutrinos}\label{sec:supernova}

Core collapse supernova explosions are the last process in the evolution of 
massive ($>8$M$_{\rm sun}$) stars.  Working their way successively through periods of 
predominantly hydrogen fusion, helium fusion, and so on, eventually silicon 
fusion starts  making iron.  
Once an iron core has formed, no more energy can be
released via its fusion into still-heavier elements, and the hydrodynamic balance 
between gravity and stellar burning is finally and catastrophically 
disrupted.   The sudden gravitational collapse of their iron cores -- 
each one of which then goes on to form either a neutron star or a black hole -- is 
the main source of energy from this type of supernova explosion.
The energy released by a supernova is estimated to be $\sim 3 \times 10^{53}$~ergs, 
making it one of the most energetic phenomena in the universe.
Since neutrinos interact weakly with matter, almost 99\% of the released energy
from the exploding star is carried out by neutrinos. As a result, 
the detection of supernova neutrinos gives
direct information of  energy flow during the explosion.
The neutrino emission from a core collapse supernova starts with a short 
($\sim$10~millisecond) burst phase of electron captures 
($p + e^- \rightarrow n + \nu_e$) called the neutronization burst, 
which releases about $10^{51}$~ergs.  
Following that, the majority of the burst energy is released
by an accretion phase ($< \sim$1~second) and a cooling phase 
(several seconds) in which all three types of
neutrinos (including anti-neutrinos) are emitted.

The observation of a handful (25 in total) of supernova burst neutrinos from SN1987a by
the Kamiokande, IMB, and Baksan experiments proved the basic scenario of 
the supernova explosion was correct. 
However, more than two decades later the detailed 
mechanism of explosions is still not known. 
Several groups 
around the world are trying to explode supernovae in computer 
simulations. However, they have not been completely successful yet.
It seems that some physical processes are lacking in the present 
simulations. 
In order to fully understand supernova explosions,
it is necessary to detect many more supernova neutrinos. 
If a supernova explosion occurs halfway across our galaxy, 
the Hyper-Kamiokande detector would detect approximately 170,000$\sim$260,000 neutrino
events. This very large statistical sample should at last reveal the detailed
mechanism of supernova explosions.

Supernova neutrinos also give us an opportunity to investigate unknown 
properties of neutrinos. Because of quite high densities at the supernova
core, neutrino-neutrino self-interaction is not negligible.
As described later in this section, the neutrino mass hierarchy could be 
determined using the time variation 
of the energy spectrum.  The 
very short rise time of the outbreak of the burst enables us to discuss 
neutrino direct masses. Also, a detection of neutrinos from distant sources 
give us a chance to check the lifetime of neutrinos.  Indeed, by far the oldest neutrinos ever 
seen are those of SN1987a.  They were about 170,000 years old when they arrived, while
the next oldest would be the solar neutrinos, seen on Earth a mere eight minutes after their birth 
in the heart of the Sun.

Taking into account the Mikheyev-Smirnov-Wolfenstein (MSW) matter effect
through the stellar medium, the flux of each neutrino type emitted from a 
supernova is related to the originally produced fluxes  
($F^0_{\nu_e}$, $F^0_{\overline\nu_e}$ and
$F^0_{\nu_x}$, where $\nu_x$ is $\nu_{\mu,\tau}$ and 
${\overline\nu_{\mu,\tau}}$)
by the following formulas \cite{Dighe:1999id, Fogli:2004ff} :

\noindent
For normal hierarchy,
\begin{eqnarray}
F_{\overline\nu_e} &\simeq& \cos^2\theta_{12} F^0_{\overline\nu_e}+
\sin^2\theta_{12}F^0_{\nu_x}\ , 
\nonumber \\
F_{\nu_e} &\simeq & \sin^2\theta_{12} P_H F^0_{\nu_e}+
(1-\sin^2\theta_{12}P_H)F^0_{\nu_x},
\nonumber \\
F_{\nu_\mu} + F_{\nu_\tau} &\simeq & (1-\sin^2\theta_{12} P_H) F^0_{\nu_e} + (1 + \sin^2\theta_{12} P_H) F^0_{\nu_x} ,
\nonumber \\
F_{\overline\nu_\mu}+ F_{\overline\nu_\tau}  &\simeq &  (1-\cos^2\theta_{12}) F^0_{\overline\nu_e} +
        (1 + \cos^2\theta_{12}) F^0_{\nu_x} ,
\nonumber 
\end{eqnarray}

\noindent
and, for inverted hierarchy,
\begin{eqnarray}
F_{\overline\nu_e} &\simeq & \cos^2\theta_{12} P_H F^0_{\overline\nu_e}+
(1-\cos^2\theta_{12} P_H)F^0_{\nu_x}\ , 
\nonumber \\
F_{\nu_e} &\simeq& \sin^2\theta_{12} F^0_{\nu_e}+
\cos^2\theta_{12}F^0_{\nu_x} ,
\nonumber \\
F_{\nu_\mu} + F_{\nu_\tau} &\simeq & (1-\sin^2\theta_{12}) F^0_{\nu_e} + (1 + \sin^2\theta_{12}) F^0_{\nu_x} ,
\nonumber \\
F_{\overline\nu_\mu}+ F_{\overline\nu_\tau}  &\simeq &  (1-\cos^2\theta_{12} P_H) F^0_{\overline\nu_e} + (1 + \sin^2\theta_{12} P_H) F^0_{\nu_x} ,
\nonumber
\end{eqnarray}

\noindent
where $P_H$ is the crossing probability through the matter resonant layer
corresponding to $\Delta m^2_{32}$.
$P_H = 0$($P_H = 1$) for adiabatic (non-adiabatic) transition.
For $\sin^22\theta_{13} > 10^{-3}$, adiabatic transition is expected for
the matter transition in the supernova envelope.
The supernova neutrino spectrum is affected not only by stellar matter
but also by other neutrinos and anti-neutrinos at the high density core
(so-called collective effects).
The collective effects swap the $\nu_{e}$ and $\overline\nu_e$ spectra
with those of $\nu_{x}$ in certain energy intervals bounded by sharp 
spectral splits \cite{Dasgupta:2009dd}.
The combination of the collective effects and the stellar matter
effects might weaken swapping the energy spectra of 
$\nu_{e}$/$\overline\nu_e$ and $\nu_{x}$.
So, in the following description of the performance of the Hyper-K
detector, three cases are considered  in order to fully cover the 
possible variation of expectations: (1) no~oscillations,  (2)  
normal~hierarchy with $P_H=0$, and (3) inverted~hierarchy with $P_H=0$.  
An assumed oscillation parameter relevant for the description is
$\sin^2\theta_{12}=0.31$.
Concerning the neutrino fluxes and energy spectrums at the production
site, we used results obtained by the Livermore simulation \cite{Totani:1997vj}.

\begin{figure}[tbp]
  \begin{center}
    \includegraphics[width=16cm]{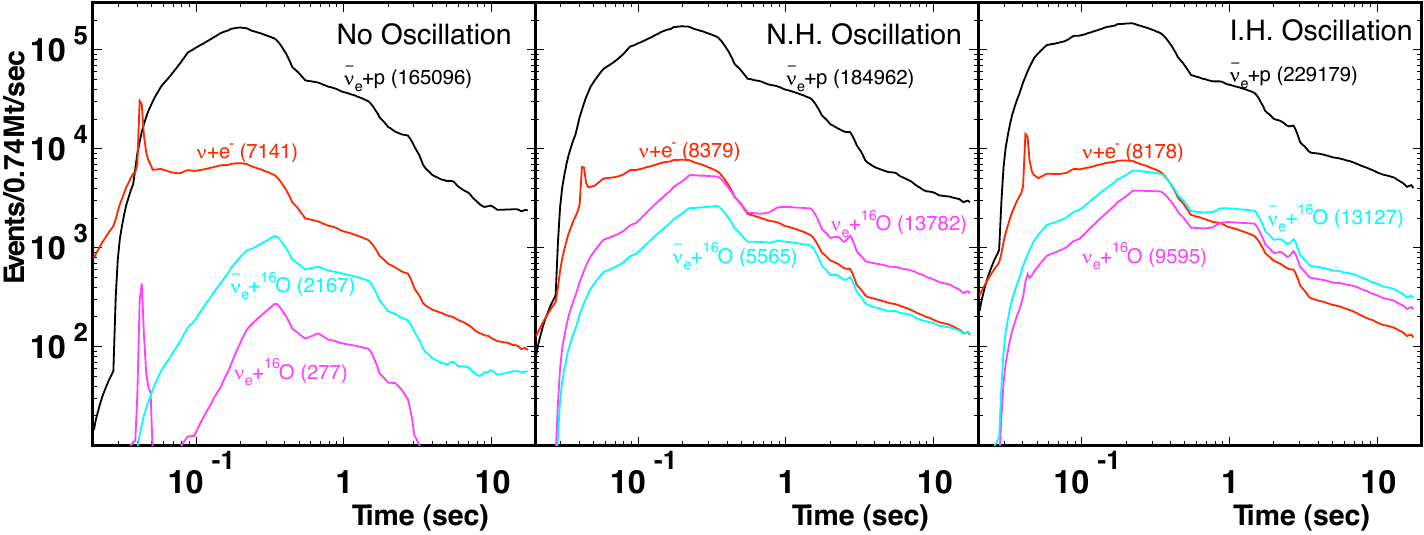}
  \end{center}
\vspace{-1cm}
  \caption{Expected time profile of a supernova at 10~kpc. 
Left, center, and right figures show profiles for no oscillation, normal 
hierarchy (N.H.), and inverted hierarchy (I.H.), respectively.
Black, red, purple, and light blue curves show event rates for
interactions of inverse beta ($\bar{\nu}_e + p \rightarrow e^+ + n$), 
$\nu e$-scattering($\nu + e^- \rightarrow \nu + e^-$), 
$\nu_e~^{16}$O CC($\nu_e + {\rm ^{16}O} \rightarrow e^- + {\rm^{16}F^{(*)}}$),
and 
$\bar{\nu}_e~^{16}$O CC
($\bar{\nu}_e + {\rm ^{16}O} \rightarrow e^+ + {\rm ^{16}N^{(*)}}$), respectively.
The numbers in parentheses are integrated
number of events over the burst. 
The fluxes and energy spectrums are from the Livermore 
simulation \cite{Totani:1997vj}}
  \label{fig:sn-rate}
\end{figure}

\begin{figure}[tbp]
  \begin{center}
    \includegraphics[width=8cm]{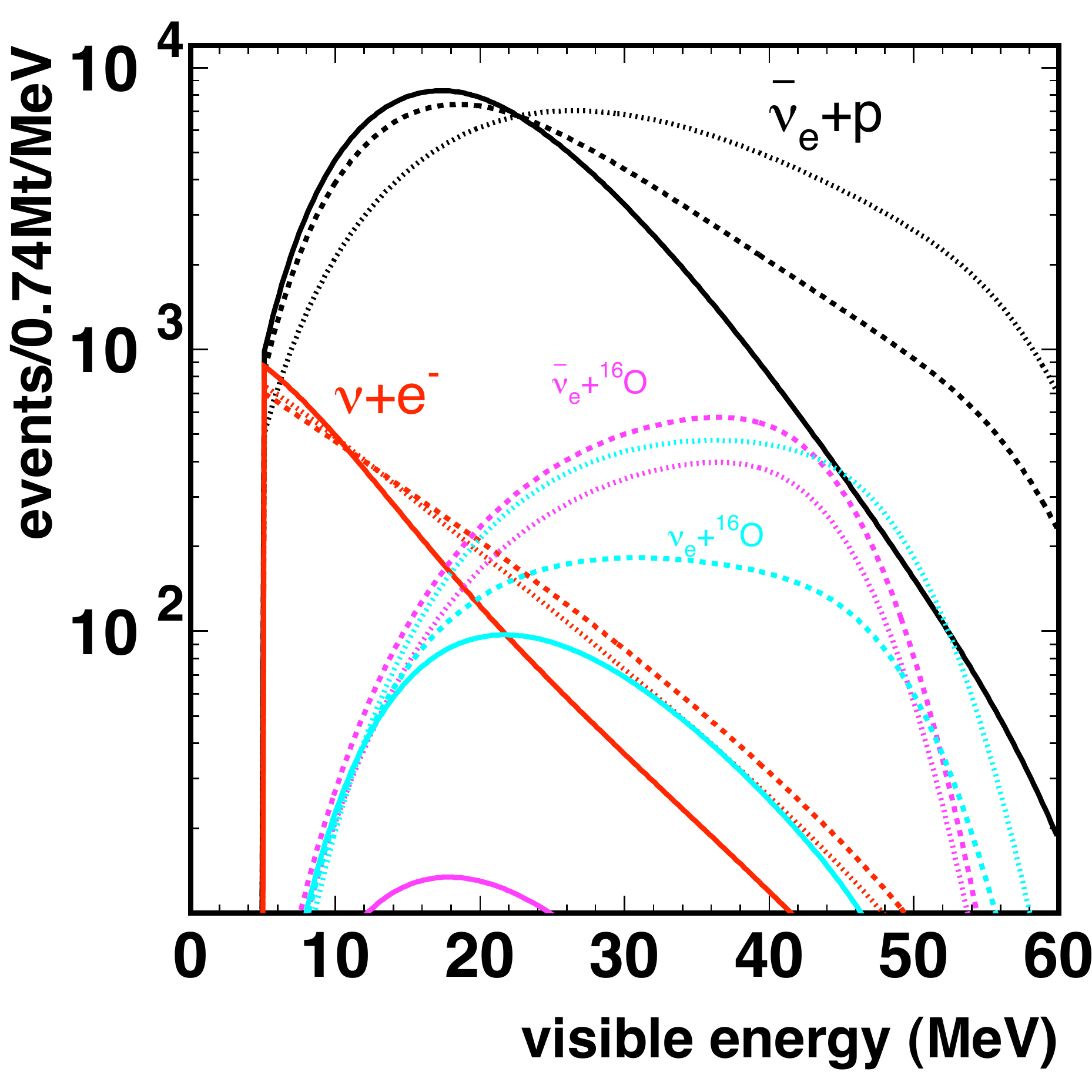}
  \end{center}
\vspace{-1cm}
  \caption{Visible energy spectrum for each interaction
for a supernova at 10~kpc. 
Black, red, purple, and light blue curves show event rates for
interactions of inverse beta, $\nu e$-scattering, 
$\nu_e ^{16}$O CC, and $\bar{\nu}_e ^{16}$O CC, respectively.
Solid, dashed, and dotted curves correspond to no oscillation, N.H., and
I.H., respectively.}
  \label{fig:sn-spec}
\end{figure}

Figure~\ref{fig:sn-rate} shows time profiles for various 
interactions expected at the Hyper-K detector, if a supernova at a
distance of 10~kiloparsecs (kpc)  is observed. This distance is a bit farther than
the center of the Milky Way galaxy at 8.5~kpc; it is chosen as being representative 
of what we might expect since a volume with a radius of 10~kpc centered at Earth 
includes about half the stars in the galaxy.
The three graphs in the figure show
the cases of no oscillation, normal hierarchy (N.H.) and 
inverted hierarchy (I.H.), respectively.
Colored curves in the figure show event rates for
inverse beta ($\bar{\nu}_e + p \rightarrow e^+ + n$), 
$\nu e$-scattering($\nu + e^- \rightarrow \nu + e^-$), 
$\nu_e~^{16}$O CC($\nu_e + {\rm ^{16}O} \rightarrow e^- + {\rm ^{16}F^{(*)}}$),
and 
$\bar{\nu}_e~^{16}$O CC
($\bar{\nu}_e + {\rm ^{16}O} \rightarrow e^+ + {\rm ^{16}N^{(*)}}$).
The burst time period is about 10~s and the peak event rate of
inverse beta events reaches about 200 kHz.
A sharp timing spike is expected for $\nu e$-scattering events at
the time of neutronization, which will be discussed in more detail later.  
Visible energy distributions of each interaction are shown in 
Fig.~\ref{fig:sn-spec}, where the visible energy is 
the electron-equivalent energy measured by a Cherenkov detector. 
The distribution of inverse beta events directly 
gives the energy spectrum of $\bar{\nu}_e$ because the energy
of the positrons ($E_{e^+}$) is simply  $E_{\nu} - 1.3~{\rm MeV}$.
The energy spectrum of $\nu$e-scattering events has enhancement at lower
energies because of the nature of the interaction.

\begin{figure}[tbp]
  \begin{center}
    \includegraphics[width=8cm]{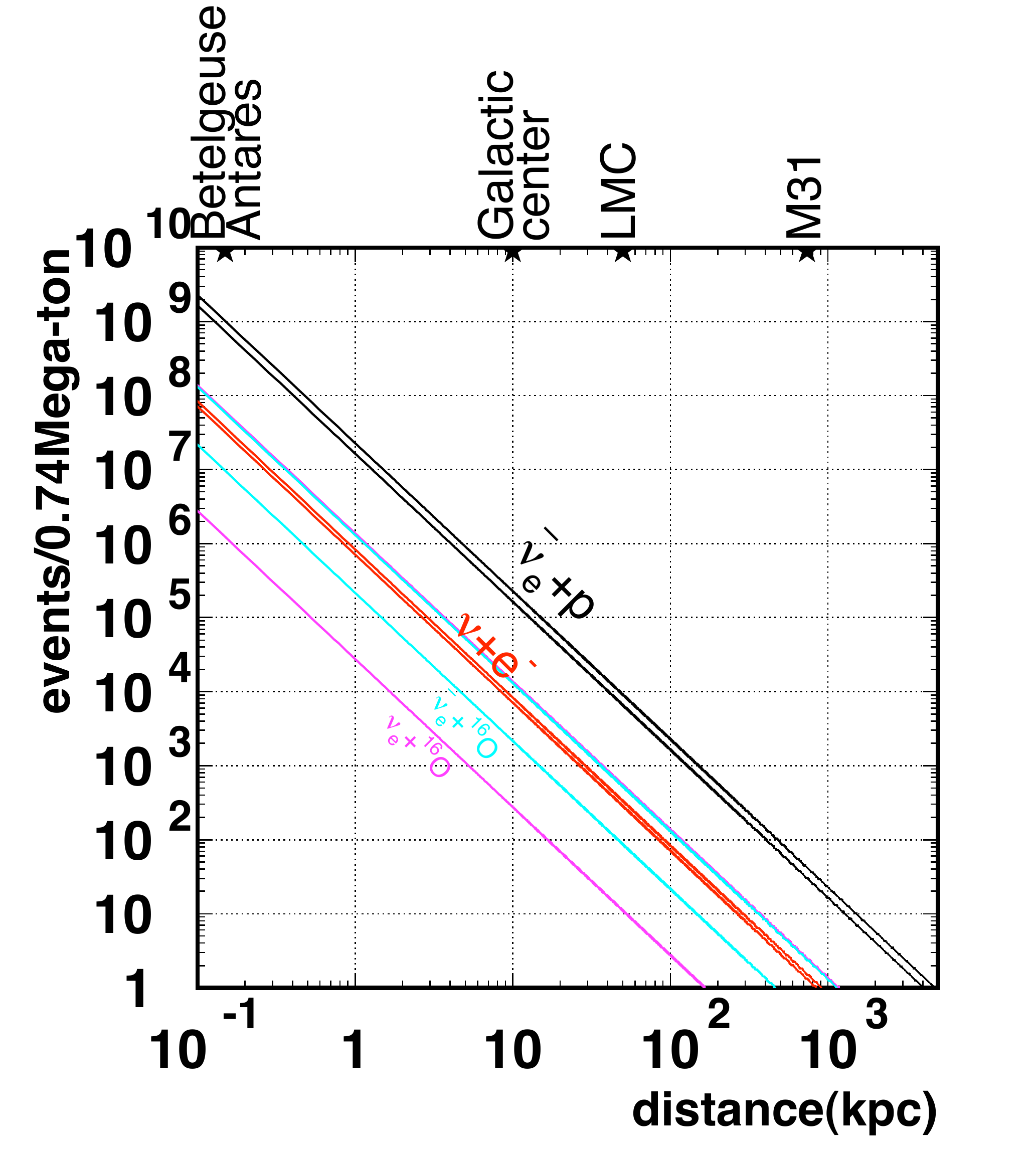}
  \end{center}
\vspace{-1cm}
  \caption{Expected number of supernova burst events for each
interaction as a function of the distance to a supernova.
The band of each line shows the possible variation due to the assumption
of neutrino oscillations.}
   \label{fig:sn-evtvsdist}
\end{figure}

Figure \ref{fig:sn-evtvsdist} shows the expected number of supernova
neutrino events at Hyper-K versus the distance to a supernova.
At the Hyper-K detector, we expect to see about 165,000$\sim$230,000 
inverse beta events, 7,000$\sim$8,000 $\nu e$-scattering events, 
300$\sim$14,000 $\nu_e ^{16}$O CC events, and 
2,000$\sim$13,000 $\bar{\nu}_e ^{16}$O CC events, in 
total 170,000$\sim$260,000 events,
for a 10~kpc supernova.
The range of each of these numbers covers possible variations due to the 
neutrino oscillation scenario (no oscillation, N.H., or I.H.).
Even for a supernova at M31 (Andromeda Galaxy), about 30$\sim$50 
events are expected at Hyper-K.
In the case of the  Large Magellanic Cloud (LMC) where SN1987a was located, 
about 7,000$\sim$10,000 events are expected.

\begin{figure}[tbp]
  \begin{center}
    \includegraphics[width=10cm]{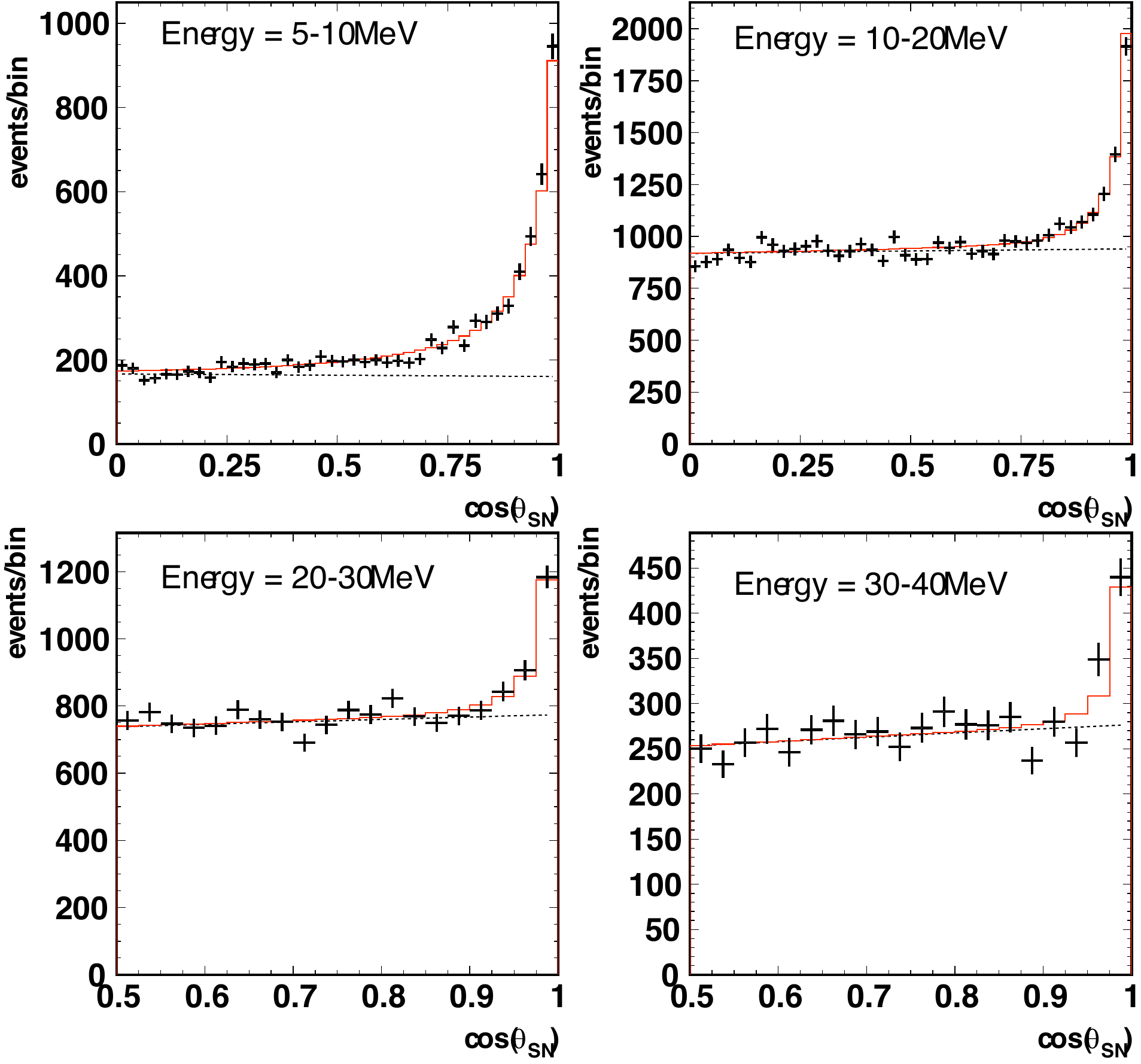}
  \end{center}
\vspace{-1cm}
  \caption{Angular distributions of a simulation of
a 10~kpc supernova. The plots show a visible energy range of 
5-10~MeV (left-top), 
10-20~MeV (right-top), 20-30~MeV (left-bottom), and 30-40~MeV (right-bottom).
The black dotted line and the red solid histogram (above the black dotted line)
are fitted contributions of inverse beta and $\nu e$-scattering events.
Concerning the neutrino oscillation scenario, the $no~oscillation$ case is shown
here.}
  \label{fig:sn-cossn}
\end{figure}

Figure \ref{fig:sn-cossn} shows expected angular distributions with respect 
to the direction of the supernova for four visible energy ranges.
The inverse beta events have a nearly isotropic angular distribution.
On the other hand, $\nu e$-scattering events have a strong peak
in the direction coming from the supernova.
Since the visible energy of $\nu e$-scattering events are lower than
the inverse beta events, the angular distributions for lower energy
events show more enhanced peaks.
The direction of a supernova at 10~kpc can be reconstructed with an accuracy of
about two degrees according to these angular distributions. 
In Hyper-K, we can statistically extract an energy distribution of 
$\nu_e + \nu_X$  ($X=\mu,\tau)$ events using the angular
distributions in much the same way as solar neutrino signals are separated from background
in Super-K. 
Although the effect of neutrino oscillations must be
taken into account, the $\nu_e + \nu_X$ spectrum gives another handle for 
discussing the temperature of neutrinos.
Note that inverse beta events directly provide a very precise measurement of the temperature
of $\bar{\nu}_e$. Hyper-K will be able to evaluate the temperature difference 
between $\bar{\nu}_e$ and $\nu_e + \nu_X$.  This would be a valuable input to model builders.

\begin{figure}[tbp]
  \begin{center}
    \includegraphics[width=7cm]{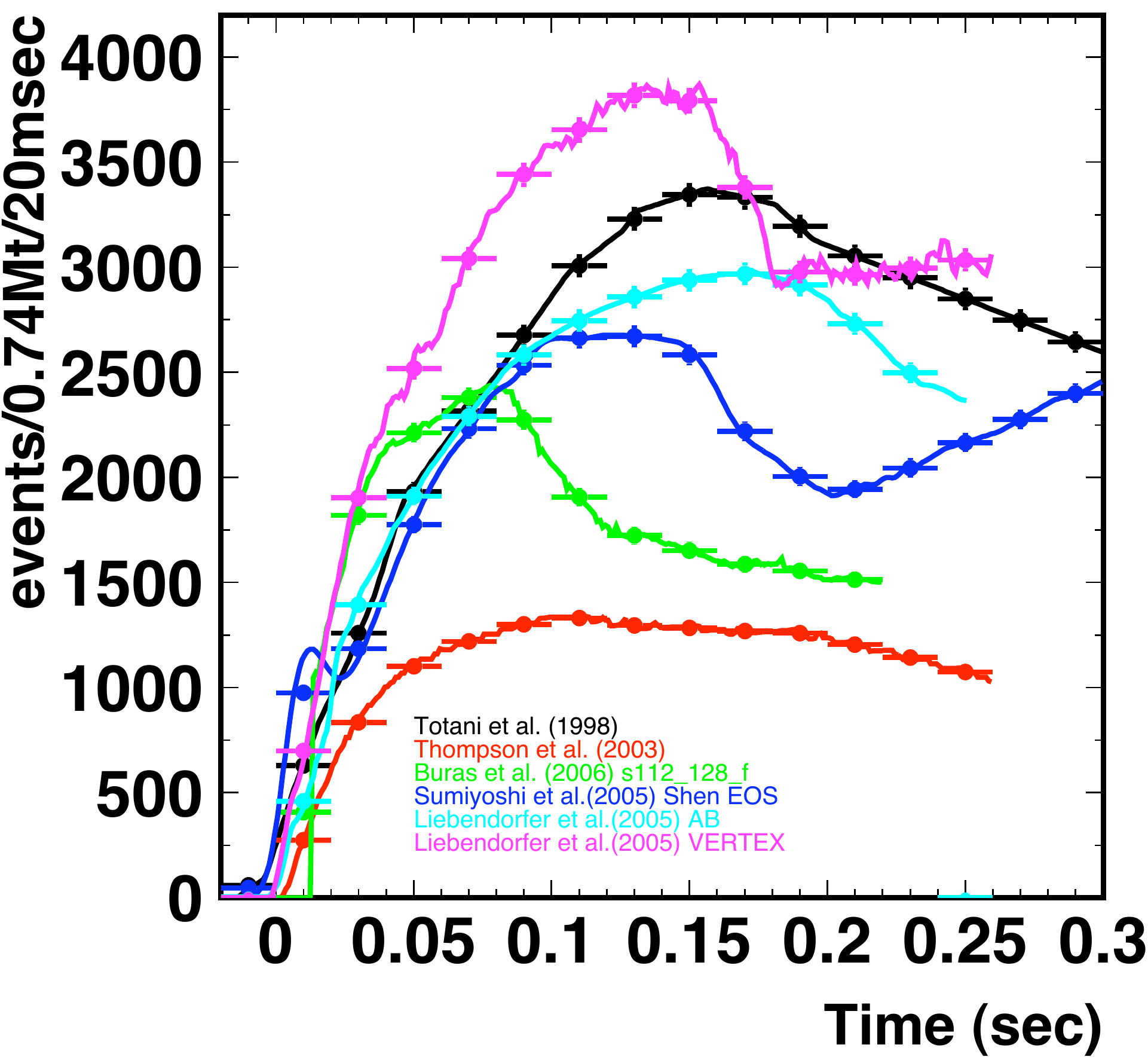}
    \includegraphics[width=7cm]{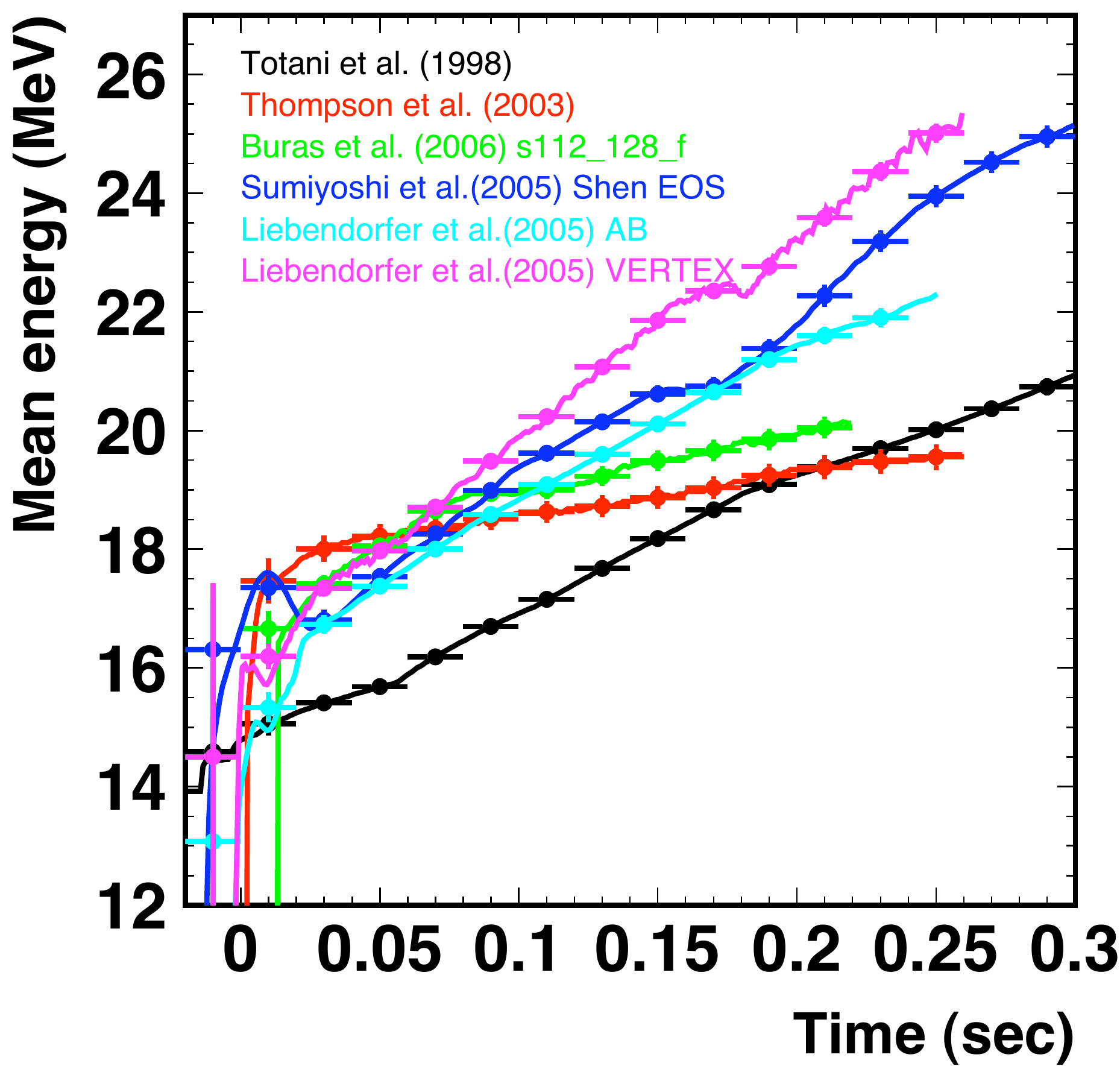}
  \end{center}
\vspace{-1cm}
  \caption{Inverse beta event rate (left) and mean energy of $\bar{\nu}_e$ (right) 
predicted by supernova simulations \cite{Totani:1997vj,Thompson:2002mw,Buras:2006aa,Sumiyoshi:2005ri,Liebendoerfer:2003es} for the first 0.3~seconds after the onset 
of a 10~kpc distant burst.}
  \label{fig:sn-model} 
\end{figure}

Figure \ref{fig:sn-model} shows inverse beta event rates and mean $\bar{\nu}_e$
energy distributions predicted by various 
models \cite{Totani:1997vj,Thompson:2002mw,Buras:2006aa,Sumiyoshi:2005ri,Liebendoerfer:2003es} for the 
first 0.3~sec after the onset of a burst. 
The statistical error of Hyper-K is much smaller than the difference
between the models, and so Hyper-K should give crucial data for comparing  
model predictions.
The left plot in Fig.~\ref{fig:sn-model} shows that
about 300-1000 events are expected in the first 20 millisecond bin.
This means that the onset time can be determined with an accuracy of about
0.03 ms. This is precise enough to allow examination of the infall of the core 
in conjunction with the signals of neutronization
(see below) as well as possible data from future gravitational wave detectors.

We can also use the sharp rise of the burst to make a measurement of the absolute mass of
neutrinos. Because of the finite mass of neutrinos, their arrival times will  
depend on their energies.
This relation is expressed as
\begin{eqnarray}
\Delta t = 5.15~ {\rm msec} \left( \frac{D}{10~{\rm kpc}} \right)
\left( \frac{m}{1~{\rm eV}} \right) ^2 \left( \frac{E_\nu}{10~{\rm MeV}} 
\right) ^{-2}  
\end{eqnarray}
where $\Delta t$ is the time delay with respect to that assuming
zero neutrino mass, $D$ is the distance to the supernova, $m$ is the
absolute mass of a neutrino, and $E_\nu$ is the neutrino energy.
Totani~\cite{Totani:1998nf} discussed Super-Kamiokande's sensitivity to neutrino mass 
using the energy dependence of the rise time; 
scaling these results to the much larger statistics provided by Hyper-K, we expect 
a sensitivity of 0.5$\sim$1.3~eV for the absolute
neutrino mass \cite{Totani:2005pv}.
Note that this measurement of the absolute neutrino mass does not depend on
whether the neutrino is a Dirac or Majorana particle.

\begin{figure}[tbp]
  \begin{center}
    \includegraphics[width=8cm]{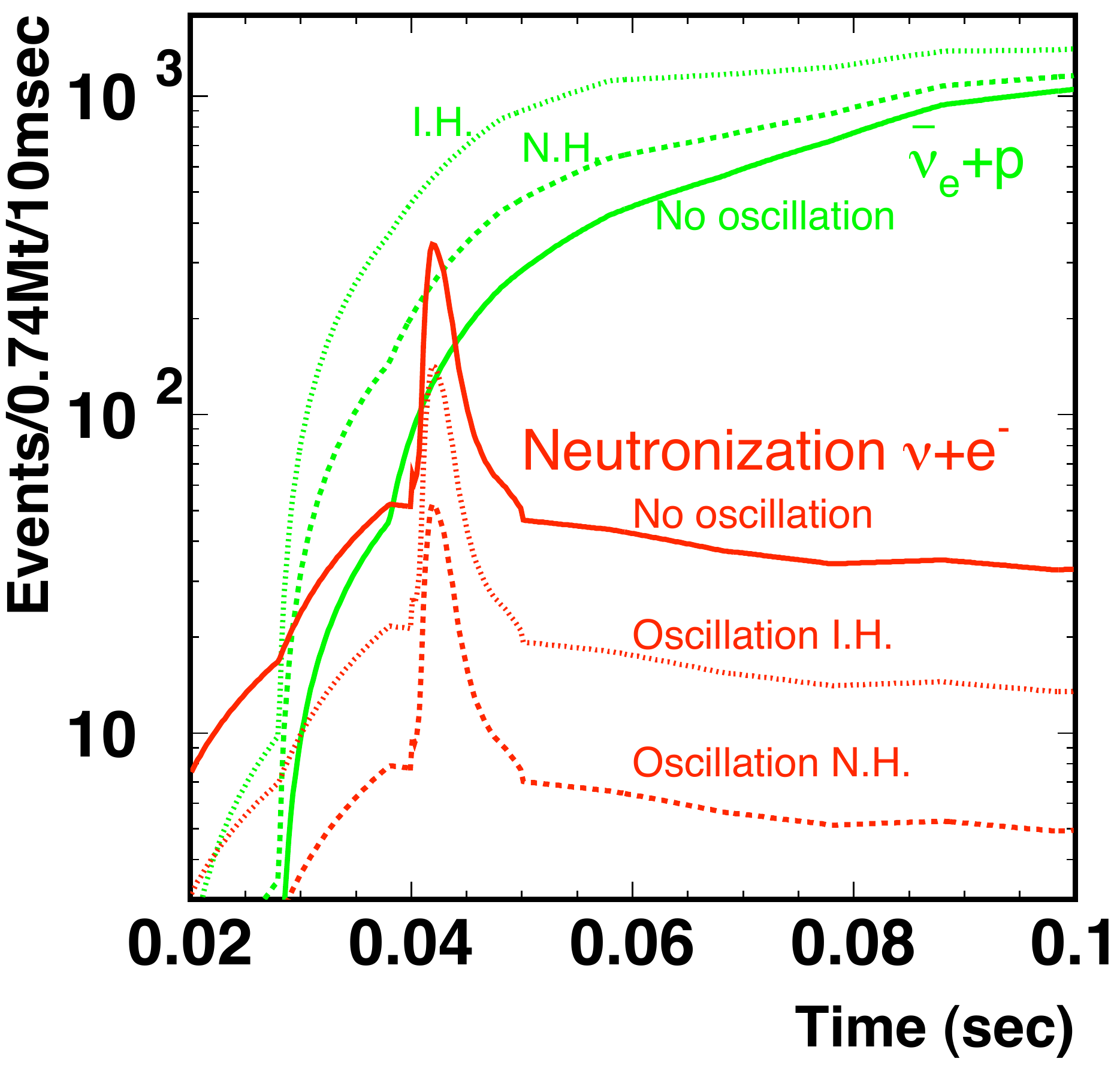}
  \end{center}
\vspace{-1cm}
  \caption{Expected event rate at the time of neutronization burst
for a supernova at 10~kpc.
Red and green show event rates for $\nu e$-scattering and 
inverse beta events, respectively. 
Solid, dotted, and dashed curved indicate the neutrino oscillation
scenarios of no oscillation, N.H., and I.H., respectively.}
  \label{fig:sn-neutronization}
\end{figure}

A sharp timing peak at the moment of neutronization is expected in 
$\nu e$-scattering events as shown in Fig. \ref{fig:sn-neutronization}.
The expected number of $\nu e$-scattering events at the neutronization
burst is $\sim$20, $\sim$56, and $\sim$130 for N.H., I.H., and no oscillation,
respectively, for a supernova at 10~kpc.
Although the number of inverse beta events is $\sim$345 (N.H.), 
$\sim$700 (I.H.), and $\sim$190(no oscillation)
in the 10~ms bin of the neutronization burst, the
number of events in the direction of the supernova is typically 1/10
of the total events. 
So, the ratio of signal events ($\nu e$-scattering) to other events
(inverse beta) is expected to
be about 20/33~(N.H.), 52/70~(I.H.) and 130/19~(no oscillation).
Thus, the $\nu e$ scattering events can be identified with high statistical 
significance thanks to the directionality of $\nu e$-scattering.

\begin{figure}[tbp]
  \begin{center}
    \includegraphics[width=16cm]{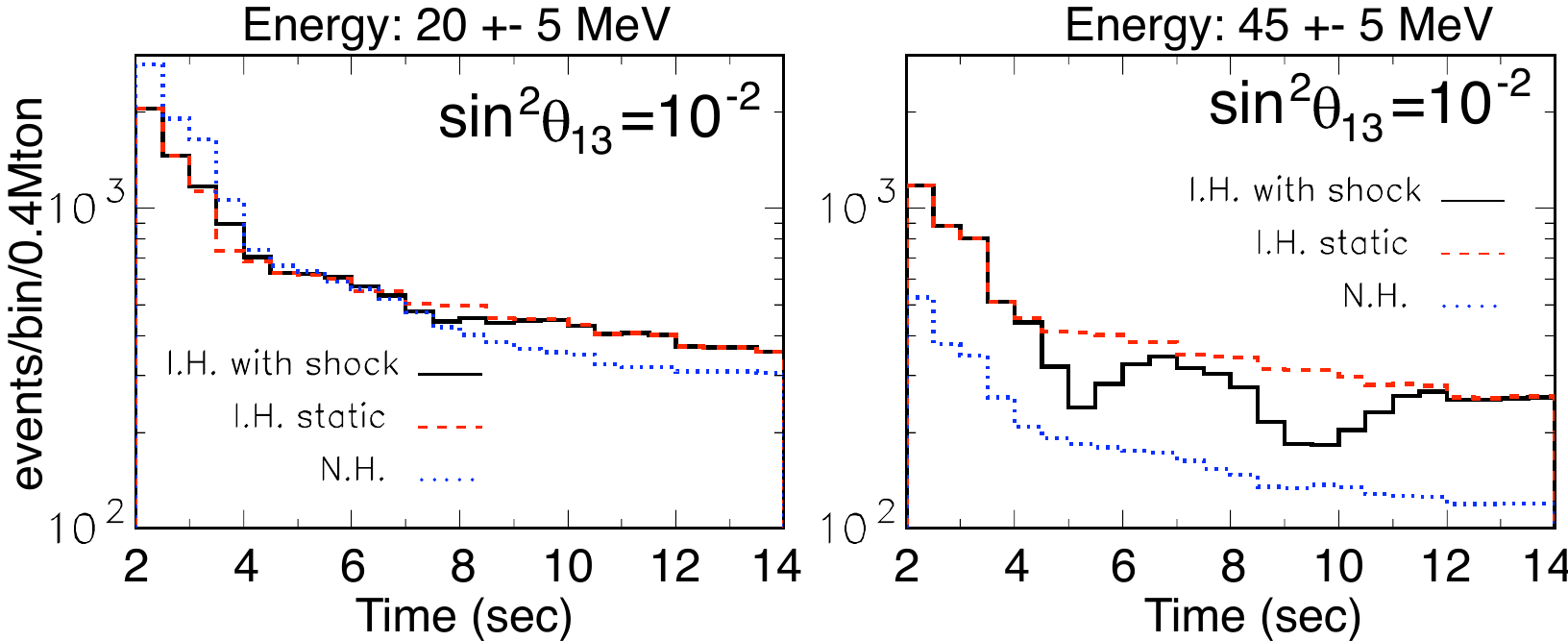}
  \end{center}
\vspace{-1cm}
  \caption{Time variation of the neutrino event rate affected by 
neutrino conversion by matter due to shock wave  propagation 
(reproduced from \cite{Fogli:2004ff}).
Left (right) plot shows the time variation of inverse beta events for
the energy range of 20$\pm$5~MeV (45$\pm$5~MeV).
Solid black, dashed red, and blue dotted histograms show the event rates
for I.H. with shock wave propagation, I.H. with static matter density profile, 
and N.H., respectively.
It has been assumed that $\sin^2\theta_{13}=10^{-2}$.}
  \label{fig:sn-nuosc}
\end{figure}

Neutrino oscillations could be studied using supernova neutrino events.
There are many papers which discuss the possibility of extracting signatures  
of neutrino oscillations free from uncertainties of 
supernova models \cite{Dighe:2003vm,Tomas:2004gr,Fogli:2004ff,Barger:2005it,Mirizzi:2006xx,EstebanPretel:2007yu,Choubey:2007ga,Skadhauge:2006su,Baker:2006gm,Dasgupta:2008my}.
One big advantage of supernova neutrinos over other neutrino sources (solar,
atmospheric, accelerator neutrinos) is that they inevitably pass through
very high density matter on their way to the detector.
This gives a sizable effect in the time variation of the energy spectrum
even for small $\sin^2\theta_{13}$ \cite{Schirato:2002tg, Fogli:2004ff, Tomas:2004gr}.
As an example, figures from the paper by Fogli et al. \cite{Fogli:2004ff}
are shown in Fig. \ref{fig:sn-nuosc}.
The propagation of the supernova shock wave causes time variations in the matter
density profile through which the neutrinos must travel. 
Because of neutrino conversion by matter, 
there might be a bump in the time variation of the inverse beta event rate for a particular energy range (i.e., 45$\pm$5 MeV as shown in Fig.~\ref{fig:sn-nuosc}(right)) while no change is 
observed in the event rate near the spectrum peak 
(i.e., 20$\pm$5 MeV as shown in Fig.~\ref{fig:sn-nuosc}(left)).
This effect is observed only in the case of inverted mass hierarchy;  
 this is one way in which the mass hierarchy could be determined by a supernova burst.

\begin{figure}[tbp]
  \begin{center}
    \includegraphics[width=7cm]{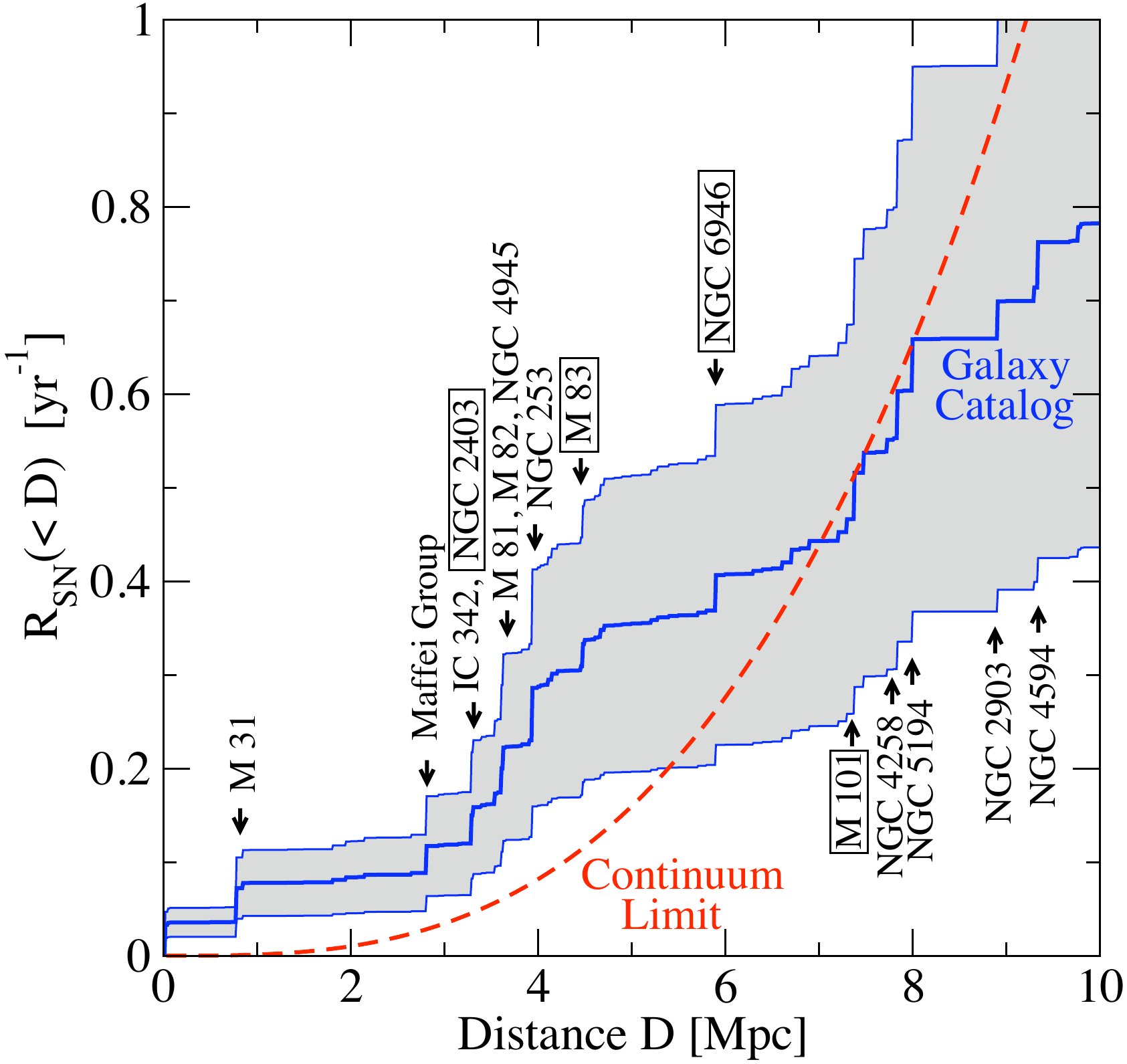}
\hspace{3mm}
    \includegraphics[width=7cm]{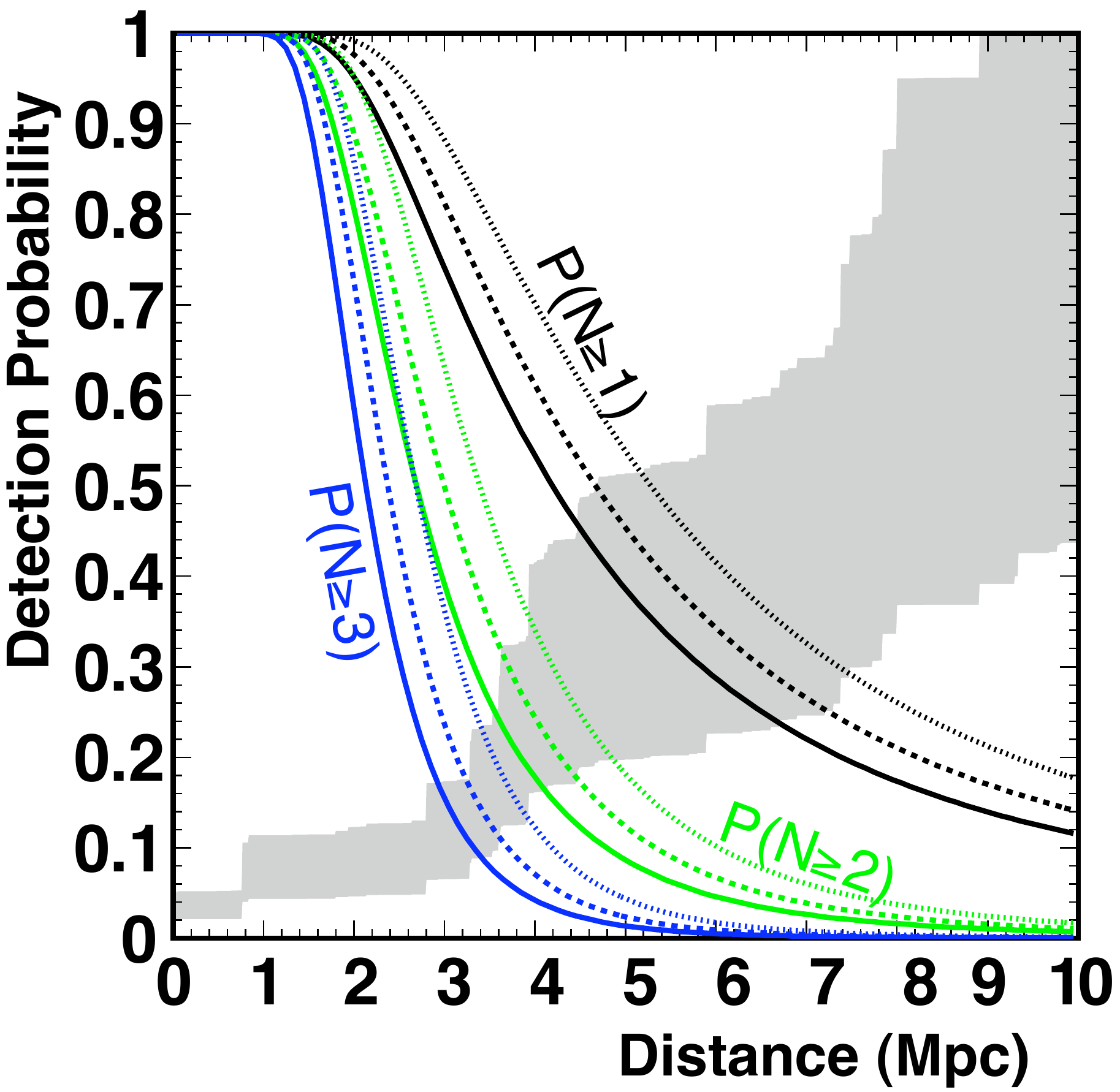}
  \end{center}
\vspace{-1cm}
  \caption{(Left) Cumulative calculated supernova rate versus distance for 
supernovae in nearby galaxies (reproduced from \cite{Ando:2005ka}).
(Right) Detection probability of supernova neutrinos versus distance at
Hyper-K assuming a 0.56~Megaton fiducial volume and 10~MeV threshold for this 
analysis.
Black, green, and blue curves show the detection efficiency resulting in requiring
at least or equal to one, two, and three events per burst, respectively.
Solid, dotted, and dashed curves are for neutrino oscillation scenarios of
no oscillation, N.H., and I.H., respectively.}
  \label{fig:sn-nearby}
\end{figure}

In Hyper-K, it could be possible to detect burst neutrinos from
supernovae in nearby galaxies.
As described above, we expect to observe a very large number of neutrino events from a 
galactic supernova.
However, galactic supernovae are expected to happen once per 30-50 years.
So, we cannot count on seeing many galactic supernova bursts.
In order to examine a variety of supernova bursts, supernovae from nearby
galaxies are useful even though the expected number of detected 
events from any single such burst are small.
Furthermore, in order to fully understand the spectrum of supernova relic neutrinos
(see next sub-section), collecting an energy spectrum without the complications of 
varying red-shift effects is highly desirable. 
The supernova events from nearby galaxies provide a reference energy spectrum
for this purpose. 
The supernovae in nearby galaxies was discussed by 
S. Ando, J. F. Beacom and H. Y{\"u}ksel \cite{Ando:2005ka}; 
a figure from their paper is shown in Fig.~\ref{fig:sn-nearby}(left).
It shows the cumulative supernova rate versus distance and indicates that if 
Hyper-Kamiokande can see signals out to 4~Mpc then we could  
expect a supernova about every three years.
It should also be noted that the paper says recent, more sensitive 
astronomical observations indicate that the true nearby supernova rates are 
probably about 3 times 
higher than this conservative calculation.
Figure~\ref{fig:sn-nearby}(right) shows detection probability versus
distance for the Hyper-K detector.
In this estimate, energy of neutrino events is required to be more than
10~MeV and the vertex position of the events should be within the fiducial 
volume (0.56 Megatons).
If we require the number of events to be more than or equal to one(two), the detection
probability is 52$\sim$69\%(17$\sim$33\%) for a supernova at 4~Mpc.
If we can use a tight timing coincidence with other types of supernova sensors (e.g. gravitational
wave detectors), we should be able to identify even single supernova neutrinos.
Assuming the observed supernova rate in nearby galaxies, we expect to collect about
10-20 supernova neutrino events from them during 20 years of operating Hyper-K.

\subsubsection{Supernova relic neutrinos}\label{sec:relic-supernova}

There are about 10$^{20}$ stars in the universe ($\sim$10$^{10}$ galaxies 
in the universe, 
and each galaxy has  about 10$^{10}$ stars). 
Because about 0.3\% of the stars have masses larger 
than 8 times the solar mass, it is estimated that 10$^{17}$ supernova 
explosions have occurred over the entire history of the universe. 
This means that on average one supernova explosion has been occurring 
every second somewhere in the universe. The neutrinos produced by all of  
the supernova explosions since the beginning of the universe are called 
supernova relic neutrinos (SRN). They must fill the present universe and 
their flux is estimated to be a few tens/cm$^2$/sec. If we can detect 
these neutrinos, it is possible to explore the history of how heavy 
elements have been synthesized since the onset of stellar formation. 
\begin{figure}[tbp]
  \begin{center}
    \includegraphics[width=8cm]{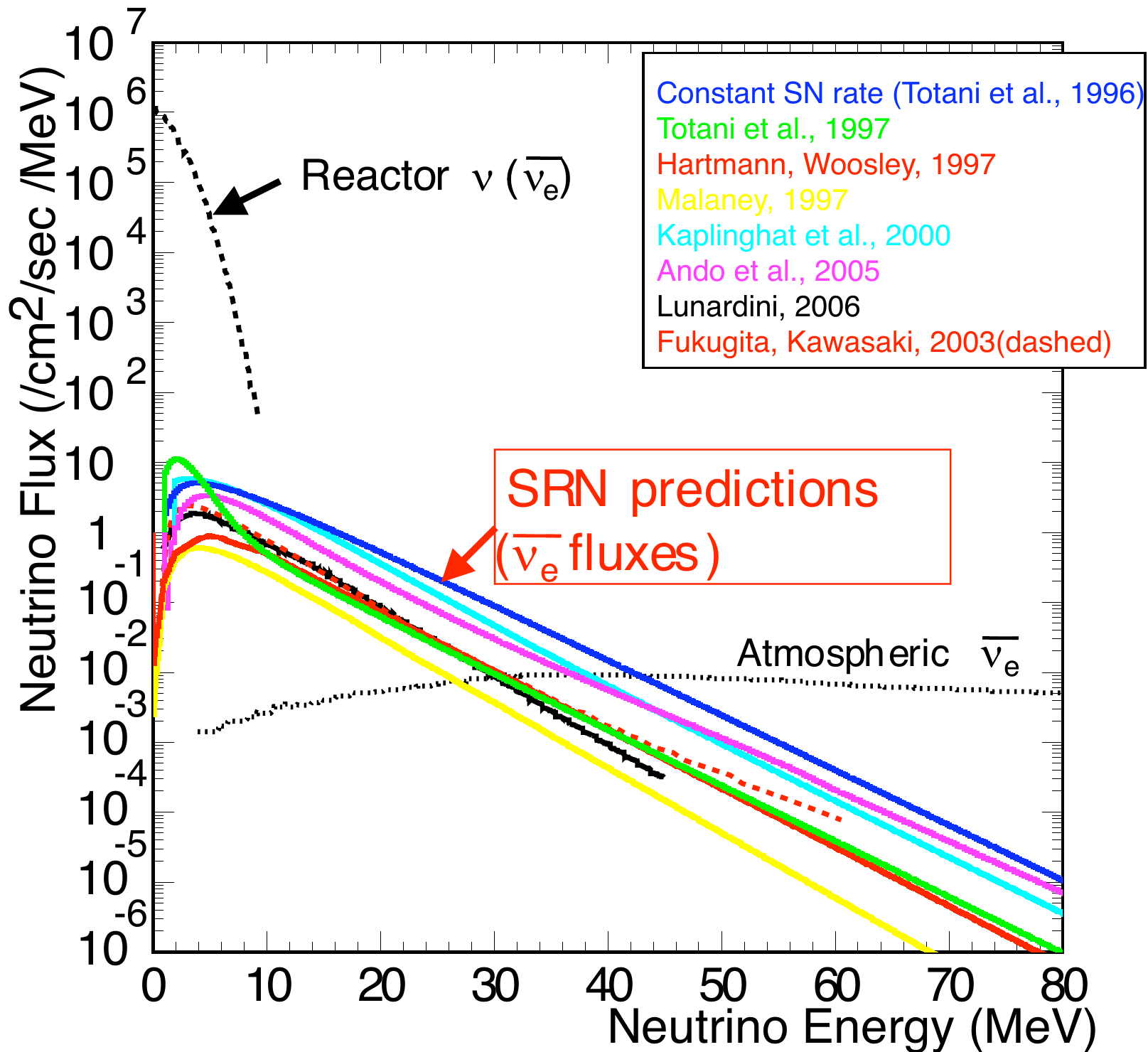}
  \end{center}
\vspace{-1cm}
  \caption{Predictions of the supernova relic neutrino (SRN) spectrum.
Fluxes of reactor neutrinos and atmospheric neutrinos are also shown.}
  \label{fig:sn-srn-prediction}
\end{figure}
Figure \ref{fig:sn-srn-prediction} shows the SRN spectra predicted by 
various models.
Although searches for SRN have been conducted at large underground 
detectors, no evidence of SRN signals has yet been obtained because of the 
small flux of SRN.
The expected inverse beta 
($\bar{\nu}_e p \rightarrow e^+ n$) event rate at Super-Kamiokande is 0.8-5 events/year 
above 10~MeV, but because of the large number of spallation products and the
low energy atmospheric neutrino background (decay electrons from muons below Cherenkov threshold produced by atmospheric
muon neutrinos, the so-called invisible muon background), SRN signals have not yet been 
observed at Super-Kamiokande.
In order to reduce background, lower the energy threshold, individually identify true 
inverse beta events by tagging their neutrons, and 
thereby positively detect SRN signals at Super-Kamiokande, a project to add
0.1\% gadolinium to the tank (the GADZOOKS! project) was 
proposed by J.F.Beacom and M.R.Vagins \cite{Beacom:2003nk}; very active 
R\&D work for the project is ongoing. The first observation of the SRN 
could be made by the GADZOOKS! project, 
but in order to measure the spectrum of the SRN and analyze the 
history of the universe we need a megaton-scale detector.

\begin{figure}[tbp]
  \begin{center}
    \includegraphics[width=8cm]{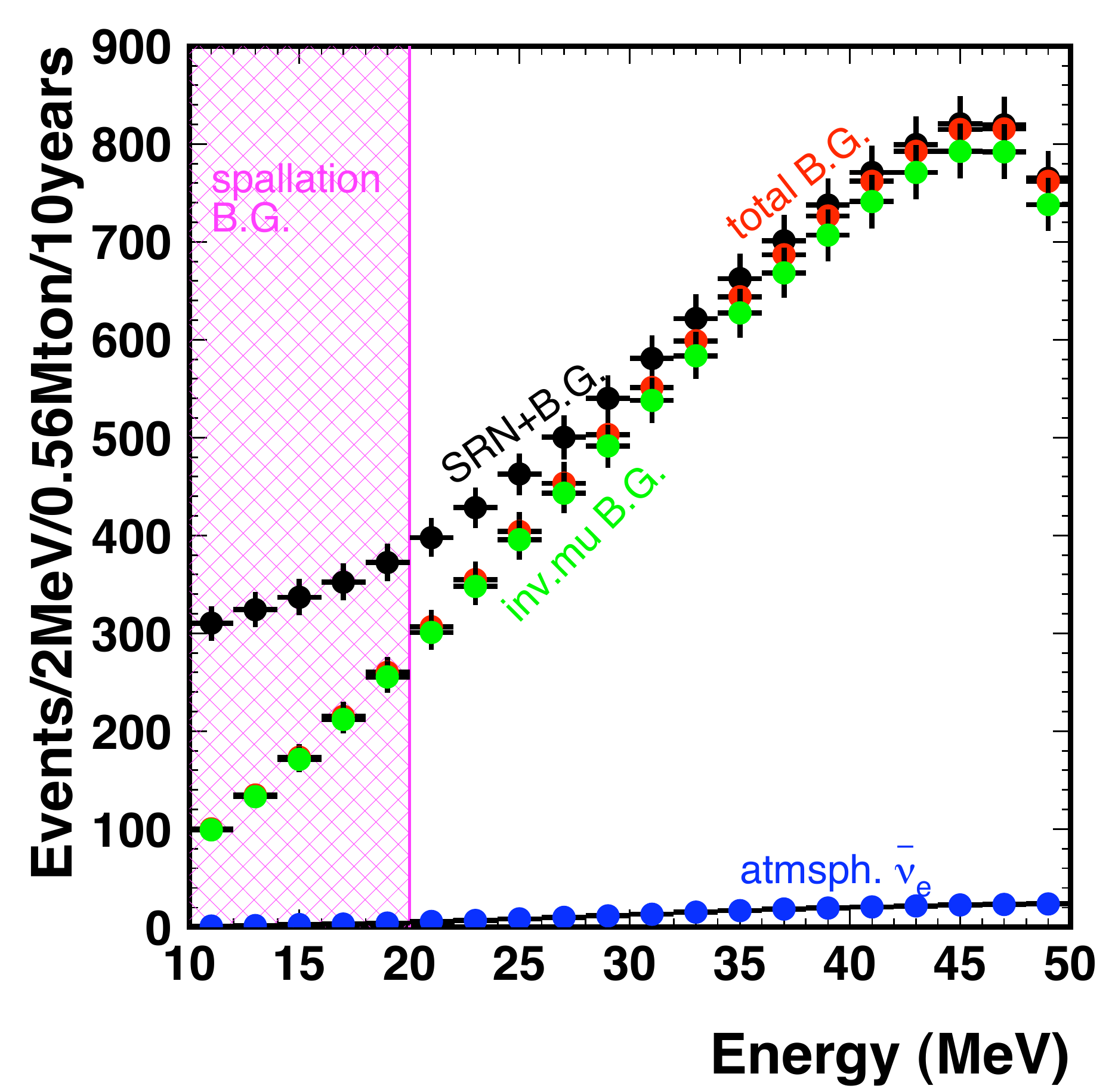}
  \end{center}
\vspace{-1cm}
  \caption{Expected spectrum of SRN signals at Hyper-K with 10 years
of livetime without tagging neutrons. The black dots show 
signal+background (red component).
Green and blue show background contributions from the invisible muon and 
$\nu_e$ components of atmospheric neutrinos.
Without tagging neutrons, spallation background dominates below 20~MeV.
The SRN flux prediction by Ando et al. \cite{Ando:2003aa} is used, and
a signal selection efficiency of 90\% is assumed.}
  \label{fig:srn-no-n-tag}
\end{figure}

Figure~\ref{fig:srn-no-n-tag} shows expected SRN signals at Hyper-K with
10~years'  livetime without tagging neutrons. Because of the high background rate below
20~MeV from spallation products, the detection of SRN signals is limited to above 20~MeV, while 
above 30~MeV the atmospheric neutrino backgrounds completely overwhelm the signal. 
The expected number of SRN events in $E=20-30$~MeV is about 310/10yrs assuming the flux
prediction of Ando et al. \cite{Ando:2003aa} and an event selection efficiency of 90\%.
The number of background events from atmospheric neutrinos 
(invisible muon and $\bar{\nu}_e$) is estimated to be 2200/10yrs.
So, it is possible to detect SRN signals with high statistical significance.

However, it is important to be able to measure the SRN spectrum down to $\sim$10~MeV in order
to explore the history of supernova bursts back to the epoch of red shift ($z$)  $\sim$1.
Therefore, in the following discussion the expected SRN signal with gadolinium neutron 
tagging is considered.

\begin{figure}[tbp]
  \begin{center}
    \includegraphics[width=6cm]{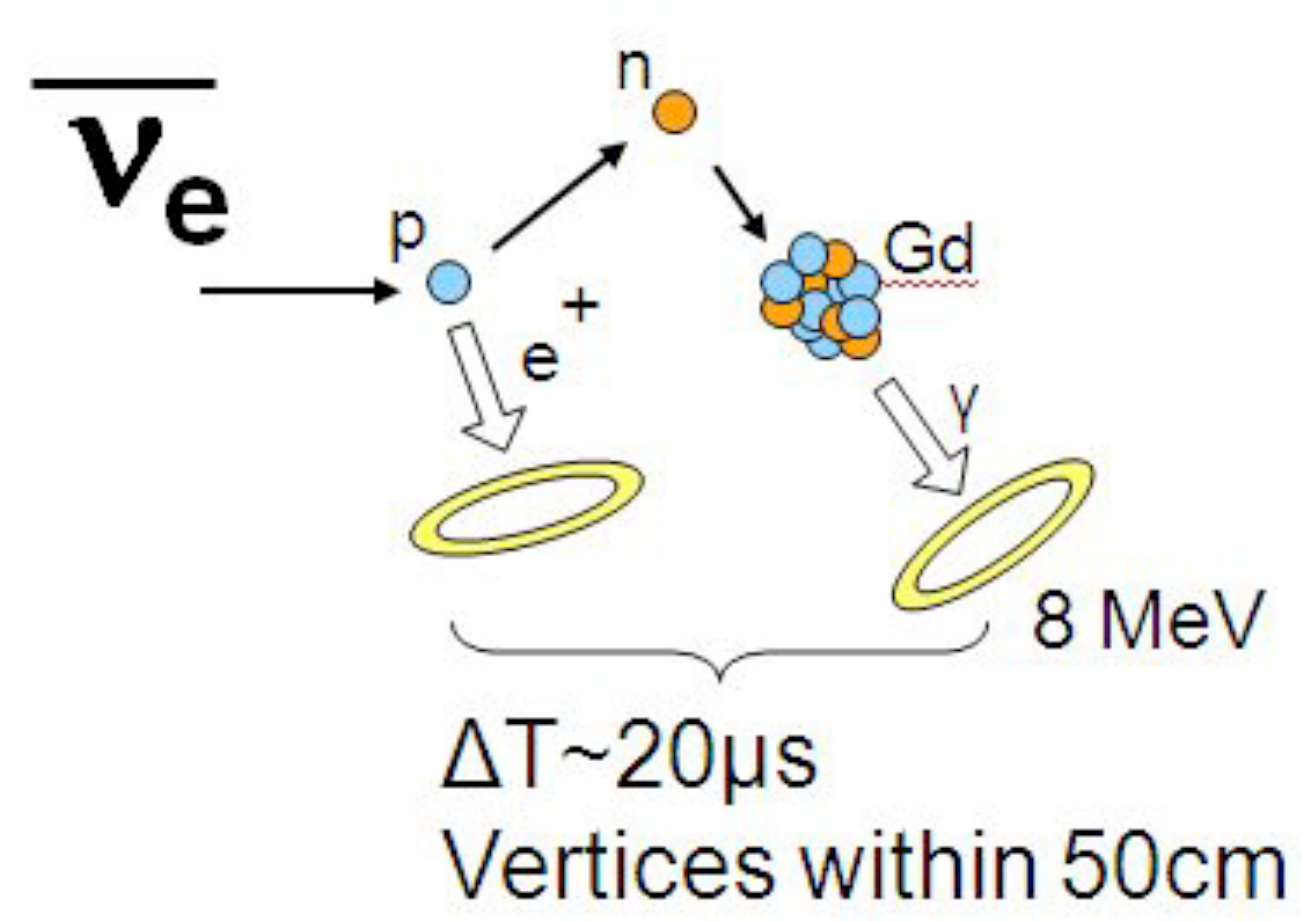}
    \includegraphics[width=6cm]{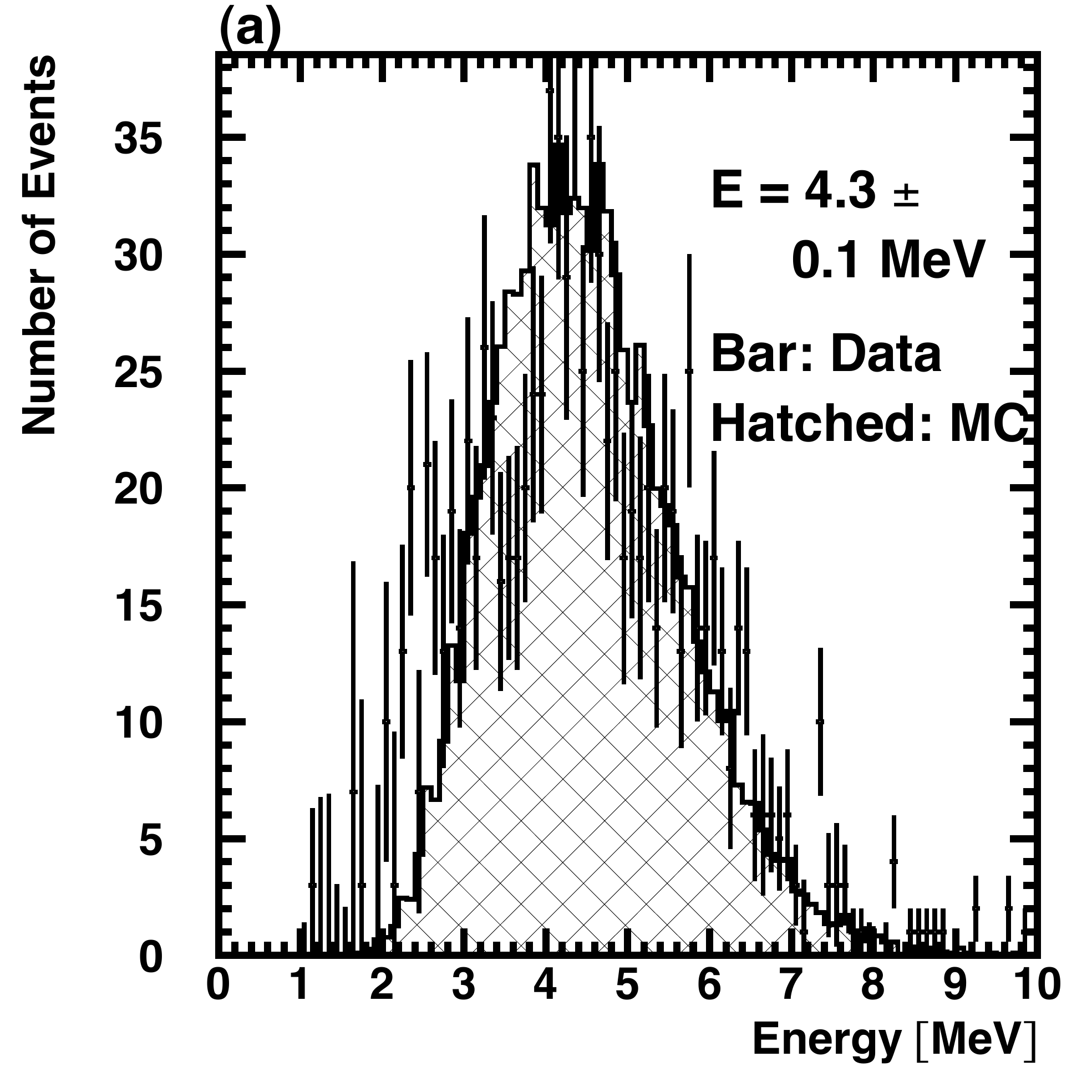}
  \end{center}
\vspace{-1cm}
  \caption{The left plot shows how to tag neutrons produced by the inverse beta
reaction, while the right plot shows the electron-equivalent energy spectrum of 
the Gd(n,gammas)Gd gamma cascade as measured in the Super-Kamiokande 
detector \cite{Watanabe:2008ru}. }
  \label{fig:srn-ntag}
\end{figure}

\begin{figure}[tbp]
  \begin{center}
    \includegraphics[width=8cm]{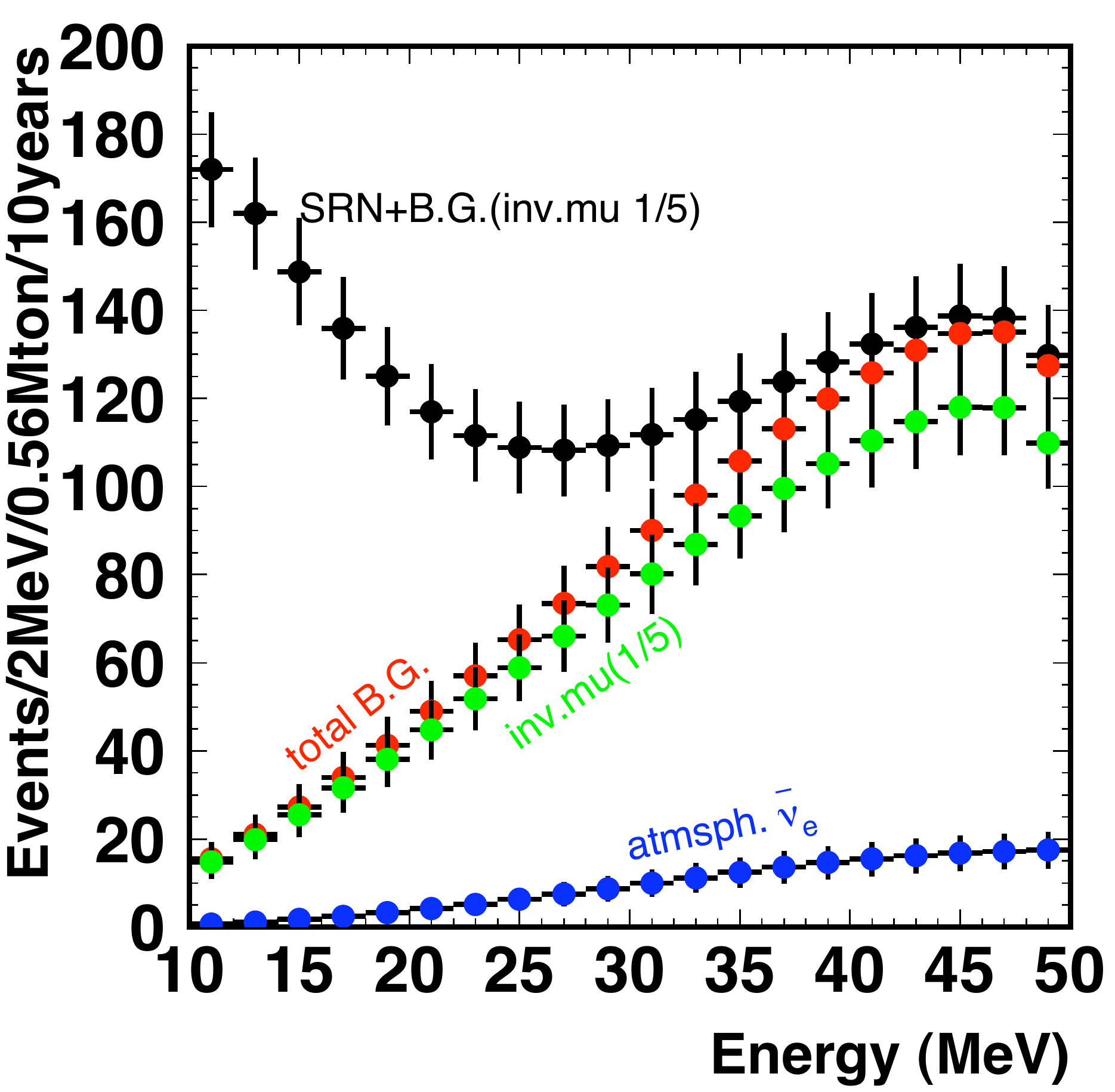}
  \end{center}
\vspace{-1cm}
  \caption{Expected spectrum of the SRN signals at Hyper-K with 10 years
of livetime. The black dots show signal+background (red component).
Green and blue show background contributions from the invisible muon and 
$\nu_e$ components of atmospheric neutrinos.
The SRN flux prediction of Ando et al. \cite{Ando:2003aa} was used, and 
a 67\% detection efficiency of 8~MeV gamma cascades and
a factor of 5 reduction in the invisible muon background were assumed.}
  \label{fig:srn-megaton}
\end{figure}

Inverse beta reactions can be identified by coincident detection of both positron and
delayed neutron signals, and requiring tight  spatial and temporal correlations between them. 
With 0.1\% by mass of gadolinium (Gd) dissolved in the water, neutrons are captured on 
gadolinium with about 90\% efficiency; the excited Gd nuclei then de-excite by emitting 8~MeV 
gamma cascades. 
The time correlation of about 20~$\mu$sec between the positron and
the Gd(n,gammas)Gd cascade signals, and the vertex correlation within about
50~cm are strong indicators of a real inverse beta event.  Requiring both correlations (as well as requiring the prompt event to be Cherenkov-like and the delayed event to be isotropic) can be 
used to reduce background of spallation products by 
many orders of magnitude while also reducing invisible muon backgrounds by about a factor of 5.
Since the 8~MeV gamma cascade of Gd(n,gammas)Gd reaction produces 
multiple gammas, its visible energy is much lower than 8~MeV.  This was 
measured to be 4.3~MeV as shown in Fig.~\ref{fig:srn-ntag}(right).
In order to reconstruct vertex and energy of such events,
about 20\% photocoverage is necessary.
Figure~\ref{fig:srn-megaton} shows the expected observation of the SRN
spectrum in Hyper-Kamiokande. 
In this plot, an SRN flux prediction by Ando et al. \cite{Ando:2003aa} was used, and
a detection efficiency of 8~MeV gamma cascades of 67\% (90\% for
capture efficiency and 74\% for event selection in which
vertex coincidence of less than 2~m, time correlation less than 60~$\mu$sec, 
energy of delayed signal more than 3~MeV, and spherical pattern of the 
Gd gamma event were required) and
a reduction factor of the invisible muon background of 5 were assumed.
The expected number of SRN events in the energy range of 10-30~MeV is 
about 830 with 10 years of live time.
These large statistics are sufficient to start discussing the supernova history
of the universe.  In addition, by comparing of the total SRN flux with optical supernova rate 
observations, a determination of the fraction of failed (optically dark) supernova explosions, 
currently unknown but thought to occur in not less than 5\% and perhaps as 
many as 50\% of all explosions, will be possible.

\begin{figure}[tbp]
  \begin{center}
    \includegraphics[width=8cm]{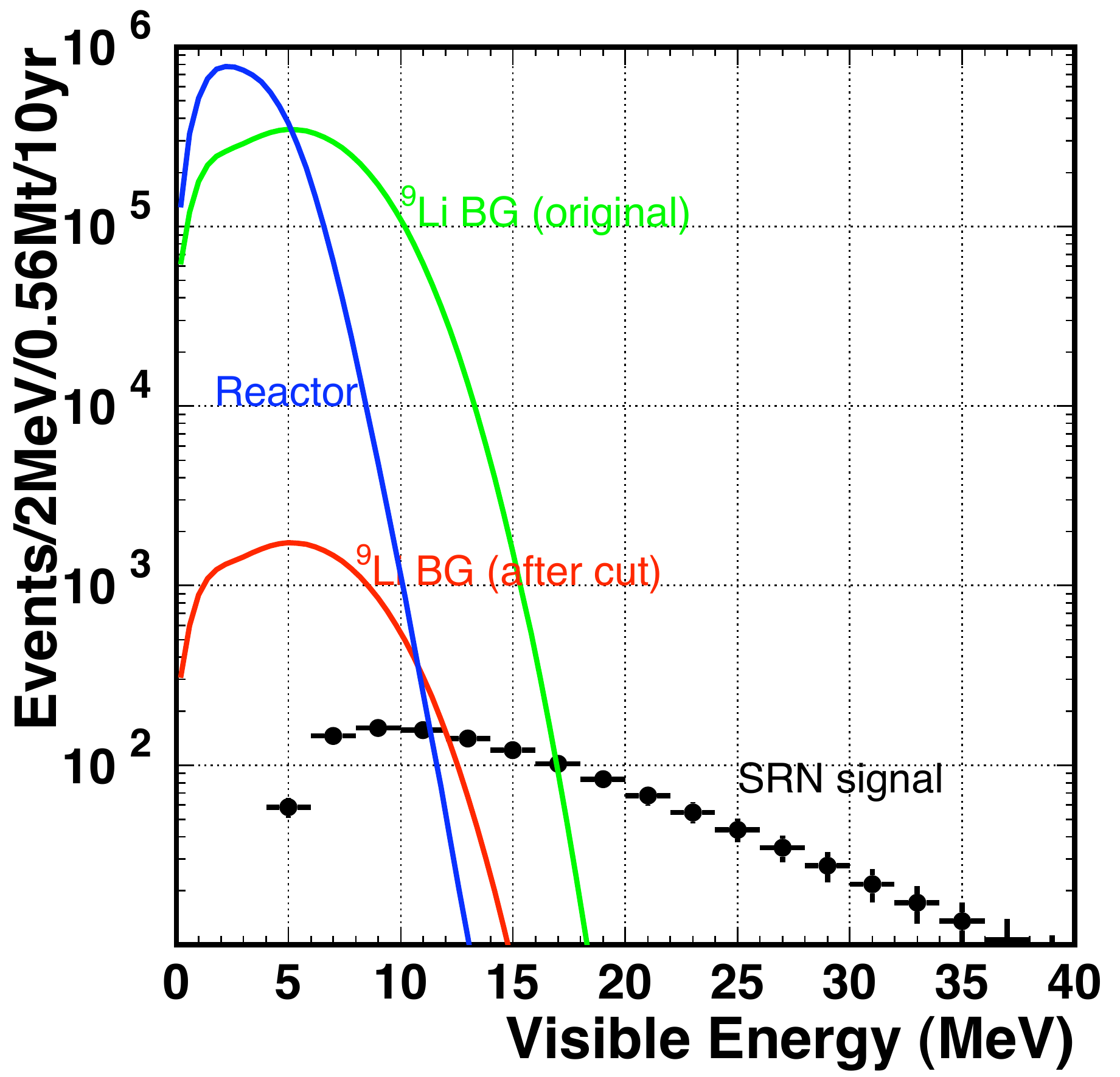}
  \end{center}
\vspace{-1cm}
  \caption{Green (red) curve shows the estimated $^9$Li production rate
before (after) applying cuts based on a correlation with cosmic ray muons.
Blue shows estimated background from reactor neutrinos at full intensity.
Black data points show expected SRN signal based on the
flux prediction by Ando et al. \cite{Ando:2003aa}.}
  \label{fig:srn-bg}
\end{figure}

Possible backgrounds to the SRN search are (1) chance coincidences, 
(2) spallation products with accompanying neutrons, and (3) the resolution tail
of the reactor neutrinos. 
For (1) chance coincidences, 
the possible source of the prompt event is the
spallation products. By requiring time coincidence, vertex correlation and
energy and pattern of the delayed event, the chance coincidence rate can
be reduced below the level of the expected SRN signal by reducing the
radioactivity of PMTs and their protective cases by a factor of about 5.
For (2) spallation products with accompanying neutrons, the 
only possible spallation product is  $^9$Li and an
estimation by a GEANT4 simulation is shown in Fig.~\ref{fig:srn-bg}.
Because of the short half-life of $^9$Li ($\tau_{1/2}$=0.18sec), a high
rejection efficiency of $\sim$99.5\% is expected.
With this expectation, the $^9$Li background is less than the signal level 
above 12~MeV; this could be lowered by further development of the
background reduction technique.
For (3) the resolution tail of the reactor neutrinos, the estimated background rate
is about 380(80)/10~years above 10~MeV (11~MeV) as shown in Fig.~\ref{fig:srn-bg}
with full reactor intensity. 

%% file: physics-astro/astrophys.tex
\clearpage

\subsubsection{Solar flare}
Solar flares are the most energetic bursts which occur in the solar surface.
Explosive release of energy stored in solar magnetic fields is caused by 
magnetic reconnections, resulting in plasma heating, particle accelerations, 
and emission of synchrotron X-rays or charged particles from the solar 
surface. 
In a large flare, an energy of 10$^{33}$ ergs is emitted over 10's of 
minutes, and the accelerated protons can reach energies greater than 10 GeV. 
Such high energy protons can produce pions by nuclear interactions 
in the solar atmosphere. Evidence of such nuclear interactions 
in the solar atmosphere are obtained via observations of solar neutrons, 
2.2 MeV gamma rays from neutron captures on protons, nuclear de-excitation 
gamma rays, and possible $>100$ MeV gamma rays from neutral pion decays. 
Thus, it is likely that neutrinos are also emitted by the decay of mesons 
following interactions of accelerated particles. 
Detection of neutrinos from a solar flare was first discussed in 1970's 
by R.Davis~\cite{dav, bacall_sf}, but no significant signal has yet been 
found~\cite{aglietta,hirata}. There have been some estimates of the number 
of neutrinos which could be observed by large water Cherenkov 
detectors~\cite{fargion,kocharov}. According to \cite{fargion}, 
about 20 neutrinos will be observed at Hyper-Kamiokande during a solar flare 
as large as the one in 20 January 2005, although the expected numbers 
have large uncertainties.  Therefore, regarding solar flares 
our first astrophysics goal is to discover solar flare neutrinos with Hyper-K.
This will give us important information about the mechanism of the particle 
acceleration at work in solar flares. 

\subsubsection{Indirect dark matter search}
The dark matter problem in the universe is one of the hottest topics 
in both particle physics and astrophysics. One possible solution 
to this conundrum is to introduce stable, weakly interacting massive 
particles (WIMPs) -- which feel only the weak interaction and gravity -- 
with sufficient mass to solve the structure formation problem of the universe.
Therefore, many experiments are trying to detect WIMPs directly or indirectly 
all over the world. Hyper-K will be able to perform indirect WIMP searches 
as was done in Super-K~\cite{tanaka} with improved sensitivity. 
In indirect searches, signals are from neutrinos emitted by WIMP 
annihilations which occur in the Sun, the Earth, or in the halo, 
where WIMPs are trapped by gravitational potentials. 
In this study, we focus on the WIMP signal from the Sun. 

\begin{figure}[htb]
\begin{center}
\includegraphics[height=20pc]{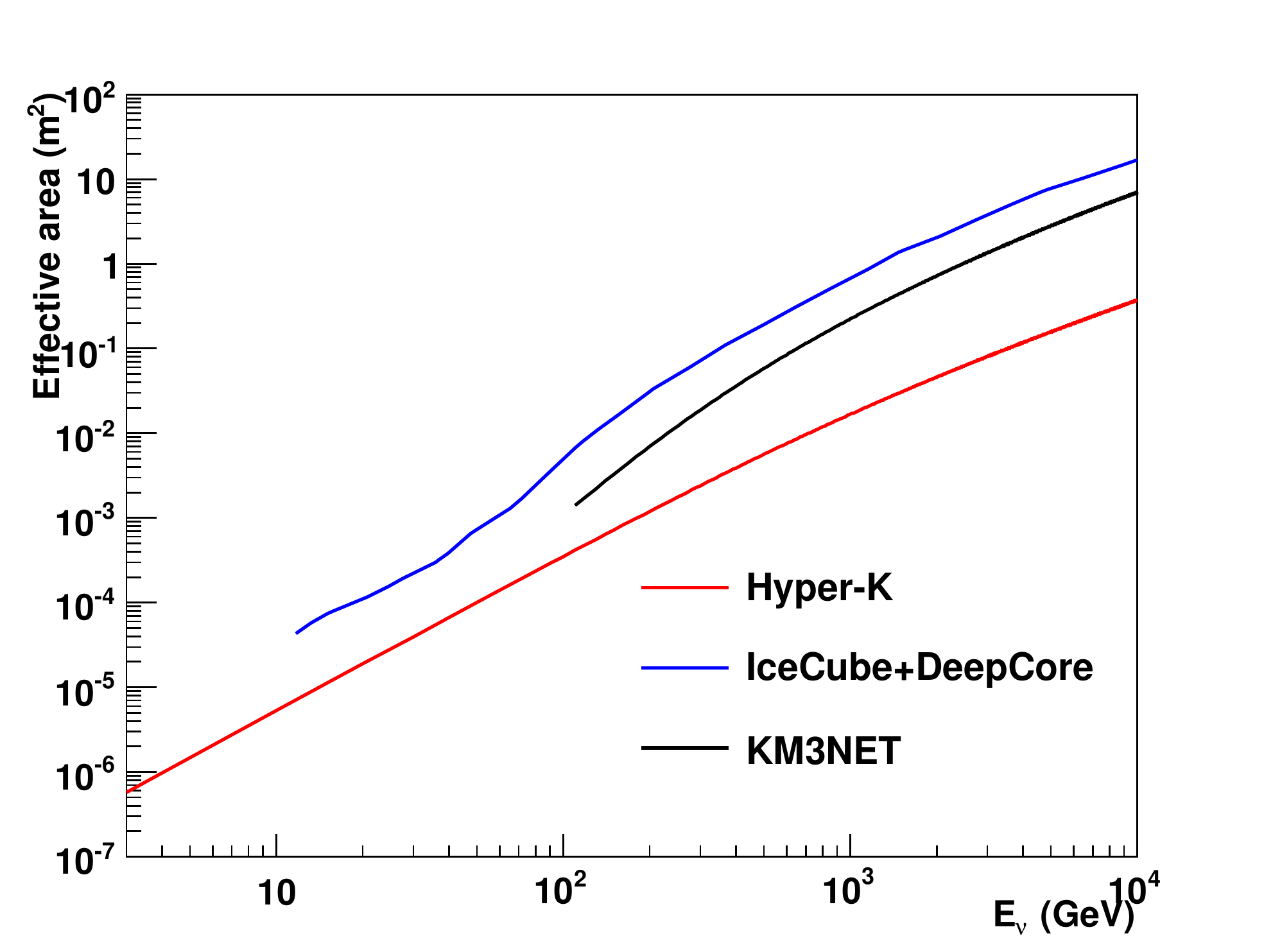}
\end{center}
\caption{The effective area for upward-going muon detection. 
The red line shows Hyper-K, while the blue and black lines show 
Ice-Cube+DeepCore (at trigger level)~\cite{icdc_aeff} and  KM3NET~\cite{gabici} .}
\label{fig:aeff}
\end{figure}

These neutrinos interact with detector targets 
or rocks around detectors, 
and the secondary particles such as muons are detected. 
To distinguish those signal muons from cosmic-ray muons, 
only upward-going muons are selected.  The detector volume for neutrino 
searches using upward-going muons can be expressed as a neutrino effective 
area ($A_{eff}$), which is defined as:
\begin{equation}
A_{eff} = \sigma_{\nu-\text{rock}} \times R_{\mu} \times A_{HK}
\end{equation}
where $\sigma_{\nu-\text{rock}}$ is the cross section of neutrino-rock 
interaction, $R_{\mu}$ is the muon range, and $A_{HK}$ is the detector area 
which in this case is 4550~cm$\times$24750~cm$\times$2=22000~m$^2$. 
Figure~\ref{fig:aeff} shows the calculated effective area of Hyper-K 
as a function of neutrino energy. Here we reasonably assume the momentum 
threshold for muon detection in Hyper-K is similar to that of Super-K 
(1.7~GeV/$c$). As shown in Fig.~\ref{fig:aeff}, Hyper-K has a better 
sensitivity below a few tens of GeV than km$^3$-scale detectors 
thanks to its lower energy threshold for muon detection.

\begin{figure}[htb]
\begin{center}
\includegraphics[height=20pc]{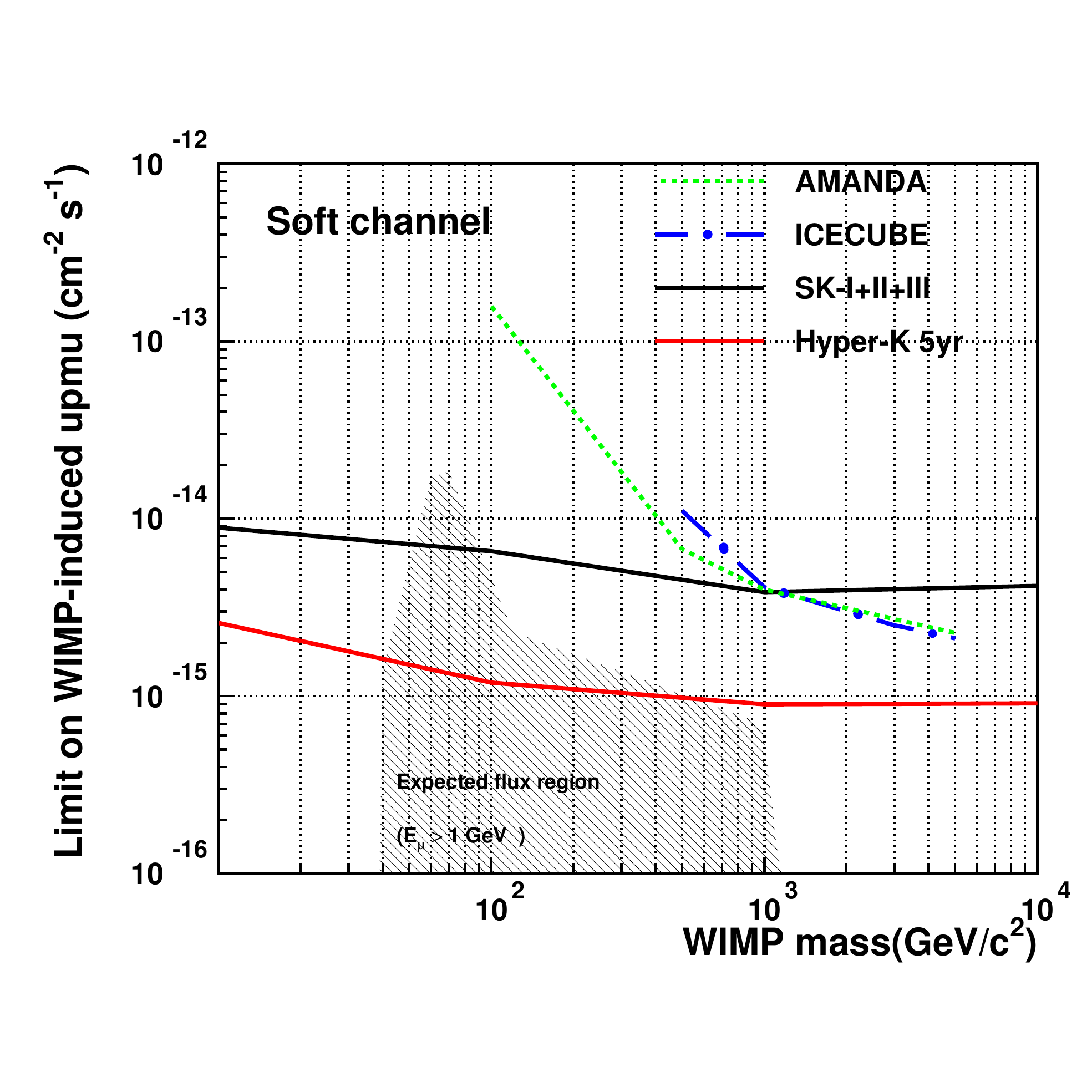}
\end{center}
\caption{Limits on the WIMP-induced upward-going muon rate as a function 
of WIMP mass are shown. Lines show AMANDA~\cite{amanda} (green dotted line), 
Ice-Cube~\cite{ic_wimp} (blue dot-dash line),
Super-K~\cite{tanaka} (black line), 
and Hyper-K (red line). The shaded region is the expected flux 
region from DARKSUSY. In this analysis, all WIMPs are assumed to annihilate 
to $b\bar{b}$.}
\label{fig:upmurate_s}
\end{figure}

\begin{figure}[htb]
\begin{center}
\includegraphics[height=20pc]{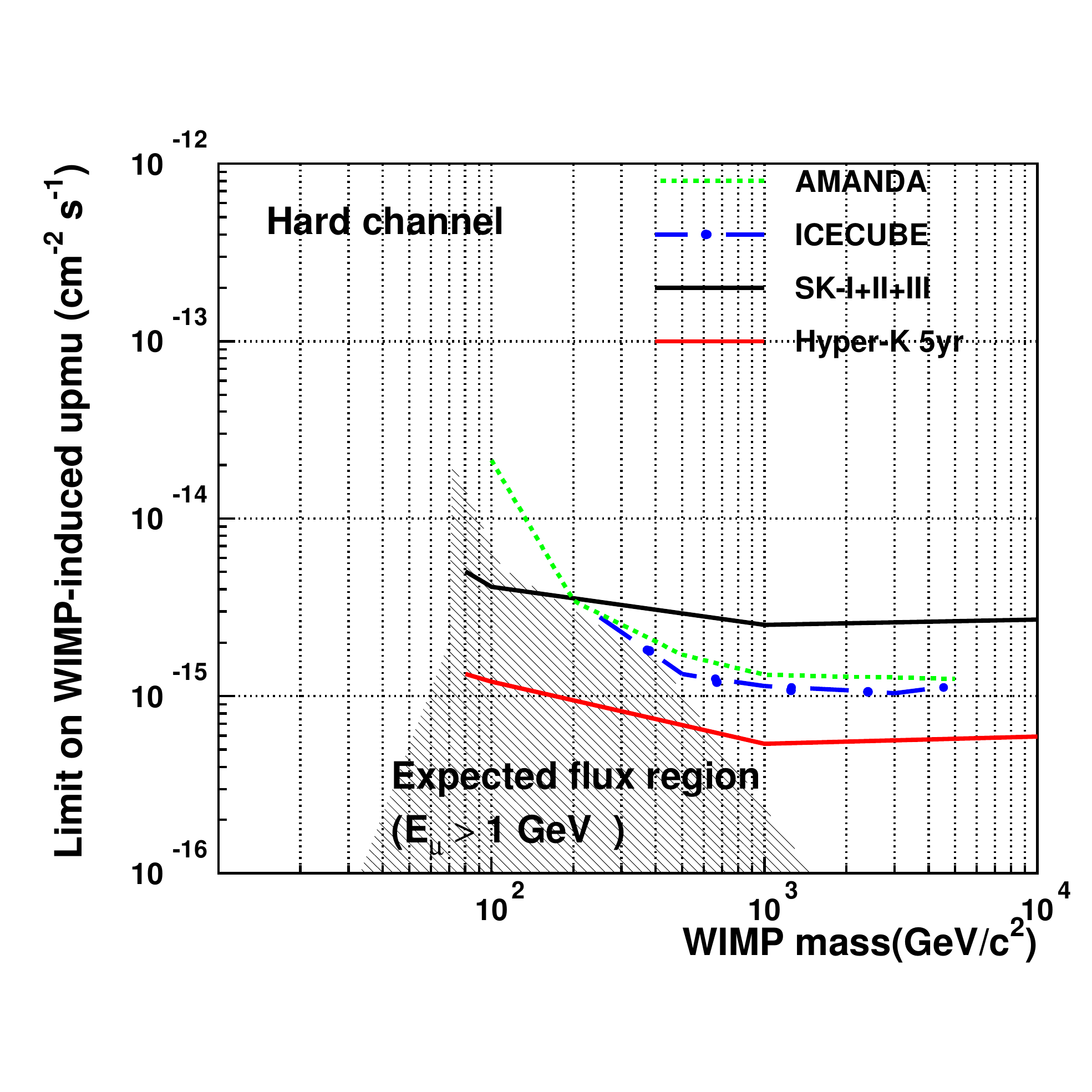}
\end{center}
\caption{Same as Fig.~\ref{fig:upmurate_s}, but in this analysis, 
all WIMPs are assumed to annihilate to $W^+W^-$.}
\label{fig:upmurate_h}
\end{figure}

\begin{figure}[htb]
\begin{center}
\includegraphics[height=20pc]{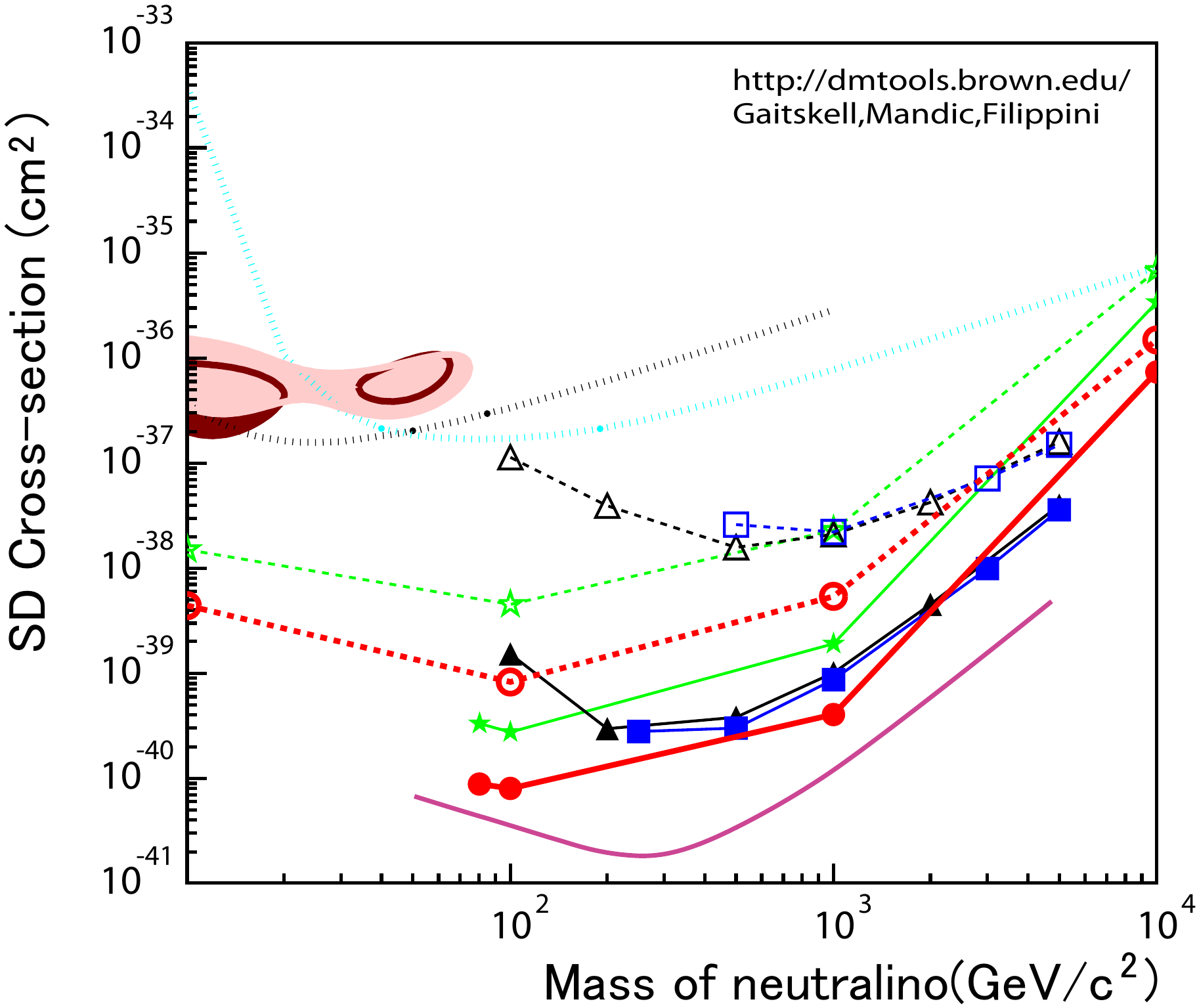}
\end{center}
\caption{Limits on WIMP-proton spin-dependent cross section as a function 
of WIMP mass. Limits from direct detection experiments: DAMA/LIBRA allowed 
region~\cite{dama}  (dark red and light red filled, with and without ion 
channeling, respectively), KIMS~\cite{kims} (light blue dotted line), 
and PICASSO~\cite{picasso} (light black dotted line) are shown. 
Limits from indirect detection (neutrino telescopes): AMANDA~\cite{amanda} 
(black line with triangles), IceCube~\cite{ic_wimp} (blue line with squares), 
and Super-K (green line with stars, green dashed line)~\cite{tanaka}.
Ice-Cube+DeepCore sensitivity is shown with the purple line~\cite{Heros:2010ss}. 
Hyper-K sensitivity is shown with the red line (soft channel) and the red dashed line (hard channel).}
\label{fig:wimp}
\end{figure}

In the Sun, WIMPs are scattered on hydrogen or heavy elements 
both with spin-dependent and spin-independent couplings. However, indirect 
searches for the annihilation signals have better sensitivity for 
the spin-dependent cross-section on hydrogen because of its dominant abundance.
In this study, the sensitivity for WIMP signals is calculated for
two cases as done in \cite{amanda,ic_wimp,tanaka}. In the first case,
all WIMPs are assumed to annihilate into $b\bar{b}$ ($\chi\chi \rightarrow
b\bar{b}$), the so-called ``soft channel'',
while in the second case, all WIMPs are assumed to annihilate into $W^+W^-$
($\chi\chi \rightarrow W^+ W^-$), the so-called ``hard channel''.
Since these channels give the softest and hardest neutrino spectra respectively,
 the actual sensitivity should lie somewhere between these two extremes.
This approach is even more sensitive than the direct searches taking place 
on the Earth. Figures~\ref{fig:upmurate_s} and \ref{fig:upmurate_h} 
show the experimental results of limit on WIMP-induced upward-going muon fluxes
as a function of WIMP mass~\cite{tanaka} and Hyper-K's expected sensitivity 
after 5 years of observation. In Fig.~\ref{fig:wimp}, Hyper-K's sensitivity 
to the WIMP-proton cross section is compared to the upper limits obtained 
from both direct and indirect dark matter searches. 
Thank to the larger effective area of Hyper-K, 5 years of exposure 
-- corresponding to 90 years of exposure for Super-K -- can improve 
the present sensitivity significantly. Another approach to search for WIMP 
annihilation neutrinos by FC events has been discussed 
elsewhere~\cite{Rott:2011fh,Cirelli:2005gh}. 
Since the neutrino energy spectrum for WIMP annihilation neutrinos shows 
a bump structure according to the WIMP mass, the reconstructed energies 
of FC events are very important to identify WIMP signals. 
In addition, because the atmospheric $\nu_{e}$ flux is much suppressed 
at higher energy (about 1/10 of $\nu_{\mu}$ at 100 GeV), 
the flavor information of FC events can be used to reduce the atmospheric 
neutrino background.  With Hyper-K's excellent capabilities of energy 
reconstruction and neutrino flavor tagging, its soft channel WIMP-proton 
cross section sensitivity could reach $10^{-39}$ cm$^2$ at 
10 GeV~\cite{Rott:2011fh}. 
The low mass region has gotten a lot of attention since the results of direct searches come out~\cite{dama,Aalseth:2011wp}.  Hence, Hyper-K's observation, which has considerably better sensitivity in this region, is likely to be
very important.
 It should be noted that the larger effective area 
and analysis improvements in the low mass region will also enhance the sensitivity
of searches for WIMP annihilation in the Earth or the galactic bulge 
as well as the Sun.

\subsubsection{Summary}
In this section, expectations of the Hyper-K detector for astrophysical
sources are discussed.

Quite high statistical observations of supernova neutrinos are expected
for galactic supernovae, e.g. 170,000$\sim$260,000 events for a 10 kpc
supernova.  Hyper-K will give us the detailed time profile and temperature
variation during the burst. The neutronization phase, the initial burst
phase which emit $\nu_e$'s, can be detected in Hyper-K.
The direction to the supernova can be determined with an accuracy of
about 2 degrees using $\nu e$-scattering events for a 10 kpc supernova.
For supernova explosions outside our galaxy, we expect about 30$\sim$50
events for M31 (Andromeda Galaxy) and 7,000$\sim$10,000 events for the
Large Magellanic Cloud (LMC).
If we could detect supernovae out to $\sim$4 Mpc, we could observe a supernova
burst once every 1$\sim$3 years with high confidence. 
Hyper-Kamiokande is able to detect at least one supernova neutrino event
for a supernova at 4 Mpc with an efficiency of 52$\sim$69\%.

We can also investigate currently unknown properties of neutrinos by studying 
supernova neutrinos.  For example, if the neutrino mass hierarchy is inverted,
the shock wave propagation may cause time variation of the event rate 
for higher energy events during the cooling phase.
The energy dependence of arrival times at the onset of the burst would tell
us the neutrino direct mass with an accuracy of 0.5$\sim$1.3 eV.

Supernova relic neutrinos (SRN) will tell us the history of massive stars in
the universe. In particular, a spectrum measurement with high statistics is
necessary because the present spectrum of SRN is the red-shifted sum of
the contribution of supernova neutrinos from every epoch of the universe.
We expect about 310 SRN events in the energy range from 20 MeV to 30 MeV
for 10 years of Hyper-K running, while the background from atmospheric 
neutrinos
will be about 2200 events without tagging neutrons.  The lower energy bound
(20 MeV here) is limited by spallation background.
It is crucial to lower the detectable energy down to 10 MeV because the
contribution from early epoch supernovae tends to be distributed at lower 
energy.
By introducing 0.1\% gadolinium to Hyper-K, a neutron coincidence signal
from the 8~MeV gamma cascade will remove spallation background.
It will enable us to lower the energy threshold down to 10 MeV and also
reduce atmospheric neutrino backgrounds by a factor of $\sim$5.
We would then expect about 830 SRN events per 10 years in E=10-30MeV
assuming 67\% efficiency for tagging neutrons.

For the solar flare neutrinos, even though there are large uncertainties 
in predictions, the existing observations of accelerated particles and 
secondary particles support the supposition that neutrinos are also produced 
during a solar flare. The first detection of solar flare neutrinos by Hyper-K 
will be a strong test of neutrino emission models. For the indirect dark matter
search, 5 years of observation by Hyper-K will yield a sensitivity to 
the WIMP-proton spin dependent cross section below $10^{-39}$ cm$^2$ around 
10 GeV for the soft channel, and $10^{-40}$ cm$^2$ around 100 GeV for 
the hard channel. It can improve the present sensitivity significantly 
for lighter WIMPs below $\sim$100 GeV.  
Other astronomical neutrino searches such as GRB neutrinos, galactic diffuse 
neutrinos, etc. can be also performed in Hyper-K with its large acceptance 
in a wide energy range as well as its excellent performance 
of event reconstruction.

%% file: physics-radiography/radiography.tex
\subsection{Neutrino geophysics}\label{section:radiography}
In this section, the radiographic measurements of the Earth with Hyper-Kamiokande will be discussed. 

\subsubsection{Geophysical motivation}

The Earth's internal structure and chemical composition have been estimated by analyzing seismic data to derive seismic velocities inside the Earth, in conjunction with many laboratory experiments and model calculations.
Until recent years, these seismic waves had been the only probe that could penetrate the Earth.
In order to construct the standard Earth model 
a number of models have been considered. PREM~\cite{PREM} and 1066A~\cite{Gilbert:1975} are two major models, both of which describe the spherically symmetric structure of the Earth. 
PREM and 1066A are both parametrized models where the seismic velocities $V_p(x)$ and $V_s(x)$ are assumed to have a certain relationship with the density $\rho(x)$ via the gravity $g$, bulk modulus $k$, and so on, and thus there is an uncertainty regarding the absolute density distribution. 
It is therefore critically important to find an independent method to directly measure the density inside the Earth. 
The question is: what kind of method can be used for this purpose? 
Drilling and core sampling enable us to directly examine material inside the Earth. 
However, considering the fact that the world's deepest such sample is 12~km deep, it is likely impossible, not to mention expensive and dangerous, to attempt to reach the core (deeper than 3000 km) by this method.

Recent observation of anti-neutrinos from the decay of radiogenic isotopes inside the Earth by the KamLAND~\cite{Araki:geonu,Gando:1900zz} and Borexino~\cite{Bellini:2010hy} experiments opened novel possibilities to investigate Earth's interior using neutrinos.
In what follows, another approach of neutrino geophysics using atmospheric neutrinos detected by Hyper-K is discussed.
The observation of neutrinos and measurement of their absorption or differential oscillations in the Earth will provide unique information about its density structure. 
Furthermore, the result will be more easily interpreted than the conventional seismological method which has intrinsic uncertainties.

We estimate the accuracy necessary to provide useful information on geophysics, taking
as  an example measuring the density of the Earth's core.
The main component of the Earth's core is assumed to be metal iron.
Seismic measurements indicate that the density of the core is 10\% less than that of pure iron~\cite{Jeanloz:1990}. 
On the other hand, siderophile elements, elements that can easily dissolve into iron, are found to have extremely low abundance in the mantle. 
Ringwood~\cite{Ringwood:1977} explained the reason why the amount of the siderophile elements is extremely low by using his model of element transportation between the mantle and the core. 
The siderophile elements that are lighter than iron can dissolve into the core iron, and as a result  their abundance in the mantle decreases.
This seemed to give an explanation of the core density being lighter than pure iron.
However, when Ringwood surveyed the amount of the siderophile elements dissolved into the core 
it was found that the core abundance of the siderophile elements could not explain its anomalous density.  The low density core remains, despite active inquiries and lively discussions, a mystery. 

Thus, a measurement of the Earth's density with the accuracy of 10\% or better 
can provide useful information on geophysics.
If the accuracy is further improved, other topics in geophysics can be pursued.
Following are examples of topics of great interest:

\begin{enumerate}
\item the density difference between the inner core and the outer core: 
this difference reflects the partition coefficient of light elements between the inner and outer cores,
which is important to understand the energy source of the geodynamo.
A measurement accuracy of 5\% will be required for this subject.
\item the value of the $C_0$ parameter of Roche's law: this parameter contains the history of the inner core~\cite{Knapmeyer20111062}.
The subject of interest is the convection of the inner core.
If this value is precisely determined, the gradient of chemical composition can be discussed.
If there is convection, the chemical composition would be homogeneous within the core, while if not it would show a gradient distribution.
\item whether the inner core is spherically or cylindrically symmetric: 
the seismic structure indicates that the inner core is cylindrically symmetric rather than spherically. 
Also, there is a report that indicates asymmetry between the east and the west hemispheres of the inner core, although it is difficult to quantitatively evaluate this difference. 
\end{enumerate}

\subsubsection{Current studies}
Neutrino radiography for surveying the internal structure of the Earth was first discussed more than 25 years ago~\cite{Volkova:1974, Nedyalkov:1981yy,DeRujula:1983ya},
and many  ideas have been proposed since then.
However, because those ideas are based on gigantic accelerators or use of  (thus far unobserved) galactic neutrinos, they are difficult to realize.
Recently, there has been an attempt to use atmospheric neutrinos  for neutrino radiography, 
based on a simulation of atmospheric neutrino events  that can be collected with the IceCube neutrino detector. 
It is found that the density difference between the core and the mantle can be determined with an  accuracy of 20\% in 10 years~\cite{GonzalezGarcia:2007gg}.
Following this calculation, the data taken with IceCube was analyzed.
Based on the one year data taken with 40 strings (about half of the full IceCube detector), 
the density difference between the core and the mantle can be confirmed at the 1$\sigma$ CL~\cite{IC-40}. 
Based on a simulation of neutrino radiography with KM3NeT~\cite{AdregadeMoura:2010zz},
it was reported that the average density of the core can be determined with accuracy of 6\% (1$\sigma$ CL) in 10 years.  

\subsubsection{Radiography with Hyper-Kamiokande}
The absorption neutrino radiography utilizes very high energy neutrinos with energies above 10~TeV, and thus a very large detector is naturally required. 
The IceCube and the KM3NeT detectors are designed specifically to detect very high energy neutrinos.
They are not optimized for detection of  low energy neutrinos 
because the optical modules (or photo sensors) are placed too sparsely.

\begin{figure}[tbp]
\includegraphics[width=0.7\textwidth]{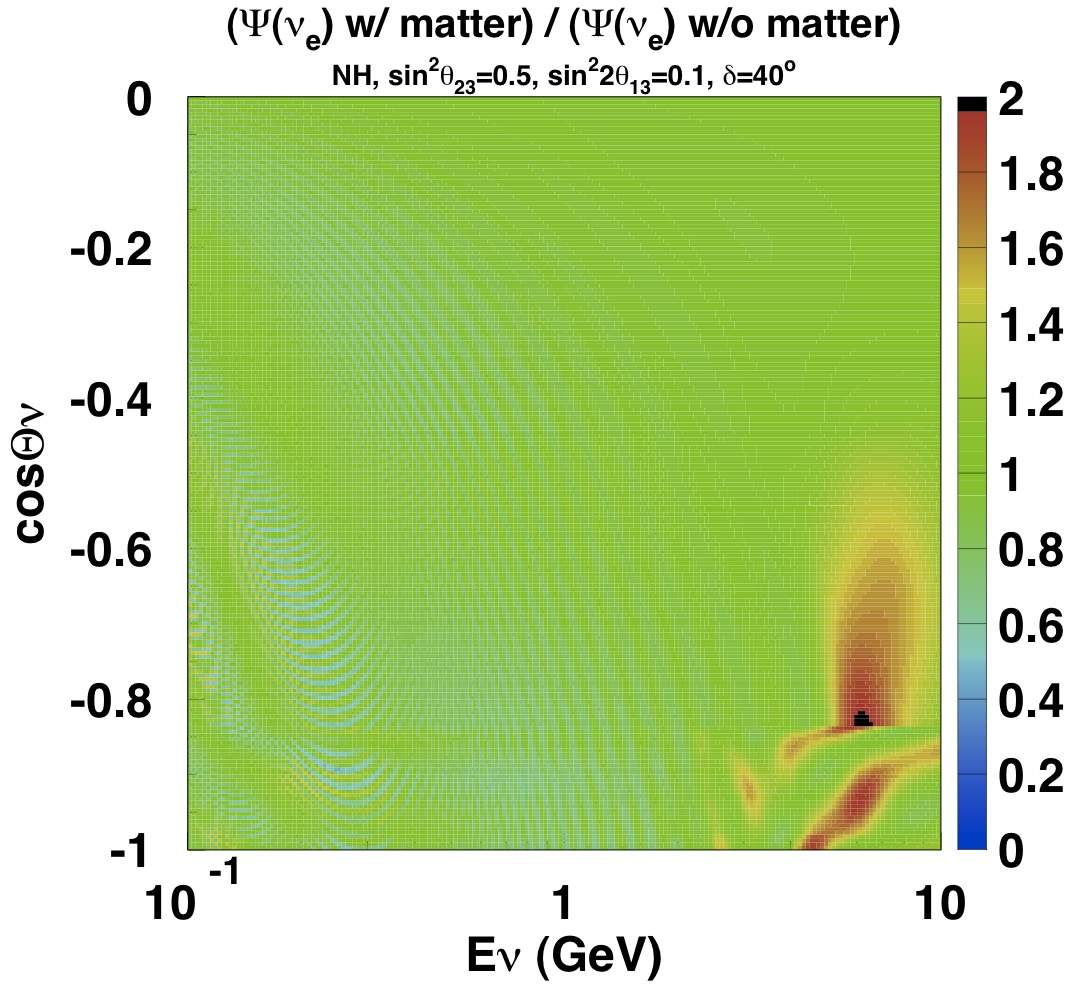}
\caption{Ratio of atmospheric $\nu_e$ flux with MSW effect to that without MSW effect,
as a function of neutrino energy and zenith angle.
$\sin^2\theta_{23}=0.5$, $\sin^22\theta_{13}=0.1$, and normal mass hierarchy are assumed.
Other parameters are listed in Table~\ref{tab:atmnu-params}.
\label{fig:radio-fig1}}
\end{figure}

Although Hyper-Kamiokande does not have such a large volume as IceCube or the KM3NeT, 
its capability to detect a large number of (lower energy) atmospheric neutrinos enables us 
to  perform the neutrino oscillation radiography using the MSW effect~\cite{Wolfenstein:1977ue,Mikheyev:1985zz,Mikheyev:1986zz}.
 Figure~\ref{fig:radio-fig1} shows the ratio of atmospheric $\nu_e$ flux with MSW effect to that without MSW effect
as a function of zenith angle and neutrino energy.
The largest change of the atmospheric neutrino oscillation probability by the MSW effect in the Earth's core can be seen in the energy range between 5 and 10 GeV as discussed in Sec.~\ref{section:atmnu}.
The density resolution is found to depend strongly  on the neutrino oscillation parameters $\theta_{23}$, $\Delta m^2_{32}$, $\theta_{13}$, and $\delta$.
The flux of neutrinos in the energy range between 5-10~GeV is more than $10^6$ times greater than the flux of such very high energy neutrinos (assuming the differential cosmic ray spectral index of 2.7). 
Given the precisely known oscillation parameters, Hyper-K would be able to perform the neutrino oscillation radiography with considerably higher statistics than these km$^3$-scale detectors based on MSW effect.

